\documentclass{aa}  

\usepackage{graphicx}
\usepackage{txfonts}
\usepackage{color}
\usepackage{diagbox}
\usepackage{lscape}

\newcommand{\mc}[3]{\multicolumn{#1}{#2}{#3}}
\newcommand{\fg}{{F-GAMMA} }
\newcommand{\frm}{\textit{Fermi}}

\begin{document}

   \title{F-GAMMA: Multi-frequency radio monitoring of \frm\ blazars}
   \subtitle{{The 2.64 to 43 GHz Effelsberg light curves from 2007--2015}}

%   \title{F-GAMMA program: multi-frequency long term monitoring of \frm\ blazars}
%   \subtitle{The Effelsberg 8-year-long 2.64 -- 43 GHz light curves}

%   \title{The F-GAMMA program: long term Effelsberg monitoring of \frm\ blazars}
%   \subtitle{The 8-year-long 2.64 -- 43 GHz light curves}

%   \title{F-GAMMA: the Effelsberg dataset of \frm\ blazars}
%   \subtitle{The 8-year-long 2.64 -- 43 GHz light curves}

%   \title{The F-GAMMA Effelsberg dataset of \frm\ blazars}
%   \subtitle{The 8-year-long 2.64 -- 43 GHz light curves}

%   \title{The F-GAMMA Effelsberg light curves of \frm\ blazars}
%   \subtitle{The 8-year-long 2.64 -- 43 GHz dataset}

%   \title{The F-GAMMA Effelsberg monitoring of \frm\ blazars}
%   \subtitle{The 8-year-long 2.64 -- 43 GHz dataset}

%   \title{The F-GAMMA 8-year 2.64 -- 43 GHz light curves or \frm\ blazars}
%   \subtitle{The 8-year Effelsberg 2.64 -- 43 GHz data}

%   \title{F-GAMMA: Multi-frequency radio monitoring of \textit{Fermi} blazars}
%   \subtitle{The 8-year Effelsberg 2.64 -- 43 GHz data}

   \author{E. Angelakis
          \inst{1}\thanks{agele@physics.auth.gr}
          \and
                  L. Fuhrmann 
          \inst{1,2}
          \and
                  I. Myserlis 
                  \inst{1}
          \and
                  J.~A.~Zensus
                  \inst{1}
          \and
                  I. Nestoras 
                  \inst{1,3}
          \and
                  V. Karamanavis 
                  \inst{1,2}
          \and
                  N. Marchili 
                  \inst{4}
          \and
                  T.~P.~Krichbaum 
                  \inst{1}
          \and
                  A. Kraus 
                  \inst{1}
          \and
                  J.~P.~Rachen 
                  \inst{5,6}
%\fnmsep
          }

   \institute{Max-Planck-Institut f\"ur Radioastronomie, Auf dem H\"ugel 69, 53121 Bonn, Germany\\
              \email{angelaki@mpifr-bonn.mpg.de}
         \and
                         Fraunhofer Institute for High Frequency Physics and Radar Techniques, Fraunhoferstra\ss e 20, 53343 Wachtberg, Germany
         \and
                         Deimos Space UK, Fermi Avenue R103, Harwell Oxford, OX11 0QR, UK
                \and
        Istituto di Astrofisica e Planetologia Spaziali Via Fosso del Cavaliere 100, 00133, Rome, Italy
%               \and
%       Astrophysical Institute, Vrije Universiteit Brussel, Pleinlaan 2, 1050 Brussels, Belgium
                \and
        Department of Astrophysics/IMAPP, Radboud University, PO Box 9010, 6500 GL Nijmegen, The Netherlands
                \and
        Max-Planck-Institute for Astrophysics, Karl-Schwarzschild-Str. 1, 85748 Garching, Germany
             %\thanks{The university of heaven temporarily does not
             %        accept e-mails}
             }

%a) Astrophysical Institute, Vrije Universiteit Brussel, Pleinlaan 2, 1050 Brussels, Belgium
%
%b) Department of Astrophysics/IMAPP, Radboud University, PO Box 9010, 6500 GL Nijmegen, The Netherlands
% 
%c) Max-Planck-Institute for Astrophysics, Karl-Schwarzschild-Str. 1, 85748 Garching, Germany

   \date{Received September 15, 1996; accepted March 16, 1997}

% \abstract{}{}{}{}{} 
% 5 {} token are mandatory
 
  \abstract
   {The advent of the \textit{Fermi} gamma-ray space telescope with its superb sensitivity, energy range, and its unprecedented capability to monitor the entire $4\pi$ sky within less than 2--3 hours, introduced new standard in time domain gamma-ray astronomy. {Among several breakthroughs, \textit{Fermi} has -- for the first time -- made it possible to investigate, with high cadence, the variability of the broadband spectral energy distribution (SED), especially for active galactic nuclei (AGN). This is necessary for understanding the emission and variability mechanisms in such systems.}
%      Among other breakthroughs, for the first time it has been possible to investigate high cadence variability of the broadband spectral energy distribution (SED), especially of active galactic nuclei, which is necessary for understanding emission and variability mechanisms. 
To explore this new avenue of extragalactic physics the \textit{Fermi}-GST AGN Multi-frequency Monitoring Alliance (\fg  ) programme undertook the task of conducting nearly monthly, broadband radio monitoring of selected {blazars, which is the dominant population of the extragalactic gamma-ray sky, from January 2007 to January 2015} . In this work we release all the multi-frequency light curves from 2.64 to 43~GHz and first order derivative data products after all necessary post-measurement corrections and quality checks.}
   {Along with the demanding task to provide the radio part of the broadband  SED in monthly intervals, the \fg programme was also driven by a series of well-defined fundamental questions immediately relevant to blazar physics. On the basis of the monthly sampled radio SEDs, the \fg aimed at quantifying and understanding the possible multiband correlation and multi-frequency radio variability, spectral evolution and the associated emission, absorption and variability mechanisms. The location of the gamma-ray production site and the correspondence of structural evolution to radio variability have been among the fundamental aims of the programme. Finally, the programme sought to explore the characteristics and dynamics of the multi-frequency radio linear and circular polarisation.}
   {The \fg ran two main and tightly coordinated observing programmes. The Effelsberg 100 m telescope programme monitoring 2.64, 4.85, 8.35, 10.45, 14.6, 23.05, 32, and 43~GHz, and the IRAM 30 m telescope programme observing at 86.2, 142.3, and 228.9~GHz. The nominal cadence was one month for a total of roughly 60 blazars and targets of opportunity. In a less regular manner the \fg programme also ran an occasional monitoring with the APEX 12 m telescope at 345~GHz. We only present the Effelsberg dataset in this paper. The higher frequencies data are released elsewhere. }
   {The current release includes 155 sources that have been observed at least once by the \fg programme. That is, the initial sample, the revised sample after the first \textit{Fermi} release, targets of opportunity, and sources observed in collaboration with a monitoring programme following up on \textit{Planck} satellite observations. For all these sources we release all the quality-checked Effelsberg multi-frequency light curves. The suite of {post-measurement} corrections and {flagging and} a thorough system diagnostic study and error analysis is discussed as an assessment of the data reliability. We also release data products such as flux density moments and spectral indices. The effective cadence after the quality flagging is around one radio SED every 1.3 months. The coherence of each radio SED is around 40 minutes. }
   {The released dataset includes more than $4\times10^4$ measurements for some 155 sources over a broad range of frequencies from 2.64 GHz to 43~GHz {obtained between 2007 and 2015}. The median fractional error at the lowest frequencies (2.64--10.45~GHz) is below 2\%. At the highest frequencies (14.6--43~GHz) with limiting factor of the atmospheric conditions, the errors range from 3\% to 9\%, respectively.}

   \keywords{Astronomical databases: miscellaneous -- Galaxies: active -- Galaxies: jets -- Radio continuum: galaxies -- Galaxies: BL Lacertae objects: general -- Galaxies: quasars: general}

   \maketitle
%
%==================================================================================================
\section{Introduction}
\label{sec:intro}
The current work constitutes the release of the first part of the \fg \citep{2016A&A...596A..45F} dataset, which includes the centimetre and subcentimetre light curves obtained with the 100 m Effelsberg radio telescope between 2007 and 2015. The \fg programme collected a vast amount of monthly monitoring data for more than 100, almost exclusively, \frm\ blazars over an unprecedentedly broad radio spectrum down to submillimetre wavelengths. The current release alone contains more than $4\times10^4$ measurements that have survived quality filtering, and naturally raises the question of what is the motivation for such a massive observational effort on blazars?  

Blazars are active galactic nuclei (AGN) with their collimated, relativistic plasma outflow (jet) in close alignment to the line of sight of the observer (a few degrees, \citealt{1978bllo.conf..328B, 1979ApJ...232...34B}). The relativistic beaming induced by this configuration causes the jet emission to outshine all other components thus making it an ideal probe of the physical conditions and processes in the exotic environments of relativistic jets. These sources exhibit an exceptionally broad spectral energy distribution (SED), which often spans 20 orders of magnitude in frequency or even more \citep[e.g.][]{2012A&A...541A.160G}, making blazars the dominant high-energy population. In the \frm/LAT third source catalogue \citep[3FGL;][]{2015ApJS..218...23A} blazars comprise 60\% of the detected sources and more than 90\% of the associated sources. The high-energy-peaked component of their characteristic double-peaked SED \citep[e.g.][]{2010ApJ...716...30A} is argued to be the signature of a photon field that is inverse Compton up-scattered by plasma producing its low-energy-peak synchrotron component. Blazars exhibit intense variability at all energies  \citep[e.g.][]{2017A&ARv..25....2P} over timescales from minutes \citep[e.g.][inferred doubling time at TeV energies of around 10 minutes]{2011ApJ...730L...8A} to several months or more \citep[e.g.][]{2016A&A...596A..45F}. They typically appear significantly polarised \citep[][]{1972ApJ...175L...7S}, especially at higher energies \citep[e.g.][]{2016MNRAS.463.3365A}, and their polarisation also undergoes intensive variability not only in terms of fraction but also polarisation plane orientation \citep[e.g.][]{2008Natur.452..966M,2015MNRAS.453.1669B}. The blazar phenomenology and the richness of the relevant jet physics becomes even more apparent with their role as complex particle accelerators and, as of recently, even confirmed neutrino emitters \citep{2018arXiv180708794I,2018arXiv180704461P} urging the re-evaluation of our understanding of their dominant emission mechanisms.  

The \fg programme was initiated with the scope to provide necessary multi-frequency radio monitoring complementary to the \frm/LAT \citep{2009ApJ...697.1071A} monitoring of the gamma-ray sky, and to study certain aspects of relevant radio physics. Among the notable advantages of radio monitoring of blazars is that the radiation in these bands originates almost entirely at the plasma jet and the contamination with other sources of radio emission is insignificant, if any. The uniqueness of the \fg programme in particular is attributed to four elements: (a) its multi-frequency character, which allows us to follow the evolution and dynamics of the radio SED and link it with underline physical processes; (b) the relatively high cadence observations, which are optimal (in most cases) for a satisfactory sampling of the spectral evolution, and most importantly, which {resolves} the inevitable degeneracies that stem from merging several evolving elements into one comparatively large telescope beam;  (c) its long duration, which allows us to acquire a firm understanding of the source behaviour at different timescales and to collect a large number of spectral evolution events that probe the emission and variability mechanisms; and finally (d) the availability of multi-frequency linear and circular
polarisation light curves \citep{2018A&A...609A..68M} which give access to
the microphysics of the emitting plasma \citep{2016Galax...4...58M,2017Galax...5...81A}. Element (c) is particularly important as it allowed us to collect a large sample of events from an otherwise inherently biased sample as we discuss later.  

The analysis that has already been carried out {for} a limited part of the dataset has led to a series of noteworthy findings.   
{An examination} of the \fg dataset with reference to the first \frm\ releases \citep{Abdo_2009ApJ_700_597A,abdo_2010ApJS_188_405A} showed that the detectability in the {giga electron volt (GeV)} energy range was an increasing function of the variability in the radio regime \citep{2016A&A...596A..45F}  as  found by other studies \cite[e.g.][with observations at 15~GHz]{2011ApJS..194...29R}. After accounting for the numerous biases affecting a flux-flux correlation analysis we found that radio and gamma-ray emissions are correlated with a significance that increases with radio frequency \citep{2016A&A...596A..45F}. This finding was interpreted as an indication that the gamma-ray emission is produced very close to (or in the same region as) the  millimetre-band emission region. {The radio spectral evolution was used to model and quantify the broadband variability mechanism} \citep{2012arXiv1202.4242A}. On the basis of the cross-correlation of radio and gamma-ray light curves of selected cases, we managed to constrain the gamma-ray emission site \citep{2014MNRAS.441.1899F,2016A&A...590A..48K}. We further developed a method that parametrises each outburst separately allowing us to account for different variability mechanisms operating in the same source at different times. This method was then used to quantify the presence of a relativistic jet in narrow-line Seyfert 1 galaxies \citep{2015A&A...575A..55A} and constrain the multi-frequency variability Doppler factors \citep{2017MNRAS.466.4625L}. Finally, assuming the radio variability to be caused primarily by traveling shocks \citep{2012arXiv1202.4242A} we constructed a realistic radio jet emission model able to reproduce the linearly and circularly polarised radiation \citep{2018A&A...609A..68M} and quantify the physical conditions and their evolution in those systems \citep{2016Galax...4...58M, 2017Galax...5...81A}.

In the following we present the \fg dataset from 2.64 to 43~GHz. The higher frequencies datasets will be released in subsequent publications. We begin with a detailed description of the sample that is included in the data release (Section~\ref{sec:sample}) and then we give a detailed description of the observations (Section~\ref{sec:obs}) and the post-measurement data treatment (Section~\ref{corrections}). The raw data are presented in Section~\ref{sec:LCs} in the form of multi-frequency light curves. Finally, in Section~\ref{sec:spind} we present spectral indices as a higher level data product. The content of this paper is strictly confined to the needs of a data release. Radio astrophysical studies and interpretation of the data will be presented in subsequent publications.

\section{Source sample}
\label{sec:sample}

As we discuss in Section~2 of \cite{2016A&A...596A..45F}, the \fg programme was optimised to complement the {\sl Fermi} blazar monitoring. Specifically, we developed the programme to quantify and understand the broadband blazar variability, localise the gamma-ray emission site, and study the evolution of conditions in the emitting elements. The initial \fg sample included 62 sources previously detected by the Energetic Gamma Ray Experiment Telescope (EGRET) with flat-spectrum radio quasars (FSRQs) comprising roughly {52\% and BL Lacs 37\% of the total}. 

With the release of the First \textit{Fermi}/LAT Source Catalogue \citep{abdo_2010ApJS_188_405A}, the sample was revised {around mid-2009} to include exclusively LAT-detected sources. The updated sample comprised a total of 65 \textit{Fermi} sources, 25 of which are already in the first sample. The new sample was chosen on the basis of the observability of the sources from Effelsberg and IRAM observatories, their variability in radio and gamma rays, whether they were monitored by other programmes and several cases of special interest such as narrow-line Seyfert 1 galaxies. The sources were observed with different cadences and priorities depending on all these parameters (cf. Sect.~\ref{sec:obs}). A considerable amount of observing effort was put on targets of opportunity. 

As a consequence, the \fg sample includes: (a) sources that have been observed only for the first 2.5 years until the sample revision, (b) sources that have been observed over the entire duration of the programme (before and after the sample revision), (c) sources that were included after the sample revision, and (d) targets of opportunity that were observed occasionally. In this work we release the data for every source with at least one measurement. The exception to this rule constitutes a sample of TeV sources \citep{2008ICRC....3.1341W} and narrow-line Seyfert 1 galaxies \citep{2015A&A...575A..55A} that will be published in separate papers. Table~\ref{tab:sample} lists all the sources in our data release and includes five sections: {(a) the main monitored sample that was finalised with the mid-2009 revision, (b) sources monitored until that revision (labelled ``old''); (c) sources observed within the F-GAMMA-\textit{Planck} satellite MoU; (d) other sources observed mostly as targets of opportunity; and (e) calibrators.  
The sample in section (c) was} the result of the partial overlap of the F-GAMMA monitoring and the regular sky-scans of ESA's \textit{Planck} satellite\footnote{https://www.esa.int/Our\_Activities/Space\_Science/Planck}. The source sample and science goals are discussed by \cite{Rachen2016}. Part of the dataset is also presented elsewhere \citep{PlanckEarly14, PlanckEarly15, Giommi2012}. 
%We include all Effelsberg data taken within the Planck MoU (also for non-F-GAMMA sources) in the present data release. 

\begin{figure}
  \resizebox{\hsize}{!}{\includegraphics[trim=20pt 20pt 30pt 30pt  ,clip, width=0.45\textwidth,angle=0]{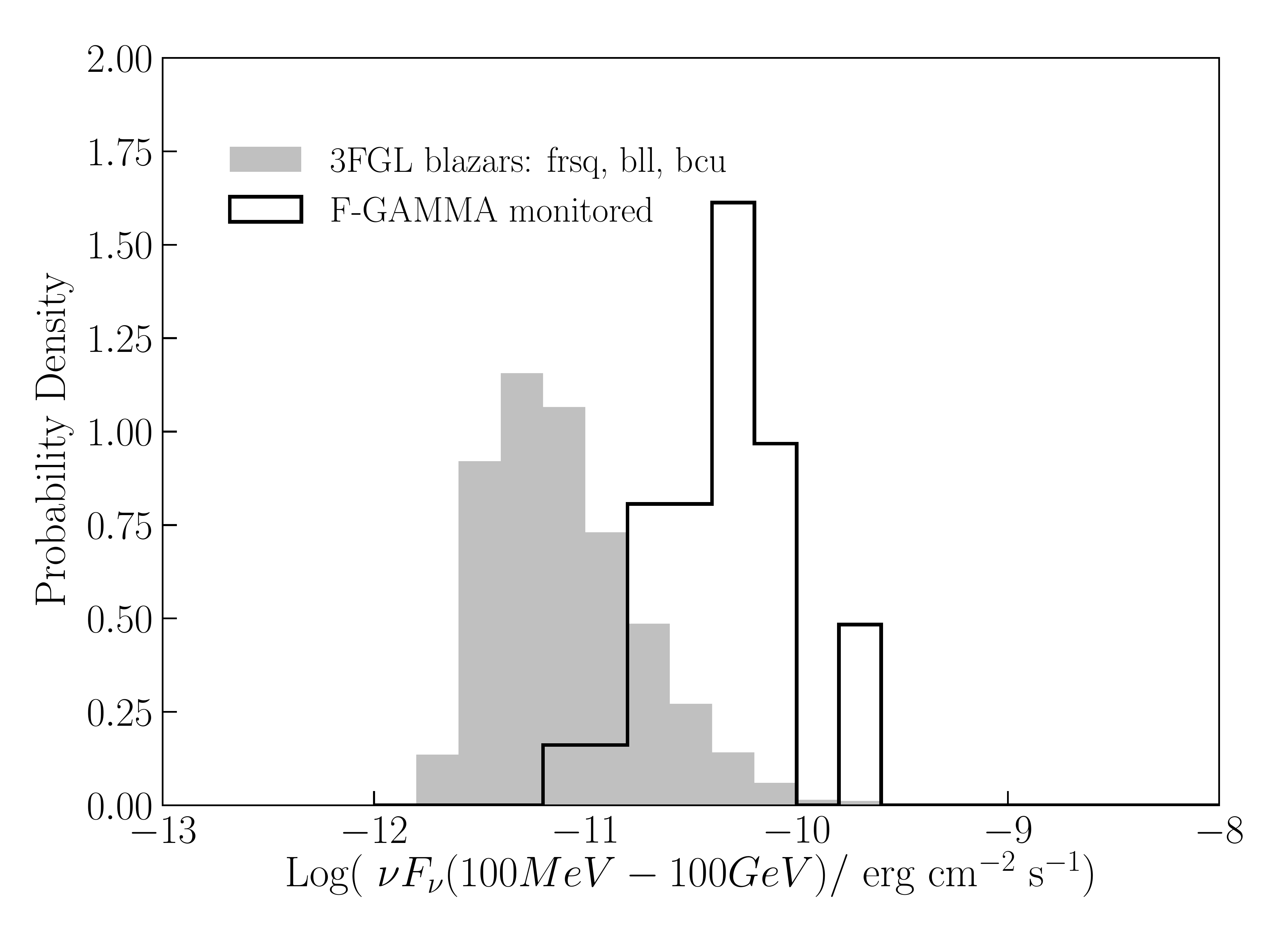}}
  \resizebox{\hsize}{!}{\includegraphics[trim=20pt 20pt 30pt 30pt  ,clip, width=0.45\textwidth,angle=0]{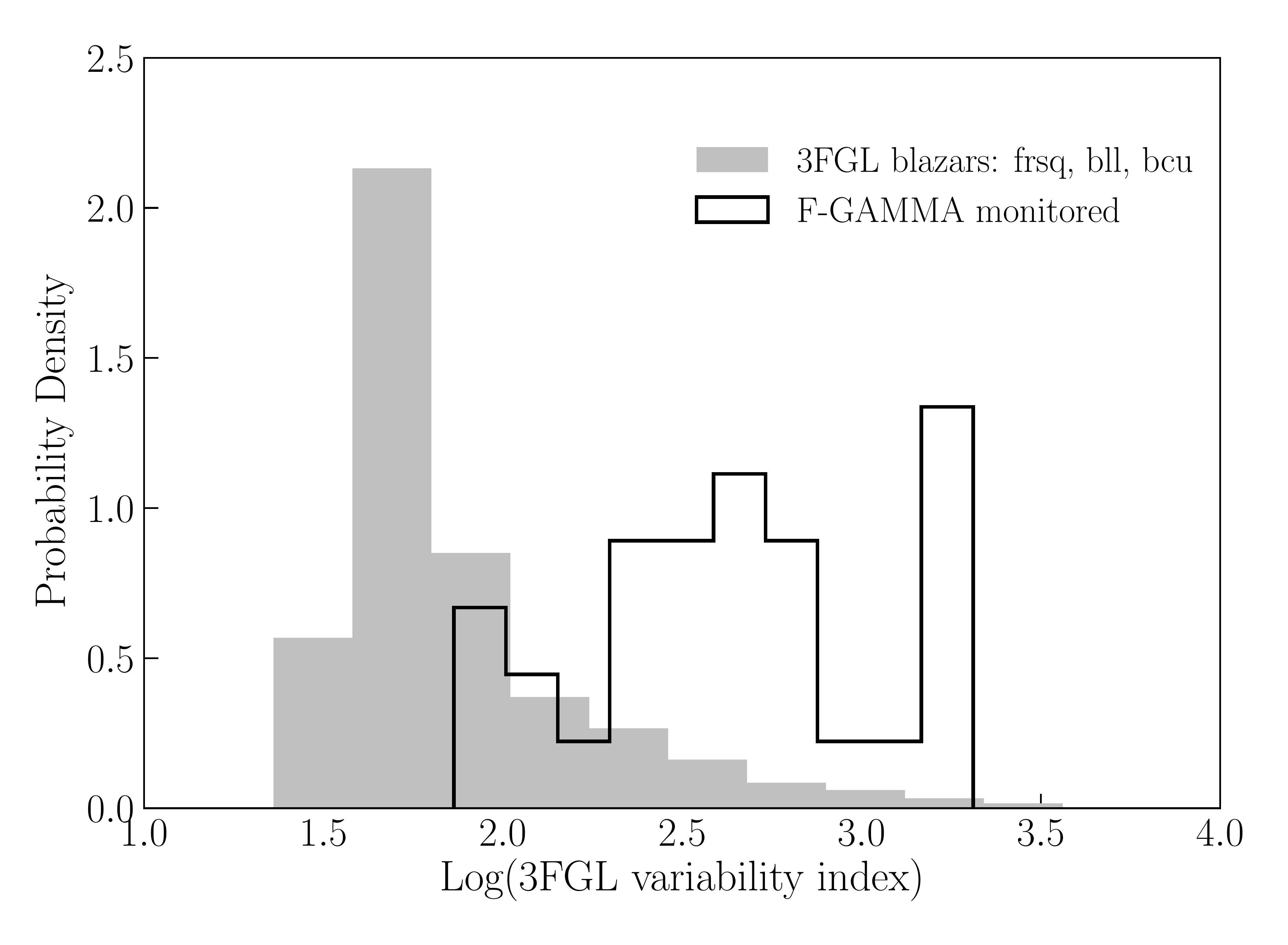}}
  \caption{
  %\blue{180301} 
  Distribution of energy flux (upper panel) and variability index (lower panel) in the band 100~MeV--100~GeV. {The grey area corresponds to all the sources in the 3FGL catalogue designated as ``fsrq'', ``bll',' and ``bcu'' while the black solid line to the \fg monitored sample}.}
\label{fig:hist_nuFnu}
\end{figure}

\begin{figure}
  \resizebox{\hsize}{!}{\includegraphics[trim=20pt 20pt 30pt 30pt  ,clip, width=0.45\textwidth,angle=0]{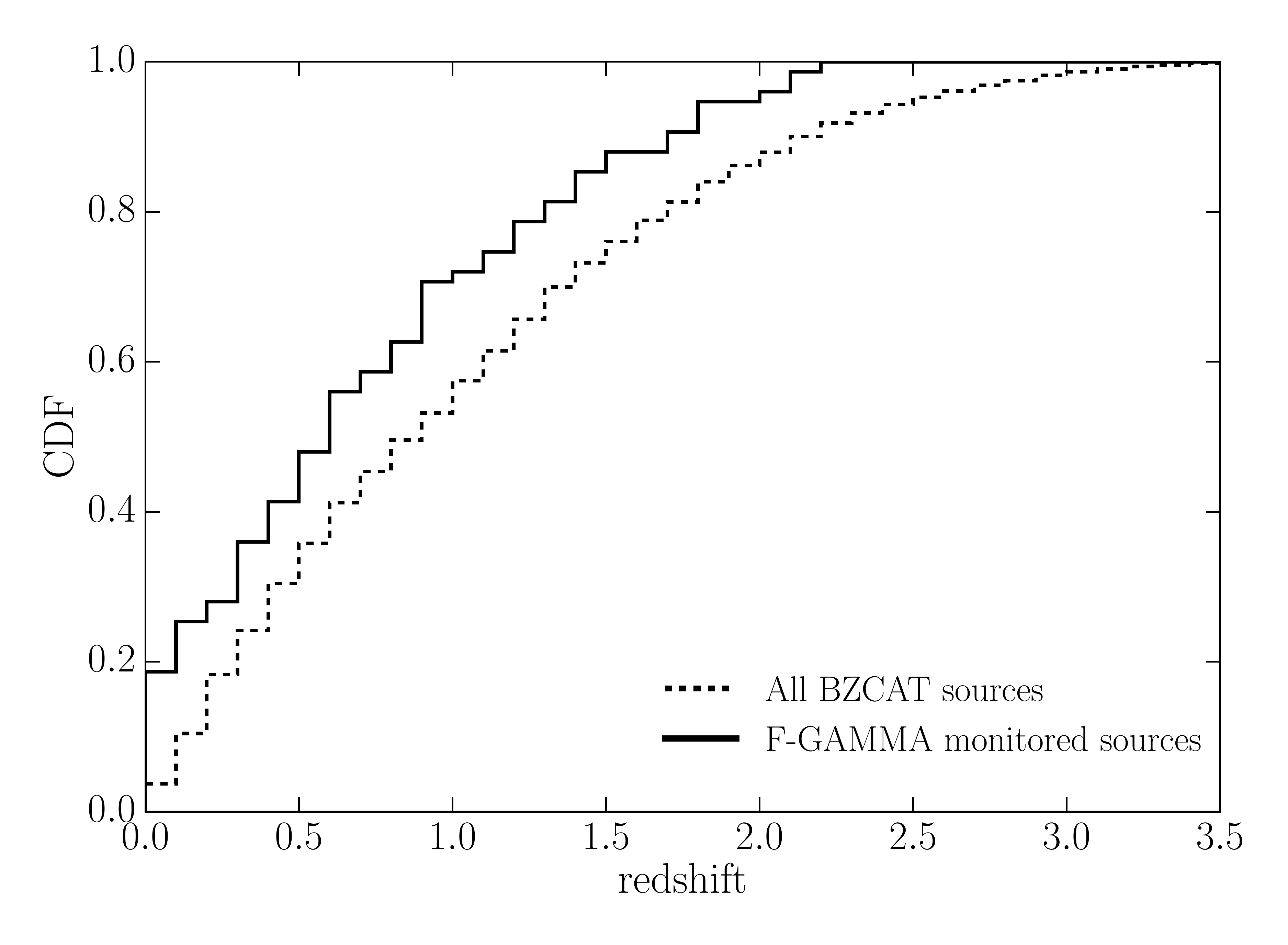}}
  \caption{
  %\blue{180620} 
  Redshift distribution for the \fg monitored sources compared to all the BZCAT sources \citep{2015ApSS.357...75M}.}
\label{fig:hist_z}
\end{figure}

Given the primary aims of the programme are to understand mechanisms and not population statistics, the \fg sample has been inevitably biased towards the brightest and most variable blazars; the former guarantee the highest quality datasets and the latter frequent outbursting events. Figure~\ref{fig:hist_nuFnu} provides a qualitative impression of how our monitored sample is representative of the entire blazar population. There we show the distributions of the energy flux in the range 100~MeV -- 100~GeV (upper panel) and the variability index (lower panel), respectively, for all the sources in the 3FGL catalogue \citep{2015ApJS..218...23A} designated as ``fsrq'', ``bll',' and ``bcu''. These plots show that the \fg monitored sample populates the upper end of high-energy distribution as well as that of the variability index. In Fig.~\ref{fig:hist_z} we compare the redshift distribution of the \fg monitored sources to that of all BZCAT catalogue sources \citep{2015ApSS.357...75M}. A two-sample KS test showed that the distribution \fg monitored sample over redshift is not qualitatively different from that of all the BZCAT sources (p-value: 0.009). From this we conclude that despite the biases in its selection the \fg sample is representative of the blazar population at least in terms of source cosmological distribution. 
%That is not the case for the 3LAC catalogued sources. The 2-sample KS gave a p-value of $2\times10^{-15}$ which must be due to overpopulating the small 3LAC redshift bins.  
% -------------------------------------------------------------------------------------------------
% -------------------------------------------------------------------------------------------------
% -------------------------------------------------------------------------------------------------
\section{Observations}
\label{sec:obs}

After the sample revision in mid-2009 the \fg adopted an optimised observing scheme for the more efficient usage of time. The 35 fastest varying sources, labelled group ``f'' (table~\ref{tab:sample}, Column 8), were observed on a monthly basis in every session. Another 30 slower variable sources were grouped in two sets of 15 sources each, labelled groups ``s1'' and ``s2', and were observed every other session.  The two groups comprised a total of 65 sources that were being monitored.       

%%..................................................................................................
%%\subsection{Receivers and observing technique} 
%\label{subsec:rx}

The observations were conducted with the secondary focus heterodyne receivers of the 100 m Effelsberg telescope (table~\ref{tab:receivers}). The systems at 4.85, 10.45, and 32\,GHz are equipped with multiple feeds allowing differential
measurements which partially remove effects of disturbing atmospheric emission or
absorption perturbations. Practically, only linear tropospheric disturbances are treated  (only partial overlap of the atmosphere columns ``seen'' by the feeds). 

Because the sources are point-like or only slightly extended for the 100 m telescope beam, the observations were conducted with ``cross-scans" i.e. by recording the antenna response while repeatedly slewing
over the source position in two orthogonal directions. One slew in one direction has been termed a ``sub-scan''.  In our case the scanning was carried out over the azimuthal (AZI) and the elevation (ELV) directions. This technique offers immediate detection of extended source structure or confusion from field sources and pointing offset
corrections. The observing cycle typically included two scans: one for telescope pointing and one as the actual measurement.
% -------------------------------------------------------------------------------------------------

\setcounter{table}{1}

\begin{table*}
\caption{Calibration parameters of the used receivers}             
\label{tab:receivers}      
\centering          
\begin{tabular}{rrrrcccccccc} 
\hline\hline       
    \mc{1}{c}{$\nu$}  &\mc{1}{c}{$\Delta\nu$} &\mc{1}{c}{$T_\mathrm{sys}$} &\mc{1}{c}{$\theta$}  &Feeds    &Polarisation   &Aperture  &\mc{1}{c}{$A_0$} &\mc{1}{c}{$A_1$} &\mc{1}{c}{$A_2$} &\mc{1}{c}{Epoch} &\mc{1}{c}{ELV$_{\mathrm{max}}$}\\    
     \mc{1}{c}{(GHz)} &\mc{1}{c}{(GHz)}       &\mc{1}{c}{(K)}              &\mc{1}{c}{(\arcsec)} &         &               &(\%)   &      &\mc{1}{c}{($10^{-4}$)}  &\mc{1}{c}{($10^{-5}$)} &      &\mc{1}{c}{(Deg)} \\    
\hline \\                   
% NOTE: this values are from the website. The Tsys is the average over channels
    2.64   &0.08  &17  &260 &1 &LCP, RCP             &58    &1.00000 &0.0000 &-0.0000  &Feb. 2007 &\ldots        \\ 
    4.85   &0.5  &27  &146 &2 &LCP, RCP              &53    &0.99500 &5.2022 &-1.2787  &Feb. 2008 &20.3          \\ 
    8.35   &1.1  &22  &82  &1 &LCP, RCP              &45    &0.99500 &4.3434 &-1.0562  &Feb. 2007 &20.6          \\ 
    10.45  &0.3  &52  &68  &4 &LCP, RCP              &47    &0.99000 &8.2490 &-1.7433  &Feb. 2007 &23.7          \\ 
    14.60  &2.0  &50  &50  &1 &LCP, RCP              &43    &0.97099 &18.327 &-2.8674  &Feb. 2007 &32.0          \\ 
    23.05  &2.0  &77  &36  &1 &LCP, RCP              &30    &0.91119 &47.557 &-6.2902  &Feb. 2007 &37.8          \\ 
    32.00  &4.0  &64  &25  &7 &LCP                   &32    &0.91612 &49.463 &-7.1292  &Feb. 2007 &34.7          \\ 
    43.00$^\mathrm{a}$  &2.8  &120 &20  &1 &LCP, RCP &19    &0.88060 &58.673 &-7.1243  &Feb. 2007 &41.2          \\\\ 
\hline                  
\end{tabular}
  \tablefoot{Entry in each column is as follows: 1: central
    frequency; 2: receiver bandwidth; 3: system temperature;
    4: FWHM; 5: number of available feeds; 6: available polarisation channels;
    7: telescope aperture efficiency at the corresponding frequency; 8, 9, and 10: parameters $A_0$ ,
                    $A_1$,  and $A_2$ defining the gain curve, respectively; 11: epoch of the gain curve observation; and 12: ELV where the
                    gain is maximised.\\
    \tablefoottext{a}{We used 43~GHz occasionally with set-ups centred at frequencies slightly different than 43~GHz within a range of a couple hundred megahertz however. For simplicity we always assume 43~GHz as the central frequency.}
        }
\end{table*}
% -------------------------------------------------------------------------------------------------

%..................................................................................................
\section{Data reduction and system diagnostics }
\label{corrections}

For the reconstruction of the true source flux density every observation was subjected to a series of post-measurement corrections. 
The median fractional effects of these corrections are reported in table~\ref{tab:corrections}. Below we discuss them in order of execution.  
% -------------------------------------------------------------------------------------------------
% the next table I produce by runnign the fit of a conastnt function on differences
% between steos f corrections. Eg. 
% fit [][0:] f(x)'LIST.13.all.sens' u :(100*(($12-$10)/$12)) via c
% 
% Th eaverage offsets for each ferquency I got them by 
% fit [][:100]f(x) 'test' u : (abs(($11))) via c
%
\begin{table}
\caption{
%\blue{-done} 
Median percentage effect of each post-measurement correction applied to
    the data for each observing frequency}    
\label{tab:corrections}     
\centering                         
\begin{tabular}{cccclc}       
\hline\hline                 
    $\nu$   &Pointing   &Opacity  &Gain curve &\mc{1}{c}{$\tau_\mathrm{z}$} &$\sigma$\\
    (GHz)   &(\%)       &(\%)     &(\%) &                  &        \\
\hline  \\                     
    2.64      &0.5 &3.5 &0.0  &0.020\tablefootmark{a} &0.003\\
    4.85      &0.4 &3.7 &0.6  &0.020\tablefootmark{a} &0.003\\
    8.35      &0.5 &3.7 &0.5  &0.020\tablefootmark{a} &0.003\\
    10.45     &1.2 &4.1 &0.6  &0.022 &0.006\\
    14.60     &1.3 &3.8 &0.3  &0.023 &0.016\\
    23.05     &1.6 &12.2 &0.7 &0.077 &0.076\\
    32.00     &3.1 &11.8 &0.6 &0.058 &0.021\\
    43.00     &5.1 &26.0 &1.1 &0.090 &0.027\\\\
\hline                             
\end{tabular}
\tablefoot{Columns: (1) observing frequency; (2), (3), and (4) median percentage effect of pointing, opacity, and gain correction, respectively; (5): average opacity at zenith; and (6): standard deviation around the mean.\\
\tablefoottext{a}{The opacity at the low end of the bandpass may be overestimated as a result of the enhanced beam side-lobes. The tabulated values should then be seen as upper limits.}}\end{table}
% -------------------------------------------------------------------------------------------------

%..................................................................................................
\subsection{Pointing offset correction}
\label{subsec:poi}

At this first stage the reduction pipeline accounts for the power loss caused by possible differences between the commanded and actual source position. The pointing offset is deduced from the difference between the expected source position and the maximisation of the telescope response. If we approximate the telescope main beam pattern with a Gaussian, and the antenna temperature observed by scanning over a direction ``$i$'' is $T_i$, then that corrected for pointing offset is  \begin{equation}
    \label{eq:Tpoi}
    T_{i,\mathrm{poi}}=T_{i}\cdot \exp\left(4\cdot\ln2\cdot \left(\frac{\Delta p_{j}}{\theta}\right)^{2}\right)
  ,\end{equation}
  where
  \[
  \begin{array}{rp{0.8\linewidth}}
    i,j&is the scanning direction indices with $i:\mathrm{ELV, AZI}$ and $j:\mathrm{AZI,ELV,}$ \\
    {\Delta p}_j &the pointing offset in $j$ direction, and\\
    \theta &the full width at half maximum (FWHM) at the observing frequency.
  \end{array}
  \]
We note that the offset in the $j$ direction is used for the correction of the measurement in the $i$ direction.   

\begin{figure} 
  \resizebox{\hsize}{!}{\includegraphics[width=0.45\textwidth,angle=0]{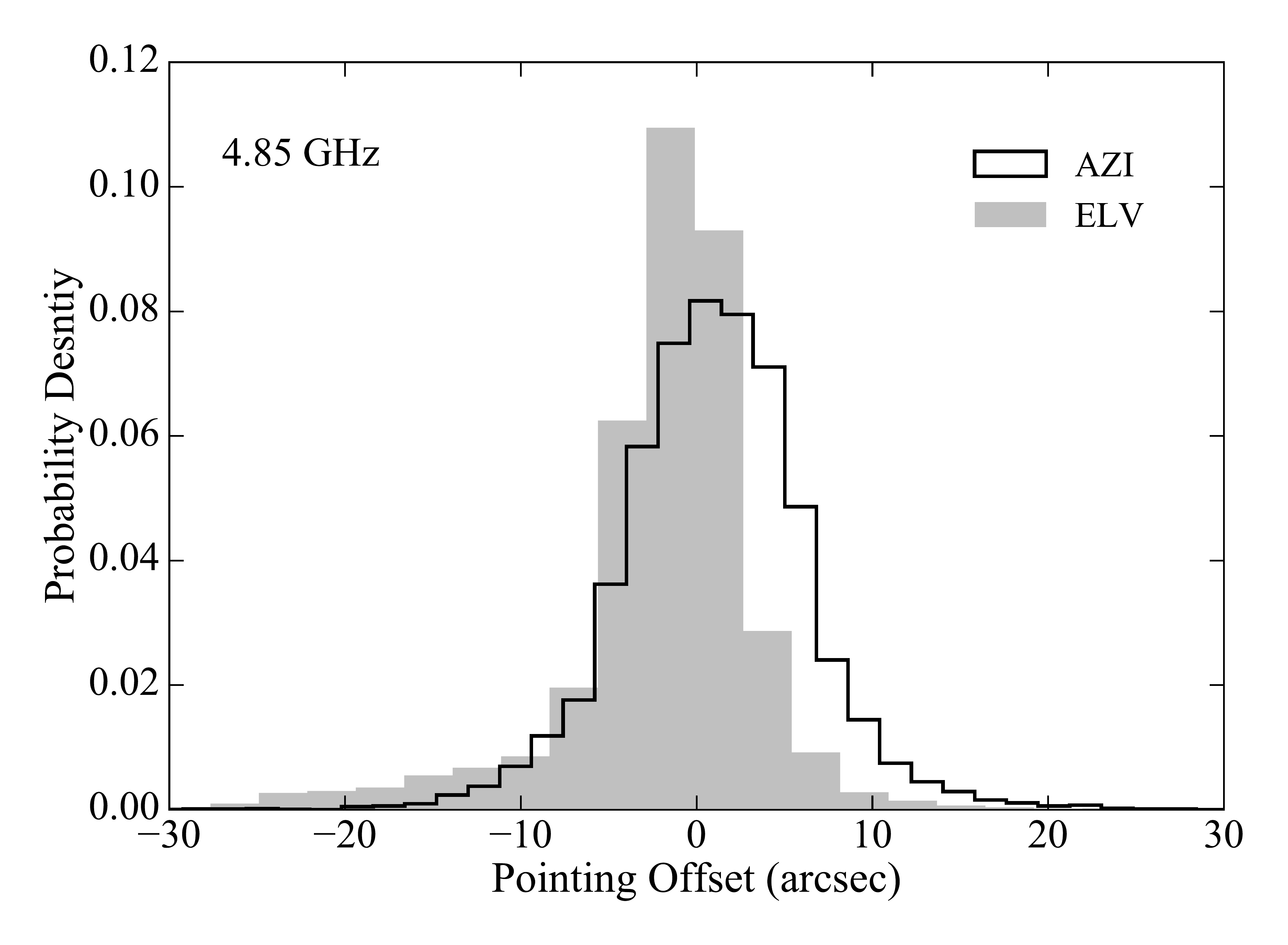}} 
%\\[-15pt]
  \resizebox{\hsize}{!}{\includegraphics[width=0.45\textwidth,angle=0]{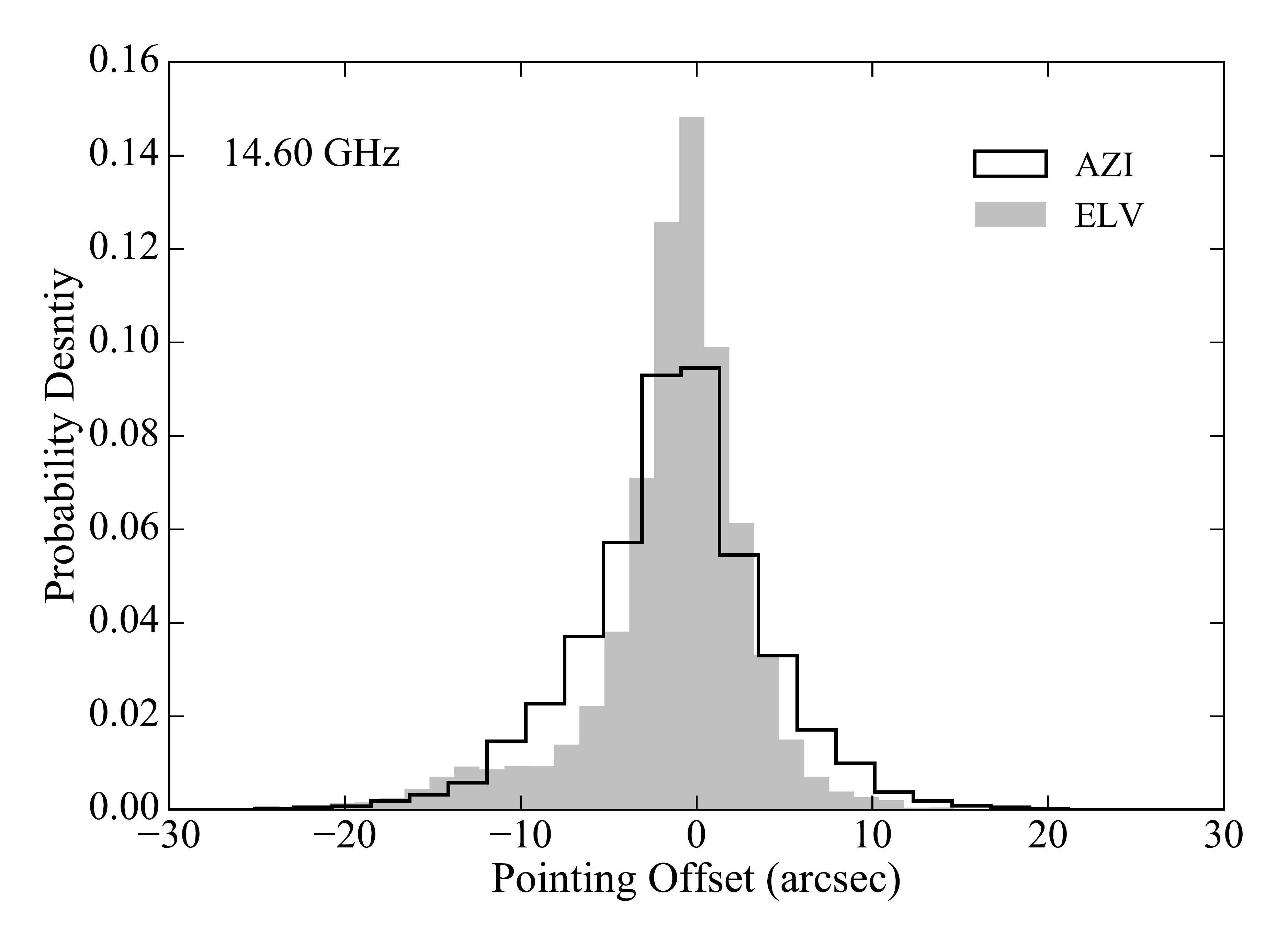}} 
%\\[-15pt]
  \resizebox{\hsize}{!}{\includegraphics[width=0.45\textwidth,angle=0]{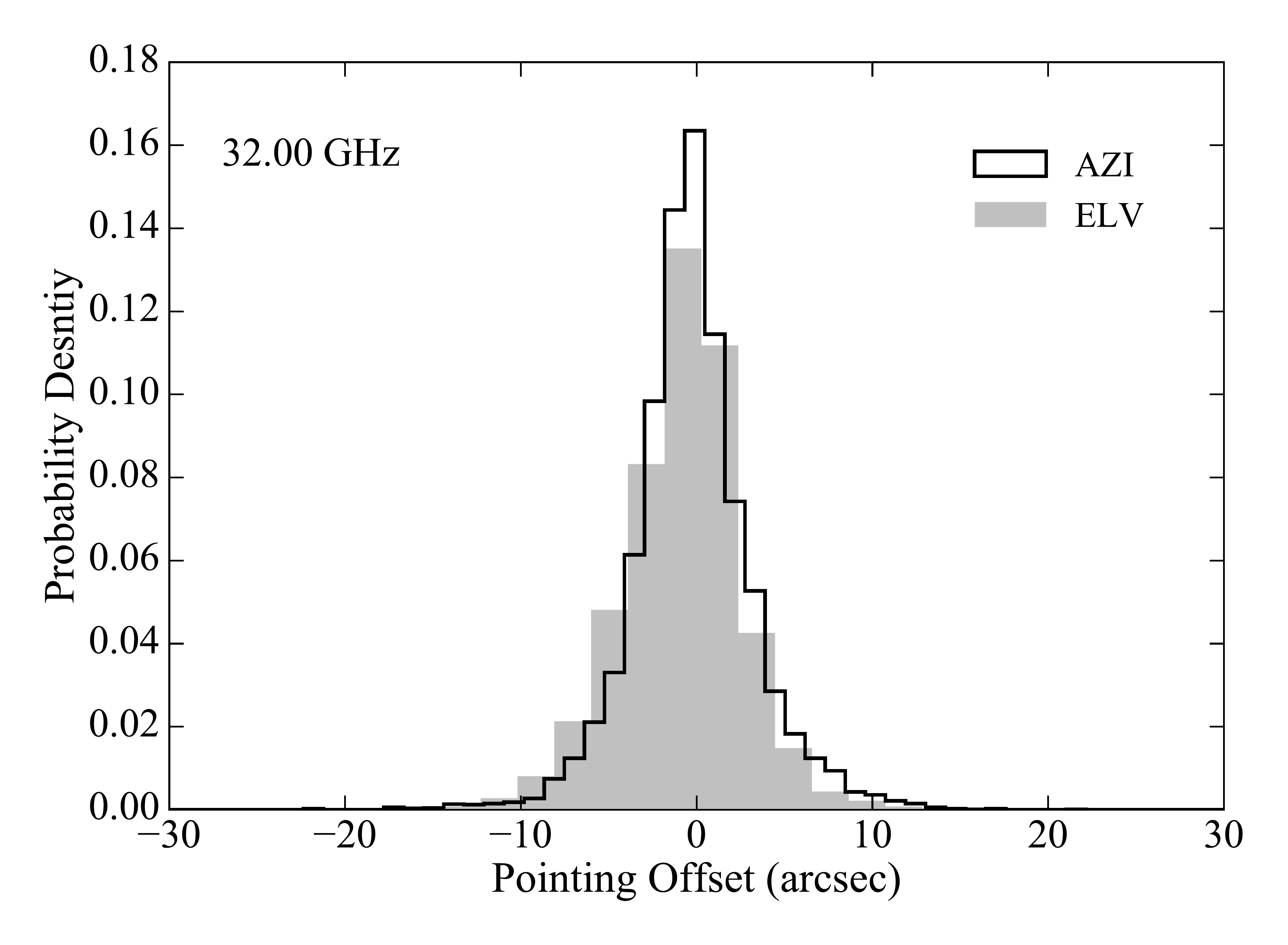}} 
  \caption{
  %\blue{-done}
  Pointing offsets for scans that have
    passed basic quality checks. {\it Upper} panel: 4.85~GHz (low band),
    {\it middle band} panel 14.60~GHz (intermediate band), and {\it lower} panel 32.00~GHz (high band). The
    black solid line corresponds to the longitudinal scan and the grey area to the
    latitudinal scan.}
  \label{fig:poi_distr}
\end{figure}

In Fig.~\ref{fig:poi_distr} we show the pointing offsets at three characteristic frequencies: 4.85\,GHz (low), 14.60\,GHz (intermediate band), and 32.00\,GHz (high band). The black solid line corresponds to the  AZI and the grey area to the ELV  scan. 
% -------------------------------------------------------------------------------------------------
% NOTE see the previous comment fro wheer are these data cominf rrom.
% -------------------------------------------------------------------------------------------------
Table~\ref{tab:poi} summarises the corresponding pointing parameters. As shown in table~\ref{tab:corrections} we conclude that this effect is of the order of a few percent at maximum {for all the receivers}.
% -------------------------------------------------------------------------------------------------
% NOTE these numbers I have fitted gaissuian with xmgrace in the 060_poi.agr etc files. 
% the data in those files are iusing the offset cut from the all.fit files. That means
% that 1. they are measueremnts that have been used and 2. they are only after rthe new
% systems started because only then we had all.fit files
\begin{table}
  \caption{
  %\blue{-done}
  Parameters of the Gaussian function fitted in the distributions of
    pointing offsets for three characteristic frequencies.}
  \label{tab:poi}  
  \centering                    
  \begin{tabular}{cccccccc} 
    \hline\hline                 
    Frequency &\mc{3}{c}{AZI Scan}& &\mc{3}{c}{ELV
      Scan}\\
\cline{2-4}
\cline{6-8}
             &N &$\Delta p$     &$\sigma$  & &N &$\Delta p$     &$\sigma$\\
             &  &(\arcsec) &(\arcsec) & &  &(\arcsec) &(\arcsec)\\
    \hline
    4.85  &9140 &1.0    &6.0 &  &9140 &$-1.1$   &8.9 \\
    14.60 &9824 &$-1.1$ &5.8 &  &9740 &$-0.8$   &5.4 \\
    32.00 &7761 &$-0.4$ &3.8 &  &7498 &$-0.6$   &4.0 \\
    \hline                                  
  \end{tabular}
  \tablefoot{The entry in each column is as follows: 1: observing frequency; 2 and 5: number of data points; 3 and 6 the median pointing offsets; and 4 and 8 the scatter of the pointing offset distribution. The latter is rather increased as compared to single frequency observations as a result of the instrumental effects caused by the usage of different receivers. }
\end{table}
% -------------------------------------------------------------------------------------------------

%..................................................................................................
\subsection{Atmospheric opacity correction}
\label{subsec:opc}
\begin{figure} 
  \resizebox{\hsize}{!}{\includegraphics[width=0.45\textwidth,angle=0]{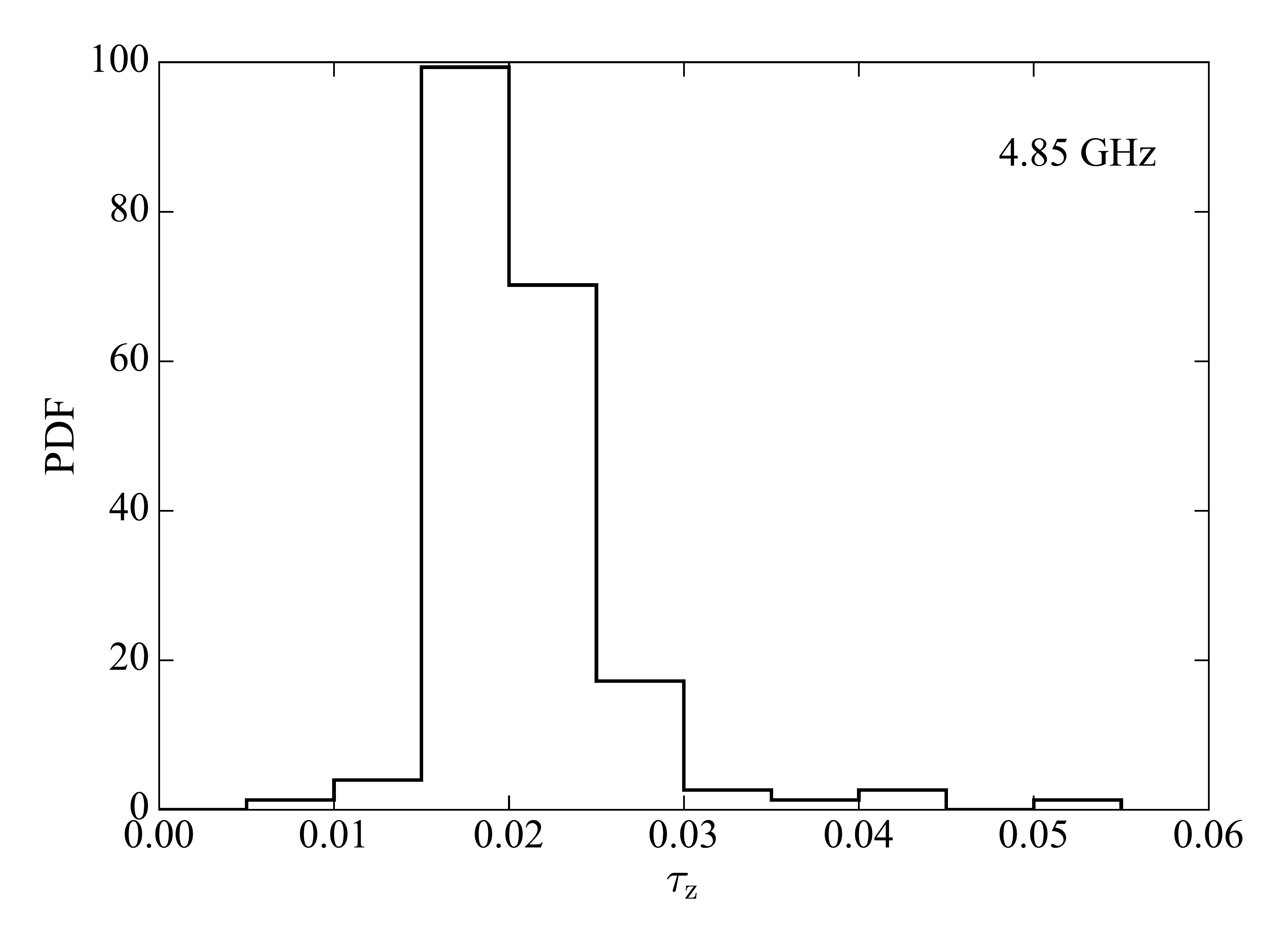}} 
%\\[-15pt]
  \resizebox{\hsize}{!}{\includegraphics[width=0.45\textwidth,angle=0]{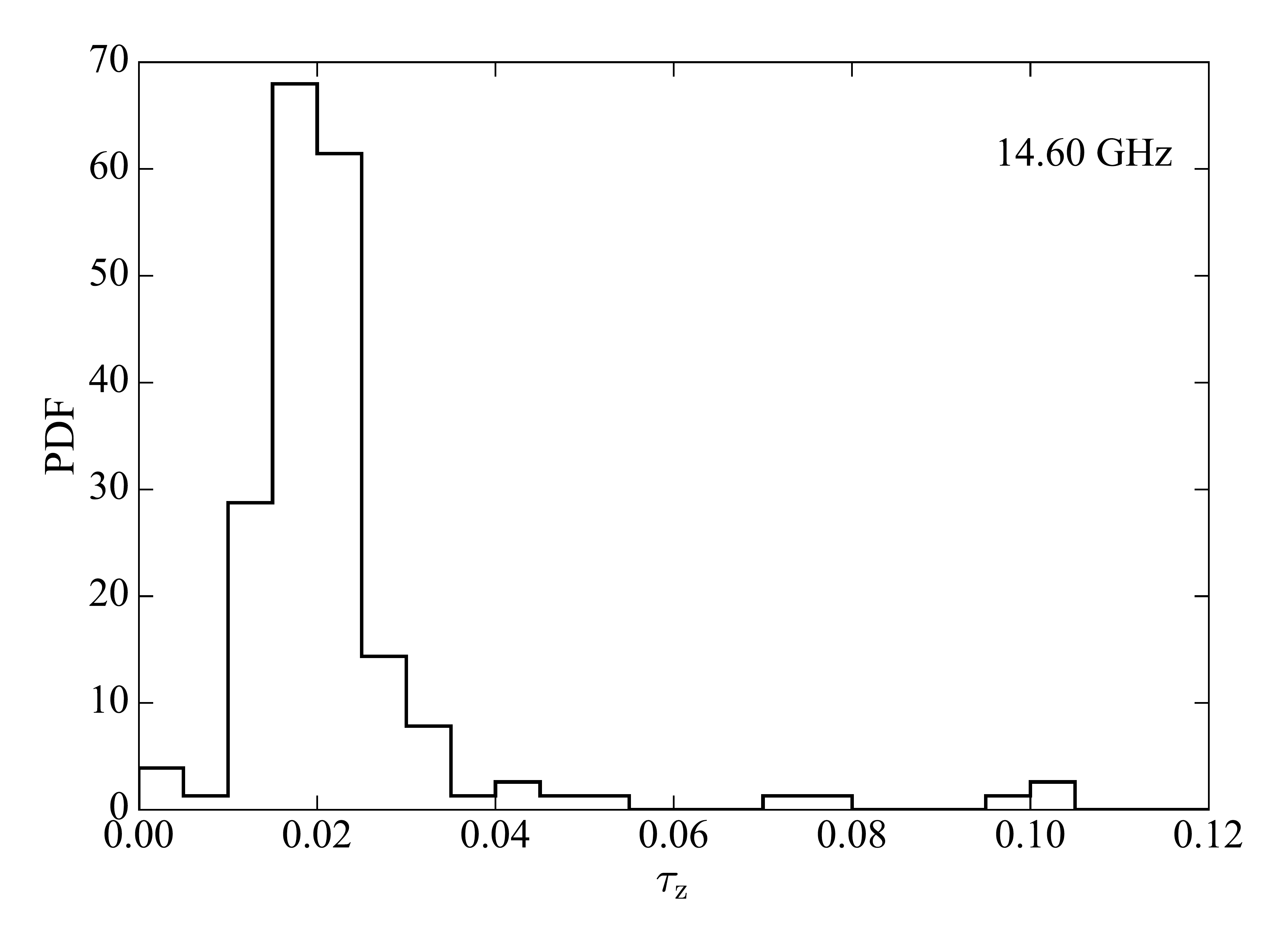}} 
%\\[-15pt]
  \resizebox{\hsize}{!}{\includegraphics[width=0.45\textwidth,angle=0]{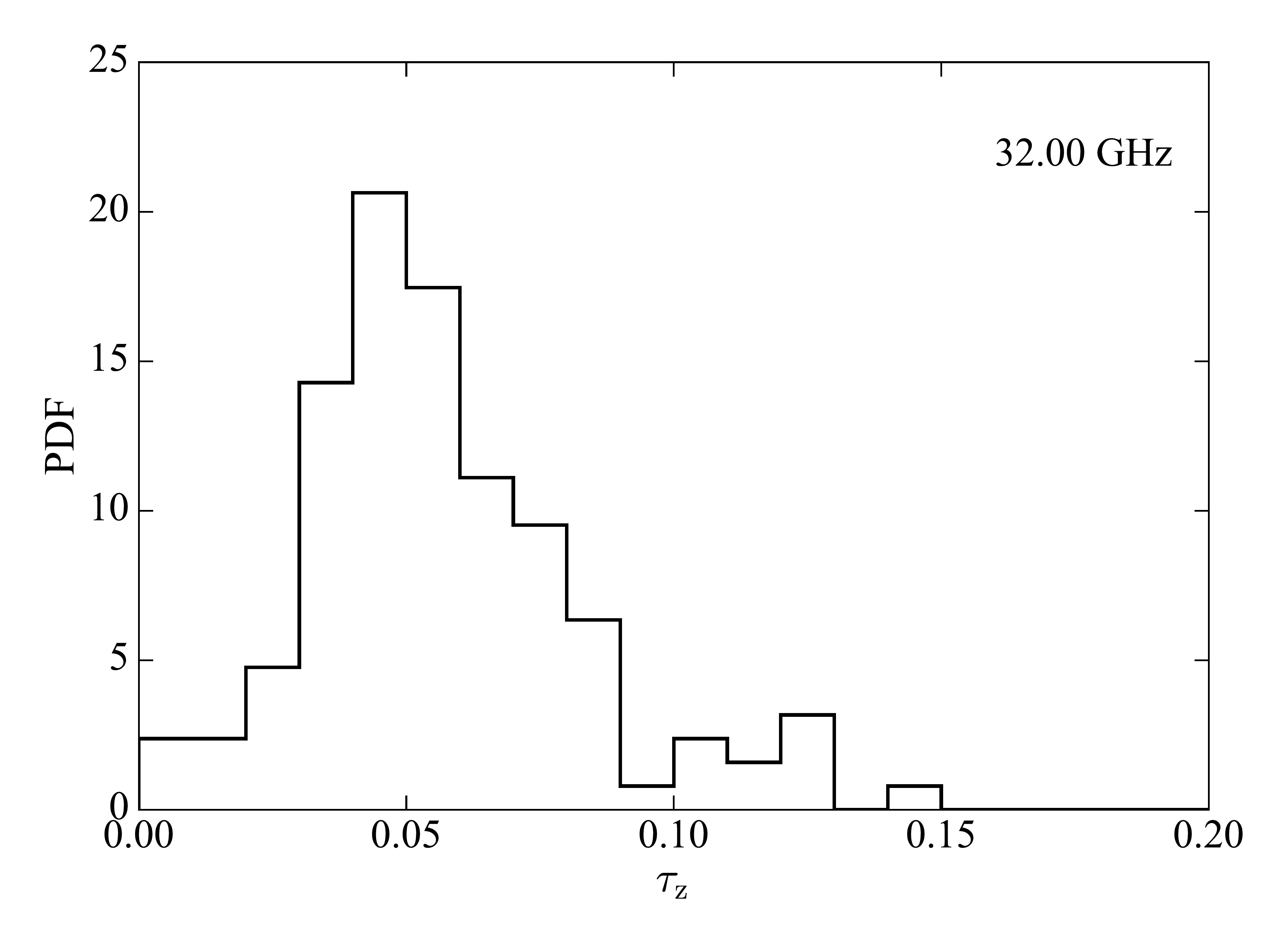}} 
  \caption{
  %\blue{-180625}
  Zenith opacity in three characteristic bands. {\it Upper} panel: 4.85~GHz (low band),
    {\it middle} panel 14.60~GHz (intermediate band), and {\it lower} panel 32.00~GHz (high band).}
  \label{fig:tau_z}
\end{figure}

This operation is correcting for the attenuation induced by the signal transmission through the terrestrial atmosphere. The opacity-corrected antenna temperature $T_\mathrm{opc}$ is computed from the observed antenna temperature $T$,  as
% -------------------------------------------------------------------------------------------------
  \begin{equation}
    \label{eq:Topct}
    T_{\mathrm{opc}}=T\cdot e^{\tau_\mathrm{atm}}
  ,\end{equation}
  where  \[
  \begin{array}{rp{0.8\linewidth}}
    \tau_\mathrm{atm}&is the atmospheric opacity at the source ELV.
  \end{array}
  \]
  The opacity $\tau_\mathrm{atm}$ at the source position is a function of its 
  ELV, and is given by  \begin{equation}
    \label{eq:tau}
    \tau_\mathrm{atm}=\tau\left(\mathrm{ELV}\right)=\tau_\mathrm{z}\cdot \mathrm{AM} = \tau_\mathrm{z}\cdot \frac{1}{\sin(\mathrm{ELV})}
  ,\end{equation}
  where
  \[
  \begin{array}{rp{0.8\linewidth}}
    %\tau_\mathrm{atm}&the atmosperic opacity at the source position\\
    \tau_\mathrm{z}&is the atmosperic opacity at the zenith and\\
    \mathrm{AM}&the air mass.\\
  \end{array}
  \]
% -------------------------------------------------------------------------------------------------
Correcting therefore for the atmospheric opacity at the source position requires the knowledge of the opacity at zenith, $\tau_\mathrm{z}$. Zenith opacity is computed from the observed dependence of $T_\mathrm{sys}$ on ELV (or equivalently on the airmass AM) and this is done for each individual session. First, a lower envelope (straight line) is fitted to the scatter plot of $T_\mathrm{sys}$ against AM. The fitted line is then extrapolated to estimate the system temperature at zero airmass. This point is subsequently connected with the system temperature of the actual measurement to compute the opacity for that scan. In Fig.~\ref{fig:tau_z} we plot the computed zenith opacity at three characteristic frequencies.  
% -------------------------------------------------------------------------------------------------
% NOTE see the previous comment fro wheer are these data cominf rrom.
% -------------------------------------------------------------------------------------------------

\begin{figure} 
  \resizebox{\hsize}{!}{\includegraphics[width=0.45\textwidth,angle=0]{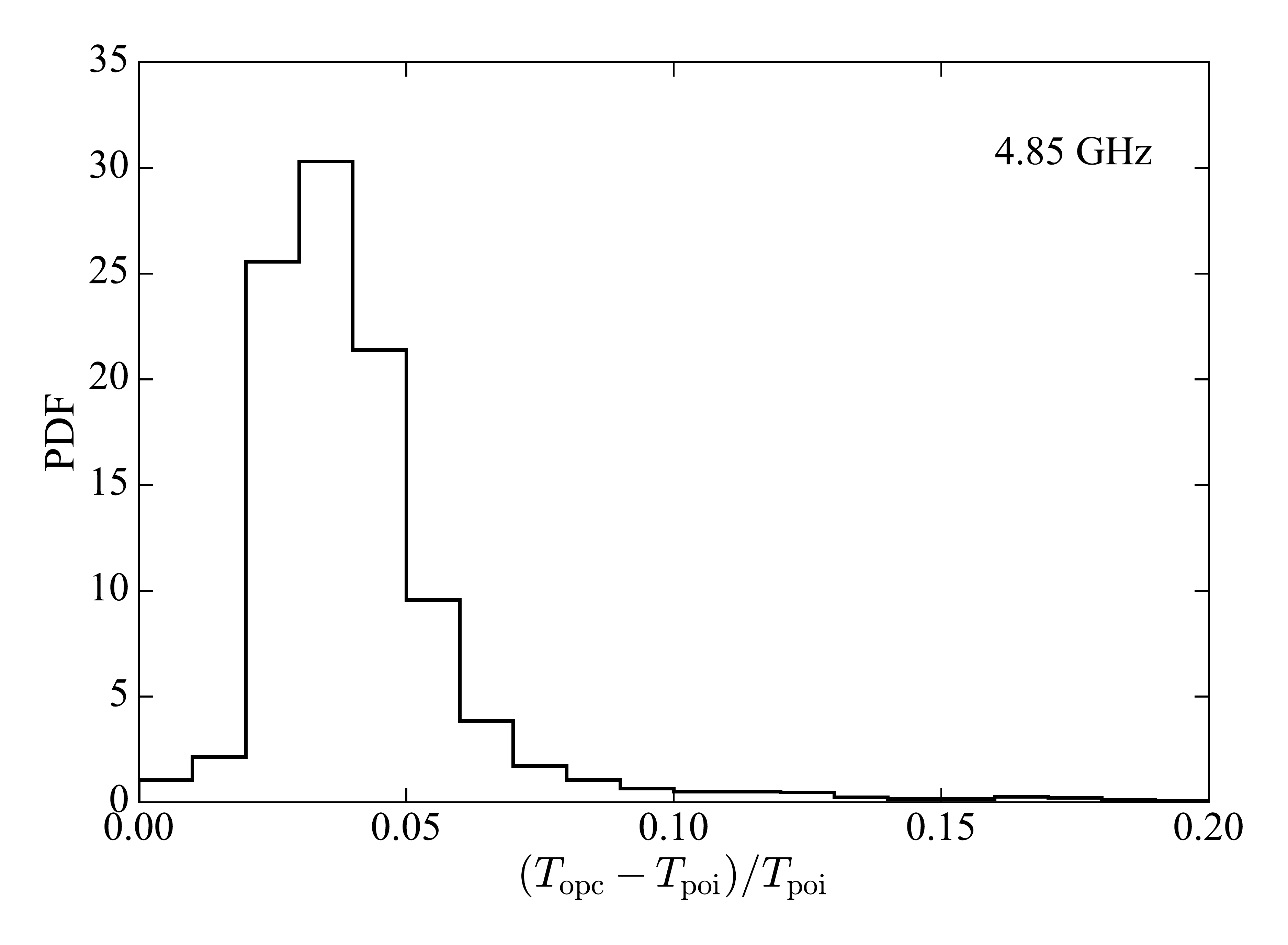}} 
%\\[-15pt]
  \resizebox{\hsize}{!}{\includegraphics[width=0.45\textwidth,angle=0]{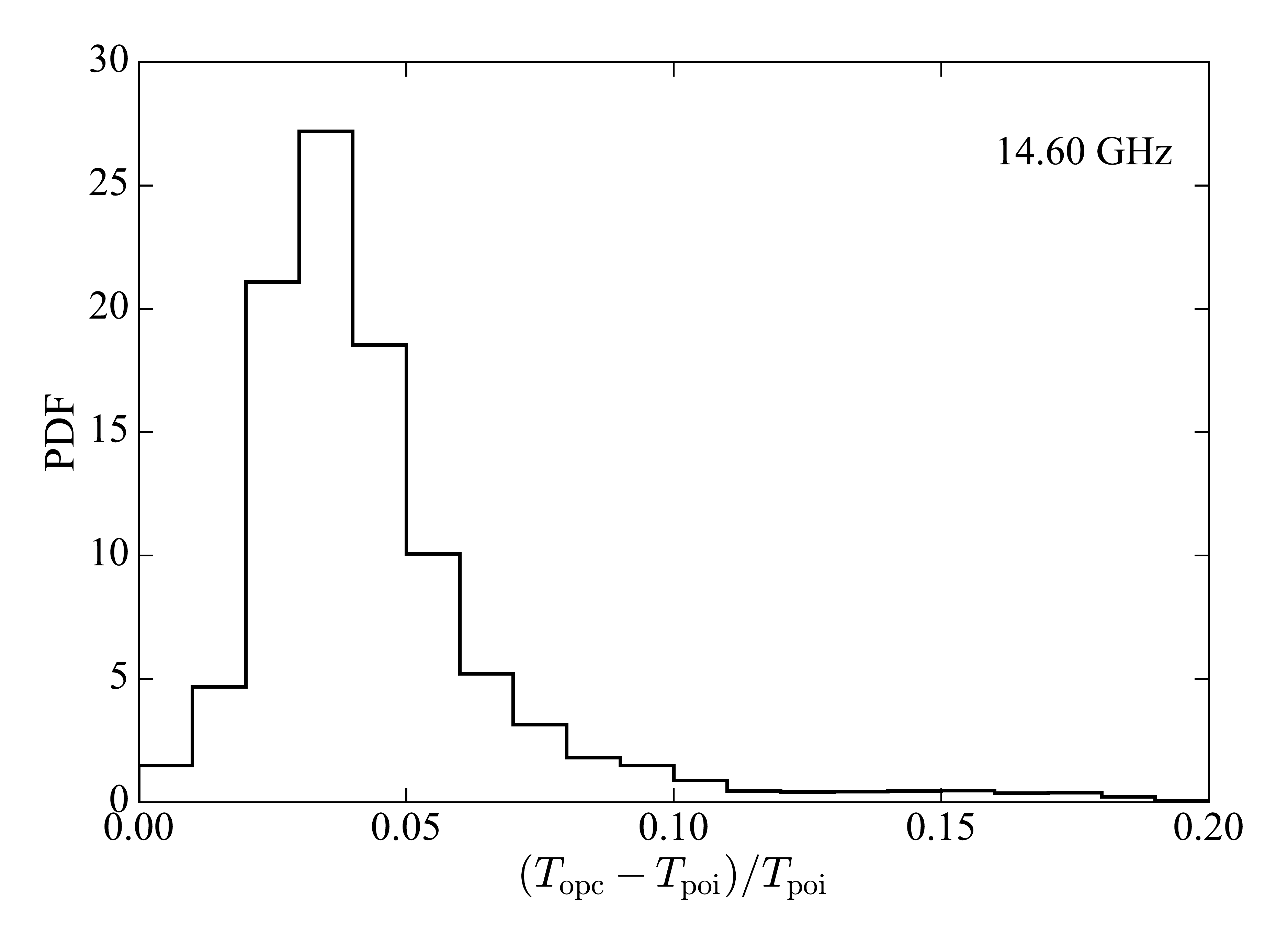}} 
%\\[-15pt]
  \resizebox{\hsize}{!}{\includegraphics[width=0.45\textwidth,angle=0]{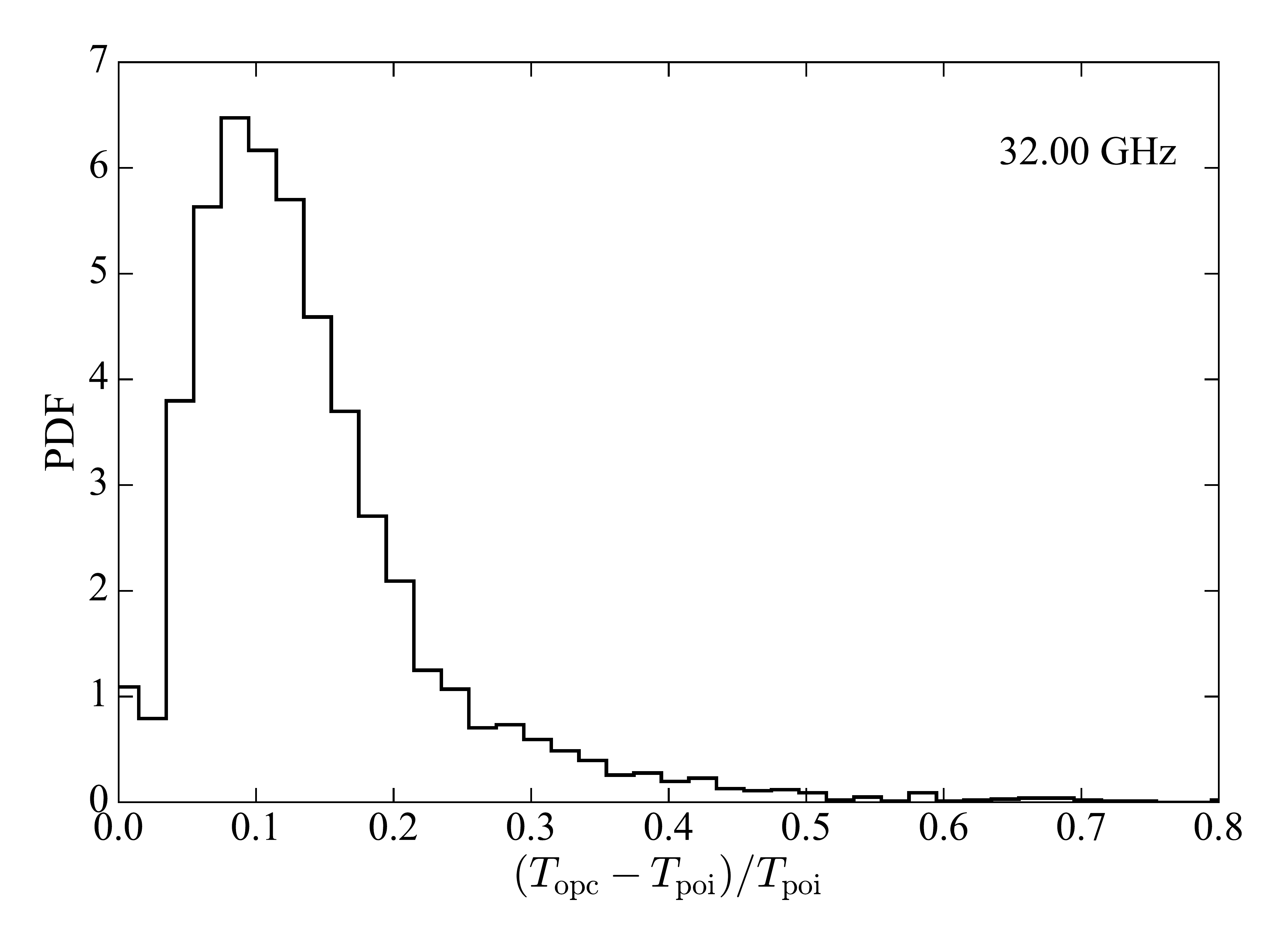}} 
  \caption{
  %\blue{-done}
  Fractional opacity correction. {\it Upper} panel: 4.85~GHz (low band),
    {\it middle} panel 14.60~GHz (intermediate band), and {\it lower} panel 32.00~GHz (high band). The x-axis is the fractional increase of the antenna temperature corrected for pointing ($T_\mathrm{poi}$) after the opacity correction has been applied ($T_\mathrm{opc}$).}
  \label{fig:opc_distr}
\end{figure}

The mean opacity at zenith for the receivers used is tabulated in table~\ref{tab:corrections}. As can be seen there, the opacity becomes particularly important towards higher frequencies. Because the opacity correction is that with the largest impact, in figure~\ref{fig:opc_distr} we present its effect at the three typical bands. In those plots we show the fractional increase of the pointing corrected antenna temperature $T_\mathrm{poi}$ when opacity correction is applied.    
% -------------------------------------------------------------------------------------------------
% NOTE see the previous comment fro wheer are these data cominf rrom.
% -------------------------------------------------------------------------------------------------
%\red{MUST DO: check the epochs with negative Tau and correct or veto them befpore data publishing.}

%..................................................................................................
\subsection{Elevation-dependent gain correction}
\label{subsec:gain}
  
\begin{figure} 
        \resizebox{\hsize}{!}{\includegraphics[]{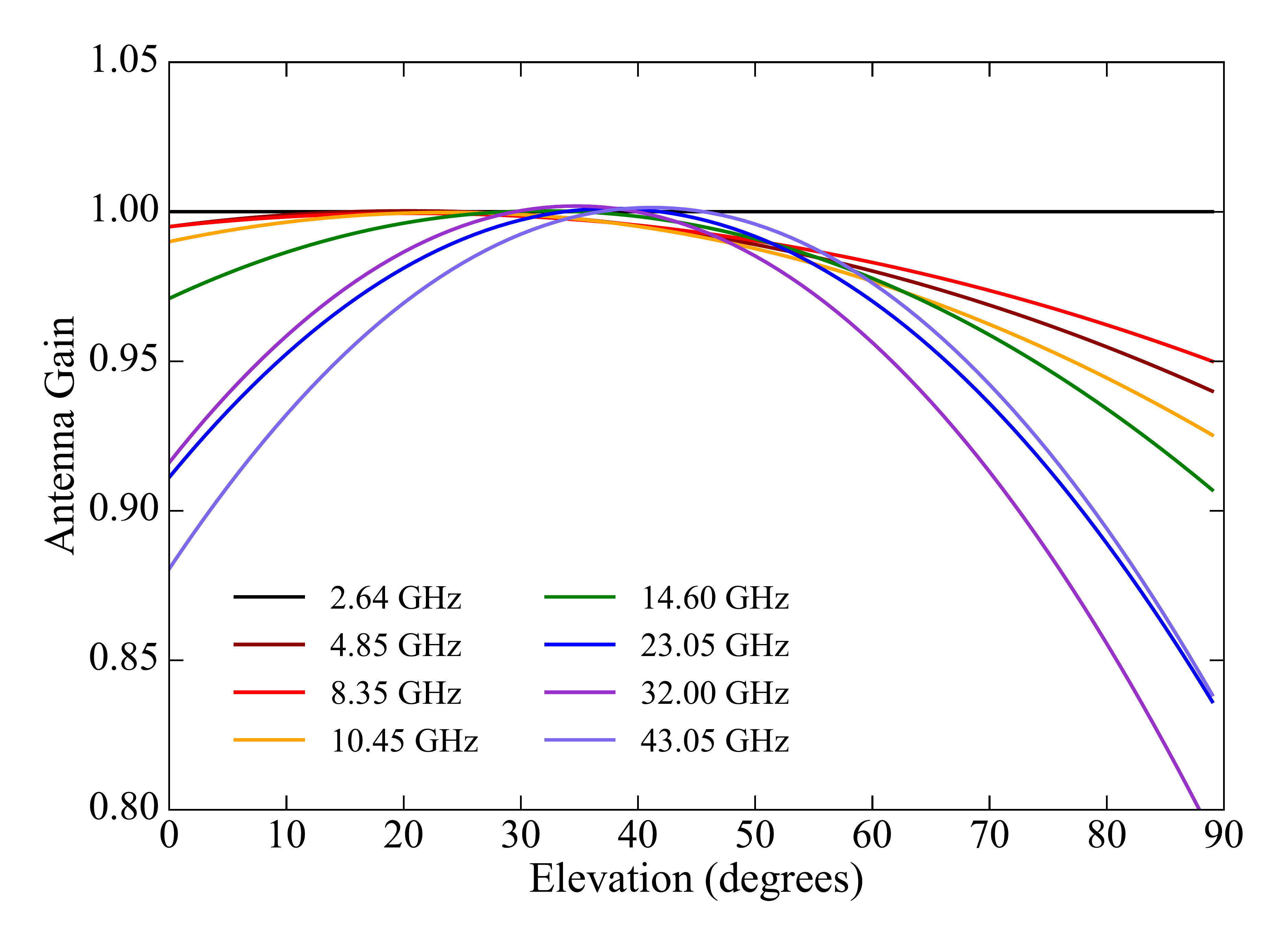}} 
        \caption{
        %\blue{-done}
        Gain curve applied at each observing frequency.}
        \label{fig:gc}
\end{figure}

This post-measurement operation accounts for losses caused by  small-scale gravity-induced departures of the geometry of the primary reflector from that of an ideal paraboloid. The power loss can be well approximated by a second order polynomial function of the source ELV. For an observed antenna temperature $T$, the corrected value, is
% -------------------------------------------------------------------------------------------------
\begin{equation}
    \label{eq:Tgc}
    T_{\mathrm{gc}}=T\cdot G^{-1}
  ,\end{equation}
  where
  \[
  \begin{array}{rp{0.8\linewidth}}
    G& is the ``gain curve'' value at the observing frequency and at the source ELV. It is given by  
    \begin{equation}
      \label{eq:gc}
      G\left(\mathrm{ELV}\right)= A_0+A_1\cdot \mathrm{ELV} +A_2\cdot \mathrm{ELV}^2
    .\end{equation}
        
  \end{array}
  \]
% -------------------------------------------------------------------------------------------------
In Fig.~\ref{fig:gc} we show the functions used for our observations. The parameters $A_0$, $A_1$, and $A_2$ are tabulated in table~\ref{tab:receivers}. As shown there, the ELV of maximum gain for the lowest frequencies tends to {be at lower elevations}. The enhancement of the beam side lobes at these frequencies imposes an overestimation of the opacity at those elevations. Hence, for low elevations the source flux densities tend to be over-corrected. As a result, because the gain curves are computed from opacity-corrected data, they tend to overestimate the gain at those lower elevations. The fractional effect of this operation is constrained to mainly less than one percent (table~\ref{tab:corrections}).
% -------------------------------------------------------------------------------------------------
% -------------------------------------------------------------------------------------------------

%..................................................................................................
\subsection{Absolute calibration} 
\label{subsec:sens} 
% -------------------------------------------------------------------------------------------------
\begin{table}
  \caption{Flux densities of the standard calibrators.}
  \label{tab:calibrators}  
  \centering                    
  \begin{tabular}{cccccc} 
    \hline\hline                 
    \diagbox[width=1.8cm]{$\nu$}{Source}    &3C\,48 &3C\,161&3C\,286&3C\,295&NGC\,7027\tablefootmark{a}\\
    \hline\\
    2.64  &9.51   &11.35    &10.69   &12.46    &3.75 \\
    4.85  &5.48   &6.62     &7.48    &6.56     &5.48 \\
    8.35  &3.25   &3.88     &5.22    &3.47     &5.92  \\
    10.45 &2.60   &3.06     &4.45    &2.62     &5.92  \\
    14.60 &1.85   &2.12     &3.47    &1.69     &5.85 \\
    23.05 &1.14   &1.25     &2.40    &0.89     &5.65 \\
    32.00 &0.80   &0.83     &1.82    &0.55     &5.49 \\
    43.00 &0.57   &0.57     &1.40    &0.35     &5.34 \\\\
    \hline                                  
  \end{tabular}
  \tablefoot{Flux densities of the calibrators are taken from
    \cite{Ott1994}, \cite{Baars1977AnA}, and Kraus priv. comm.
    \tablefoottext{a}{The flux density of NGC\,7027 is corrected for
      beam extension issues at frequencies above 10.45\,GHz and temporal evolution.}
  }
\end{table}
% -------------------------------------------------------------------------------------------------
The corrected antenna temperatures are finally converted into Jy by comparison with the observed antenna temperatures of standard targets termed primary calibrators. This operation also corrects for variations in the antenna and receiver gains between different observing epochs. The calibrators used here along with the assumed flux densities  are shown in table~\ref{tab:calibrators}. In case of multiple calibrator observations a mean correction factor was used. For NGC\,7027, our calibration procedure accounts for the temporal evolution of its spectrum \citep{2008ApJ...681.1296Z} and a partial power loss is caused by its extended structure relative to the beam size above 10.45\,GHz \cite[$7''\times10''$,][]{Ott1994}.   
        
%..................................................................................................
\subsection{Data editing and final quality check}
\label{subsec:qc}
The previously discussed reduction pipeline provides the end-to-end framework for recovering the real source flux densities from the observables. In practise, each individual sub-scan of each pointing (scan) was examined and quality checked manually by a human. The quality check protocol included various diagnostic tests at various stages of the data reduction pipeline. 
\paragraph{Sub-scans.}Each sub-scan was inspected for (a) FWHM significantly different from that expected; this could indicate source structure, field source confusion, or variable atmospheric conditions; (b) excessively large pointing offset, which could lead to irreversible power loss. (c) extraordinarily high atmospheric absorption or emission, which could cause destructive increase of noise; (d) large divergence from the mean amplitude of all sub-scans in the scan; (e) excess system temperature; and (f) possible radio frequency interference. The irreversibly corrupted sub-scans were vetoed from further analysis.
\paragraph{Sensitivity.} Highest quality observations of calibrators are clearly necessary at this stage. With human supervision this step was executed repeatedly until sensible estimates of the Jy-to-K factors were computed. This step required special care as attenuators could be activated even in the same observing session.  

\paragraph{}These quality checks were eventually followed by the ultimate test, which included two steps. First, we checked the shape of the radio SED in which every finally reduced frequency was compared against all other observing frequencies with the requirement that the line resemble a physically sensible shape. {Second, we tested whether the finally reduced radio SEDs were  following physically sensible evolution (mostly smooth)}.

%..................................................................................................
\subsection{Error budget}
\label{subsec:error_budg}

\begin{figure} 
  \resizebox{\hsize}{!}{\includegraphics[width=0.45\textwidth,angle=0]{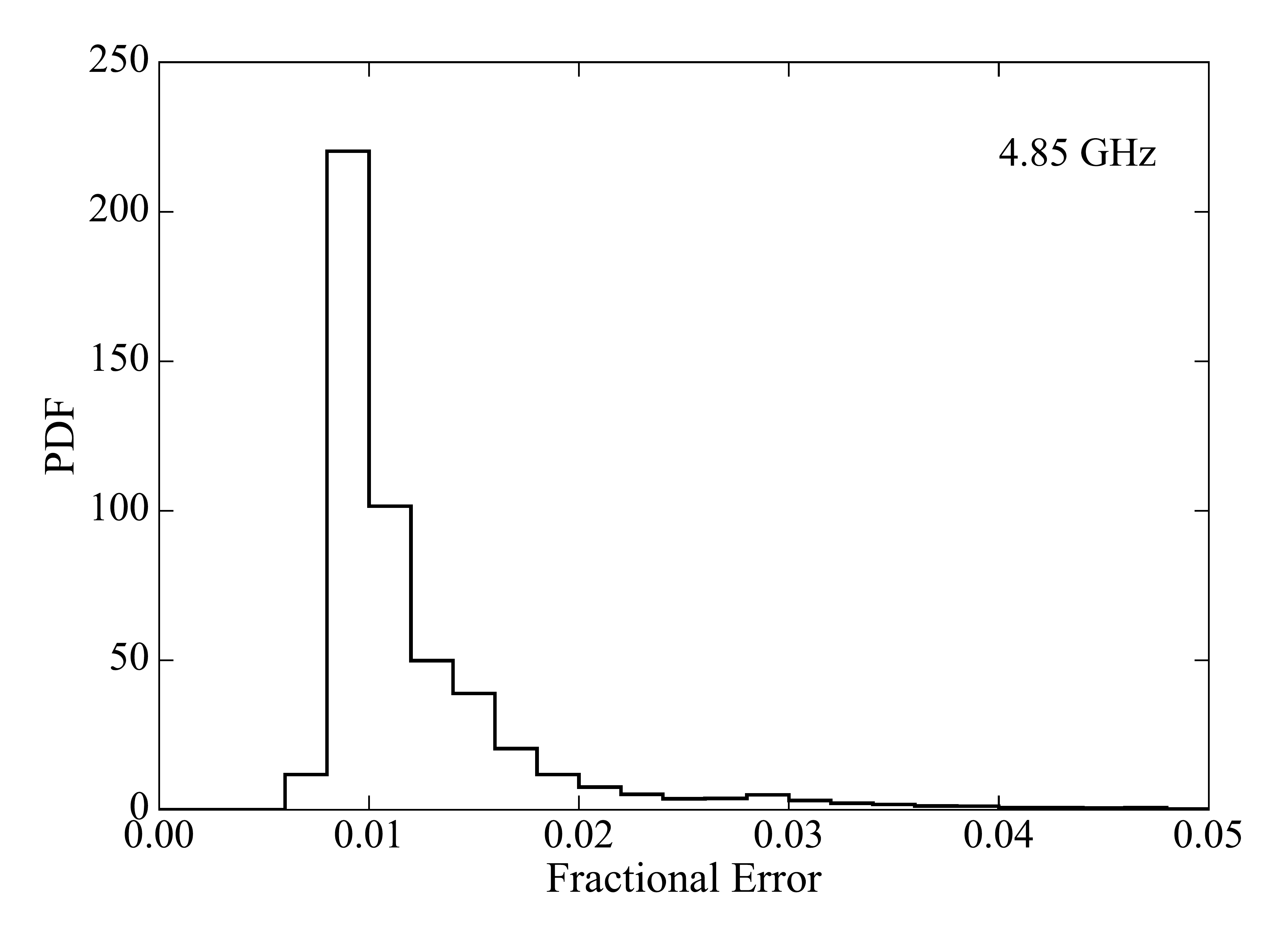}} 
%\\[-15pt]
  \resizebox{\hsize}{!}{\includegraphics[width=0.45\textwidth,angle=0]{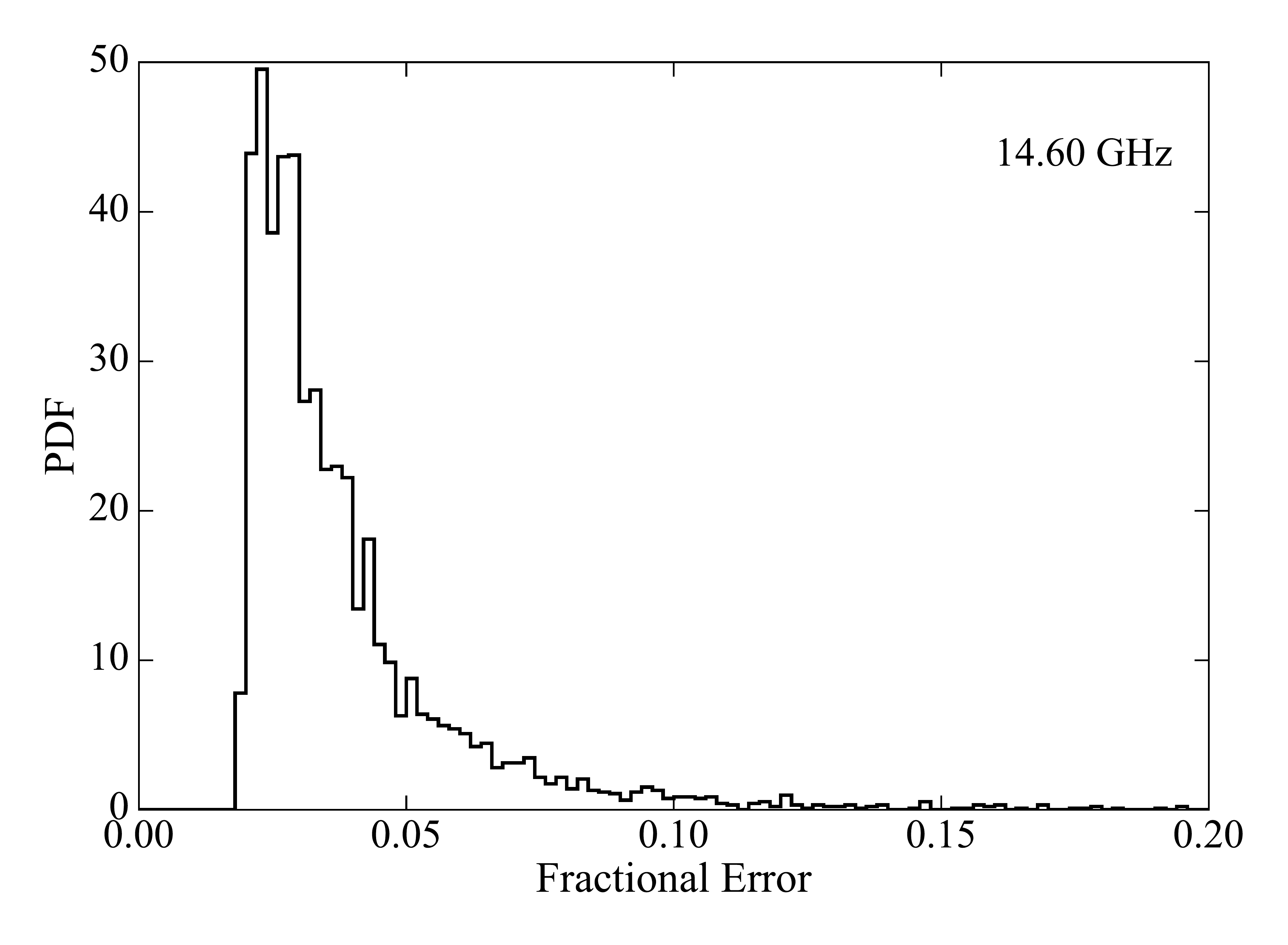}} 
%\\[-15pt]
  \resizebox{\hsize}{!}{\includegraphics[width=0.45\textwidth,angle=0]{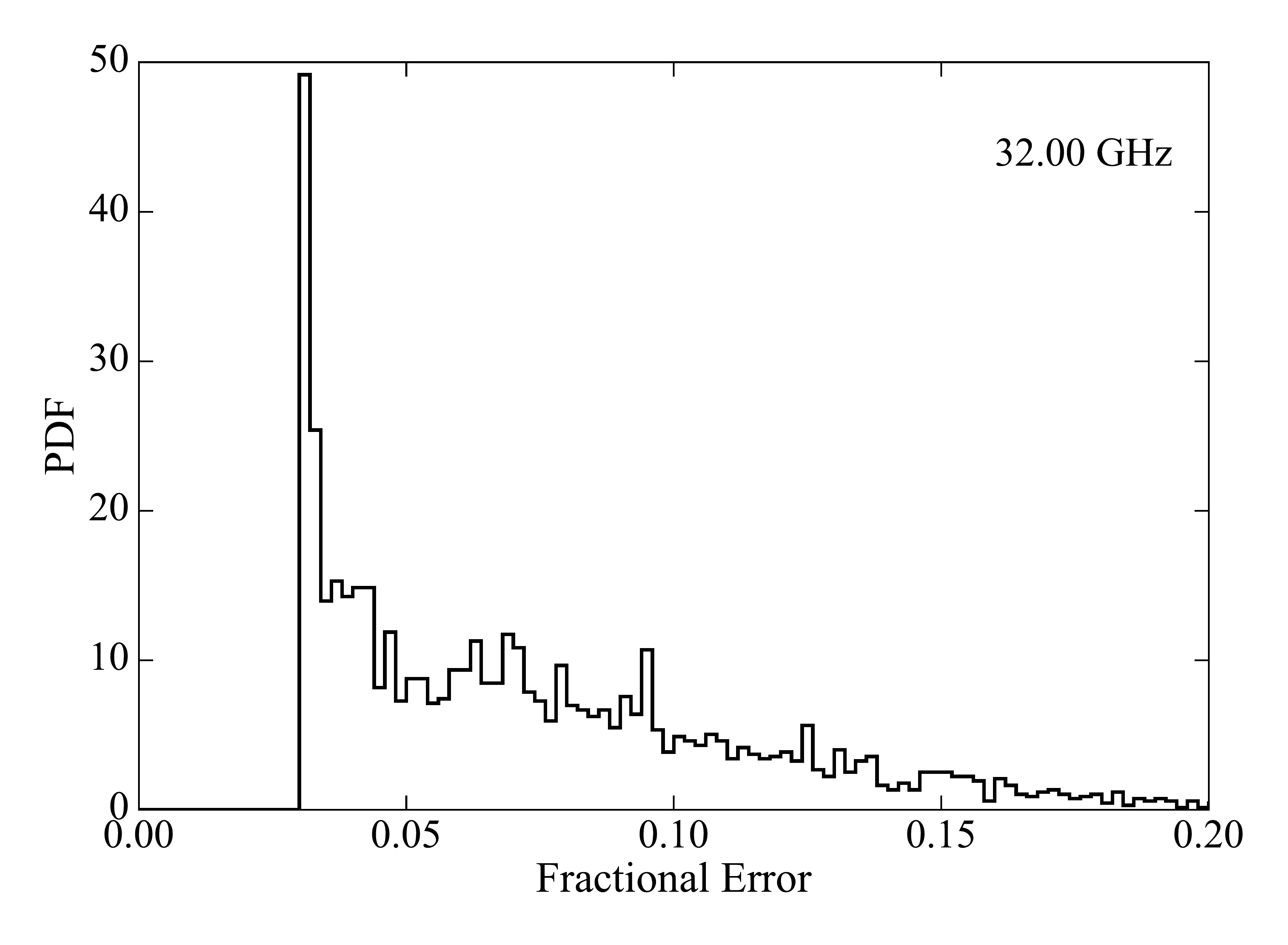}} 
  \caption{
  %\blue{180628} 
  Distribution of fractional error in three characteristic bands. {\it Upper} panel: 4.85~GHz (low band),
    {\it middle} panel 14.60~GHz (intermediate band),  and {\it lower} panel 32.00~GHz (high band).}
  \label{fig:err_distr}
\end{figure}

% sensitivity
% Tsens=T*gamma 
% Tsens_err= sqrt( (gamma * T_err)**2.0 + ( T*gamma_err )**2.0  )
%
% gain curve
% S_gc = S/G
% err_gc = err / G
%
% opacity
% flux_corr=(float(flux)*math.exp(float(tau)))
% flux_corr_err=(math.sqrt((float(flux_err)*math.exp(float(tau)))**2))
%
% pointing
%
%
% Amp_poi_LON=Amp_LON*math.exp(4*math.log(2)*Offset_LAT**2/(FWHM_tem**2))
% Amp_poi_LON_err=Amp_poi_LON*math.sqrt((Amp_LON_err/Amp_LON)**2+(8*math.log(2)*Offset_LAT/(FWHM_tem**2)*Offset_LAT_err)**2)
% Amp_poi_LAT=Amp_LAT*math.exp(4*math.log(2)*Offset_LON**2/(FWHM_tem**2))
% Amp_poi_LAT_err=Amp_poi_LAT*math.sqrt((Amp_LAT_err/Amp_LAT)**2+(8*math.log(2)*Offset_LON/(FWHM_tem**2)*Offset_LON_err)**2)
% Amp_poi=[Amp_poi_LON, Amp_poi_LAT] Amp_poi_err=[Amp_poi_LON_err, Amp_poi_LAT_err]
%
% S_poi=ea.AverageListNewNew(Amp_poi,Amp_poi_err) S_poi_err=max(S_poi[1],S_poi[6],S_poi[7])
% S_poi_err=max(S_poi[1],S_poi[6],S_poi[7])
% list_poi.append([JD,a_scn,SRC,S[5],err_final,AZI,ELV,LST,PA,S_poi[5],S_poi_err])
%
The end product of the data reduction pipeline is flux densities and their associated uncertainties, which are computed following a modified formal error propagation recipe assuming Gaussian behaviour. For the computation of the uncertainty $e_\mathrm{i}$ of a flux density measurement $S_\mathrm{i}$ the information of the entire light curve is used as follows:
% -------------------------------------------------------------------------------------------------
\begin{equation}
e_\mathrm{i}=\sqrt{{\sigma_0}^2 + (m\cdot S)^2 }
,\end{equation}
where
\[
\begin{array}{rp{0.8\linewidth}}
\sigma_0 &is the flux density independent term,\\
m &flux dependent term proportionality coefficient, amd\\
S&source mean flux density.\\
\end{array}
\]
The term $\sigma_0$, is defined as 
\begin{equation}
\sigma_0 = \max(\sigma, e)
,\end{equation}
where
\[
\begin{array}{rp{0.8\linewidth}}
  \sigma&is the standard deviation of the flux densities over the mean flux density for the corresponding
  observing session; and \\
  e&the error in the mean flux density, assuming Gaussian statistics and after propagating the error of each correction
  described in Sect.~\ref{corrections}. \\
\end{array}
\]
% -------------------------------------------------------------------------------------------------
The term $m$ is a measure of the flux-density-dependent part of the error. It is computed from the scatter of the flux density of {each one of the} calibrators which is assumed to be invariant, at least over timescales comparable to the length of the data trains discussed in this work \footnote{{The normality of the flux density distribution of each calibrator, which is expected from the assumption of intrinsically invariant flux density with the addition of random noise, has been confirmed with exhaustive tests.} }. 
The proportionality coefficient $m$ henceforth can be seen as a measure of the ``repeatability" of the system and has incorporated cumulatively all the sources of errors. In table~\ref{tab:mis} we present the mean values of $\sigma_0$ and $m$ used for each receiver. A physical interpretation of both $\sigma_0$ and $m$ can be found in \cite{angelakis2009AnA}. In Fig.~\ref{fig:err_distr} we show the distribution of the fractional error in three characteristic bands while in table~\ref{tab:median_err}  we  list the median fractional errors for all our receivers calculated from all the measurements released here. 
% -------------------------------------------------------------------------------------------------
% these numbers are calciulated in te rr_budget dir and I taek athe FLUX_avrgXXX_all and
% find th emax of th etwo thingsand then do the average
\begin{table}
  \caption{Average parameters $\sigma_0$ and $m$ of the error recipe for the different receivers.}
  \label{tab:mis}  
  \centering                    
  \begin{tabular}{lr@{\hspace{.35cm}}r@{\hspace{.35cm}}r@{\hspace{.35cm}}r@{\hspace{.35cm}}r@{\hspace{.35cm}}r@{\hspace{.35cm}}r@{\hspace{.35cm}}r} 
    \hline                 
    \hline\\
    $\nu$      & 2.64 &4.85 &8.35 &10.45 &14.6 &23.05 &32 &43\\
        $\sigma_0$ & 0.04 &0.03 &0.05 &0.06 &0.10 &0.16 &0.16 &0.22\\
        $m$        & 0.8 &0.8 &1.1 &1.2 &1.9 &2.8 &3.1 &3.5\\\\
\hline                                  
  \end{tabular}
\end{table}
% -------------------------------------------------------------------------------------------------
% -------------------------------------------------------------------------------------------------
% -------------------------------------------------------------------------------------------------
% -------------------------------------------------------------------------------------------------
\begin{table}
  \caption{Median measurement uncertainties for different frequencies.}
  \label{tab:median_err}  
  \centering                    
  \begin{tabular}{lr@{\hspace{.3cm}}r@{\hspace{.3cm}}r@{\hspace{.3cm}}r@{\hspace{.3cm}}r@{\hspace{.3cm}}r@{\hspace{.3cm}}r@{\hspace{.3cm}}r} 
    \hline                 
    \hline\\
    $\nu$      & 2.64 &4.85 &8.35 &10.45 &14.6 &23.05 &32 &43\\
        $e_i$ &0.01  &0.01  &0.02  &0.02  &0.03  &0.06  &0.07  & 0.09 \\\\
\hline                                  
  \end{tabular}
\end{table}
% -------------------------------------------------------------------------------------------------

\subsection{Noise normality}
\label{sec:noise_norm}

Throughout the preceding discussion we assumed that the noise, inevitably present in our data, behaves normally. This assumption provided the basis for a straightforward approach in the computation of the reported uncertainties in the raw and derivative quantities. We test the hypothesis and show that indeed the noise behaves in a Gaussian manner.    

\begin{figure} 
  \resizebox{\hsize}{!}{\includegraphics[trim=35pt 35pt 35pt 35pt  ,clip,   width=0.45\textwidth,angle=0]{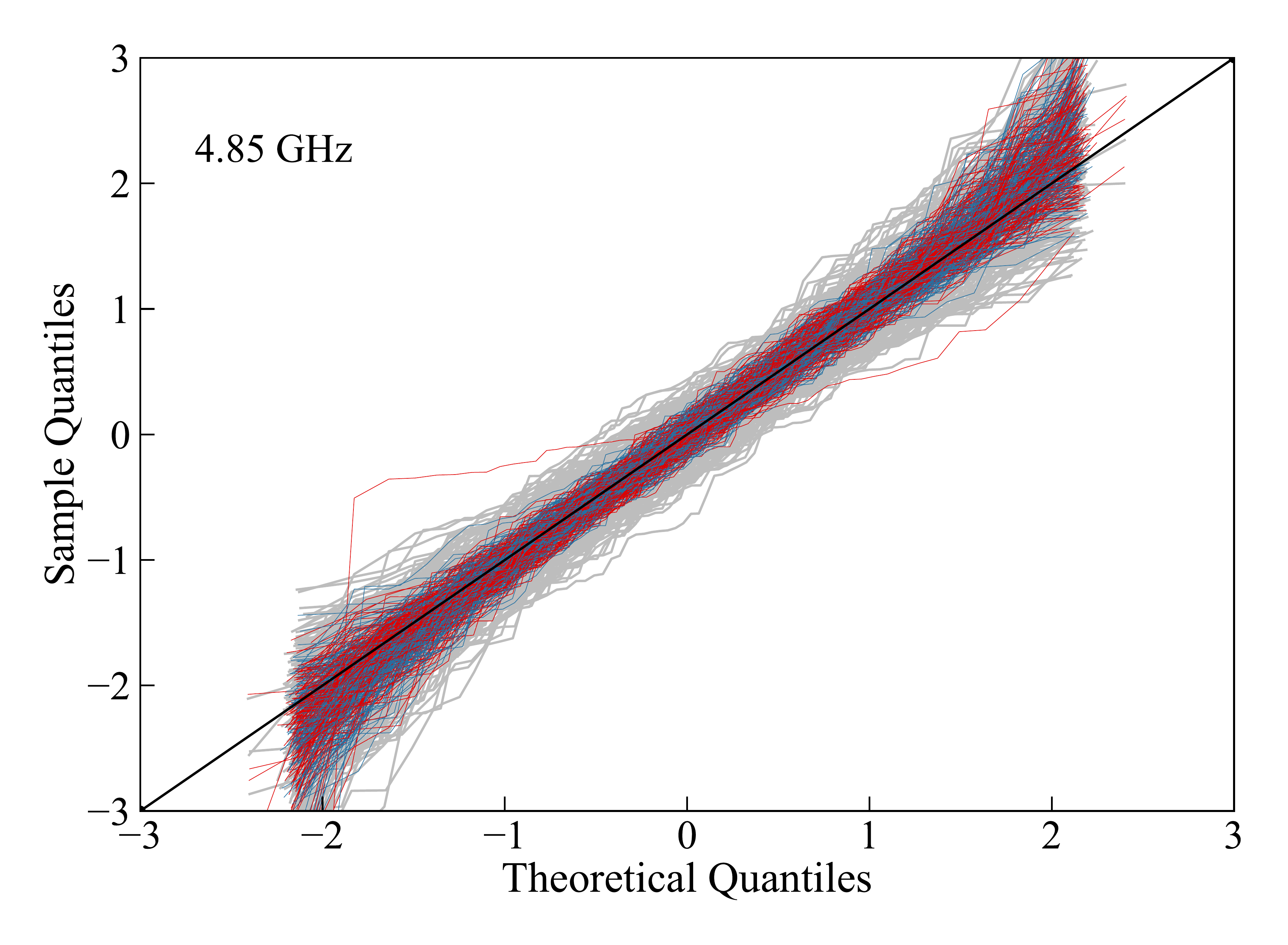}}
 
%\\[-15pt]
  \resizebox{\hsize}{!}{\includegraphics[trim=35pt 35pt 35pt 35pt  ,clip, width=0.45\textwidth,angle=0]{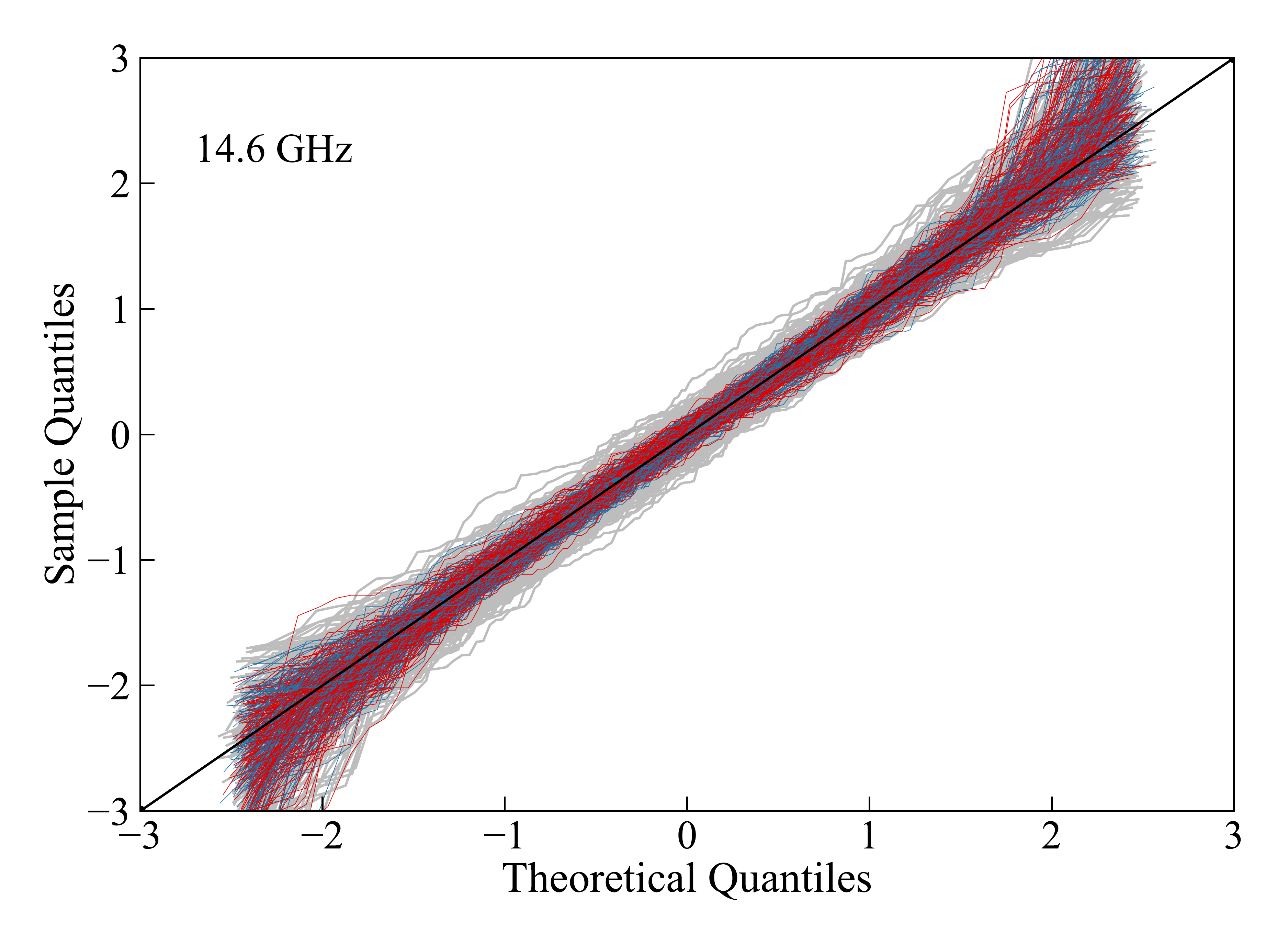}}
 
%\\[-15pt]
  \resizebox{\hsize}{!}{\includegraphics[trim=35pt 35pt 35pt 35pt  ,clip,  width=0.45\textwidth,angle=0]{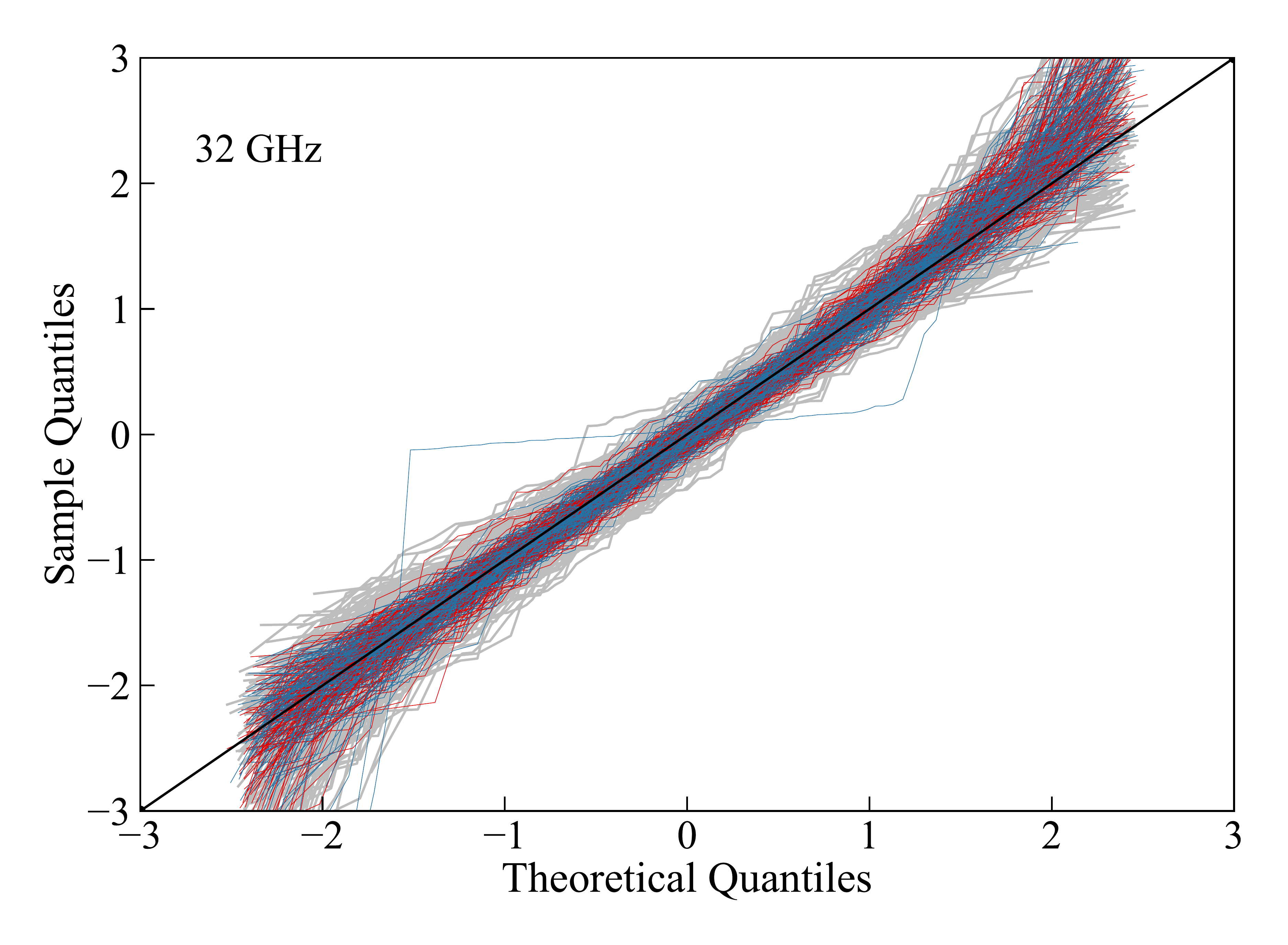}} 
  \caption{Quantile-quantile (Q-Q) probability plots in three characteristic bands. {\it Upper} panel: 4.85~GHz (low band),
    {\it middle} panel: 14.60~GHz (intermediate band), and {\it lower} panel: 32.00~GHz (high band). }
  \label{fig:qq}
\end{figure}

The normality test was run on each individual sub-scan (Sect.~\ref{sec:obs}) of one representative (in terms of tropospheric conditions) observing session. That amounts a total of several hundred datasets. In Fig.~\ref{fig:qq} we first show the quantile-quantile (Q-Q) probability plot at the three characteristic receivers for a visual inspection of the normality. Each dataset is first shifted to its average (hence it centres at zero) and is having its standard deviation normalised to unity.
% Th enormalisation is done by dividing each data point bu the stabdard deviation
These transformations allow us to compare the Q-Q plots of all the different datasets and make the interpretation of the Q-Q plots easier. An ideal dataset of a perfectly normal distribution would be described by the $y=x$ line, which in Fig.~\ref{fig:qq} is plotted as a black solid line. Each one of the blue and red lines comprises the Q-Q plot of one dataset. The red and blue lines correspond to the brightest and weakest sources, respectively, crudely classified by comparison to the median of all datasets. 

Evidently, the departure from normality is rather insignificant. For each dataset we create a mock sample from an ideal Gaussian distribution by randomly selecting the same number of data points as those in the observed dataset. These mock Q-Q lines are indicated in grey. Figure~\ref{fig:qq} makes it immediately clear then that the sampling alone can account for the departure from normality manifested as a spread of the Q-Q plots. 

For the quantification of normality we ran a D'Agostino's K$^2$ test of the hypothesis that the distribution of a given dataset is Gaussian. For a p-value threshold of 0.05 the hypothesis at 4.85~GHz cannot be rejected for more than 94\% of the cases. For the same p-value threshold this fraction is 85\% at 14.6~GHz and 89\% at 32~GHz practically proving the validity of the hypothesis that the noise is Gaussian.

% -------------------------------------------------------------------------------------------------
% -------------------------------------------------------------------------------------------------

%==================================================================================================
\section{Raw data: Multi-frequency light curves }
\label{sec:LCs}

As we discussed in Sect.~\ref{sec:sample} the purpose of the current work is the release of the Effelsberg 2.64 -- 43 GHz \fg dataset. For each source the multi frequency light curves are available on-line in the form of table~\ref{tab:example_lc}. In table~\ref{tbl:data} (also available on-line) we provide median flux densities and basic descriptive characteristics of the light curves. 

For all the sources that have been monitored by the \fg programme, i.e. tagged as ``f'', ``s1'', ``s2'', ``old'' , or F-GAMMA-\textit{Planck} MoU sources (Sect.~\ref{sec:obs}), we also present their light curves in Fig.~\ref{fig:sample_pg1} -- \ref{fig:sample_pg18}. The lower panel in each frame shows the evolution of the three-point low $\alpha_{2.64}^{8.35}$, and mid-band $\alpha_{8.35}^{14.6}$ spectral indices defined by $S\propto \nu^{\alpha}$. The mean data rate\footnote{For $N$ measurements with the first at JD$_0$ and the last at JD$_N$,  the mean data rate is computed as JD$_N-$JD$_0$/$(N-1)$.} in these light curves is around 1.3 months at the lower frequencies, 1.7 months at higher frequencies, and 3.6 months at 43 GHz. Because these values refer to post-quality check data products (not to those observed) the departure from the nominal one month cadence is mostly due to data quality filtering. %Similar plots for all the sources in table \ref{tab:sample} are \red{available online}. 
% -------------------------------------------------------------------------------------------------
\begin{table}
  \caption{
  %\blue{ea-180305:} 
  Example light curve file available on-line.}
  \label{tab:example_lc}  
  \centering                    
  \begin{tabular}{lrrrr} 
      \hline\hline                 
      Source    &\mc{1}{c}{JD} &\mc{1}{c}{$\nu$} &\mc{1}{c}{$S$}   &\mc{1}{c}{err}\\
                &   &\mc{1}{c}{(GHz)} &\mc{1}{c}{(Jy)}  &\mc{1}{c}{(Jy)}\\
      \hline\\
   J0050$-$0929 &2454953.964 &4.85    &1.199  &0.011 \\
   J0050$-$0929 &2454983.966 &4.85    &1.210  &0.346 \\
   J0050$-$0929 &2455001.709 &4.85    &0.995  &0.008 \\
   J0050$-$0929 &2455046.586 &4.85    &0.823  &0.011 \\
   J0050$-$0929 &2455072.561 &4.85    &0.773  &0.007 \\
   \mc{1}{c}{\ldots} &\mc{1}{c}{\ldots}   &\mc{1}{c}{\ldots}  &\mc{1}{c}{\ldots}     &\mc{1}{c}{\ldots}\\ \\     
\hline                                  
  \end{tabular}
\end{table}
% -------------------------------------------------------------------------------------------------

%..................................................................................................
\subsection*{Event rates and duty cycle}
\label{sec:rates}

Our variability analysis, which will be presented in a subsequent publication, showed that for practically all the frequencies our main target group (``f'' , ``s1'' , and ``s2'' sources, Fig.~\ref{fig:sample_pg1} -- \ref{fig:sample_pg18}) display one outbursting event per year. Exceptionally 43~GHZ gives much lower event rates ($\sim 0.8$~yr$^{-1}$) owing to the poor effective sampling. For the most powerful events defined as those with amplitude $A>0.5 A_\mathrm{max}$, where $A_\mathrm{max}$  is the amplitude of the most powerful flare, the median rate is about 0.6 -- 0.7 events per year.     

\begin{figure} 
  \resizebox{\hsize}{!}{\includegraphics[width=0.45\textwidth,angle=0]{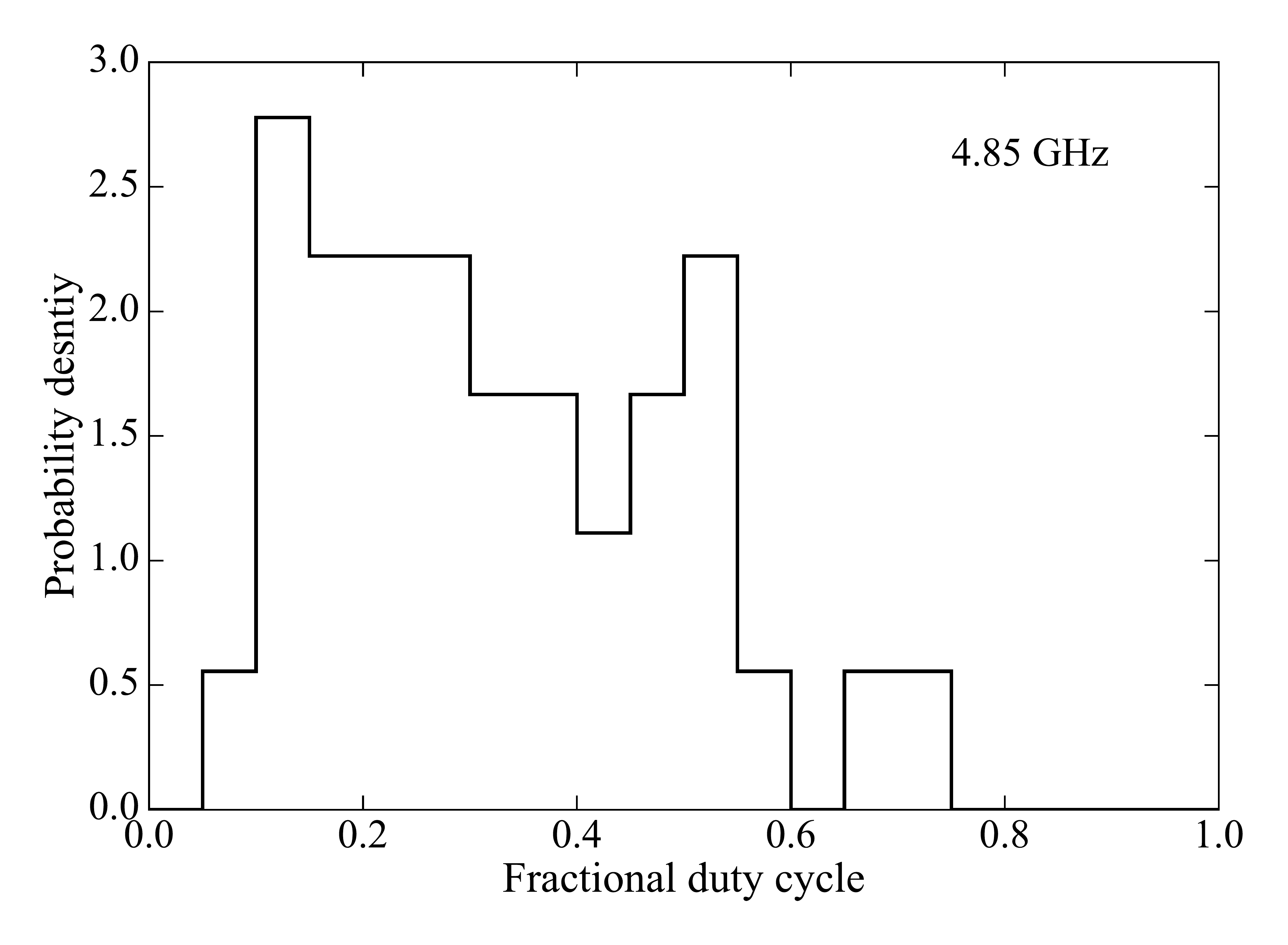}} 
%\\[-15pt]
  \resizebox{\hsize}{!}{\includegraphics[width=0.45\textwidth,angle=0]{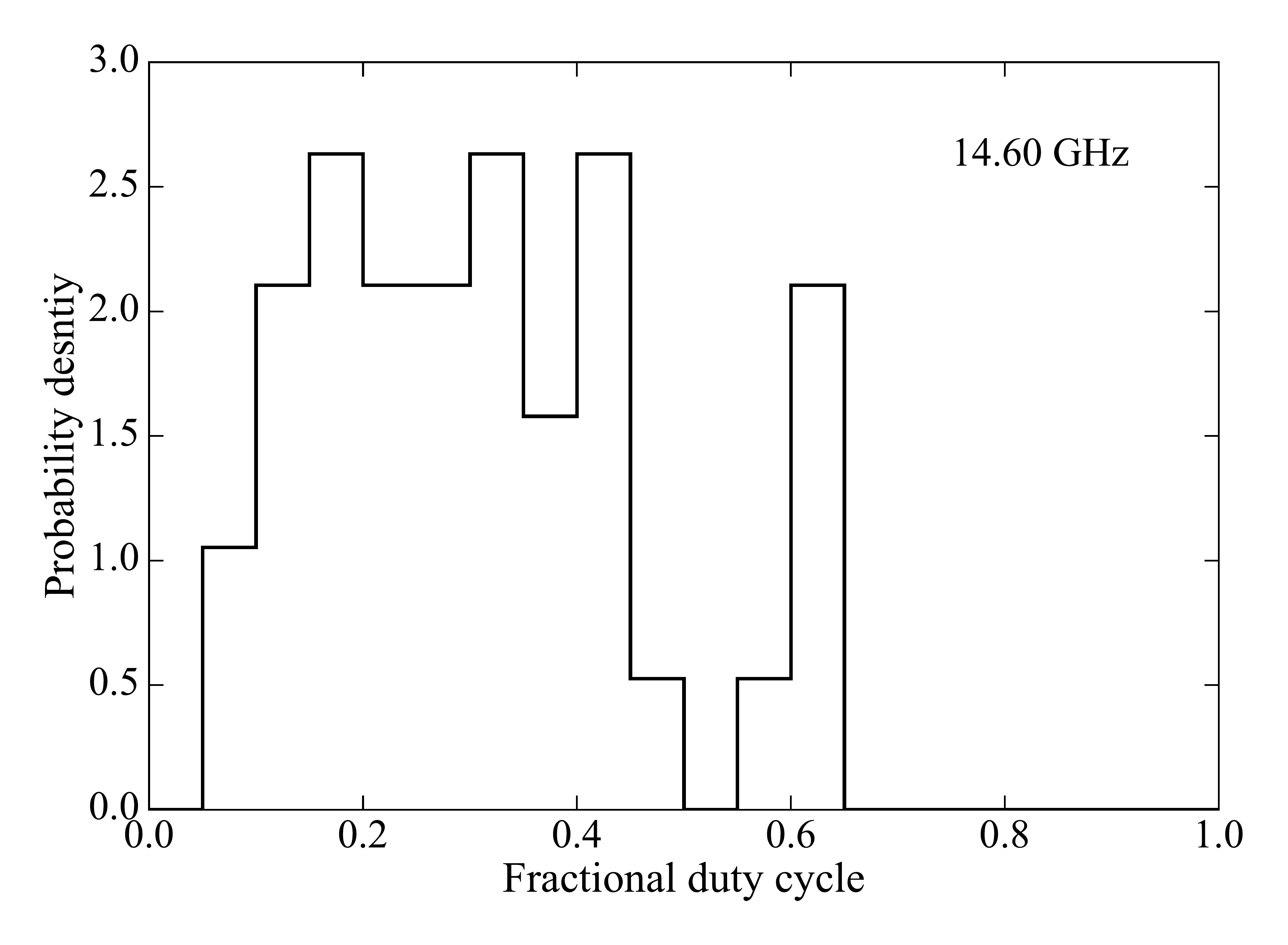}} 
%\\[-15pt]
  \resizebox{\hsize}{!}{\includegraphics[width=0.45\textwidth,angle=0]{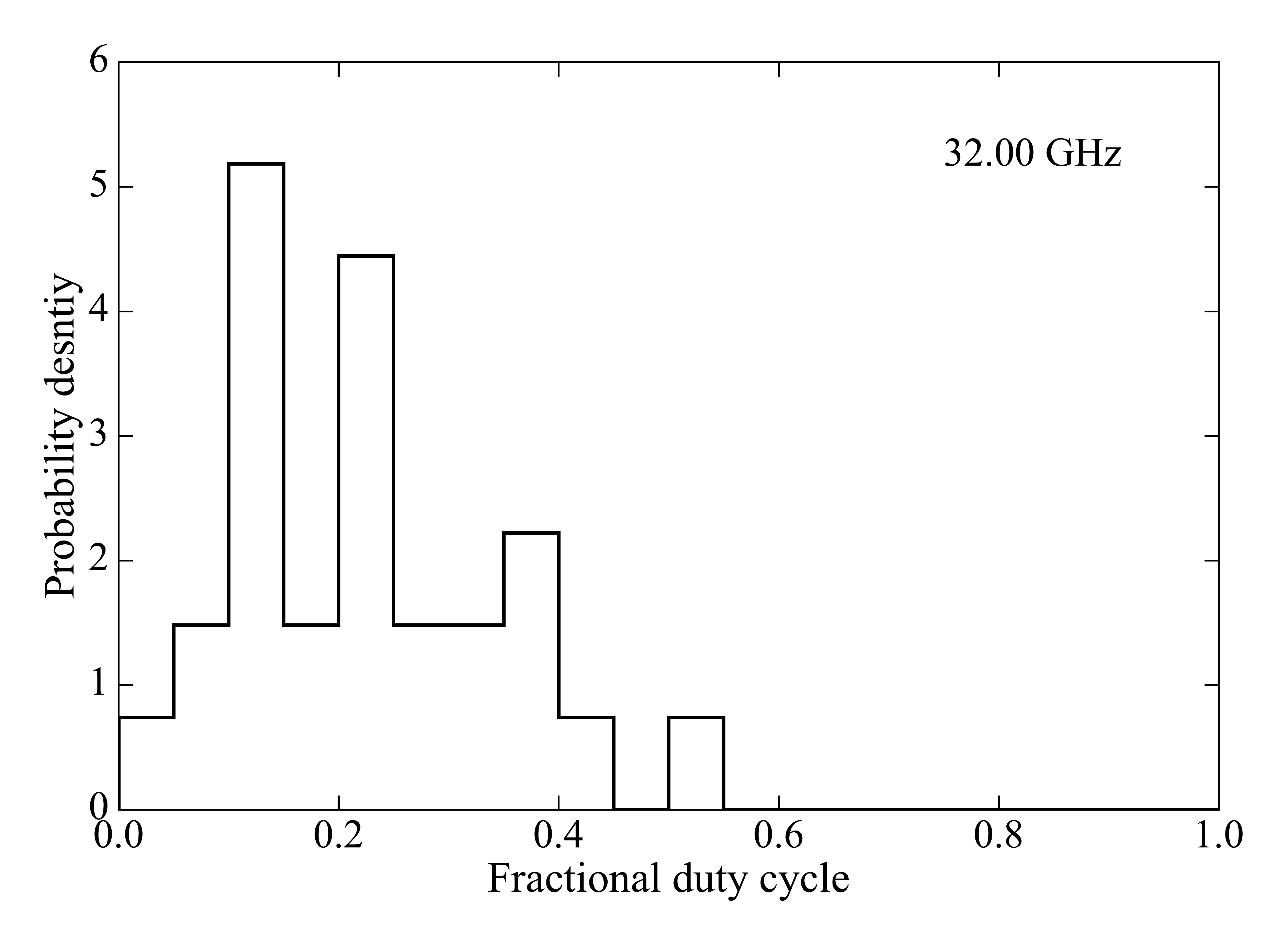}} 
  \caption{
  %\blue{-180625}
  Fractional duty cycle in three characteristic bands. {\it Upper} panel: 4.85~GHz (low band),
    {\it middle} panel 14.60~GHz (intermediate band), {\it lower} panel 32.00~GHz (high band). The bin size is set at 0.05.}
  \label{fig:DutyCycl}
\end{figure}

Beyond the frequency of event occurrence it is interesting to get a sense of the distribution of duty cycle; i.e. the fraction of observing time that the source is spending in an active state. We quantify this as the fraction of time that the source is at a phase at least half of the peak-to-peak flux density. Clearly this refers to the most luminous events. In Fig.~\ref{fig:DutyCycl} we present the distribution of the duty cycle at three characteristic frequencies. The mean and median of 0.30 and 0.33, respectively, at 4.85~GHz drop at 0.21 and 0.22 at 32~GHz; this is yet another way to show that the activity happens at progressively longer timescales as the frequency decreases.
%-------------------------------------------------------------------------------------------------
% NOTE see the previous comment fro wheer are these data cominf rrom.
% -------------------------------------------------------------------------------------------------

%..................................................................................................
\section{Data products: Spectral Indices}
\label{sec:spind}

\begin{figure} 
  \resizebox{\hsize}{!}{\includegraphics[ trim=0pt 0pt 0pt 0pt  ,clip, width=0.55\textwidth,angle=0]{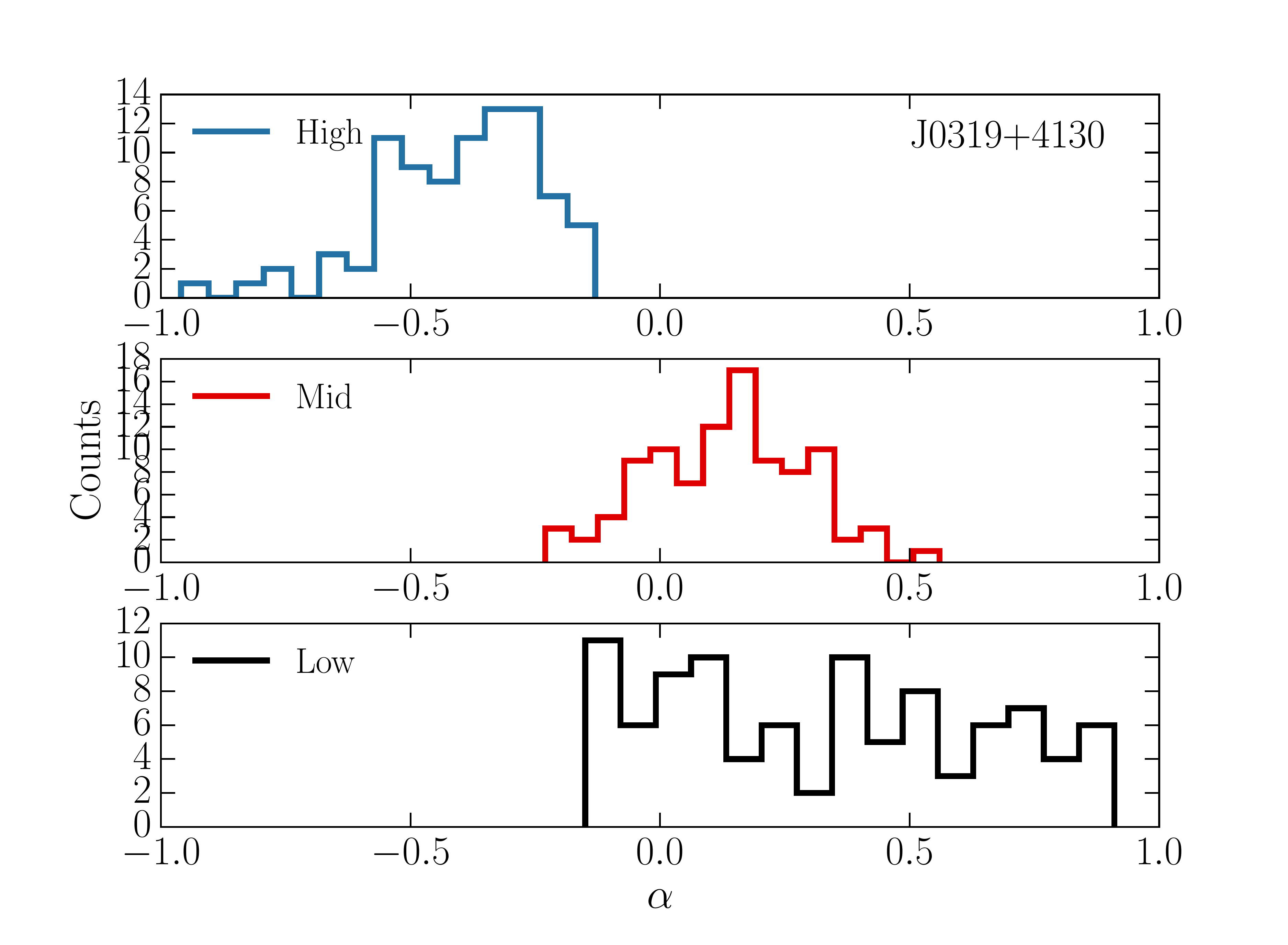}} 
%\\[5pt]

  \resizebox{\hsize}{!}{\includegraphics[ trim=0pt 0pt 0pt 0pt  ,clip, width=0.45\textwidth,angle=0]{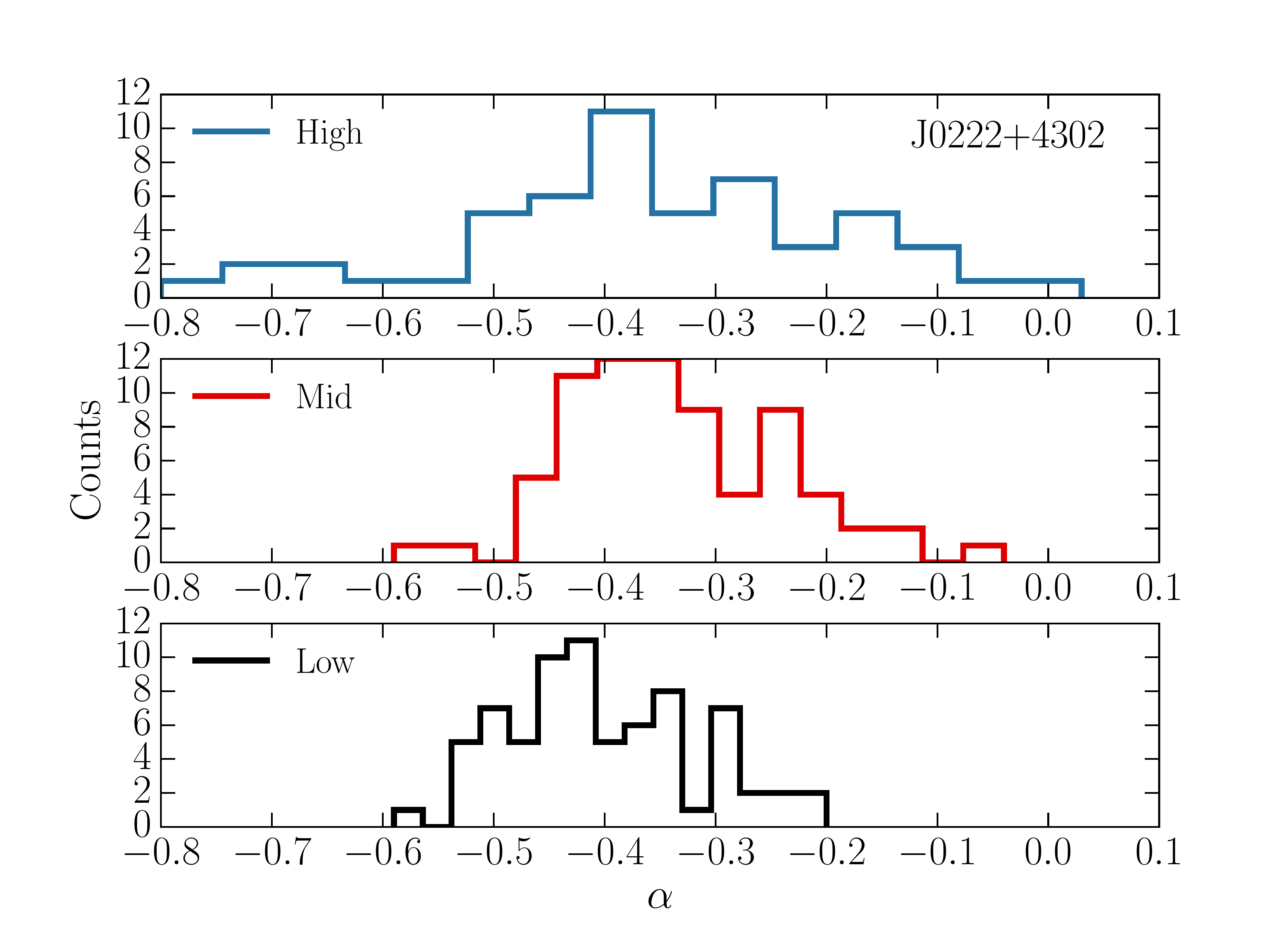}} 
  \caption{
  %\blue{ea180527-done} 
  Spectral index distribution in three sub-bands. {\it Upper} panel: a source undergoing intense spectral evolution. {\it Lower} panel: a source with an almost self-similarly variable SED. }
  \label{fig:spind_host_eg}
\end{figure}

Table~\ref{tbl:spinds} lists median and extreme values of the spectral index distributions of all the sources discussed in Sect.~\ref{sec:sample} (groups ``f'', ``s1'', ``s2'', ``old'' , and F-GAMMA-\textit{Planck} MoU) as well as targets of opportunity  for which an adequate dataset was available; in each observation band at least two frequencies were required. It is noted as we discussed earlier that practically each SED was acquired within less that one hour. For the monitored sources the median duration of an SED is between 30 and 40 minutes implying that they are practically instantaneous and hence free of variability effects. The spectral indices are computed in three bands of progressively higher frequencies: \textit{low} using 2.64, 4.85 and 8.35~GHz, the \textit{middle} band over 8.35, 10.4.5, and 14.6~GHz, and in the \textit{high} band over 14.6, 23.0, 32, and 43~GHz. The spectral index is computed with a least-squares fit of  
\begin{equation}
  S\left(\nu\right) = S_0\cdot \nu^\alpha \,
\end{equation}
to the observed flux densities. In each band, observations of at least two frequencies were required for the fit. As a measure of the uncertainty in the computation of spectral indices in table~\ref{tbl:spinds} we also list the median error, $\sigma_\alpha$. In Fig.~\ref{fig:spind_host_eg} we show example distributions of the three spectral indices in the characteristic cases of a source that undergoes spectral evolution (\textit{upper} panel) and one with an achromatically variable SED (\textit{lower} panel).    
% -------------------------------------------------------------------------------------------------
% NOTE see the previous comment fro wheer are these data cominf rrom.
% -------------------------------------------------------------------------------------------------

\begin{figure} 
  \resizebox{\hsize}{!}{\includegraphics[ trim=0pt 0pt 0pt 0pt  ,clip, width=0.45\textwidth,angle=0]{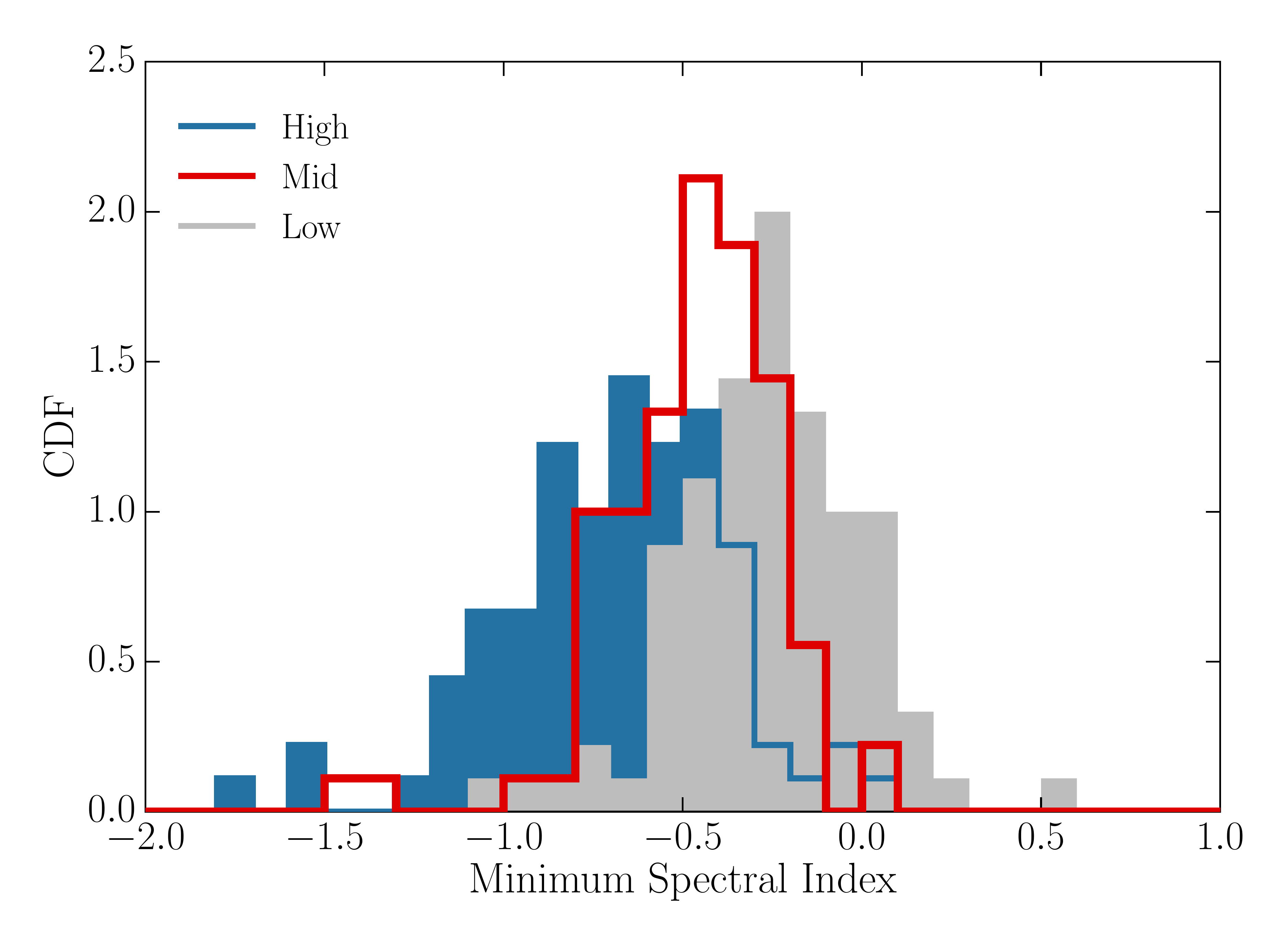}} 
%\\[5pt]

  \resizebox{\hsize}{!}{\includegraphics[ trim=0pt 0pt 0pt 0pt  ,clip, width=0.45\textwidth,angle=0]{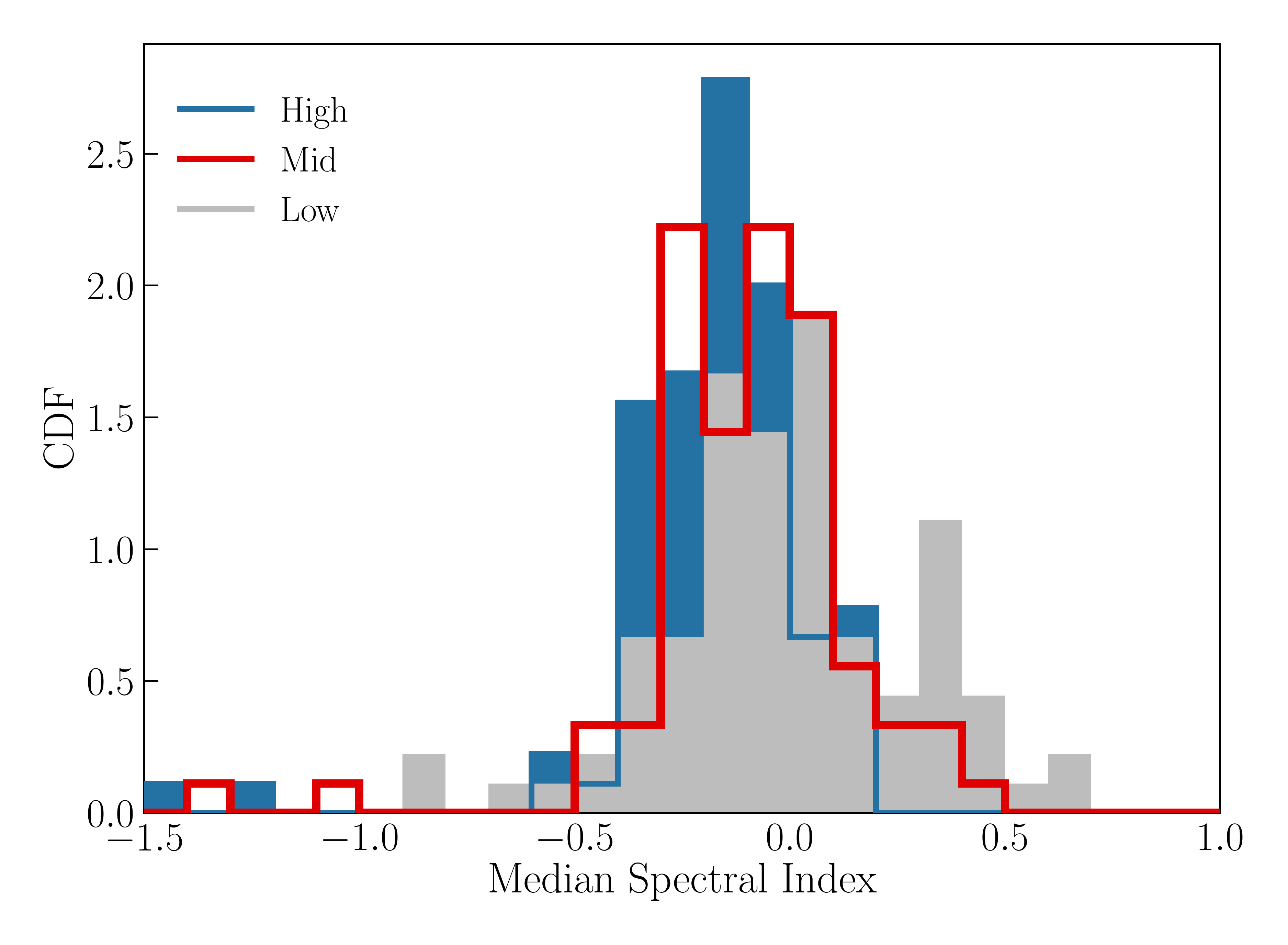}} 
%\\[5pt]

  \resizebox{\hsize}{!}{\includegraphics[ trim=0pt 0pt 0pt 0pt  ,clip, width=0.45\textwidth,angle=0]{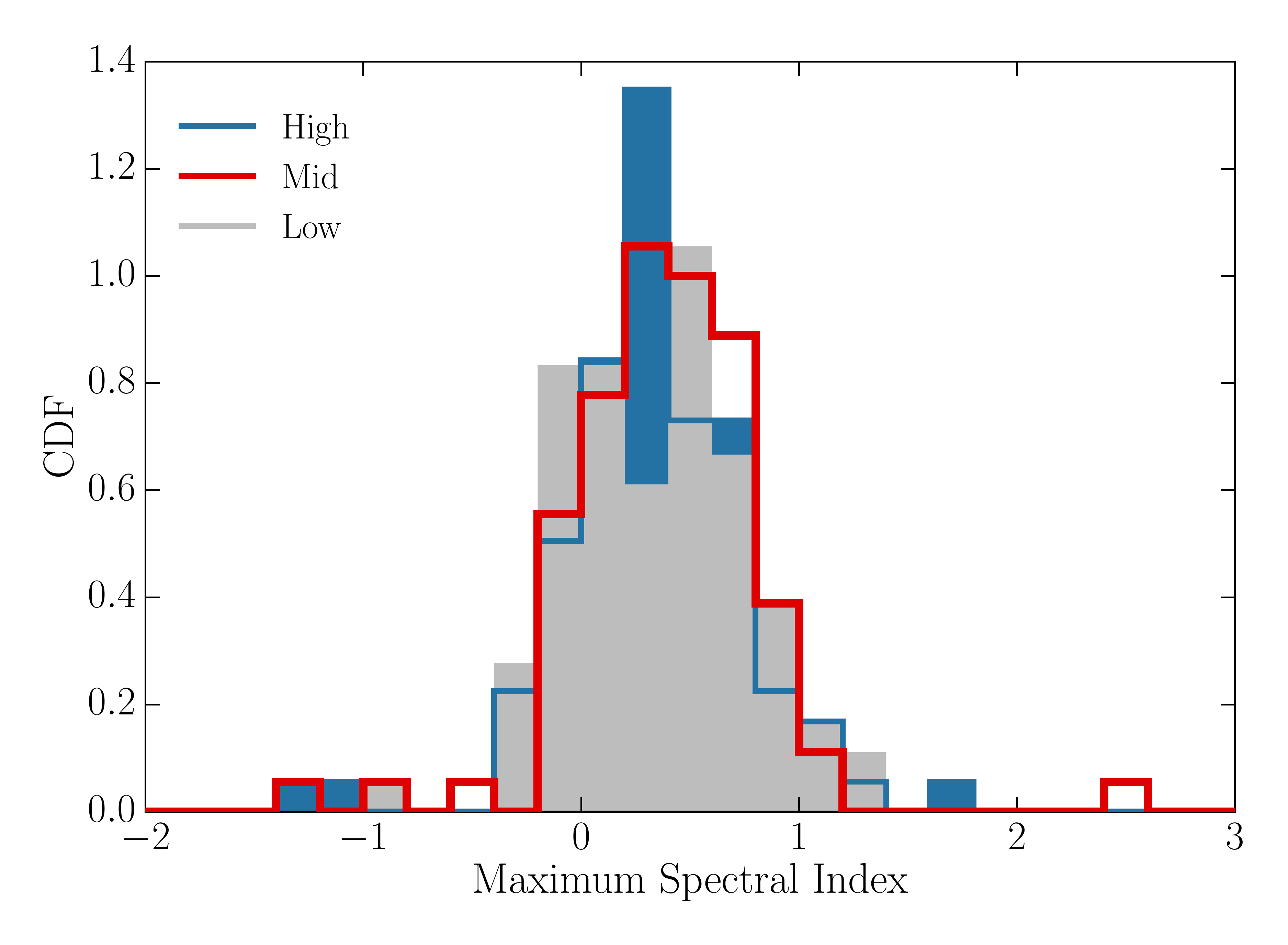}} 
  \caption{
  %\blue{ea180527-done} 
  Median (upper), minimum (middle), and maximum (lower) spectral index distribution in three sub-bands. We include sources from the groups ``f'', ``s1'', ``s2'' , and ``old''.}
  \label{fig:spind_hst}
\end{figure}

In Fig.~\ref{fig:spind_hst} we compare the distributions of the low, mid-, and high sub-band spectral indices for all the sources monitored by \fg (i.e. groups ``f'', ``s1'', ``s2'' , and ``old''). We compare separately for minimum, median, and maximum spectral indices. The median $\hat{a}$ and standard deviation $\sigma$ of the distributions are also tabulated in table~\ref{tab:spinds_mdn_min_max}.

With regard to the minimum spectral indices (upper panel in Fig.~\ref{fig:spind_hst}), the distributions of the three sub-bands are centred at $-0.26$ (low band), $-0.43$ (intermediate band), and $-0.65$ (high band), respectively. A two-sample KS test between any two of the distributions rules out the null hypothesis that they are drawn from the same parent distribution at a level above $4\sigma$. 
%(p-values for low - mid: $\sim3\times10^{-6}$, mid-high: $\sim3\times10^{-6}$ and low-high: \red{$\sim3\times10^{-14}$ (??? is this value ok???)}, respectively). 
In the case of the median spectral indices (middle panel in Fig.~\ref{fig:spind_hst}), the distinction is less significant but still present. The distributions medians are $-0.01$ (low band), $-0.07$ (intermediate band), and $-0.14$ (high band). A two-sample KS test shows that the low and the high bands are different at the level of $4\sigma$ (p-value $\sim10^{-5}$). The low sub-band median indices seem to be following a bimodal distribution. Normality tests however indicate otherwise. A D'Agostino's K$^2$ test gave a statistic of 4.2 and a p-value of 0.124 and the Shapiro-Wilk test returned a statistic of 0.97 and a p-value= 0.07 showing that the hypothesis that the distribution is normal cannot be rejected. Finally, for the maximum spectral indices (lower panel in Fig.~\ref{fig:spind_hst}) the separation disappears. The three distributions are instead centred at neighbouring medians of  $+0.34$ (low band), $+0.39$ (intermediate band), and $+0.28$ (high band). 
% -------------------------------------------------------------------------------------------------
\begin{table}
  \caption{
  %\blue{ea-180720:} 
  Median and standard deviation of the distributions of the median, minimum, and maximum spectral indices in three sub-bands.}
  \label{tab:spinds_mdn_min_max}  
  \centering                    
  \begin{tabular}{lcccccc} 
      \hline\hline             
      Spectral Index    &\mc{6}{c}{sub-band}\\
          \cline{2-7}
                               &\mc{2}{c}{Low} &\mc{2}{c}{Mid} &\mc{2}{c}{High} \\
                               &$\hat{a}$ &$\sigma$ &$\hat{a}$ &$\sigma$ &$\hat{a}$ &$\sigma$ \\
      \hline\\
   Minimum  &$-$0.26 &0.26  &$-$0.43 &0.24  &$-$0.65& 0.32 \\
   Median   &$-$0.01 &0.29  &$-$0.07 &0.25  &$-$0.14& 0.23 \\
   Maximum  &$+$0.34 &0.40  &$+$0.39 &0.44  &$+$0.28& 0.95 \\
\\\hline                                  
  \end{tabular}
\end{table}
% -------------------------------------------------------------------------------------------------
% -------------------------------------------------------------------------------------------------
% NOTE see the previous comment fro wheer are these data cominf rrom.
% -------------------------------------------------------------------------------------------------

This phenomenological discussion makes it already clear that the flat radio spectrum paradigm typically assumed for blazars is only the manifestation of an average behaviour of an otherwise intensely variable SED. The significant difference in the distributions of the low, mid-, and high bands when examining the minimum value that the spectral index, is the mere result of the spectral evolution across our observing bands. Convex spectral components are sequentially appearing in the high band and evolve towards progressively lower frequencies to dissipate as optically thin in the low end of our band. The very negative minimum values of the high-band index (blue distribution) shows that the spectral components evolve across that sub-band. The moderately negative indices in the low band (grey distribution) is caused by the fact that at those energies we observe the blend of several past components that are at different evolutionary stages making up a flat and moderately negative spectrum. Concerning the distributions of the maximum indices in each sub-band, they all populate rather positive values. This is indicative of the fact that the \fg programme has been successful in monitoring the evolution of events and this was its prime aim.     

Concerning the separation of our monitored sources in the two main classes of blazars, namely FSRQs and BL~Lacs, none of the distributions appeared significantly different from any other. FSRQs and BL~Lacs are thus indistinguishable from one another with respect to their  spectral indices. Finally, concerning the GeV energy bands we find that the source parameters are immune to the shape of the radio SED. We specifically searched for dependences of the energy flux,  GeV spectral shape, synchrotron component peak frequency, and variability index on the radio spectral index. We only find that there is a marginal indication of a relation between maximum radio spectral index in the low and middle bands and the GeV variability index. For the former case a Spearman's test gave a $\rho\sim0.59$ with a p-value of $10^{-4}$. In the latter case we found that $\rho\sim0.57$ with a p-value of $2\times10^{-4}$.
%low band  max index: SpearmanrResult(correlation=0.58518968773981761, pvalue=0.00011387852303946994)
%mid band max index : SpearmanrResult(correlation=0.56744186811084907, pvalue=0.00020304207449335423) 

To conclude, it is important to emphasise the fact that the dataset presented in this work shows that any sense of spectrum ``flatness'' is merely the result of intense spectral evolution. This is valid in the most general case {where} the evolution of several spectral {components is ultimately} integrated by the observer. It is therefore recommended to depict blazars as systems hosting intense variability in all parameter spaces (time, frequency, and intensity), rather than as simply flat radio spectra.

\section{Summary and conclusions}
{We have presented a dataset that is the result of one among the most comprehensive programmes monitoring \textit{Fermi} sources in terms of the
combination of number of sources observed, frequency coverage, regularity in the sampling, and its cadence. We summarize our findings as follows:
%
%In all the previous discussion we have presented a dataset that has been the result of one among the most comprehensive programmes monitoring \textit{Fermi} sources; in terms of a combination of number of sources observed, frequency coverage, regularity in the sampling, and its cadence. In summary,
}
   \begin{enumerate}
           \item The primary scope of the \fg programme was mostly to provide the necessary multi-frequency radio monitoring complementary to the \frm/LAT \citep{2009ApJ...697.1071A} monitoring of the gamma-ray sky and to study all the relevant radio physics. The uniqueness of the programme lies on the combination of its broadband, multi-frequency coverage, the high cadence observations, its long duration, and the availability of multi-frequency linear and circular polarisation light curves \citep{2018A&A...609A..68M}. 
        \item The \fg monitored sample is undoubtedly biased. Given its primary scope it is made of the brightest and most variable blazars. Still, at least in terms of source cosmological distribution and admixture of the two flavours of blazar AGNs (FSRQs and BLLACs), it is indeed representative of the blazar population.
           \item The current release concerns a total of 155 sources, including sources from the initial and the revised monitored sample along with targets of opportunity, primary calibrators, and sources observed within the F-GAMMA-\textit{Planck} satellite MoU. 
           \item We release the first part of the \fg \citep{2016A&A...596A..45F} dataset. This includes the light curves obtained with the 100 m Effelsberg radio telescope between 2007 and 2015 at 2.64--43~GHz. 
           \item Every data point included in this work has been subjected to a suite of post-measurement corrections: pointing, atmospheric opacity, and elevation-dependent gain. The opacity correction is that with the largest impact (up to 26\% at the highest frequency) while the other corrections are of no more than a few percent. The entirety of the dataset has been final quality checked. For possible outliers that may still be present, we found no obvious reason why they should be vetoed out of the release. Hence we kept them. The dataset that passed final quality filtering includes a total of $4\times10^4$ measurements.
           \item The reduction, final calibration, and quality checks were  carried out on the basis of the principle that each measurement is part of an evolving, broad, radio SED. 
           \item We validated the assumption, implicit in our analysis and necessary in any parametric statistical method, for the normality of the noise. The hypothesis was investigated using Q-Q plots and quantified with the D'Agostino's K$^2$ test was passed for more than 85\% of the datasets that were tested.
        \item Our variability and time series analysis, which will be presented in a forthcoming publication, shows that the median rate for the most powerful events (i.e. with amplitude $A>0.5 A_\mathrm{max}$, with $A_\mathrm{max}$ the amplitude of the most powerful flare) is about 0.6 -- 0.7 events per year.     
        \item Concerning the duty cycle, i.e the fraction of observing time that the source is spending in an active state, we find the median of 0.33 at 4.85~GHz drops to 0.22 at 32~GHz.
\item Our dataset -- in any of its possible representations (radio SEDs or light curves) -- makes it immediately apparent that the assumption of a flat radio spectrum (especially for blazars) is an oversimplification of an otherwise extremely dynamic system. It is substantially more advisable to refer to systems undergoing extreme spectral evolution the integral of which is what an observer simplistically depicts as a flat spectrum.           
   \end{enumerate}
{The released quality checked light curves provide a unique coverage of the radio part of the SED and are brought to the community for further analyses. The \fg studies have already been and will be presented elsewhere. A similar data release of the millimeter and sub-mm datasets will follow. }

%==================================================================================================
\begin{acknowledgements}
  Based on observations with the 100\,m telescope of the MPIfR (Max-Planck-Institut f\"ur Radioastronomie). I.M., I.N. and V.K. were funded by the International Max Planck Research School (IMPRS) for Astronomy and Astrophysics at the Universities of Bonn and Cologne. The authors wish to thank the anonymous journal referee and the internal MPIfR referee Dr J. Liu for a careful reading and constructive comments. The \fg programme based its ephemeris calculations on numerical routines from the scientific-grade XEphem astronomical software \cite{2011ascl.soft12013D}. This research has made use of the NASA/IPAC Extragalactic Database (NED), which is operated by the Jet Propulsion Laboratory, California Institute of Technology, under contract with the National Aeronautics and Space Administration. 
\end{acknowledgements}

\begin{appendix} %First online appendix
%==================================================================================================

%==================================================================================================
\section{{Source sample}}
\label{app:sampl}

{Table~\ref{tab:sample} lists all the sources included in the current data release.}

%==================================================================================================
\section{Multi-frequency light curves}
\label{app:lcs}

\begin{figure*}[p]
\centering
\begin{tabular}{cc}
\includegraphics[trim=60pt 30pt 100pt 50pt  ,clip, width=0.49\textwidth,angle=0]{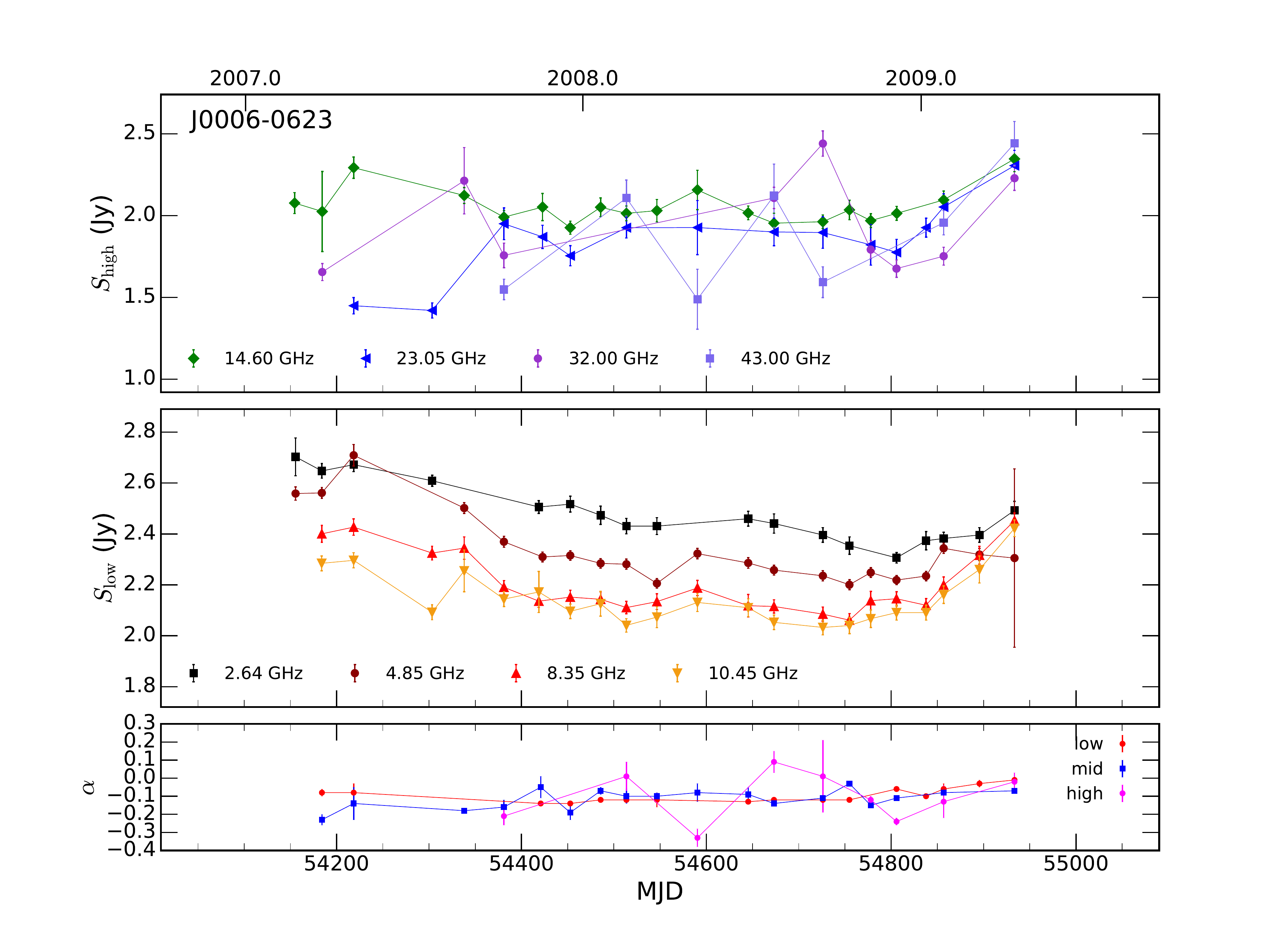}&
\includegraphics[trim=60pt 30pt 100pt 50pt  ,clip, width=0.49\textwidth,angle=0]{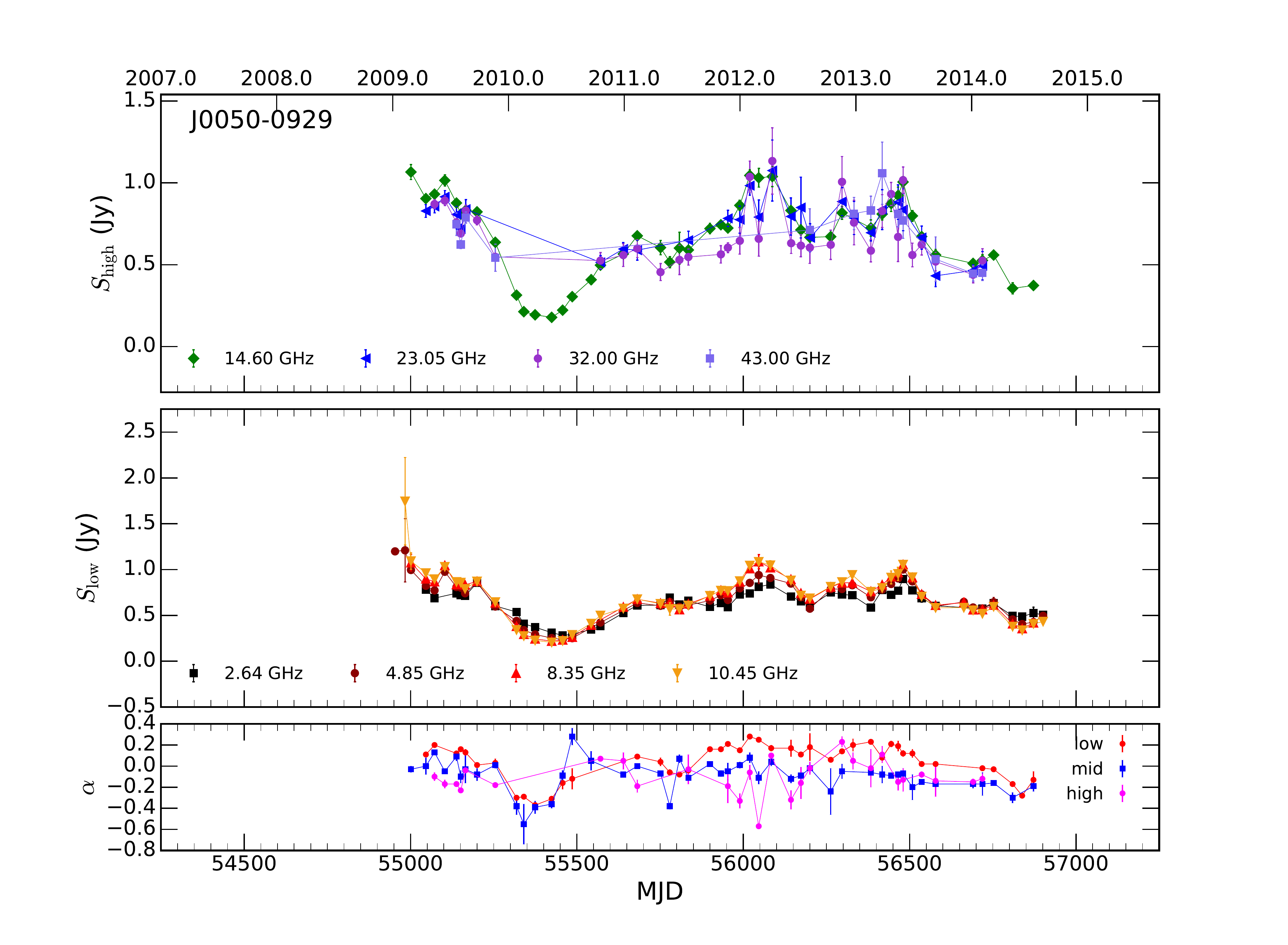}\\
\\[10pt]
\includegraphics[trim=60pt 30pt 100pt 50pt  ,clip, width=0.49\textwidth,angle=0]{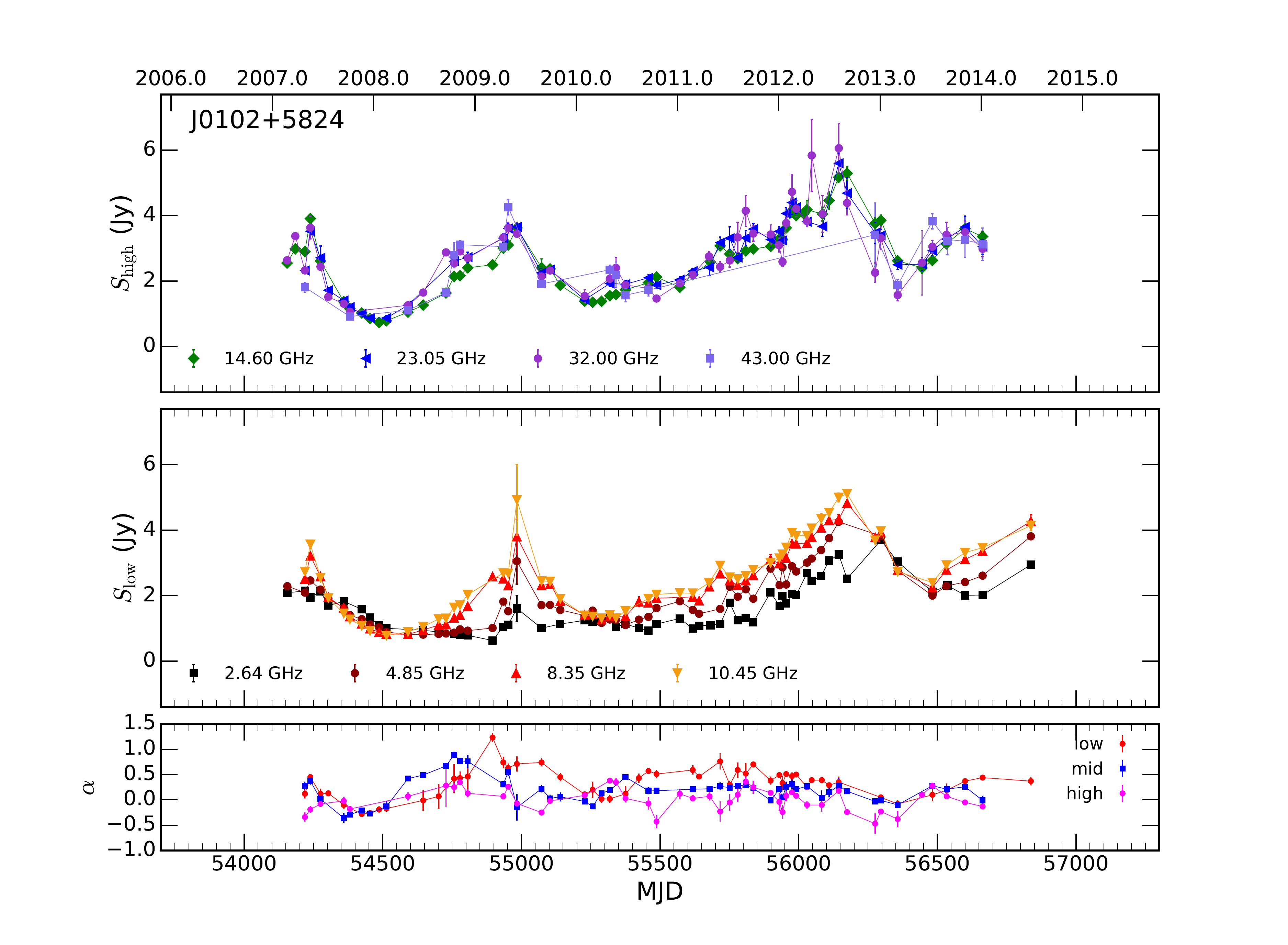}&
\includegraphics[trim=60pt 30pt 100pt 50pt  ,clip, width=0.49\textwidth,angle=0]{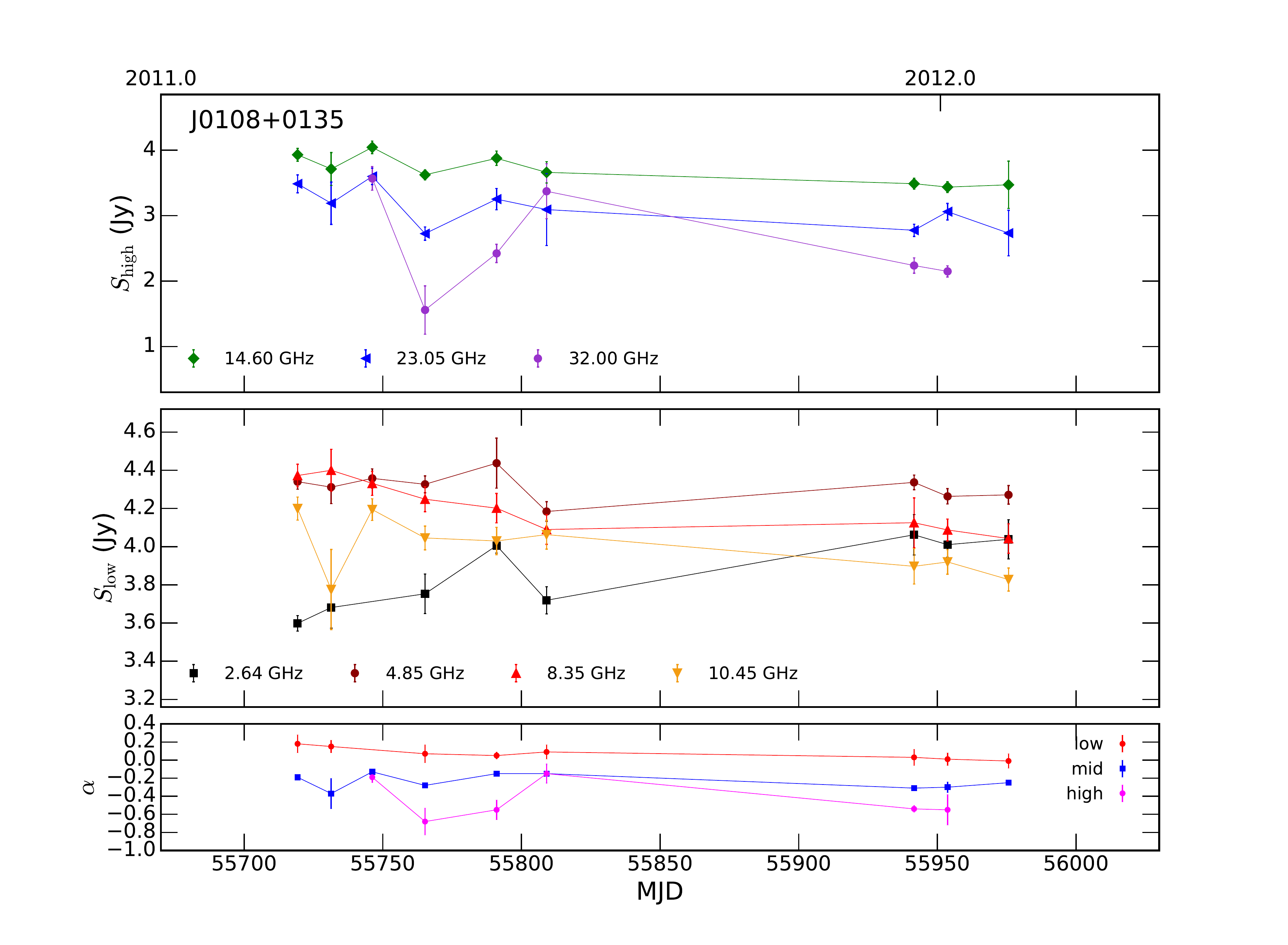}\\
\\[10pt]
\includegraphics[trim=60pt 30pt 100pt 50pt  ,clip, width=0.49\textwidth,angle=0]{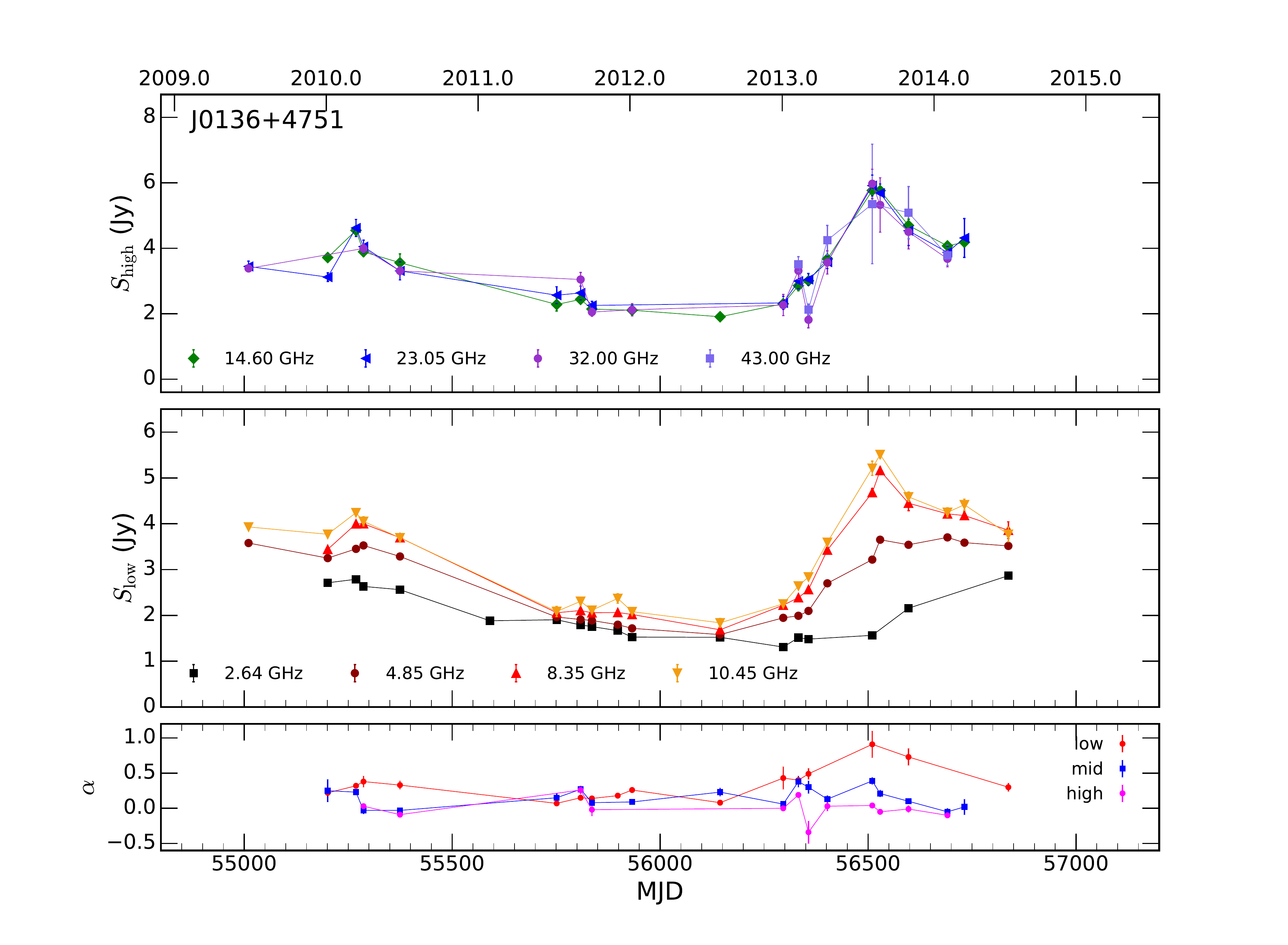}&
\includegraphics[trim=60pt 30pt 100pt 50pt  ,clip, width=0.49\textwidth,angle=0]{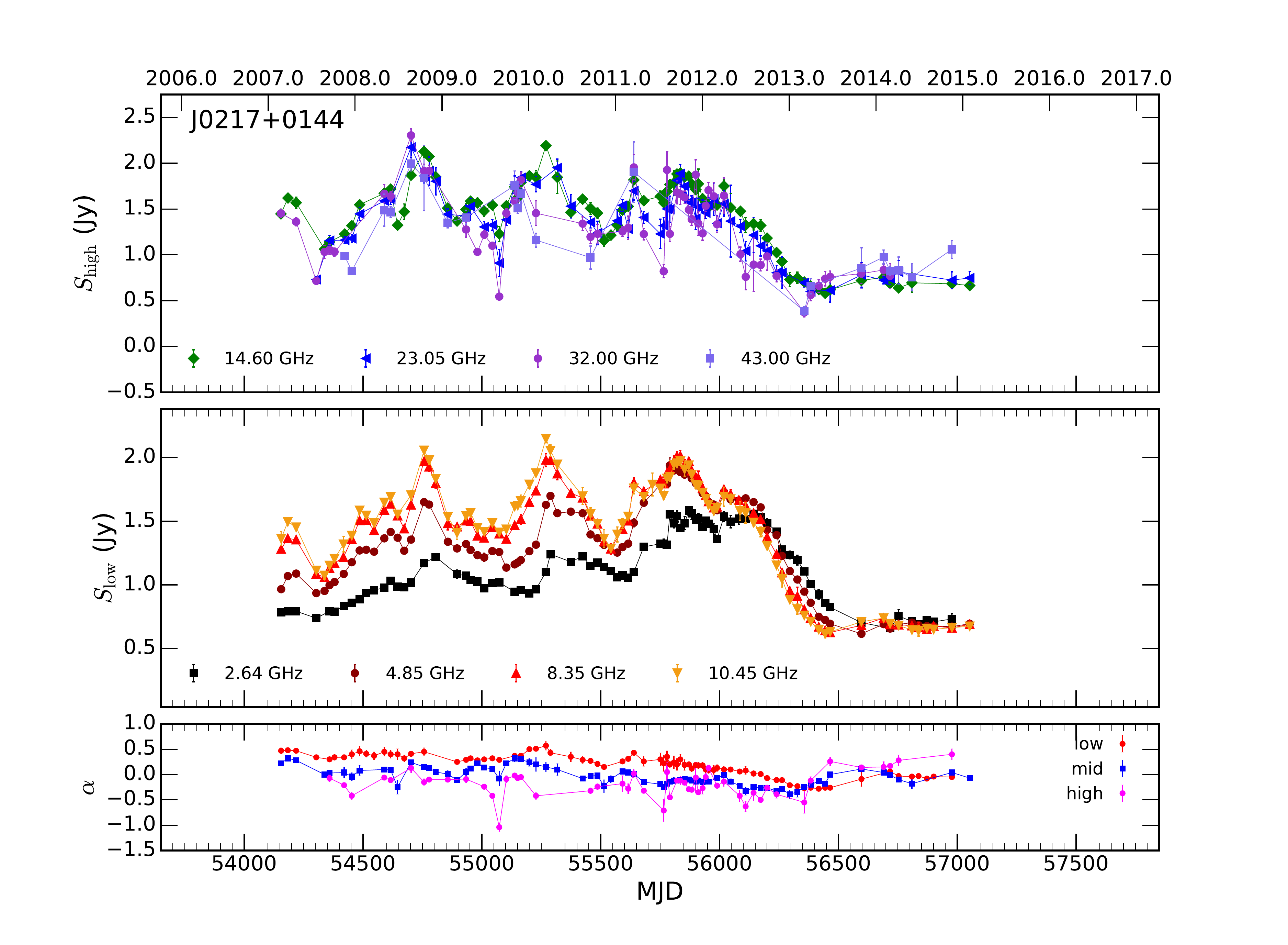}\\
\\[10pt]
\end{tabular}
\caption{Multi-frequency light curves for all the sources monitored by the \fg programme (``f'', ``s1'', ``s2'', and ``old'') and the F-GAMMA-\textit{Planck} MoU. The lower panel in each frame shows the evolution of the low (2.64, 4.85 and 8.35~GHz), mid-band (8.35, 10.45 and 14.6~GHz), and high-band (14.6, 23.05, 32, 43~GHz) spectral index. Only spectral index estimates from at least three frequencies are shown. Connecting lines have been used to guide the eye. }
\label{fig:sample_pg1}
\end{figure*}

\begin{figure*}[p]
\centering
\begin{tabular}{cc}
\includegraphics[trim=60pt 30pt 100pt 50pt  ,clip, width=0.49\textwidth,angle=0]{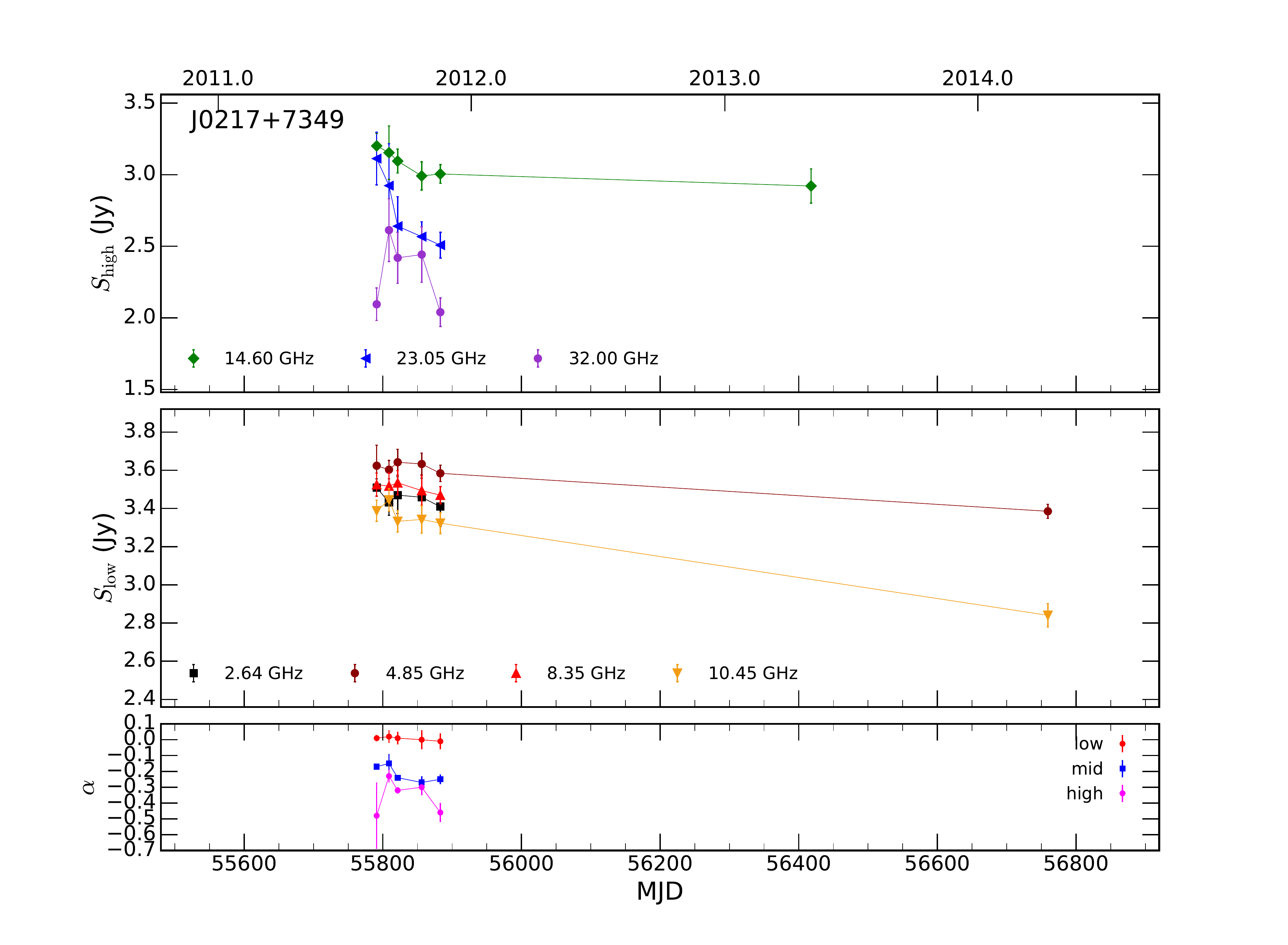}&
\includegraphics[trim=60pt 30pt 100pt 50pt  ,clip, width=0.49\textwidth,angle=0]{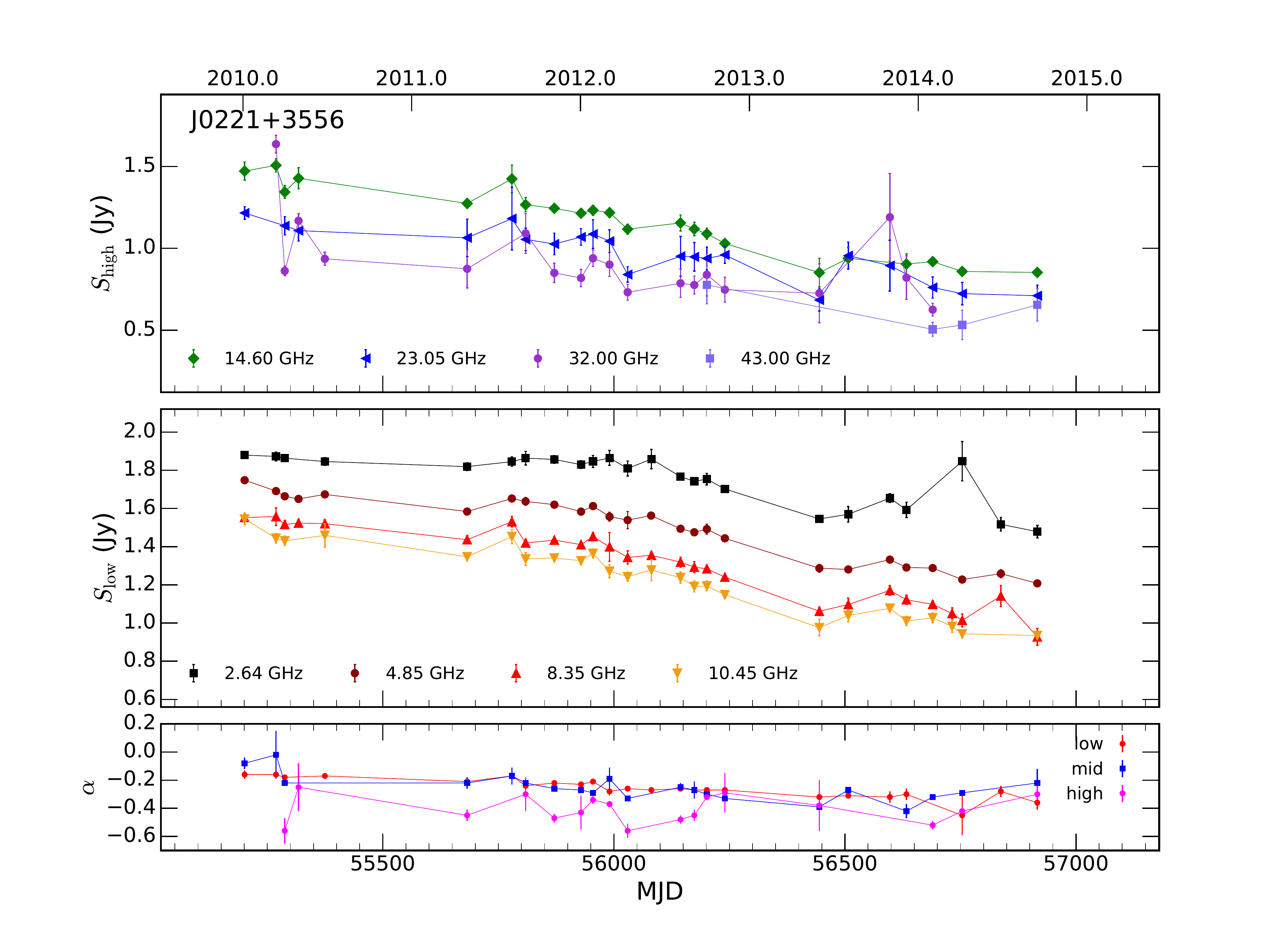}\\
\\[10pt]
\includegraphics[trim=60pt 30pt 100pt 50pt  ,clip, width=0.49\textwidth,angle=0]{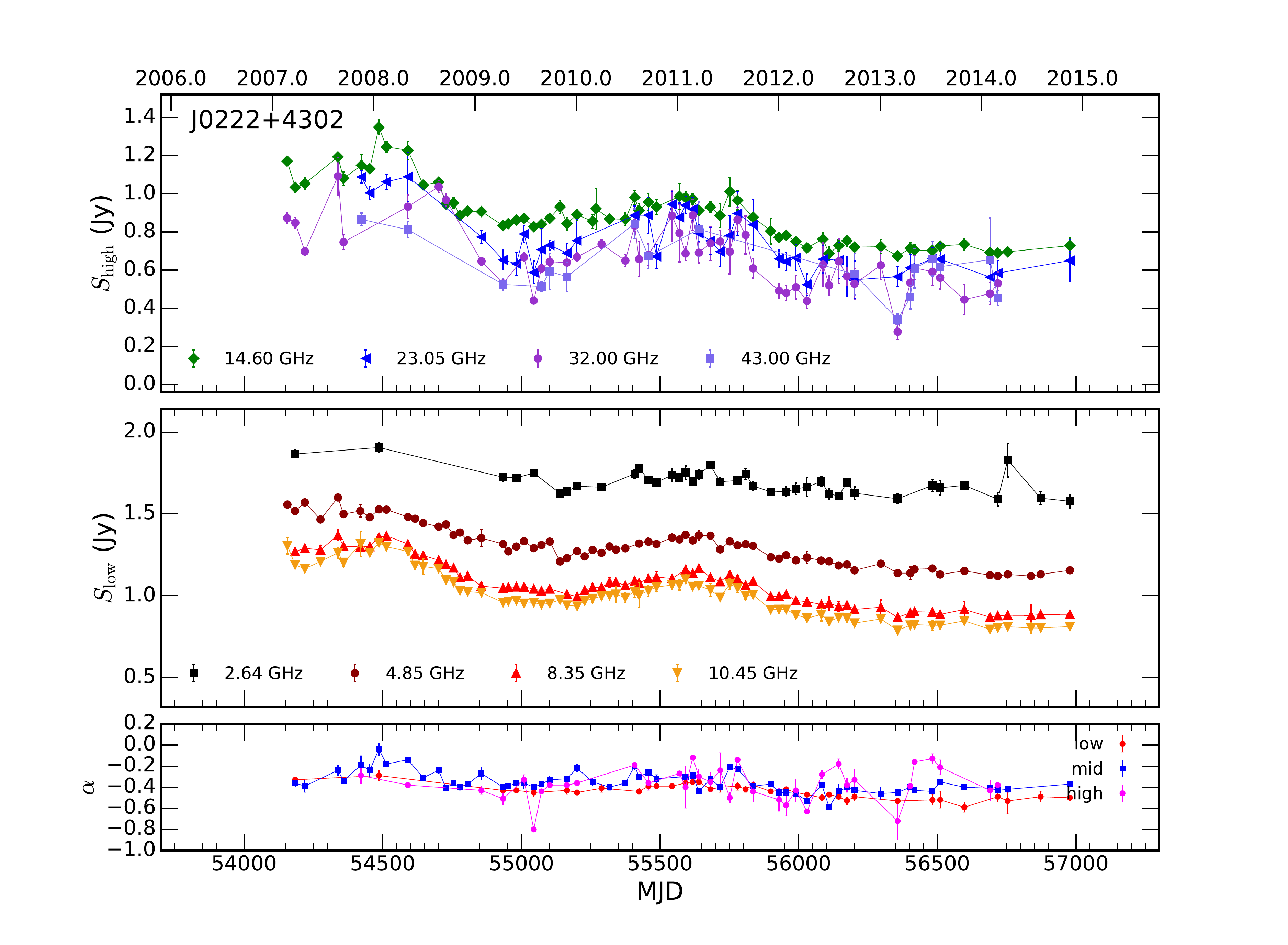}&
\includegraphics[trim=60pt 30pt 100pt 50pt  ,clip, width=0.49\textwidth,angle=0]{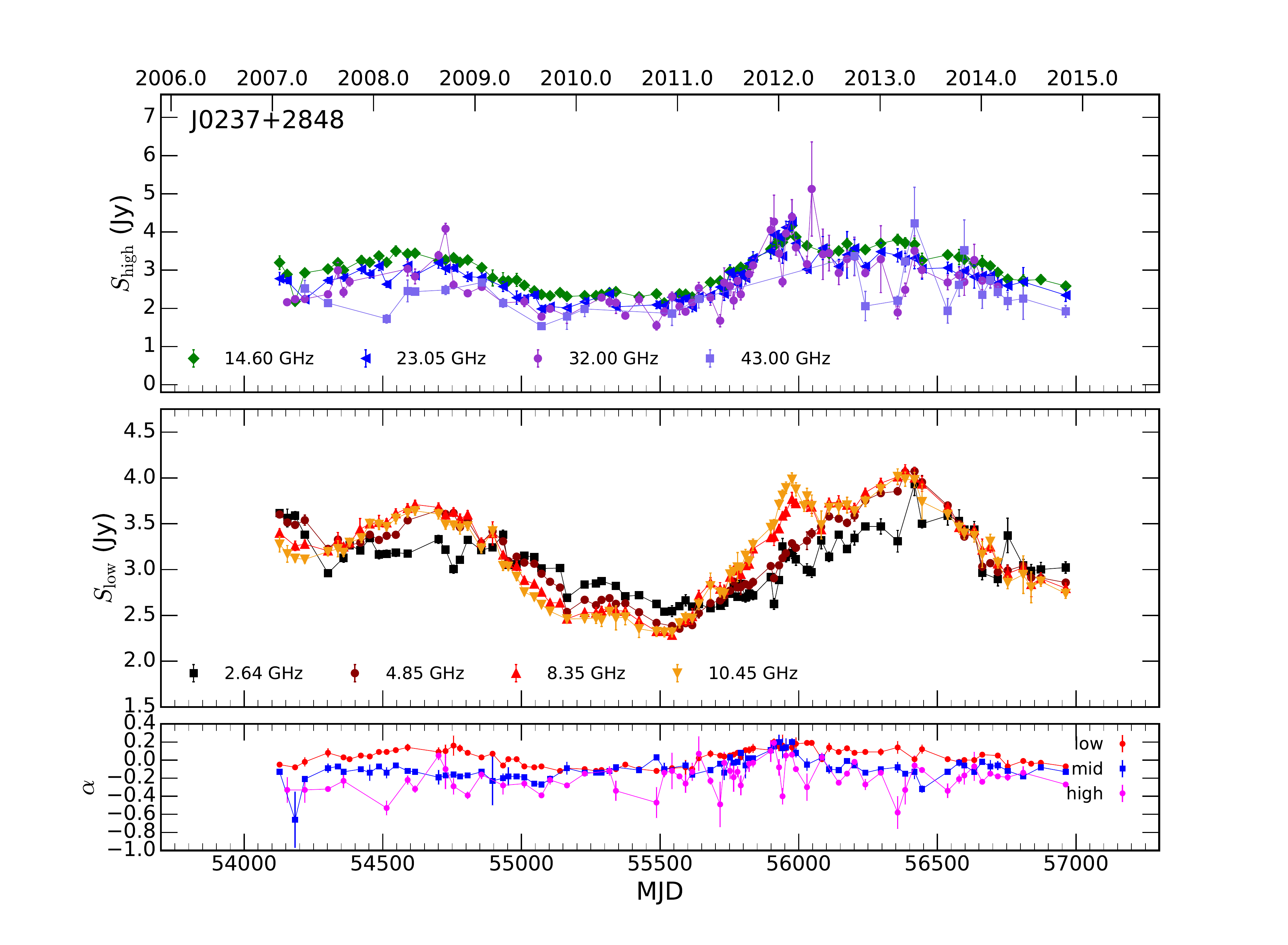}\\
\\[10pt]
\includegraphics[trim=60pt 30pt 100pt 50pt  ,clip, width=0.49\textwidth,angle=0]{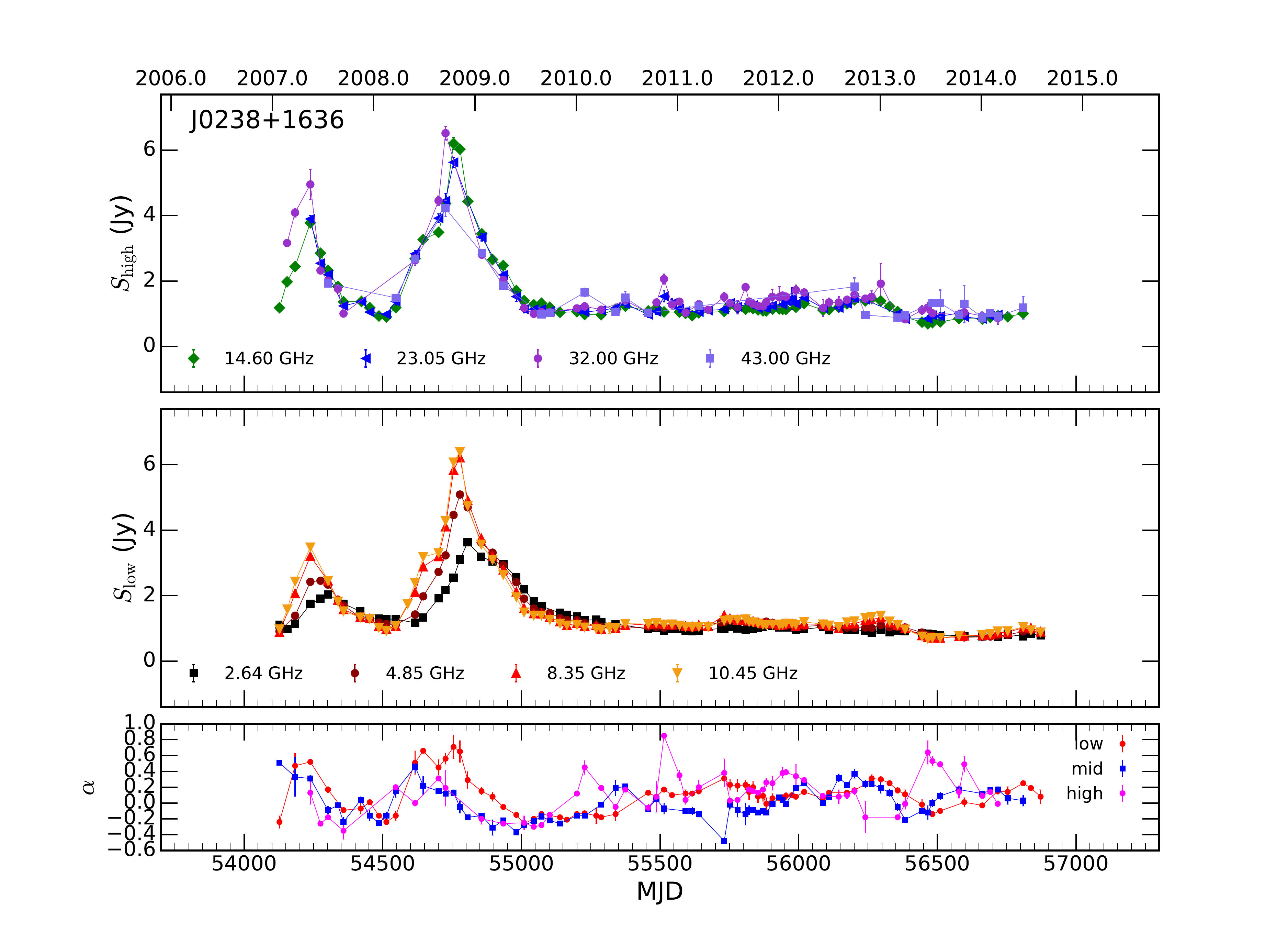}&
\includegraphics[trim=60pt 30pt 100pt 50pt  ,clip, width=0.49\textwidth,angle=0]{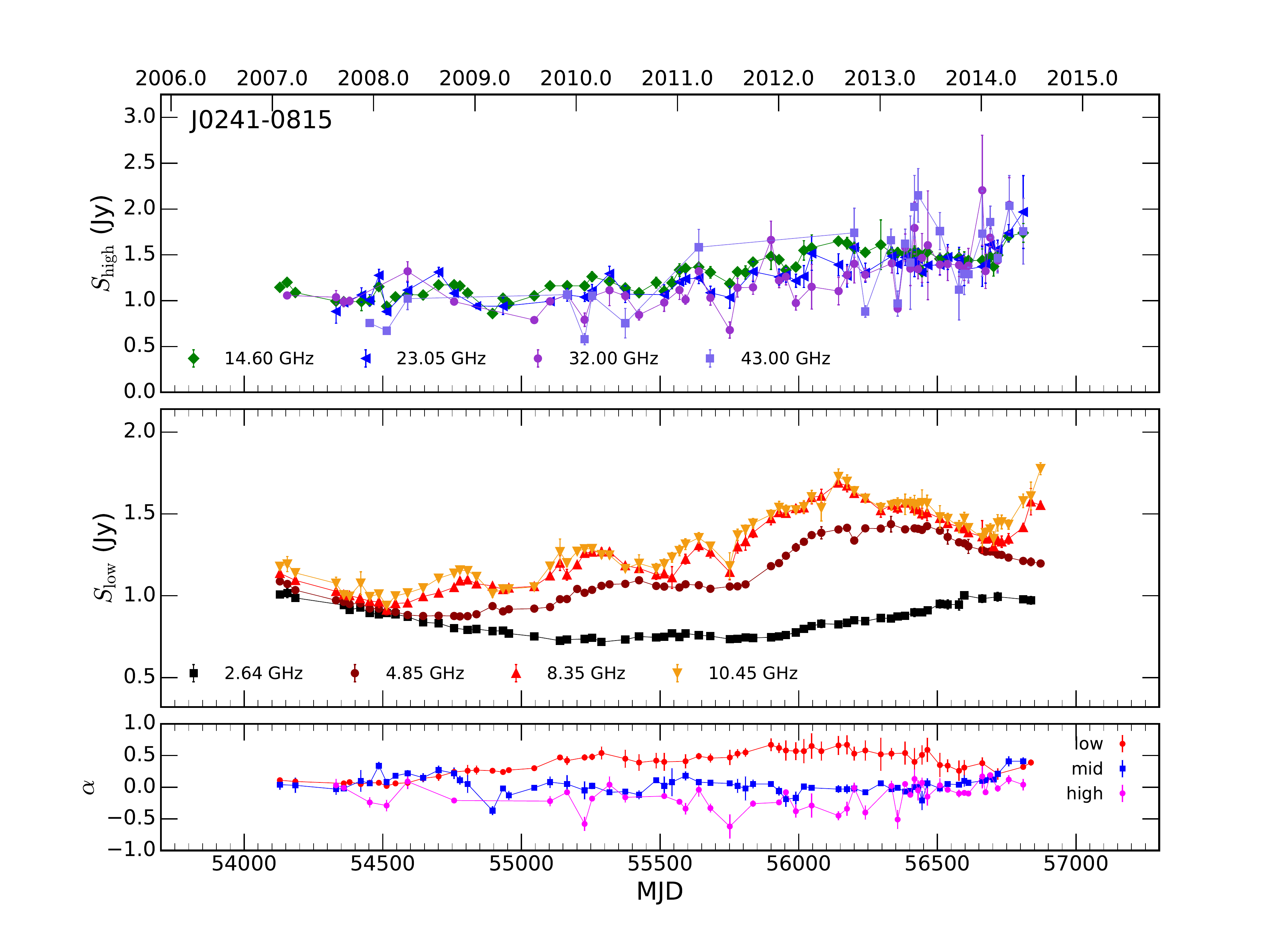}\\
\\[10pt]
\end{tabular}
\caption{Multi-frequency light curves for all the sources monitored by the \fg programme (``f'', ``s1'', ``s2'', ``old'') and the F-GAMMA-\textit{Planck} MoU. The lower panel in each frame shows the evolution of the low (2.64, 4.85 and 8.35~GHz) and mid-band (8.35, 10.45 and 14.6~GHz) and high-band (14.6, 23.05, 32, 43~GHz) spectral index. Only spectral index estimates from at least three frequencies are shown. Connecting lines have been used to guide the eye. }
\label{fig:sample_pg2}
\end{figure*}

\begin{figure*}[p]
\centering
\begin{tabular}{cc}
\includegraphics[trim=60pt 30pt 100pt 50pt  ,clip, width=0.49\textwidth,angle=0]{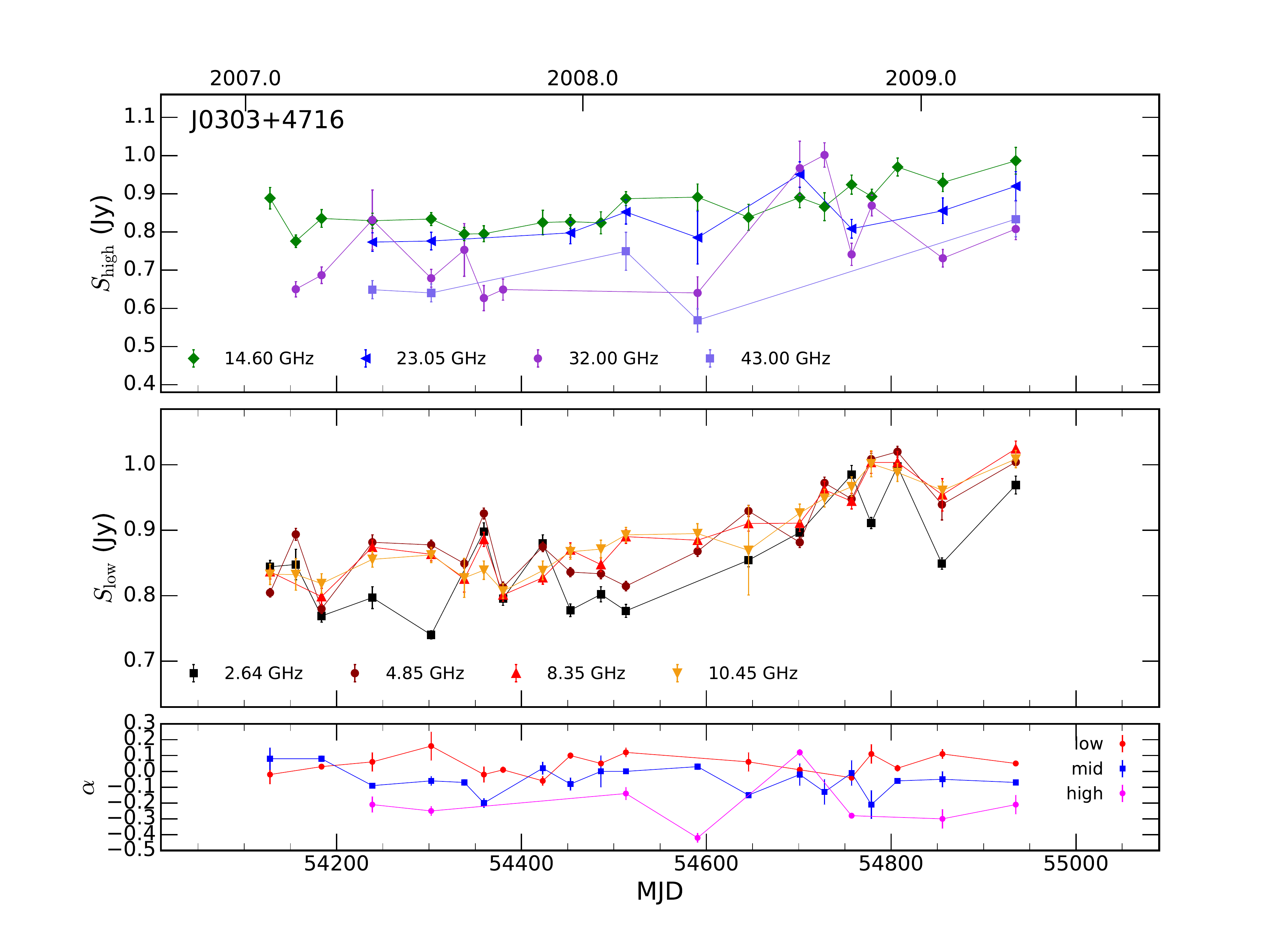}&
\includegraphics[trim=60pt 30pt 100pt 50pt  ,clip, width=0.49\textwidth,angle=0]{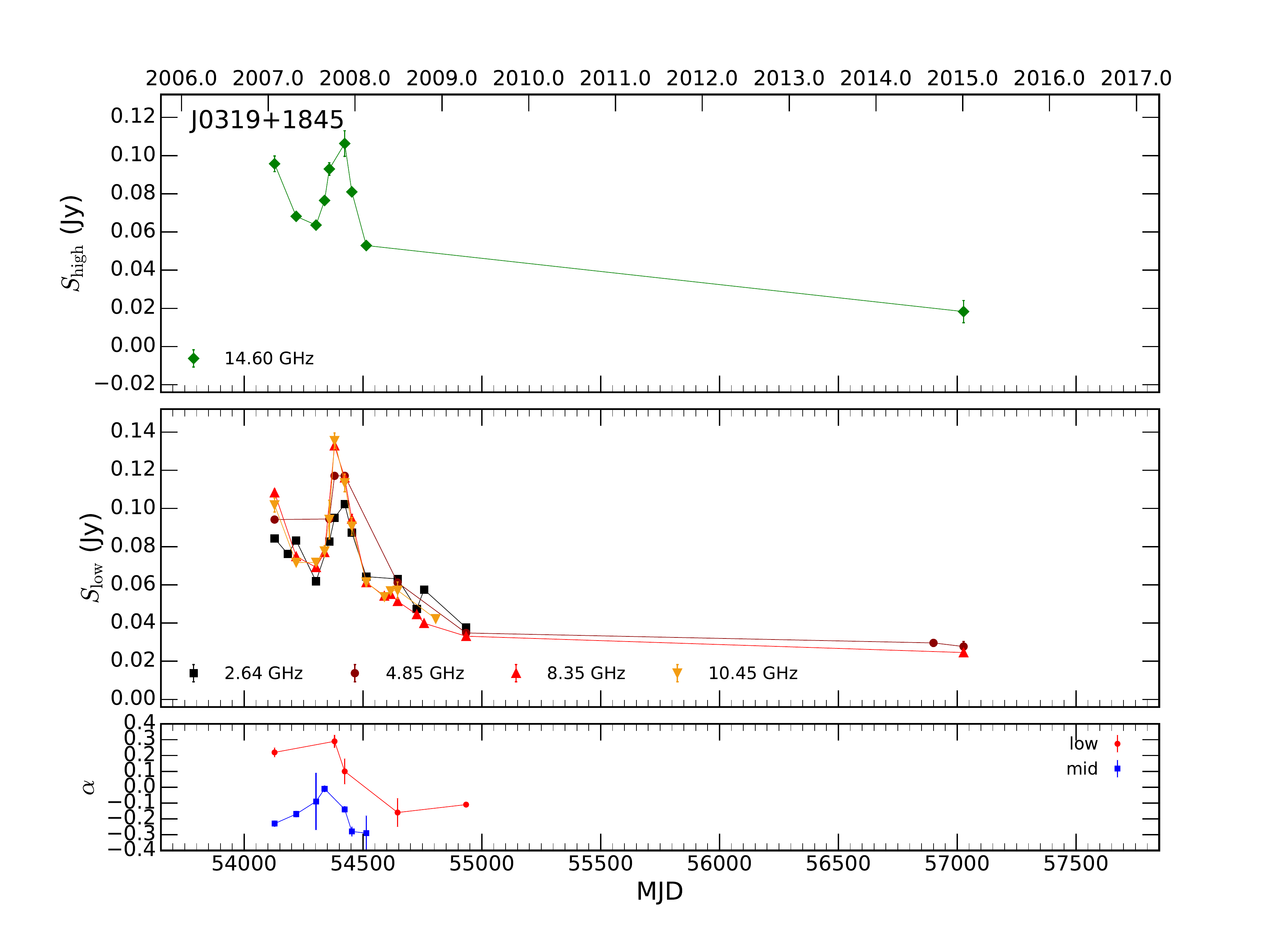}\\
\\[10pt]
\includegraphics[trim=60pt 30pt 100pt 50pt  ,clip, width=0.49\textwidth,angle=0]{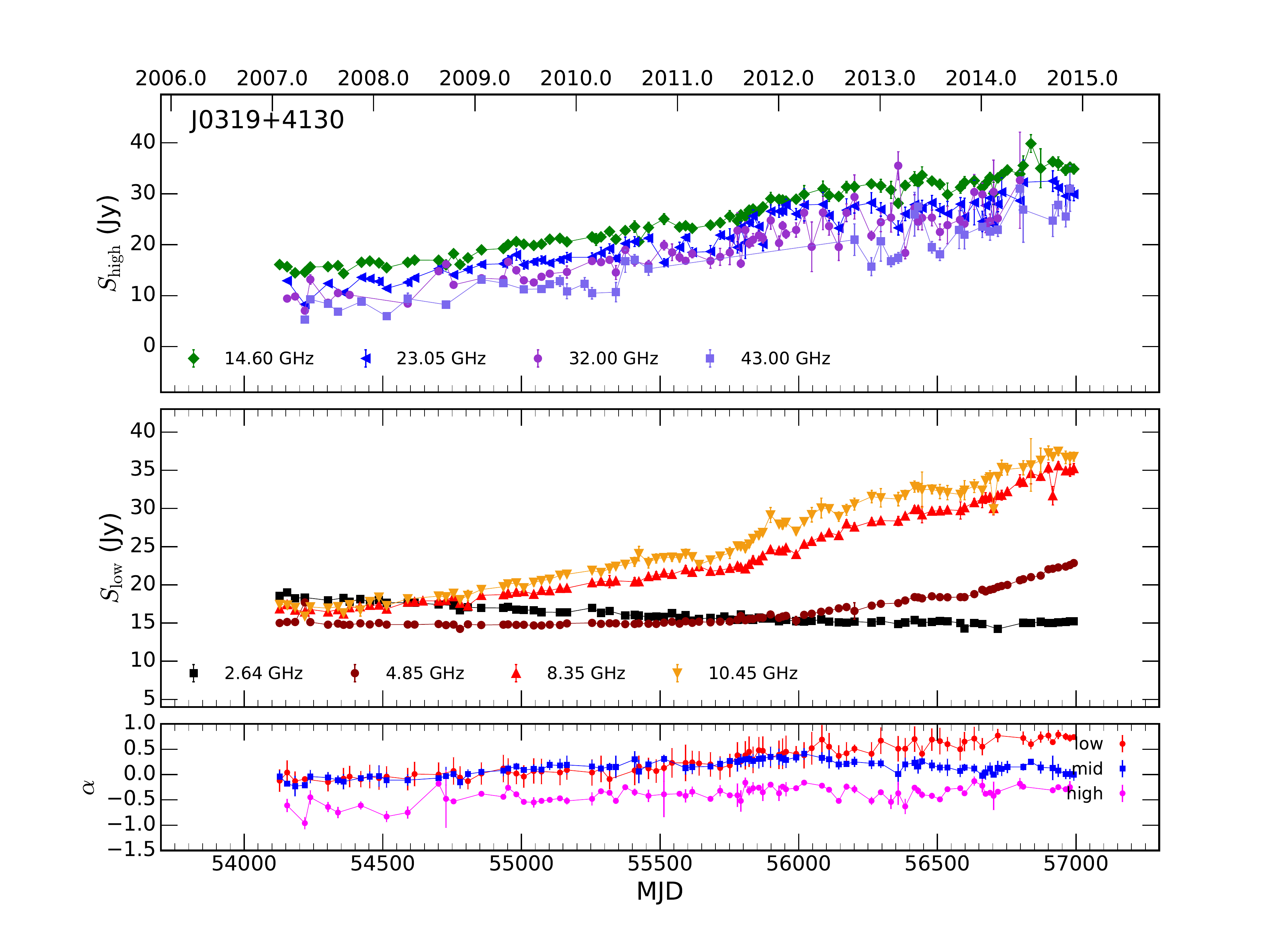}&
\includegraphics[trim=60pt 30pt 100pt 50pt  ,clip, width=0.49\textwidth,angle=0]{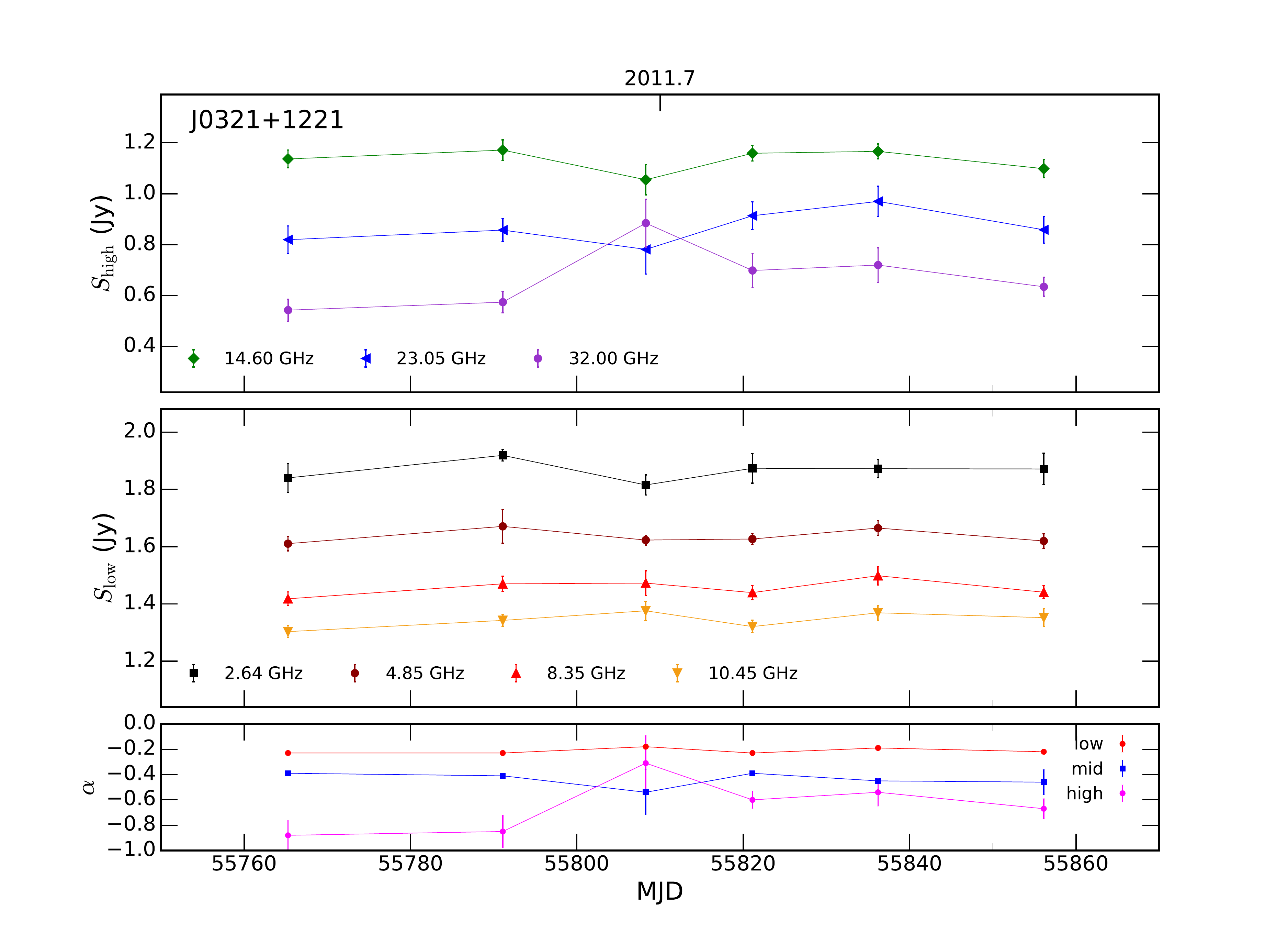}\\
\\[10pt]
\includegraphics[trim=60pt 30pt 100pt 50pt  ,clip, width=0.49\textwidth,angle=0]{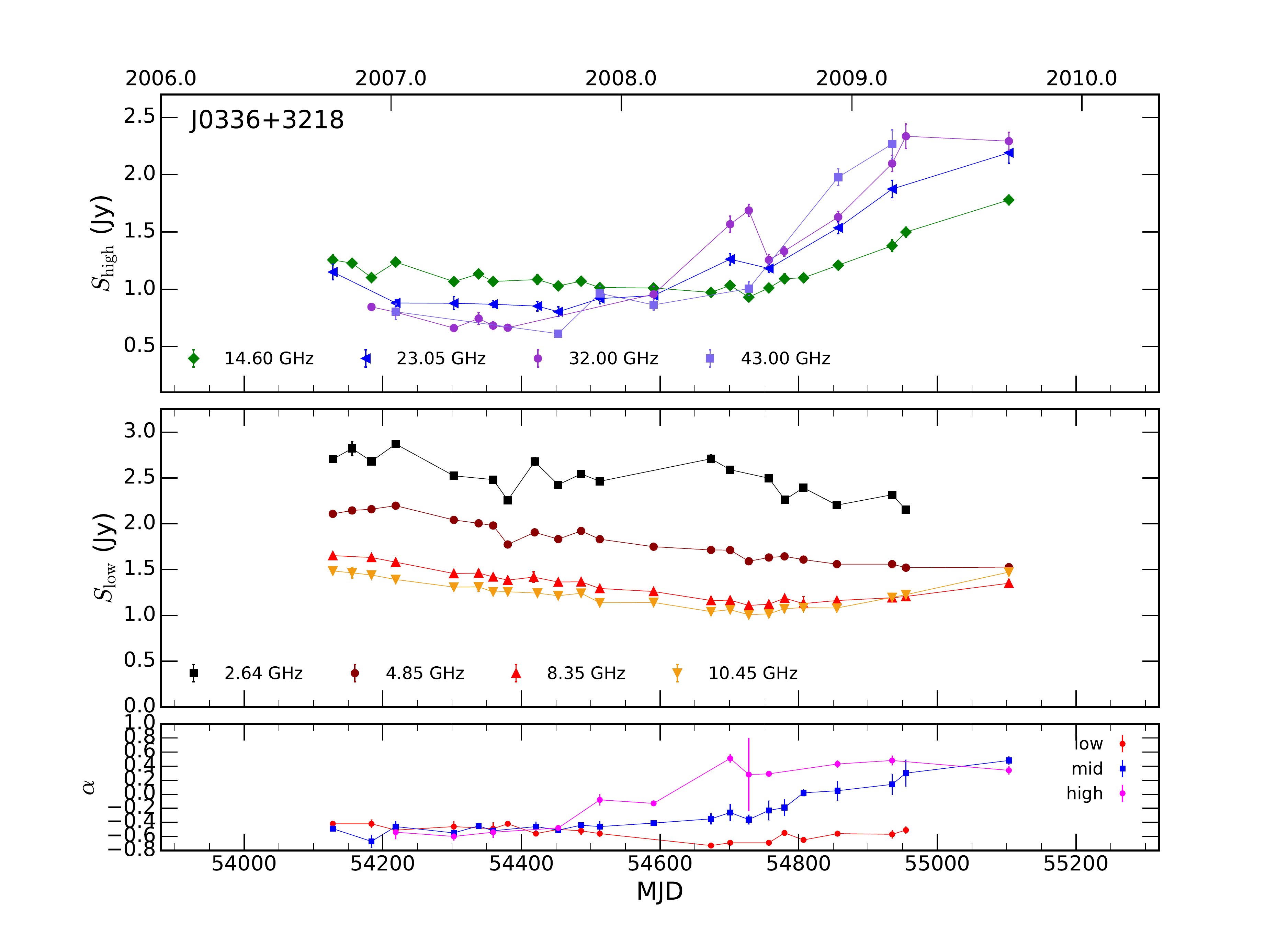}&
\includegraphics[trim=60pt 30pt 100pt 50pt  ,clip, width=0.49\textwidth,angle=0]{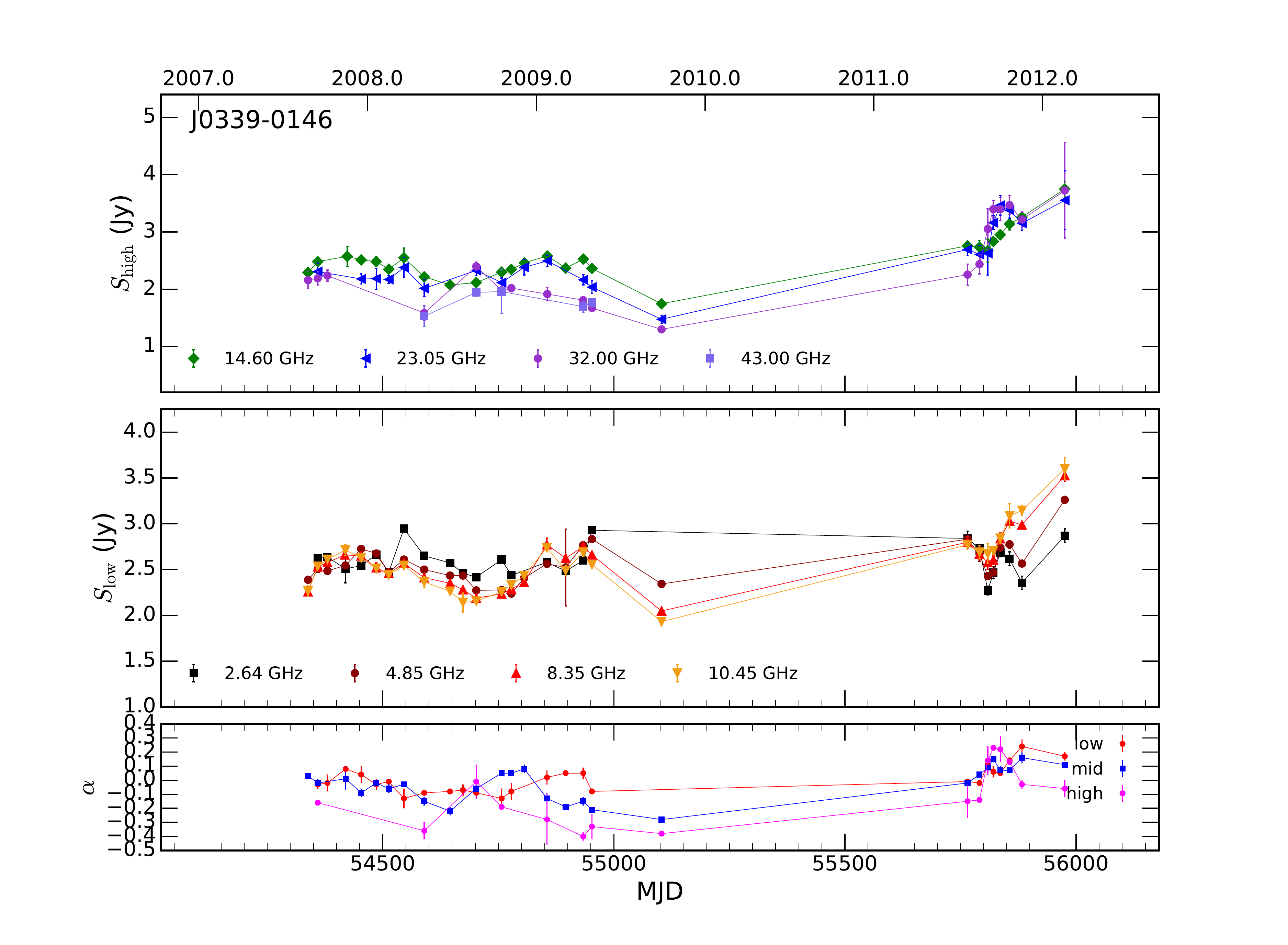}\\
\\[10pt]
\end{tabular}
\caption{Multi-frequency light curves for all the sources monitored by the \fg programme (``f'', ``s1'', ``s2'', ``old'') and the F-GAMMA-\textit{Planck} MoU. The lower panel in each frame shows the evolution of the low (2.64, 4.85 and 8.35~GHz) and mid-band (8.35, 10.45 and 14.6~GHz) and high-band (14.6, 23.05, 32, 43~GHz) spectral index. Only spectral index estimates from at least three frequencies are shown. Connecting lines have been used to guide the eye. }
\label{fig:sample_pg3}
\end{figure*}

\begin{figure*}[p]
\centering
\begin{tabular}{cc}
\includegraphics[trim=60pt 30pt 100pt 50pt  ,clip, width=0.49\textwidth,angle=0]{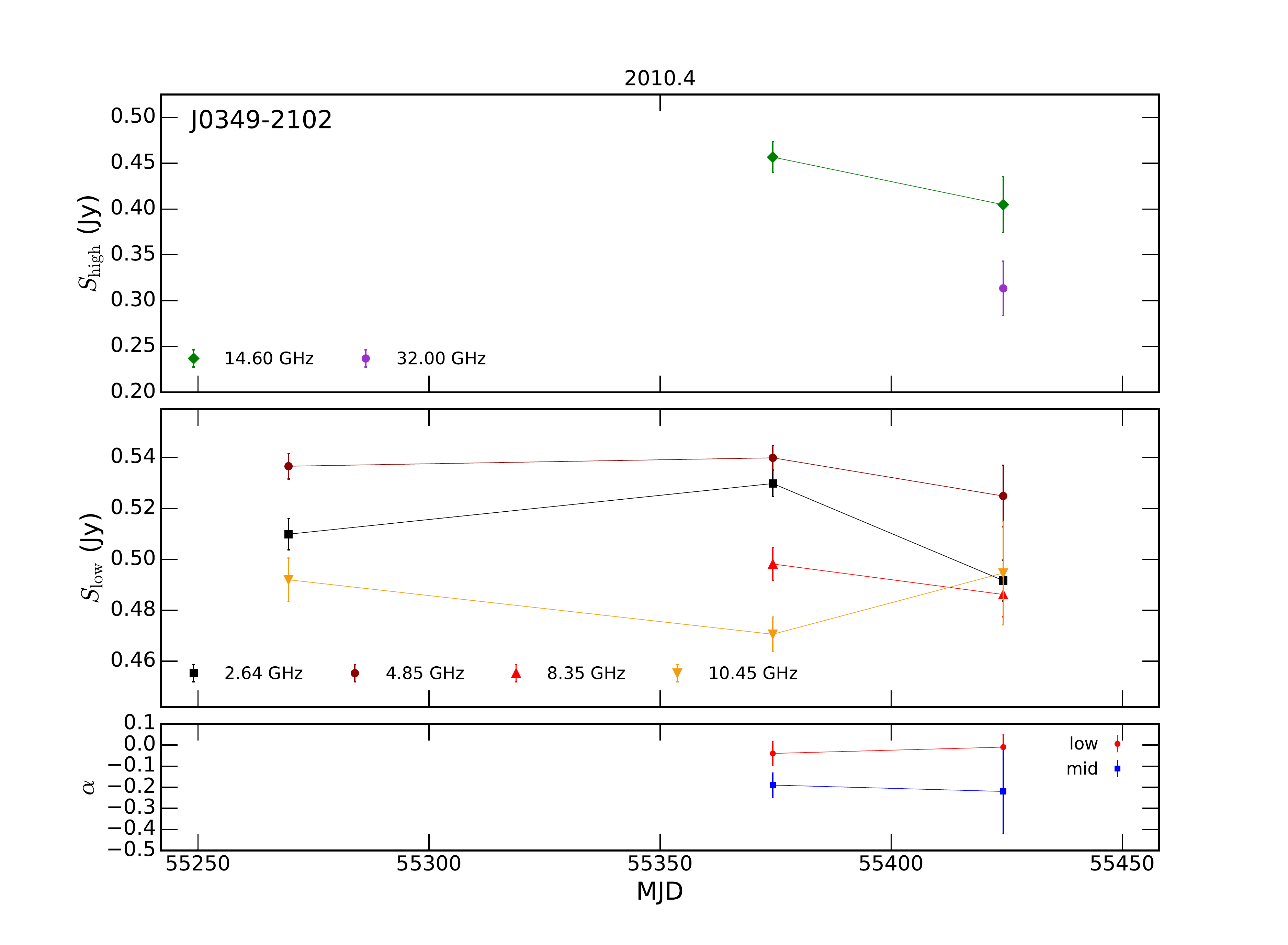}&
\includegraphics[trim=60pt 30pt 100pt 50pt  ,clip, width=0.49\textwidth,angle=0]{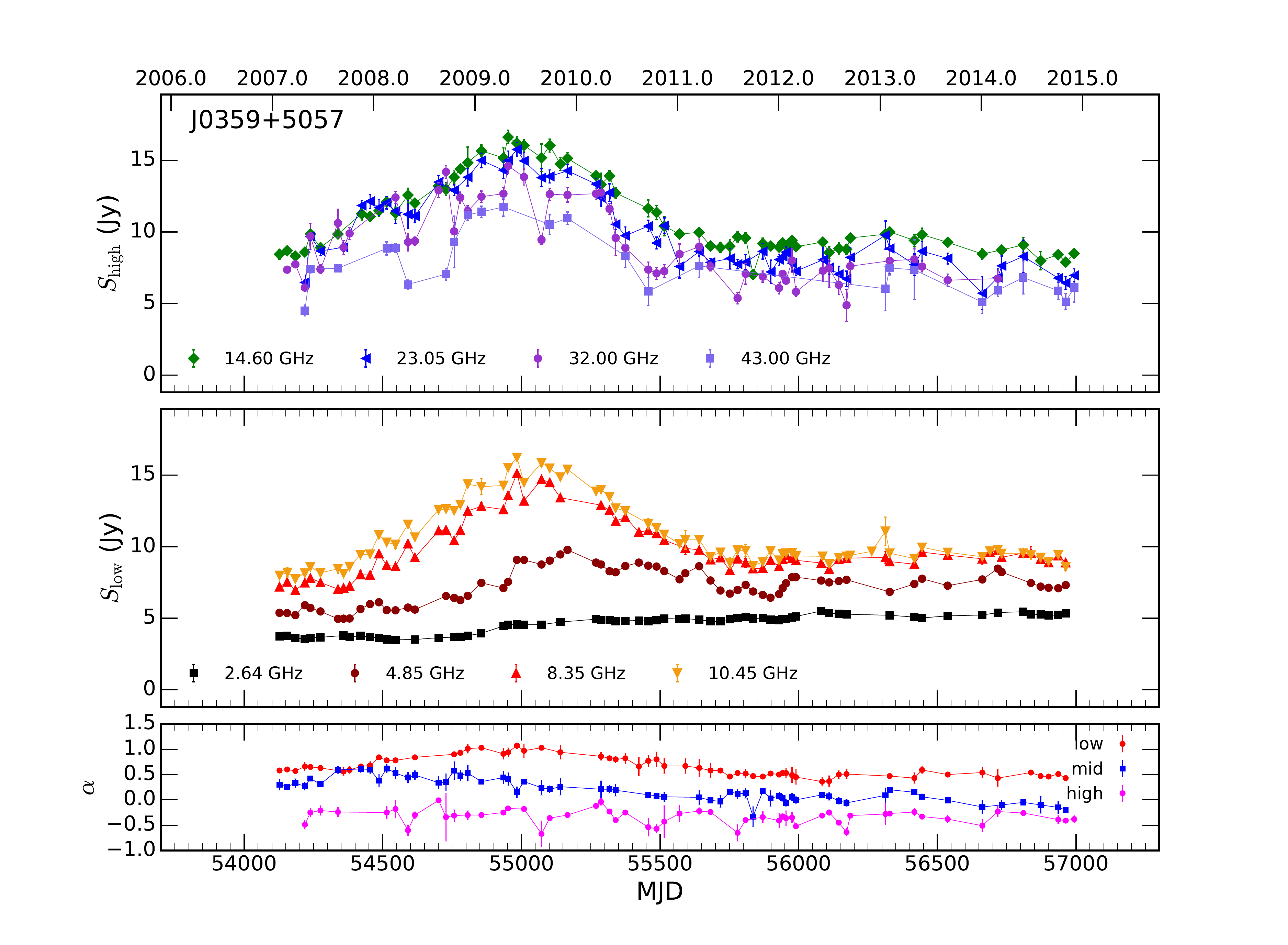}\\
\\[10pt]
\includegraphics[trim=60pt 30pt 100pt 50pt  ,clip, width=0.49\textwidth,angle=0]{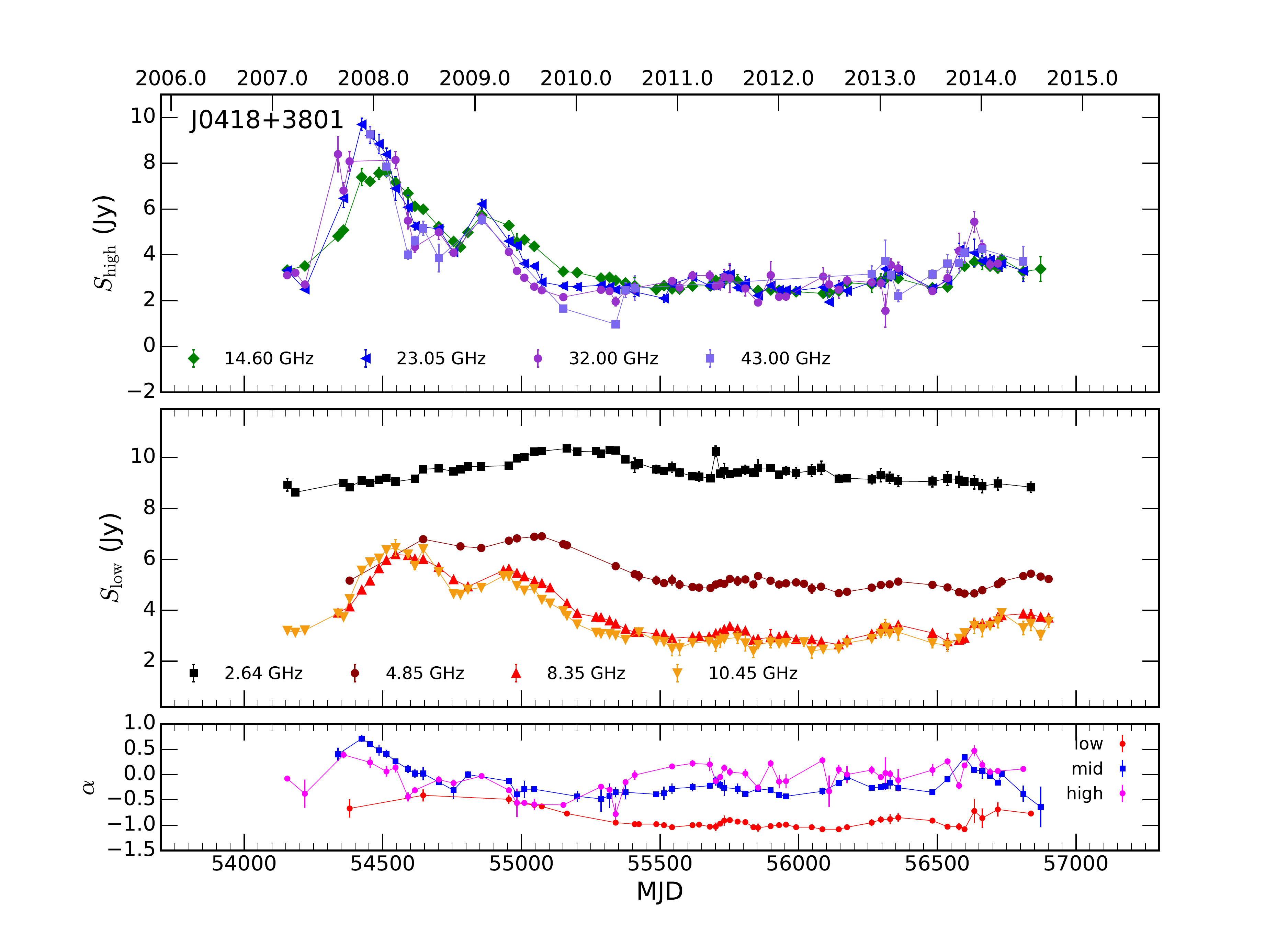}&
\includegraphics[trim=60pt 30pt 100pt 50pt  ,clip, width=0.49\textwidth,angle=0]{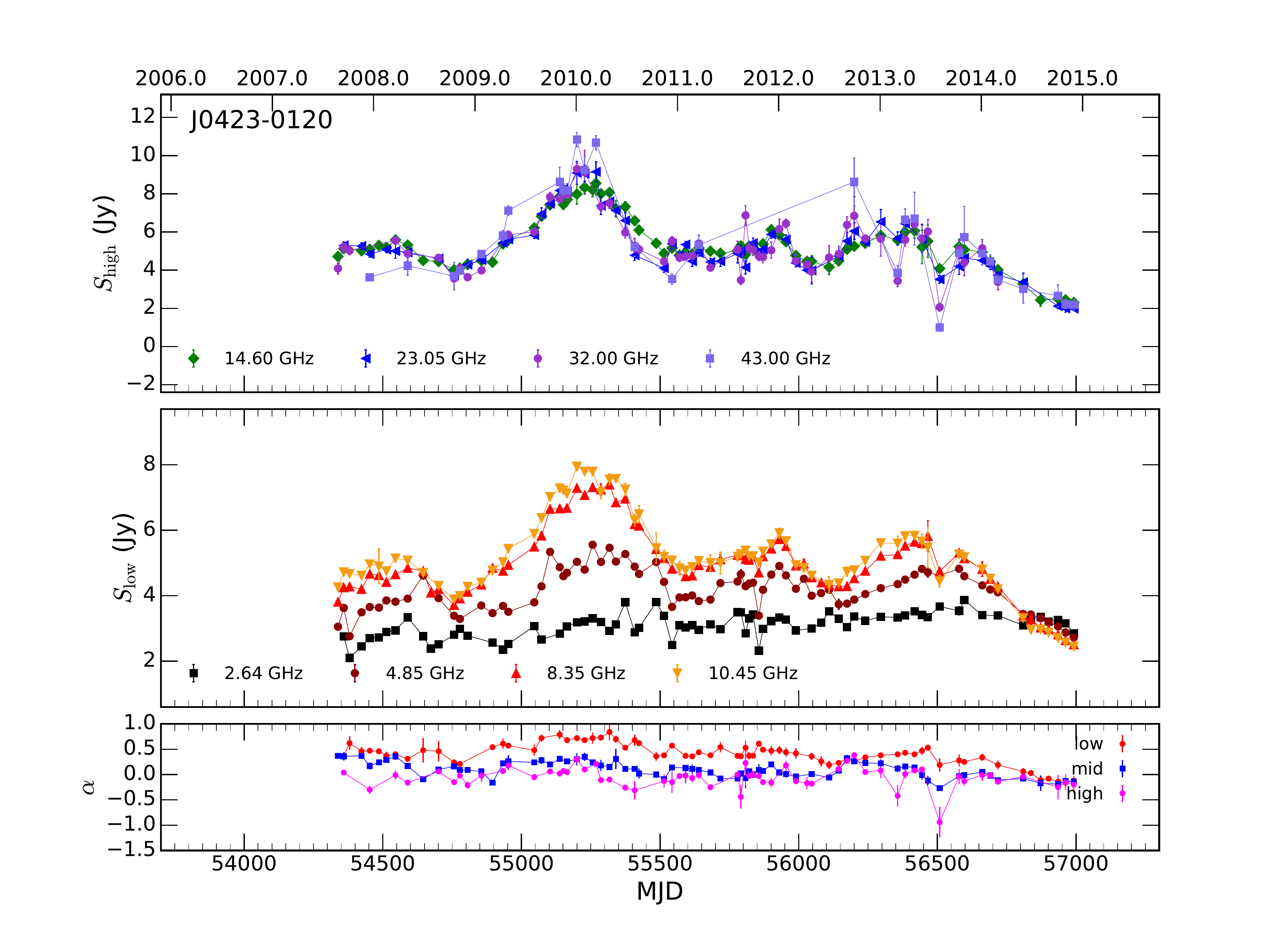}\\
\\[10pt]
\includegraphics[trim=60pt 30pt 100pt 50pt  ,clip, width=0.49\textwidth,angle=0]{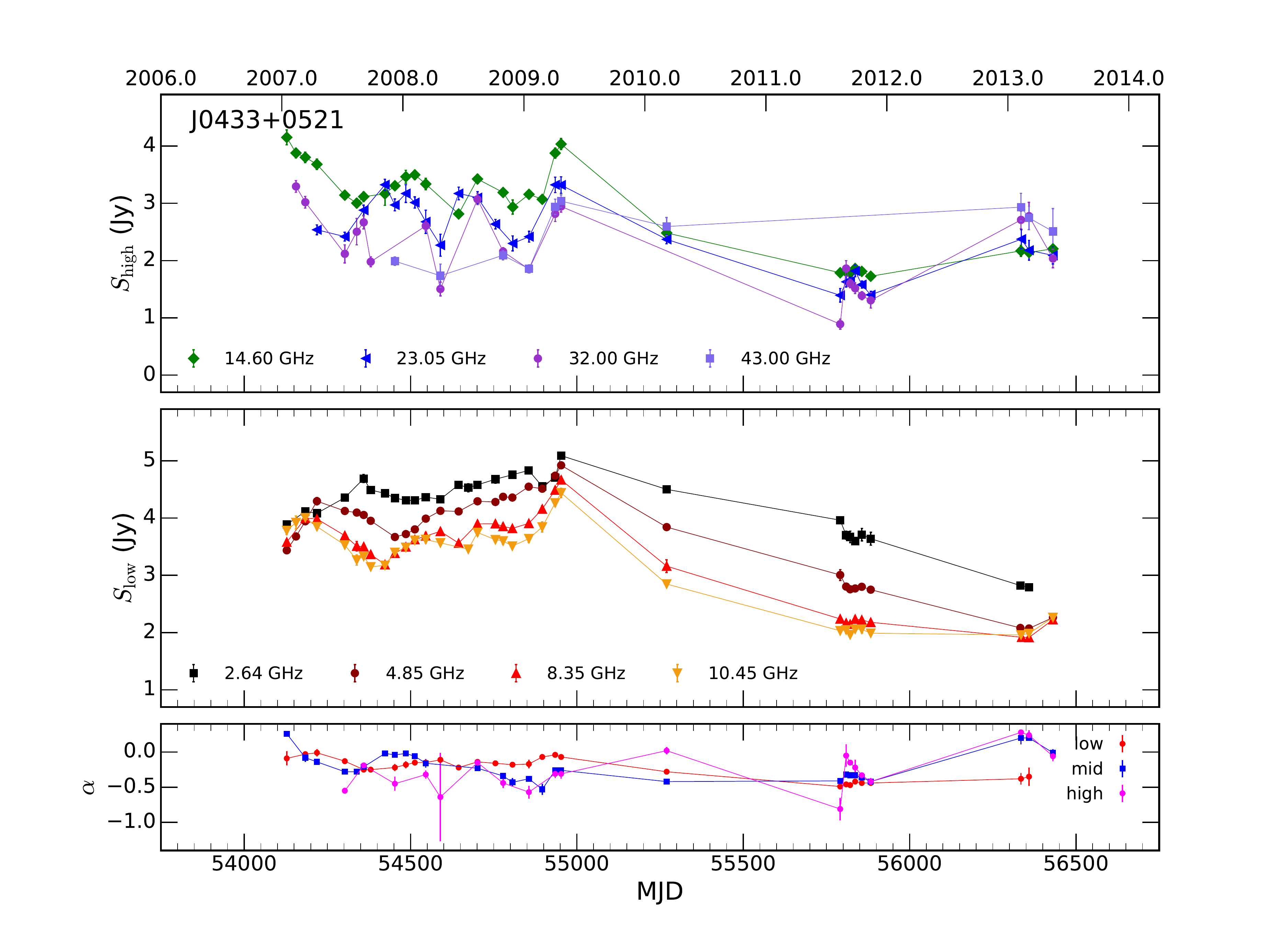}&
\includegraphics[trim=60pt 30pt 100pt 50pt  ,clip, width=0.49\textwidth,angle=0]{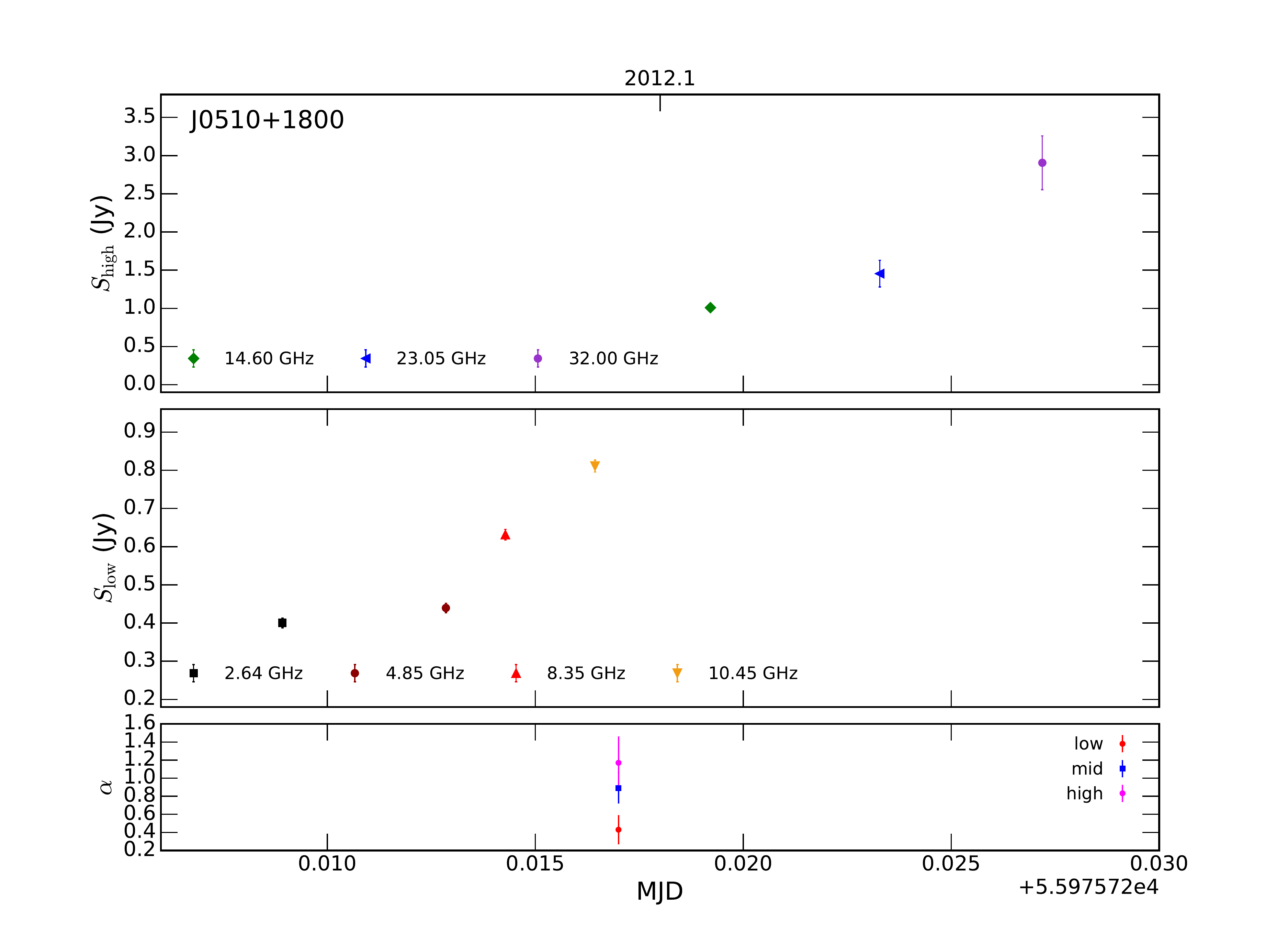}\\
\\[10pt]
\end{tabular}
\caption{Multi-frequency light curves for all the sources monitored by the \fg programme (``f'', ``s1'', ``s2'', ``old'') and the F-GAMMA-\textit{Planck} MoU. The lower panel in each frame shows the evolution of the low (2.64, 4.85 and 8.35~GHz) and mid-band (8.35, 10.45 and 14.6~GHz) and high-band (14.6, 23.05, 32, 43~GHz) spectral index. Only spectral index estimates from at least three frequencies are shown. Connecting lines have been used to guide the eye. }
\label{fig:sample_pg4}
\end{figure*}

\begin{figure*}[p]
\centering
\begin{tabular}{cc}
\includegraphics[trim=60pt 30pt 100pt 50pt  ,clip, width=0.49\textwidth,angle=0]{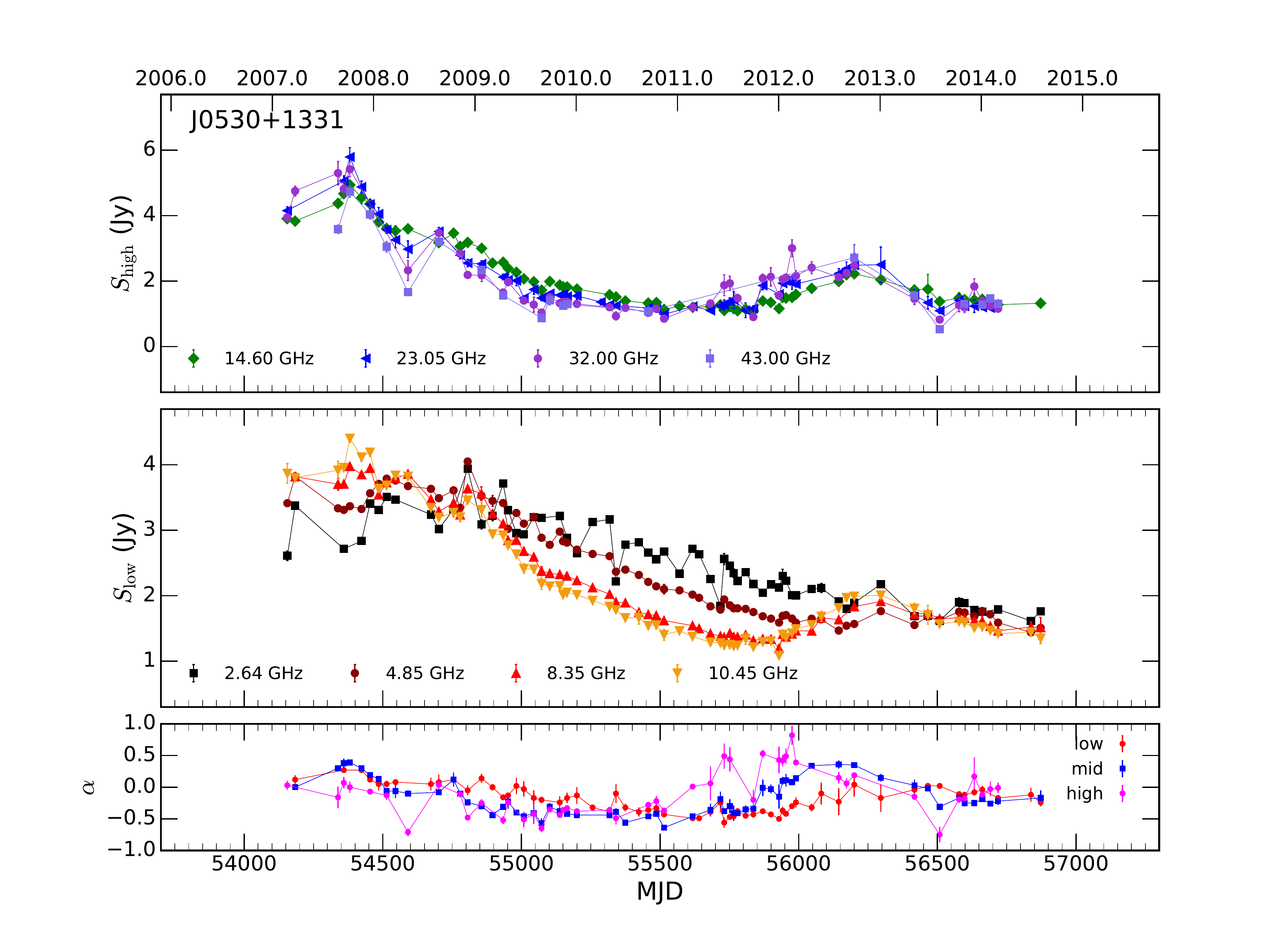}&
\includegraphics[trim=60pt 30pt 100pt 50pt  ,clip, width=0.49\textwidth,angle=0]{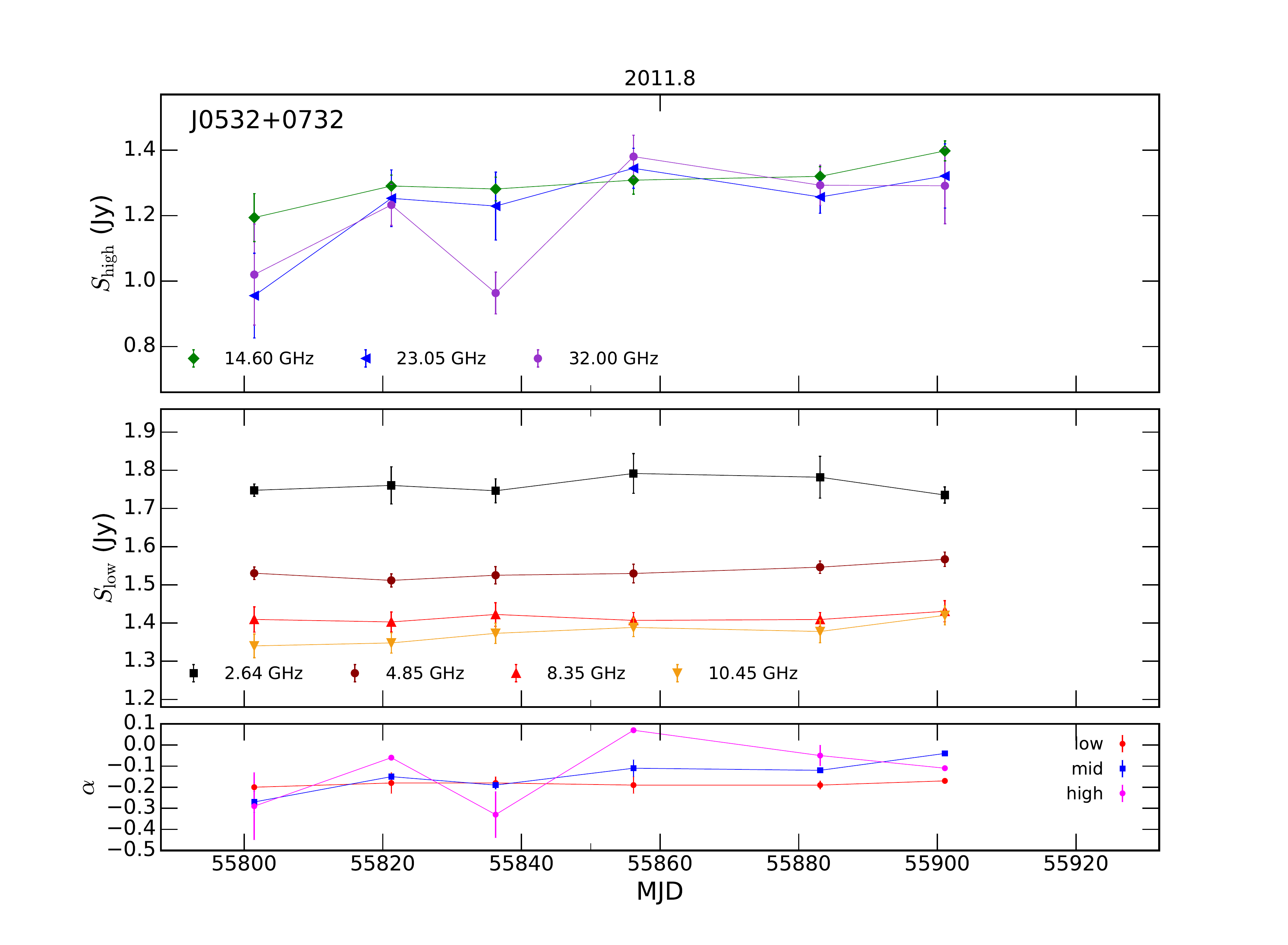}\\
\\[10pt]
\includegraphics[trim=60pt 30pt 100pt 50pt  ,clip, width=0.49\textwidth,angle=0]{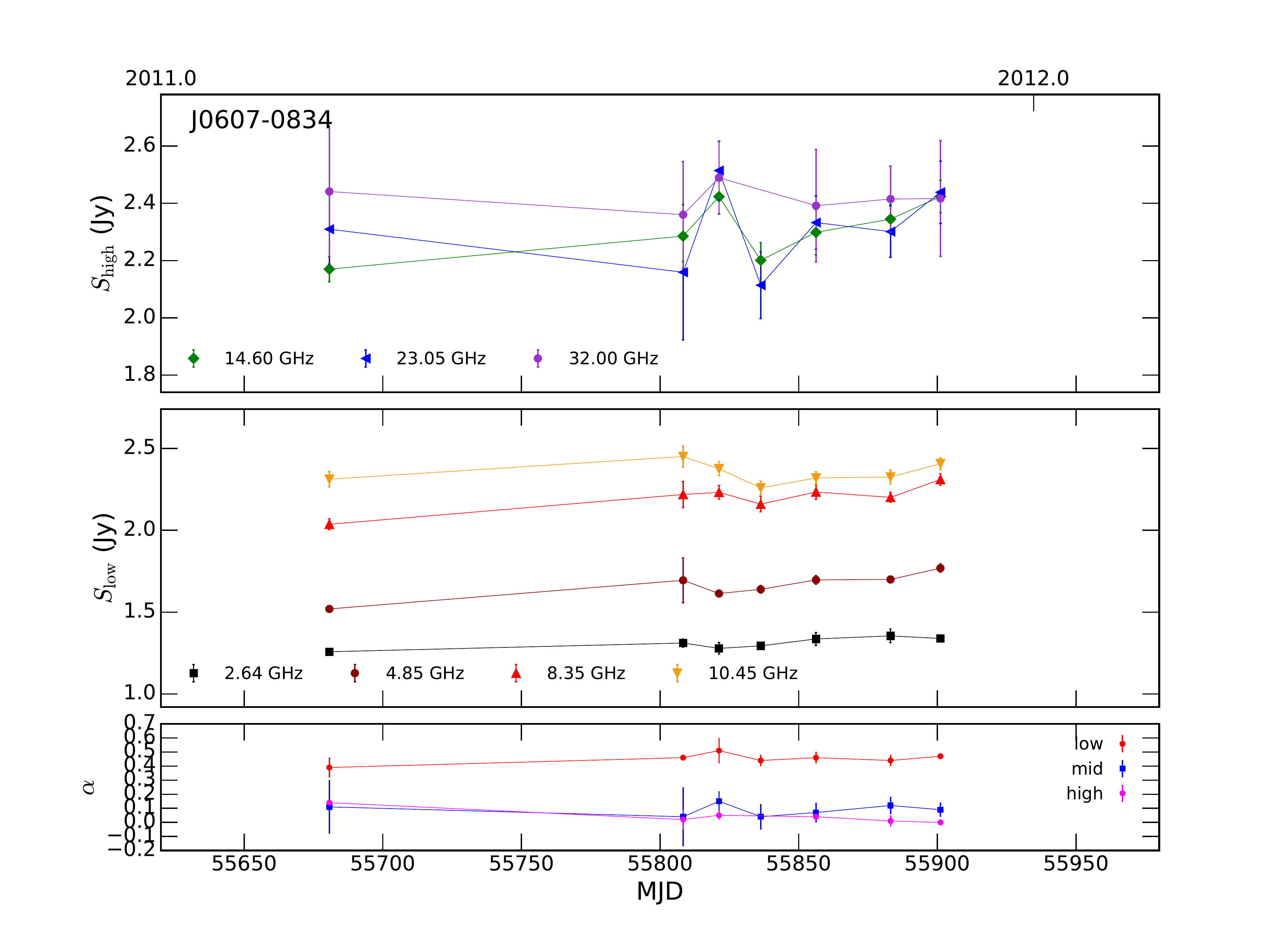}&
\includegraphics[trim=60pt 30pt 100pt 50pt  ,clip, width=0.49\textwidth,angle=0]{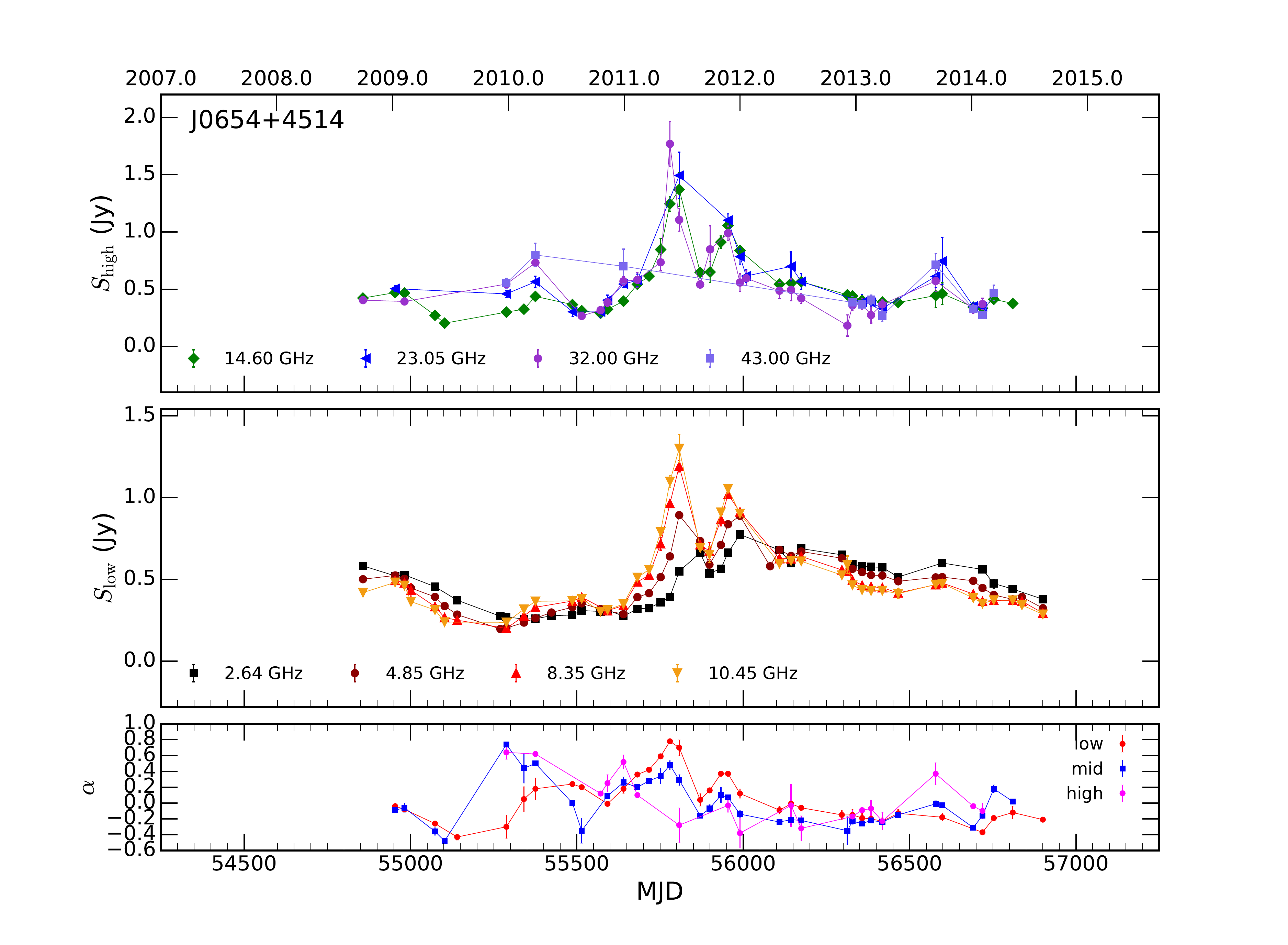}\\
\\[10pt]
\includegraphics[trim=60pt 30pt 100pt 50pt  ,clip, width=0.49\textwidth,angle=0]{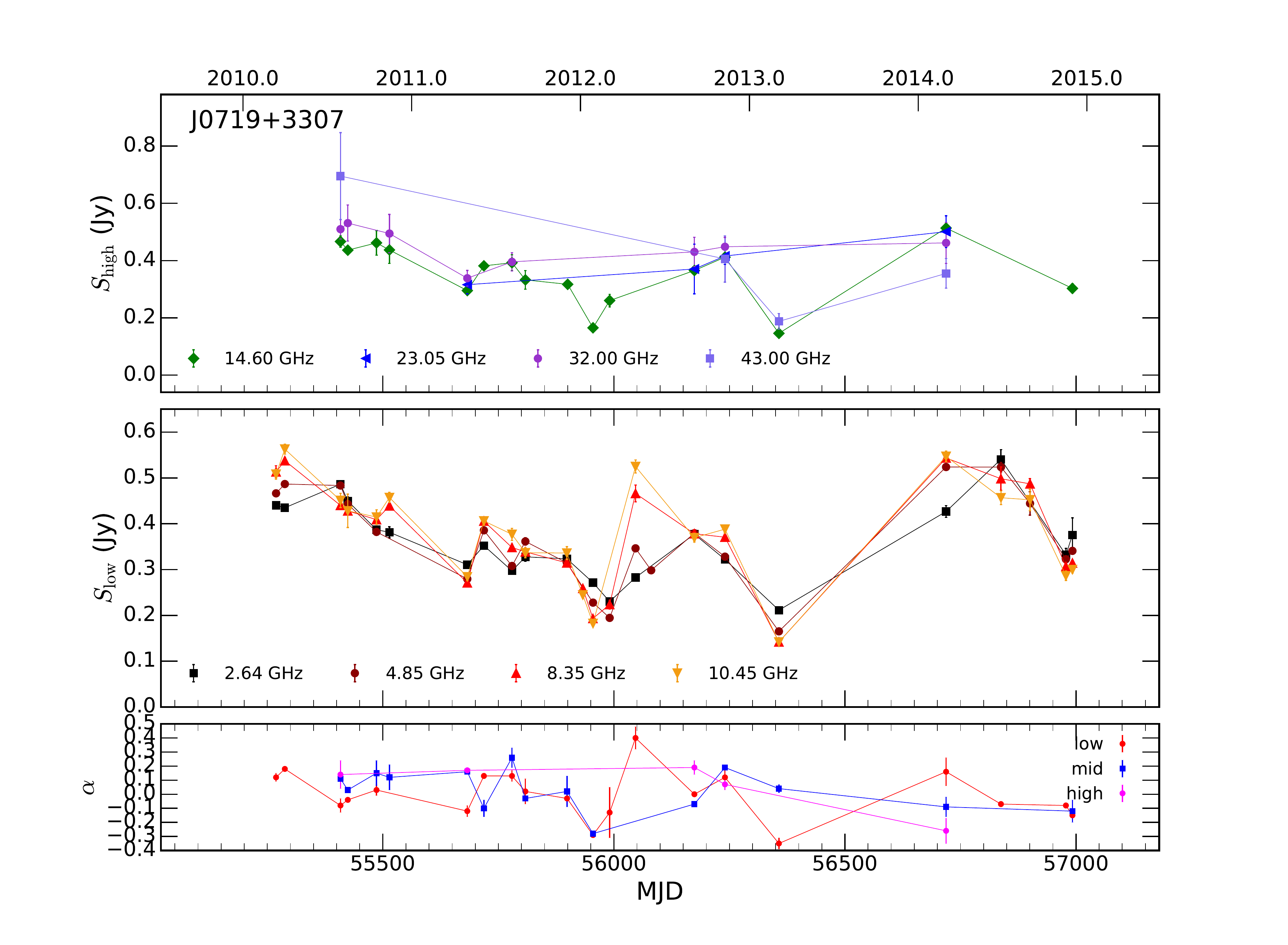}&
\includegraphics[trim=60pt 30pt 100pt 50pt  ,clip, width=0.49\textwidth,angle=0]{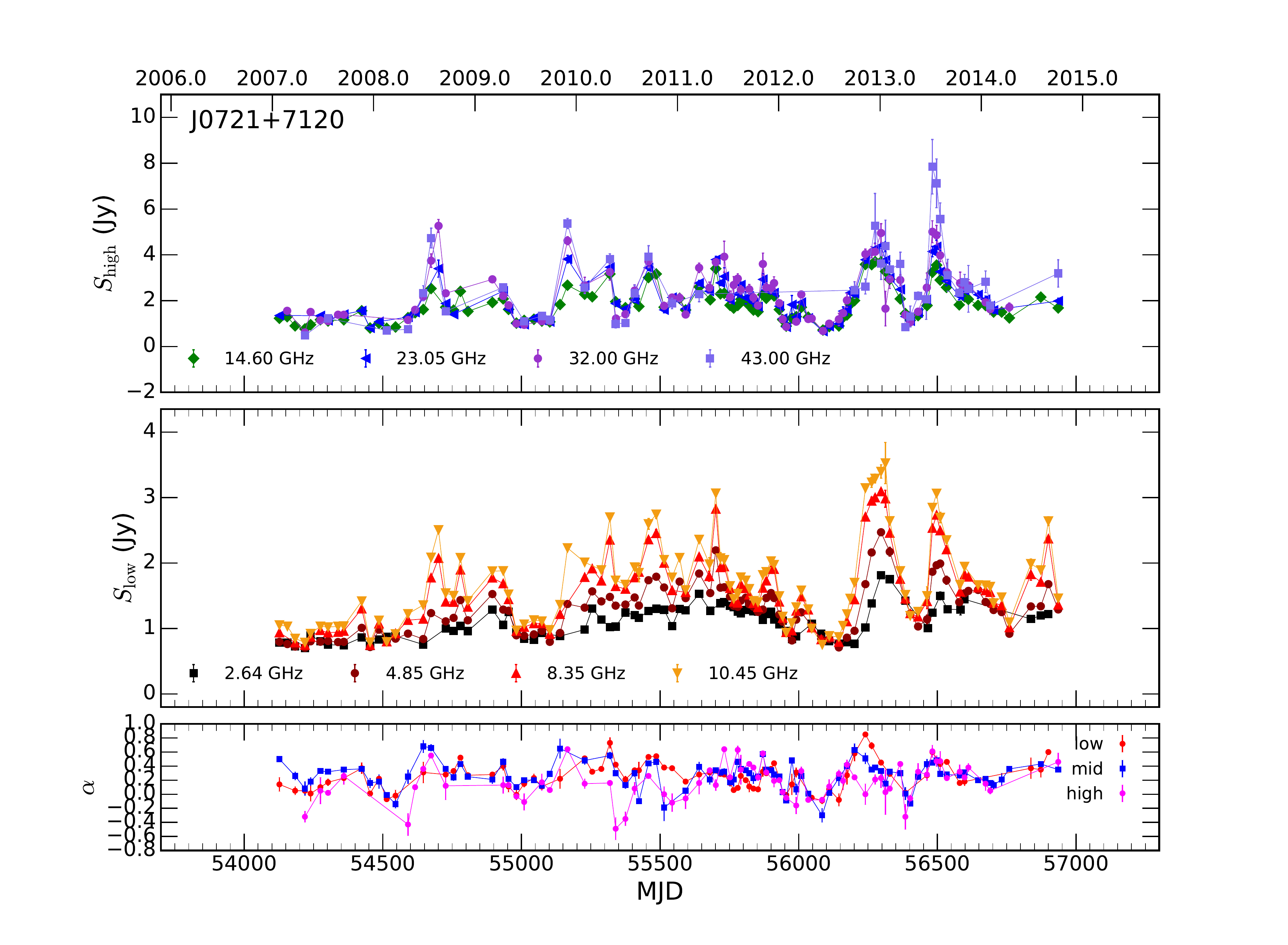}\\
\\[10pt]
\end{tabular}
\caption{Multi-frequency light curves for all the sources monitored by the \fg programme (``f'', ``s1'', ``s2'', ``old'') and the F-GAMMA-\textit{Planck} MoU. The lower panel in each frame shows the evolution of the low (2.64, 4.85 and 8.35~GHz) and mid-band (8.35, 10.45 and 14.6~GHz) and high-band (14.6, 23.05, 32, 43~GHz) spectral index. Only spectral index estimates from at least three frequencies are shown. Connecting lines have been used to guide the eye. }
\label{fig:sample_pg5}
\end{figure*}

\begin{figure*}[p]
\centering
\begin{tabular}{cc}
\includegraphics[trim=60pt 30pt 100pt 50pt  ,clip, width=0.49\textwidth,angle=0]{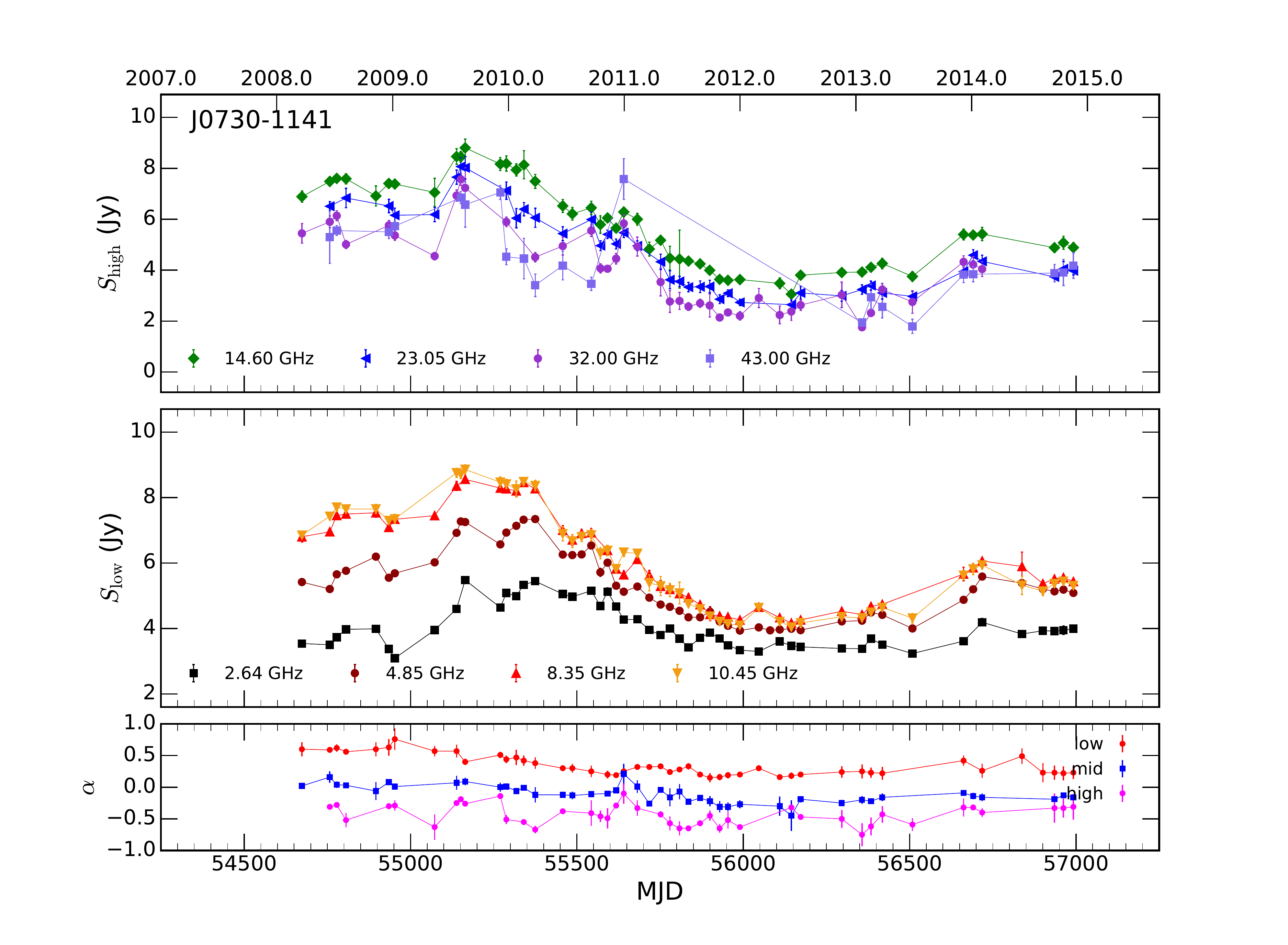}&
\includegraphics[trim=60pt 30pt 100pt 50pt  ,clip, width=0.49\textwidth,angle=0]{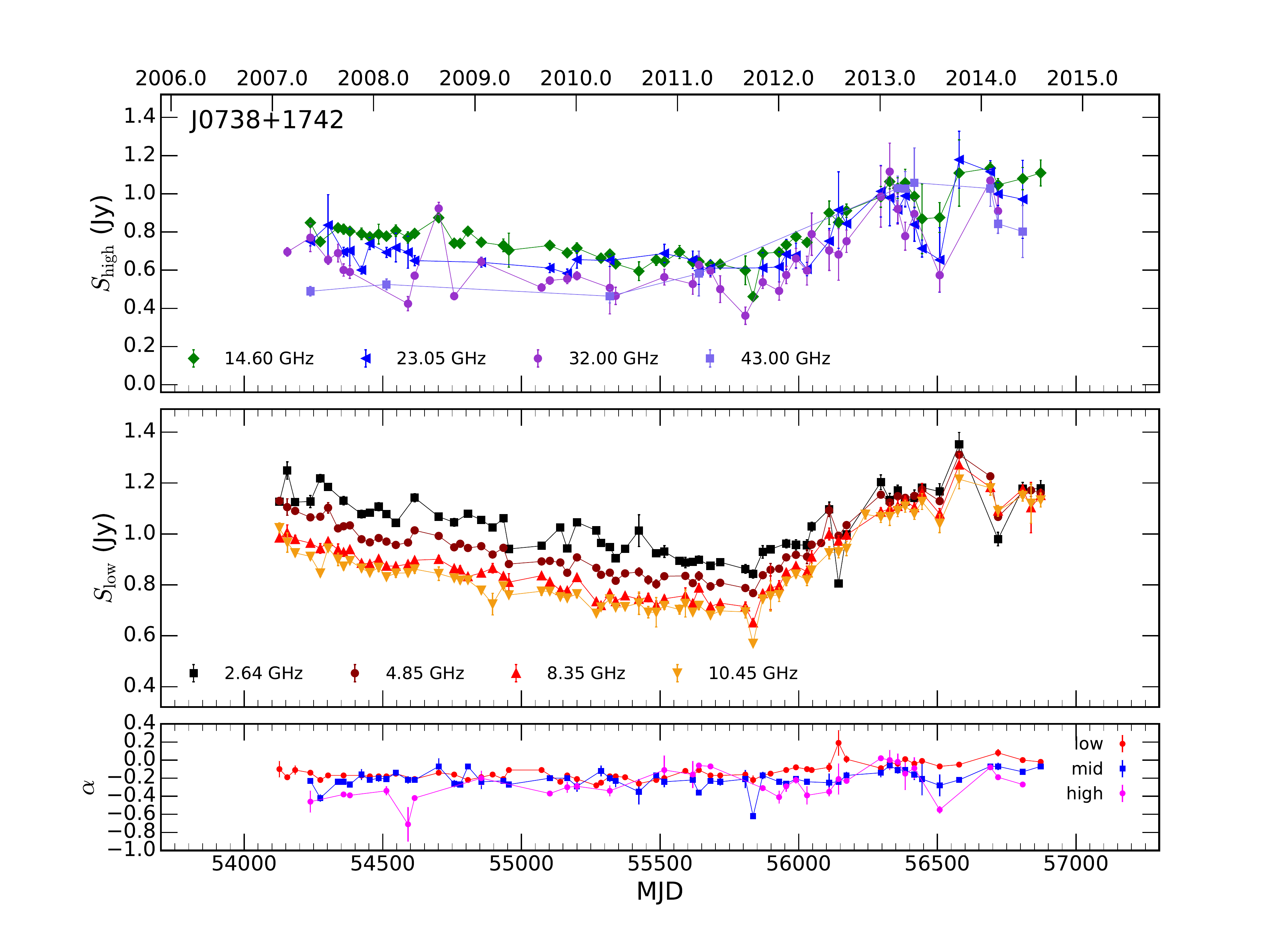}\\
\\[10pt]
\includegraphics[trim=60pt 30pt 100pt 50pt  ,clip, width=0.49\textwidth,angle=0]{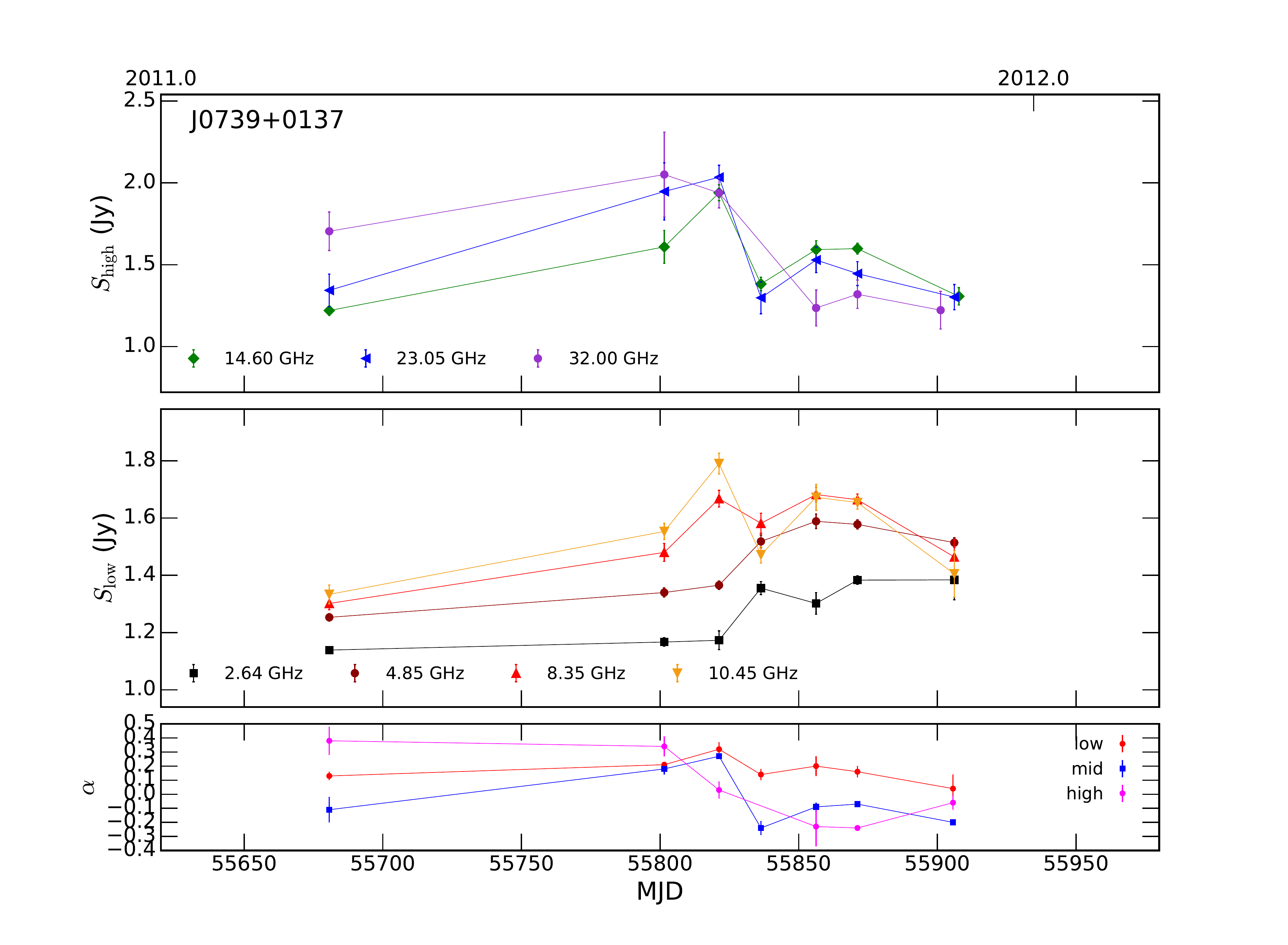}&
\includegraphics[trim=60pt 30pt 100pt 50pt  ,clip, width=0.49\textwidth,angle=0]{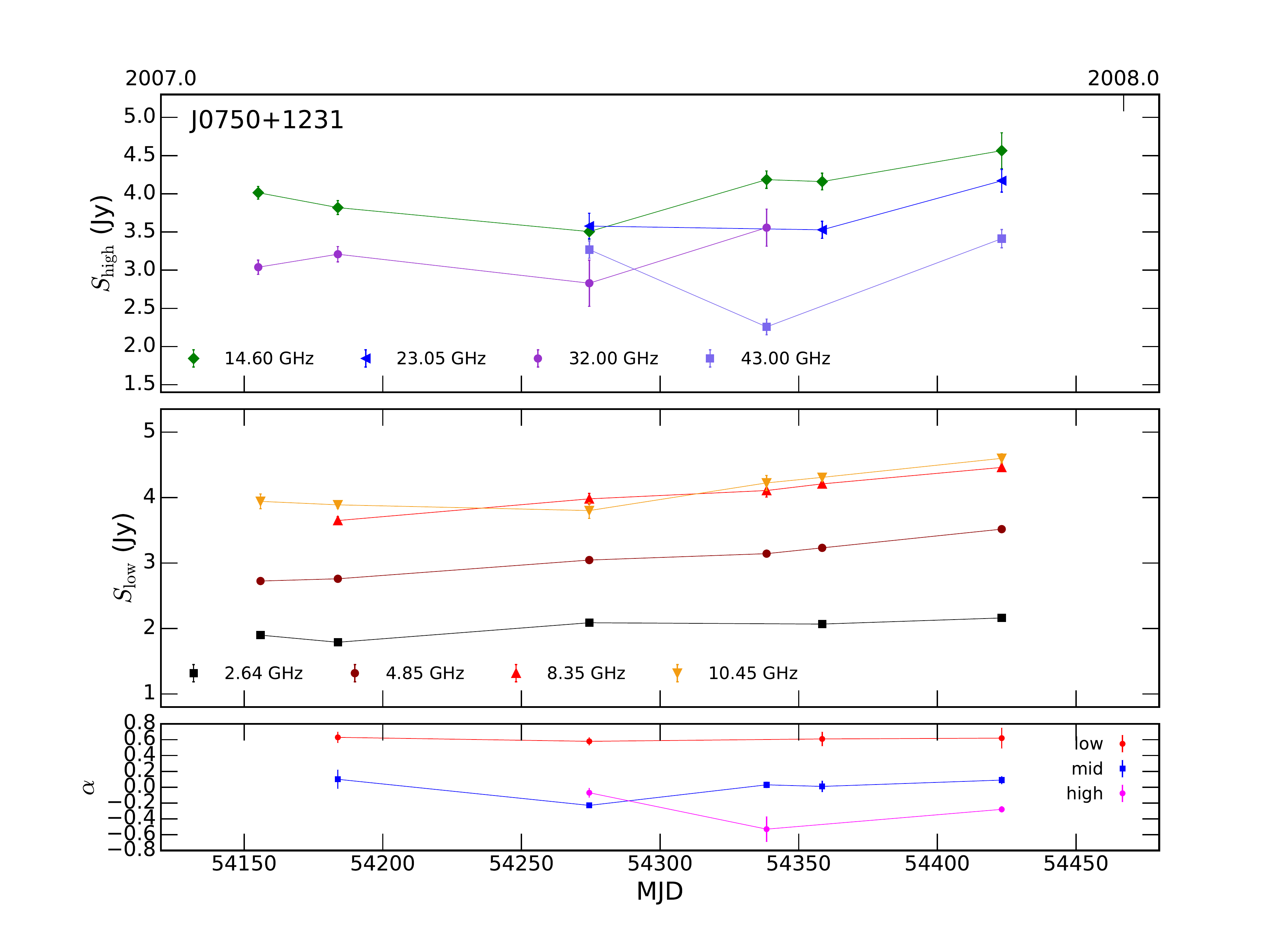}\\
\\[10pt]
\includegraphics[trim=60pt 30pt 100pt 50pt  ,clip, width=0.49\textwidth,angle=0]{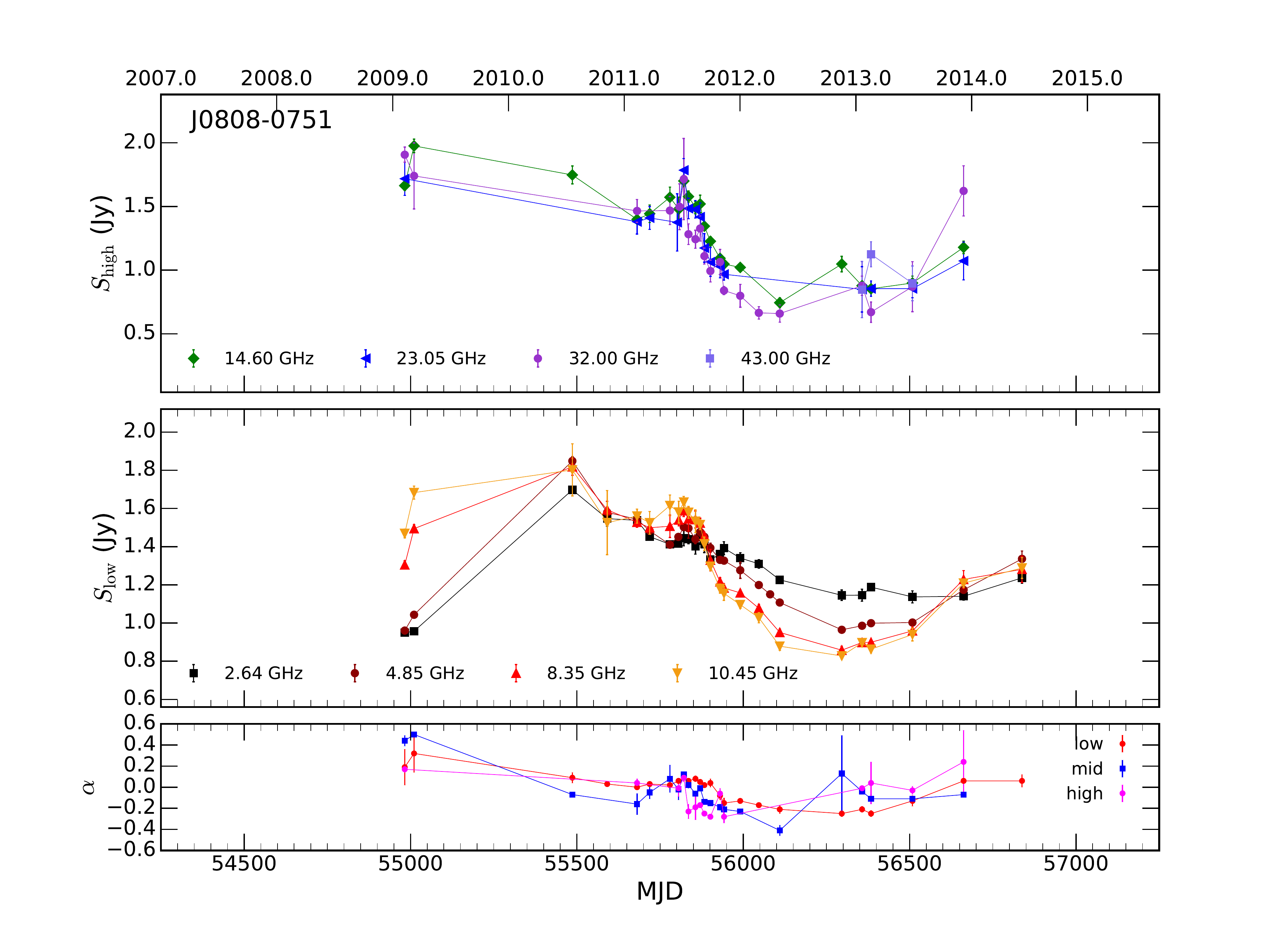}&
\includegraphics[trim=60pt 30pt 100pt 50pt  ,clip, width=0.49\textwidth,angle=0]{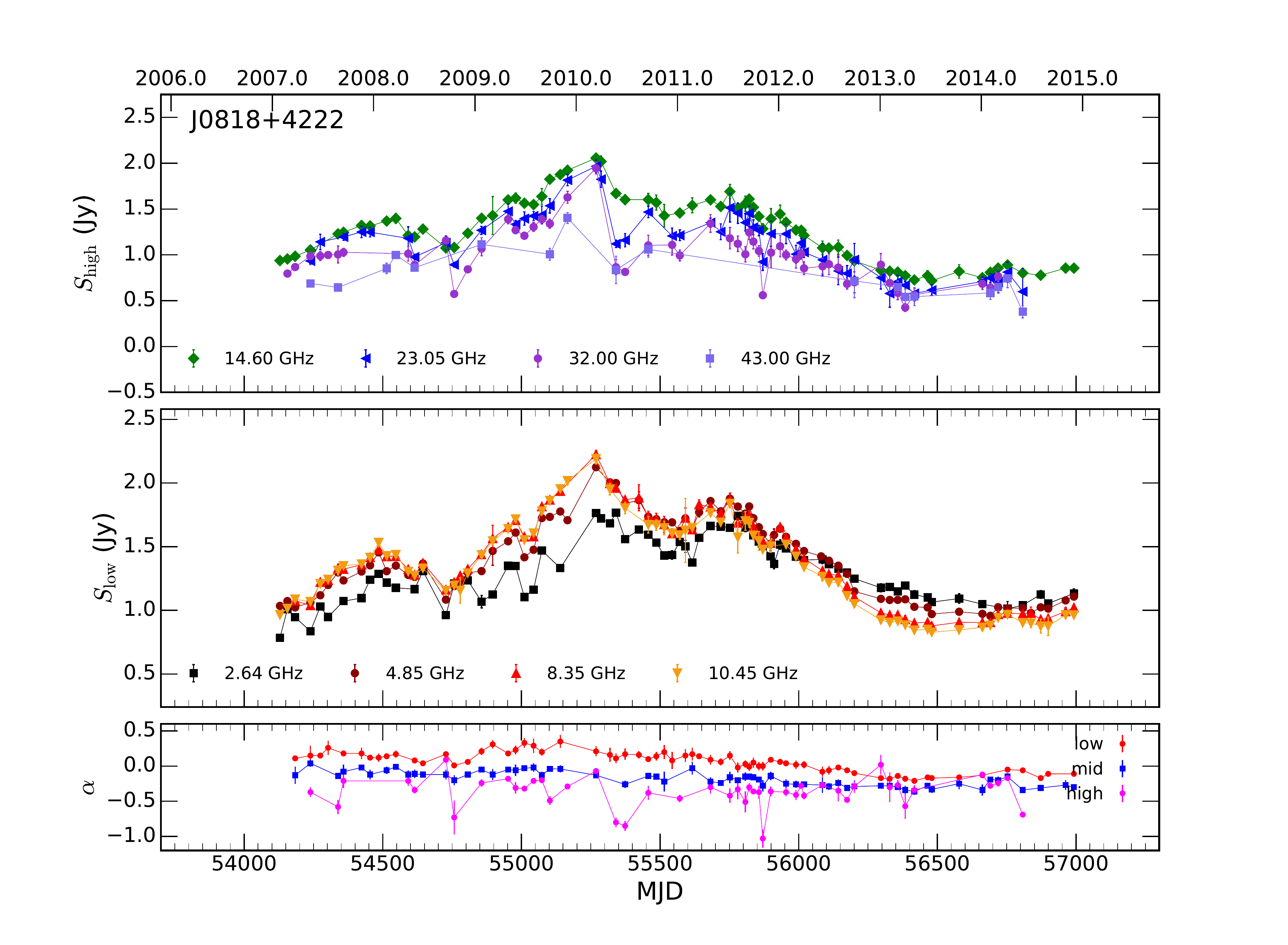}\\
\\[10pt]
\end{tabular}
\caption{Multi-frequency light curves for all the sources monitored by the \fg programme (``f'', ``s1'', ``s2'', ``old'') and the F-GAMMA-\textit{Planck} MoU. The lower panel in each frame shows the evolution of the low (2.64, 4.85 and 8.35~GHz) and mid-band (8.35, 10.45 and 14.6~GHz) and high-band (14.6, 23.05, 32, 43~GHz) spectral index. Only spectral index estimates from at least three frequencies are shown. Connecting lines have been used to guide the eye. }
\label{fig:sample_pg6}
\end{figure*}

\begin{figure*}[p]
\centering
\begin{tabular}{cc}
\includegraphics[trim=60pt 30pt 100pt 50pt  ,clip, width=0.49\textwidth,angle=0]{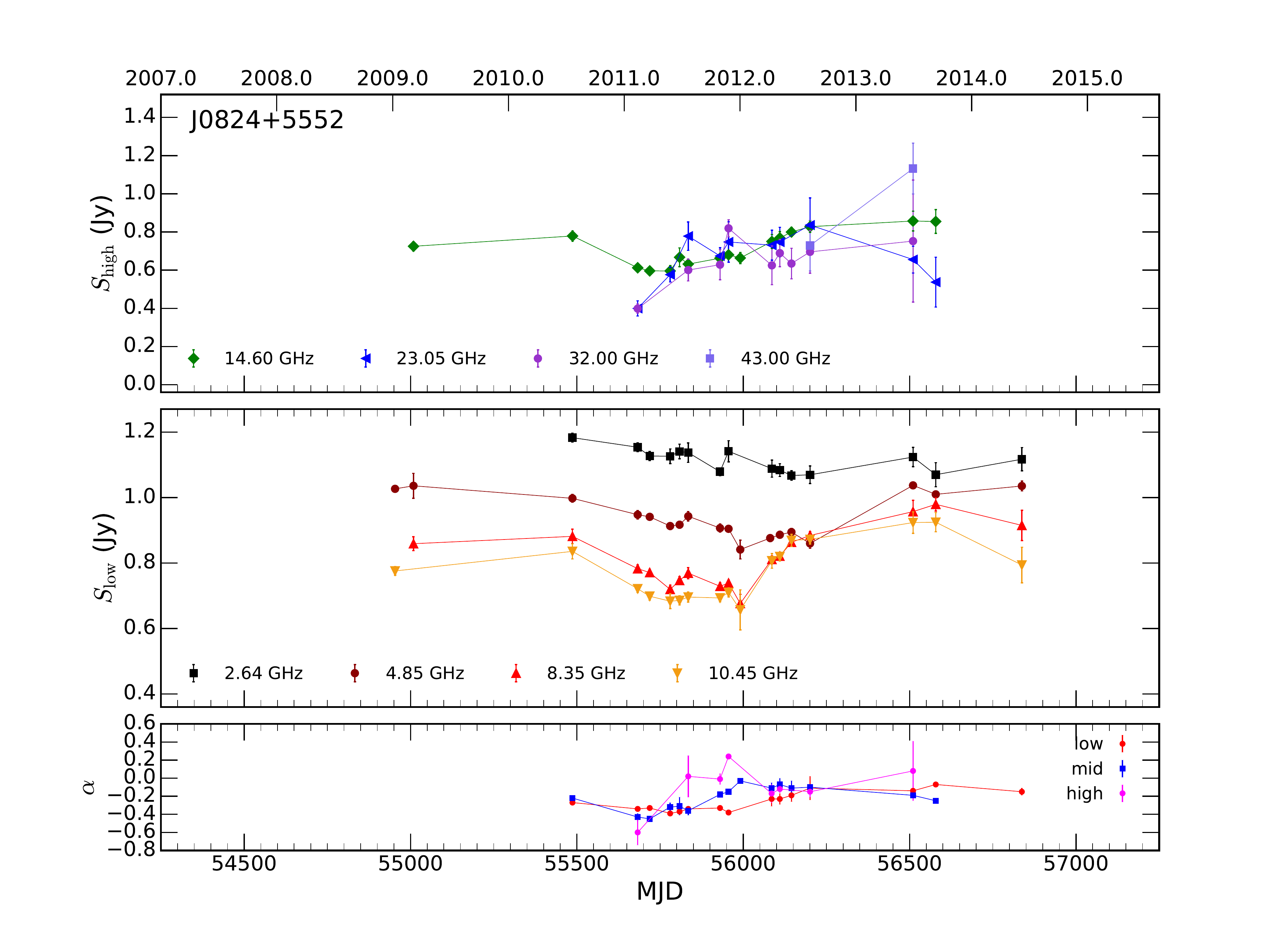}&
\includegraphics[trim=60pt 30pt 100pt 50pt  ,clip, width=0.49\textwidth,angle=0]{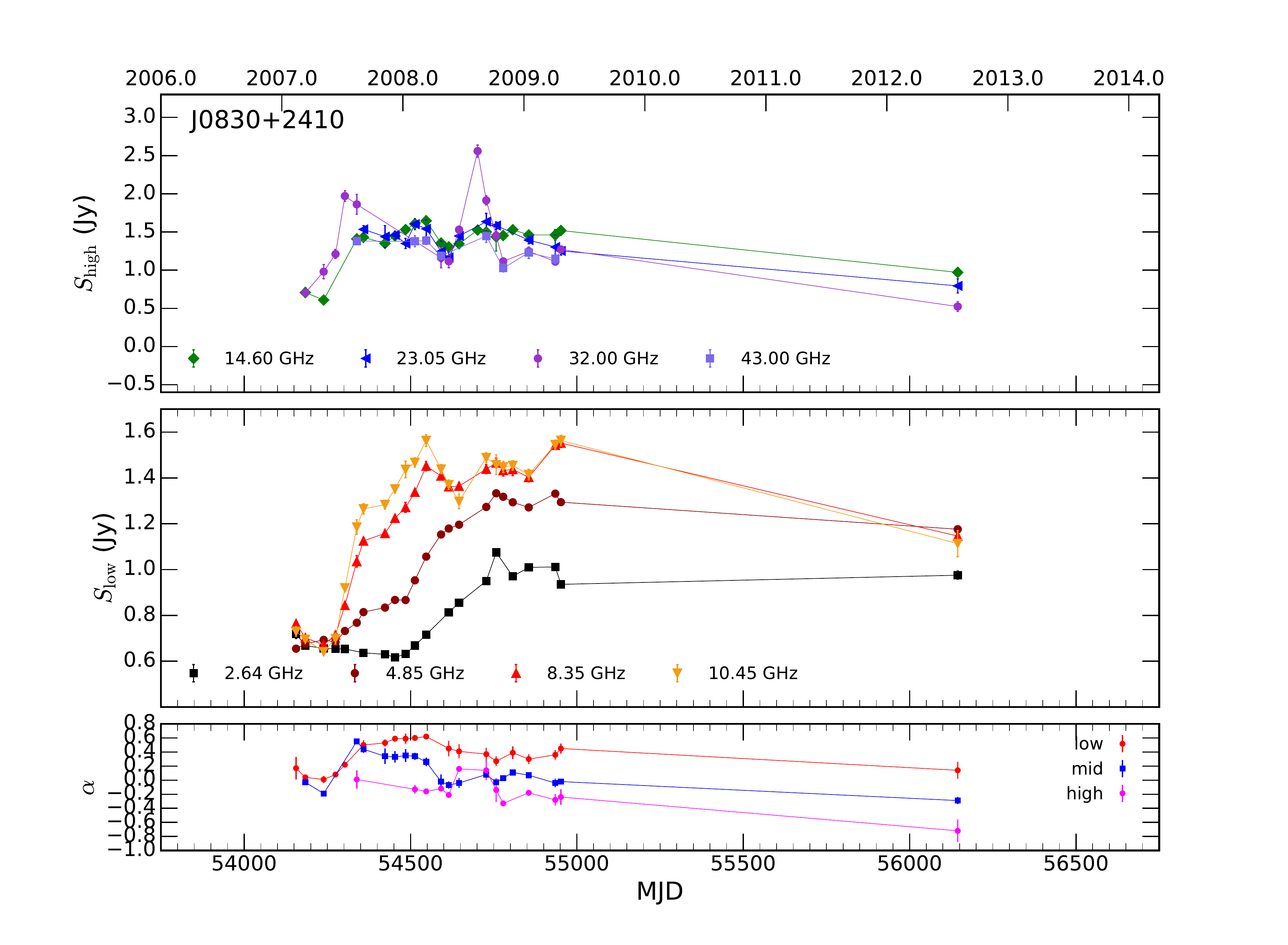}\\
\\[10pt]
\includegraphics[trim=60pt 30pt 100pt 50pt  ,clip, width=0.49\textwidth,angle=0]{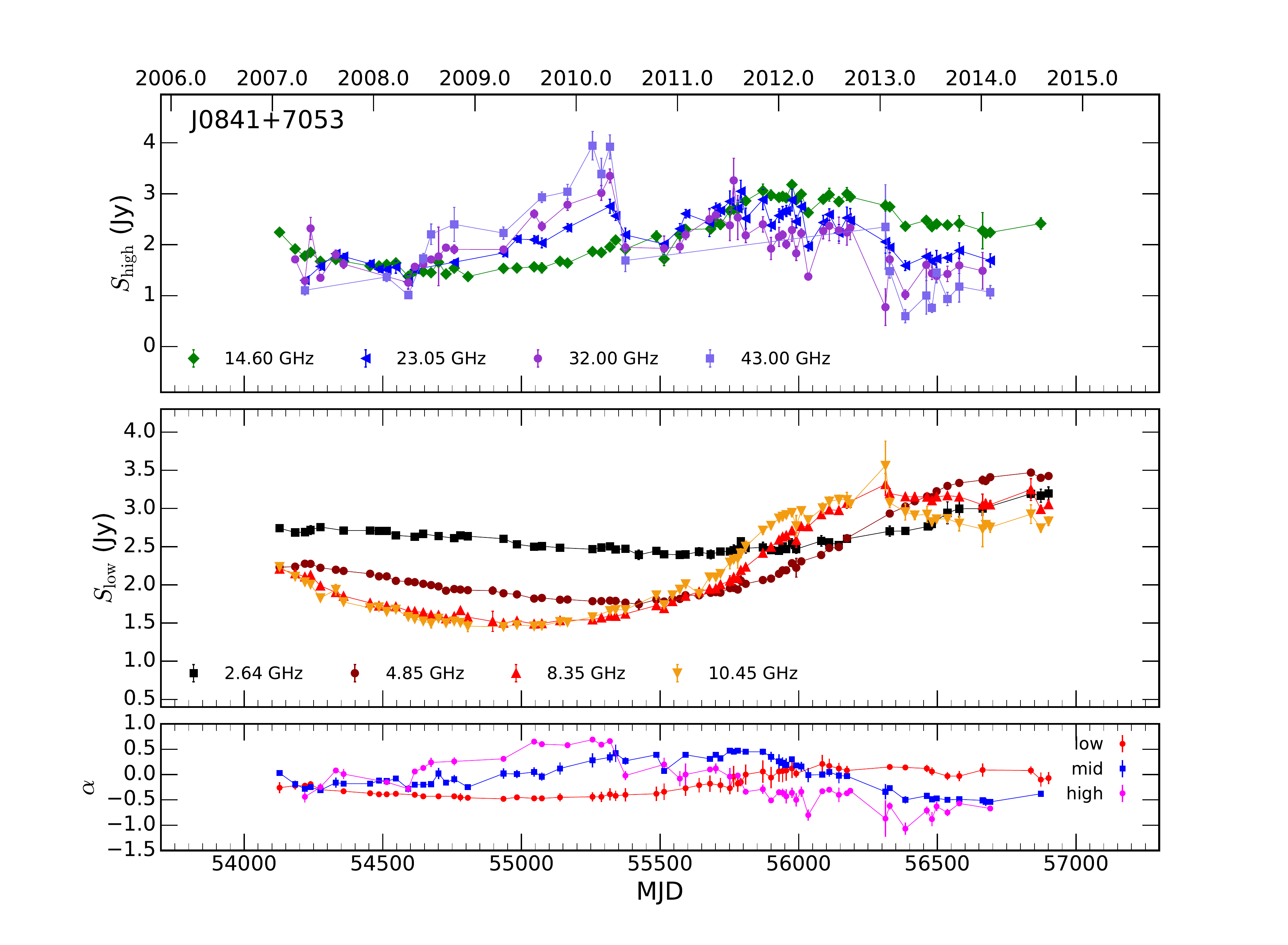}&
\includegraphics[trim=60pt 30pt 100pt 50pt  ,clip, width=0.49\textwidth,angle=0]{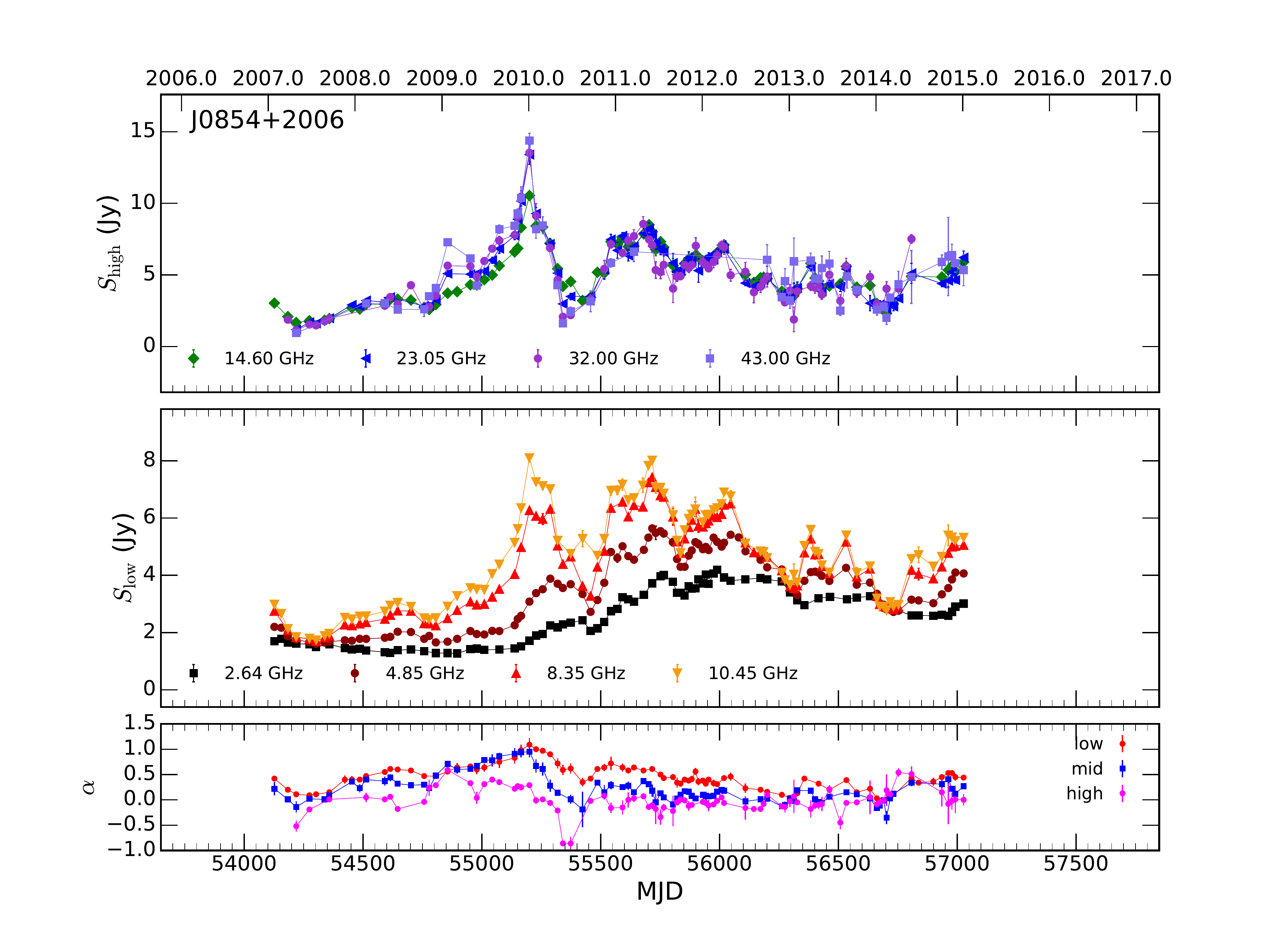}\\
\\[10pt]
\includegraphics[trim=60pt 30pt 100pt 50pt  ,clip, width=0.49\textwidth,angle=0]{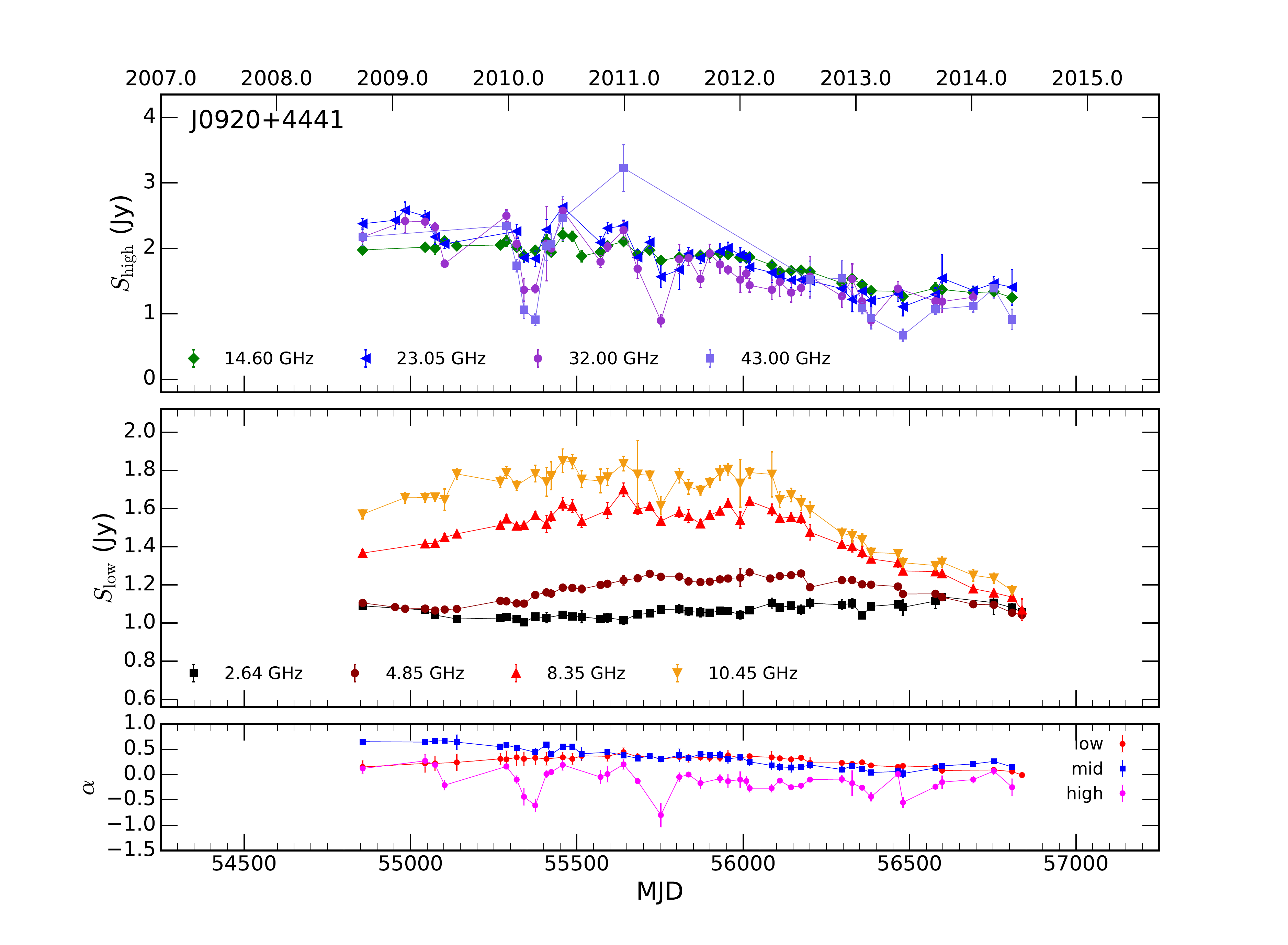}&
\includegraphics[trim=60pt 30pt 100pt 50pt  ,clip, width=0.49\textwidth,angle=0]{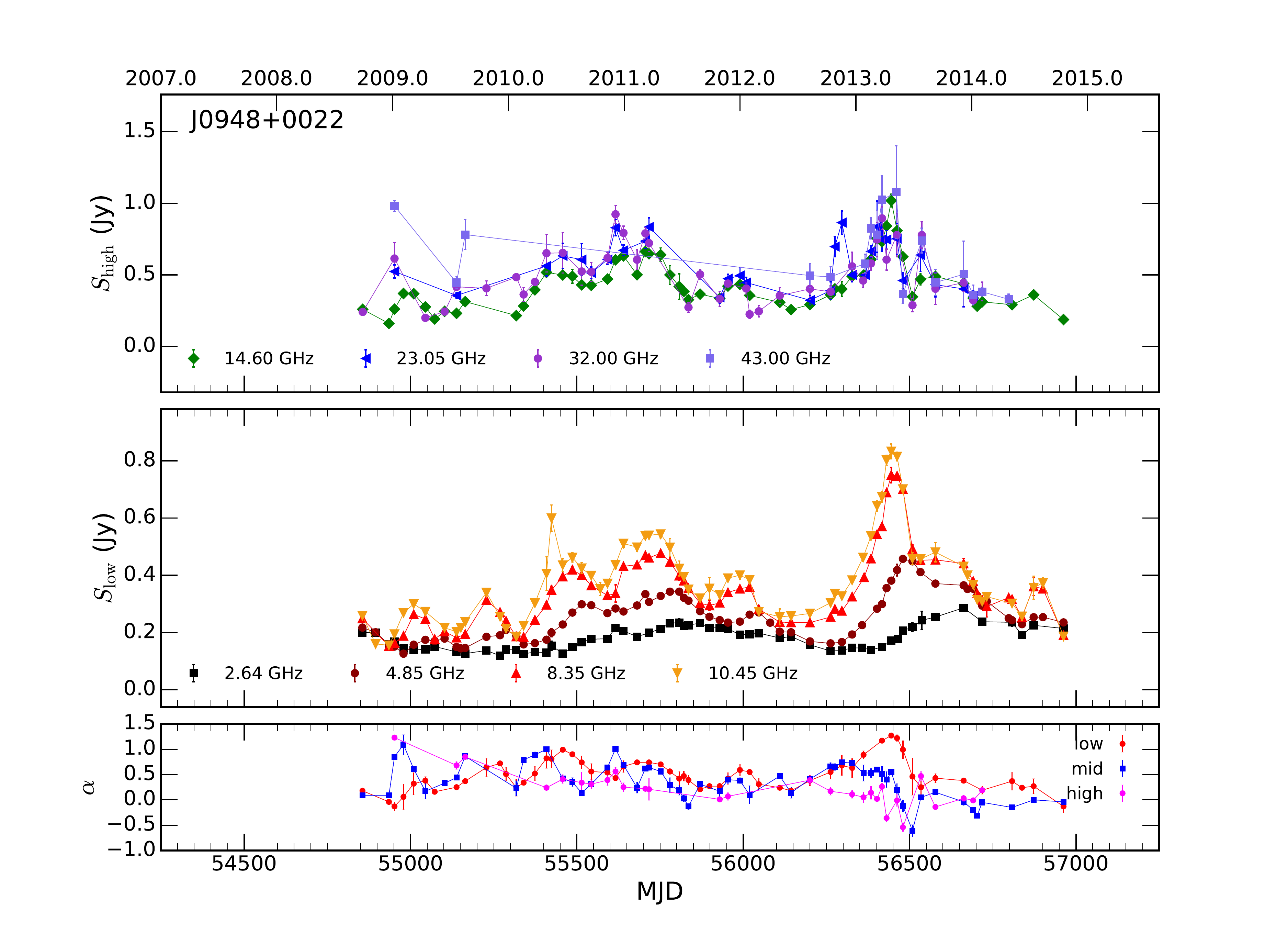}\\
\\[10pt]
\end{tabular}
\caption{Multi-frequency light curves for all the sources monitored by the \fg programme (``f'', ``s1'', ``s2'', ``old'') and the F-GAMMA-\textit{Planck} MoU. The lower panel in each frame shows the evolution of the low (2.64, 4.85 and 8.35~GHz) and mid-band (8.35, 10.45 and 14.6~GHz) and high-band (14.6, 23.05, 32, 43~GHz) spectral index. Only spectral index estimates from at least three frequencies are shown. Connecting lines have been used to guide the eye. }
\label{fig:sample_pg7}
\end{figure*}

\begin{figure*}[p]
\centering
\begin{tabular}{cc}
\includegraphics[trim=60pt 30pt 100pt 50pt  ,clip, width=0.49\textwidth,angle=0]{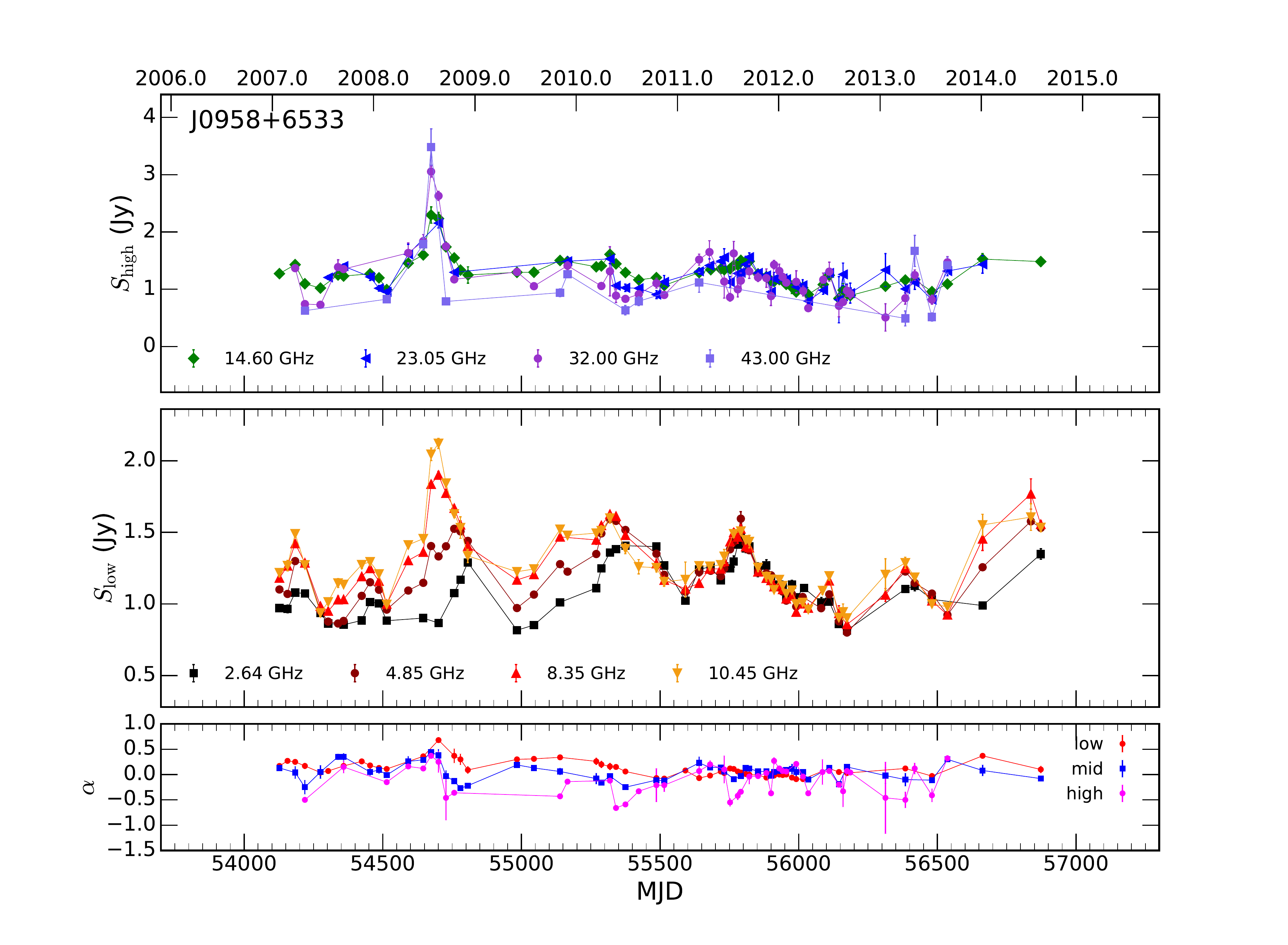}&
\includegraphics[trim=60pt 30pt 100pt 50pt  ,clip, width=0.49\textwidth,angle=0]{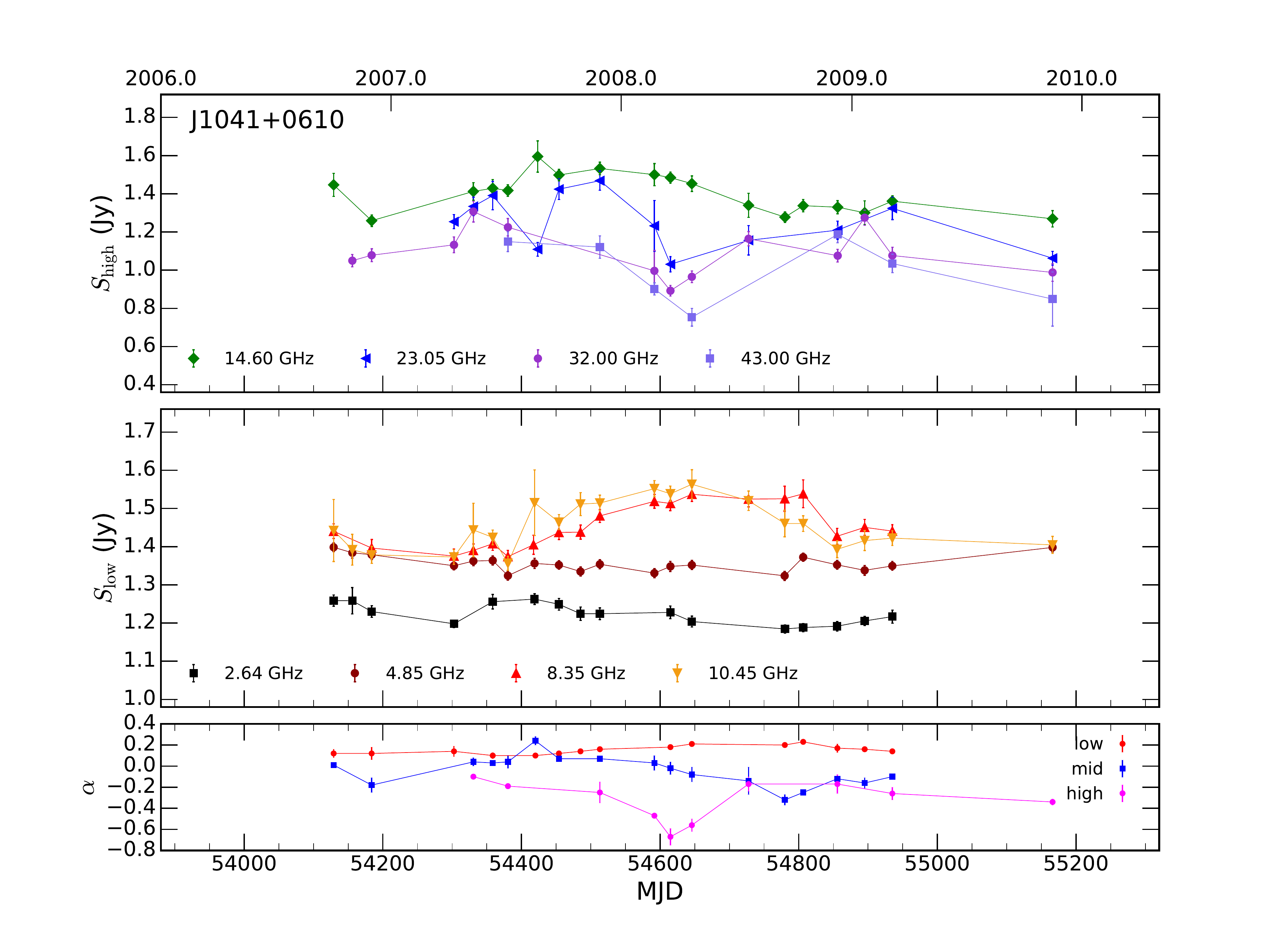}\\
\\[10pt]
\includegraphics[trim=60pt 30pt 100pt 50pt  ,clip, width=0.49\textwidth,angle=0]{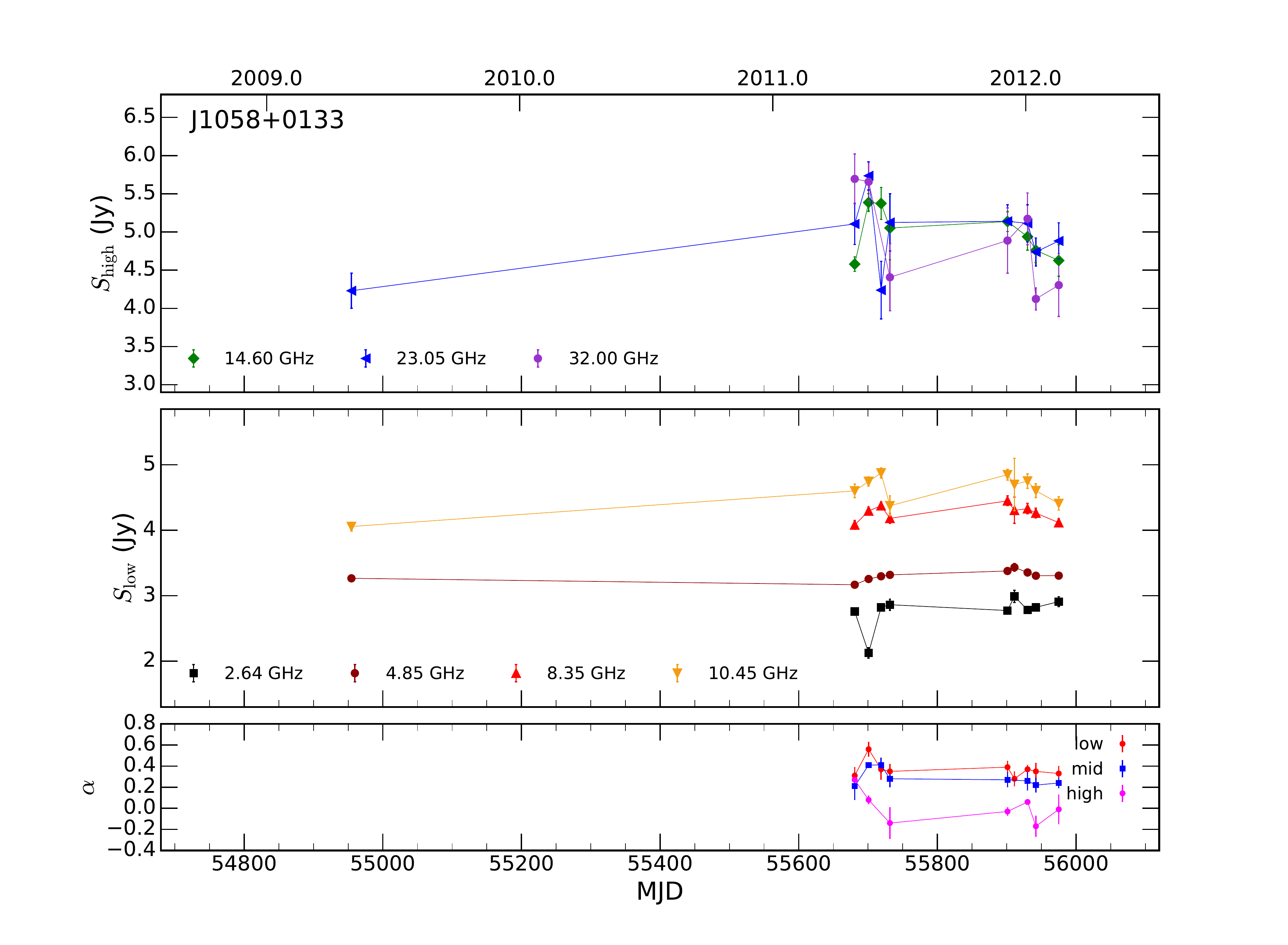}&
\includegraphics[trim=60pt 30pt 100pt 50pt  ,clip, width=0.49\textwidth,angle=0]{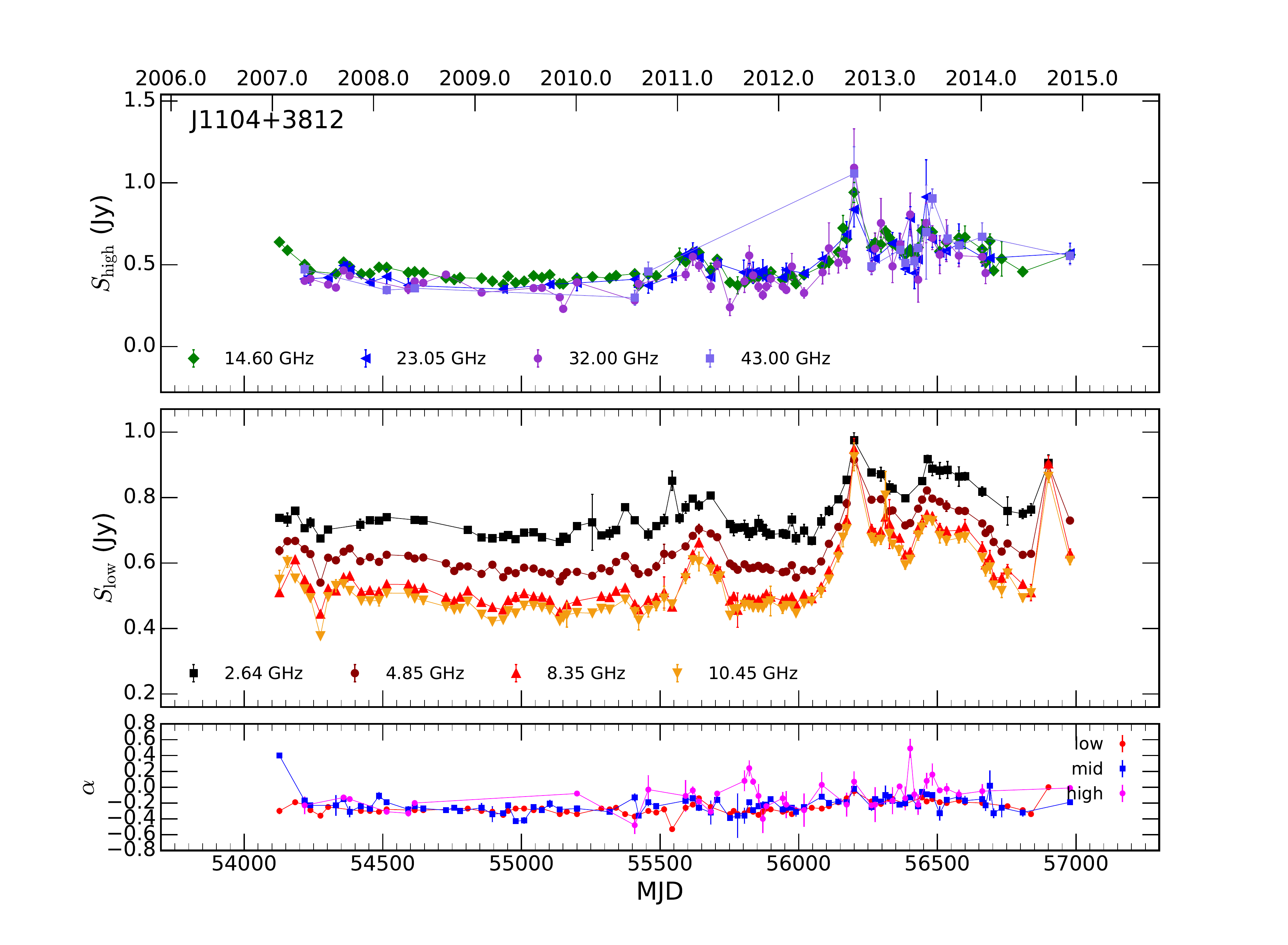}\\
\\[10pt]
\includegraphics[trim=60pt 30pt 100pt 50pt  ,clip, width=0.49\textwidth,angle=0]{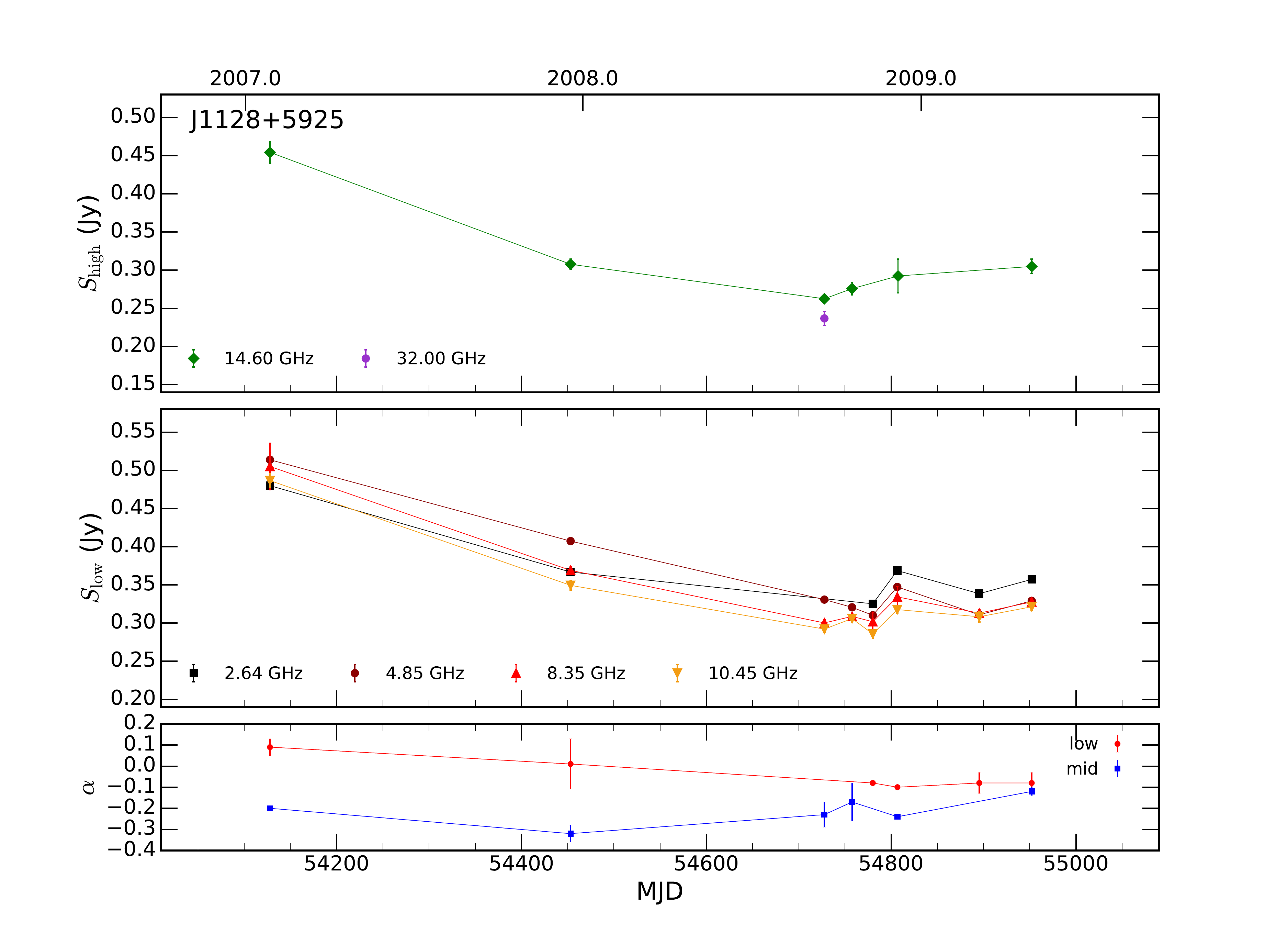}&
\includegraphics[trim=60pt 30pt 100pt 50pt  ,clip, width=0.49\textwidth,angle=0]{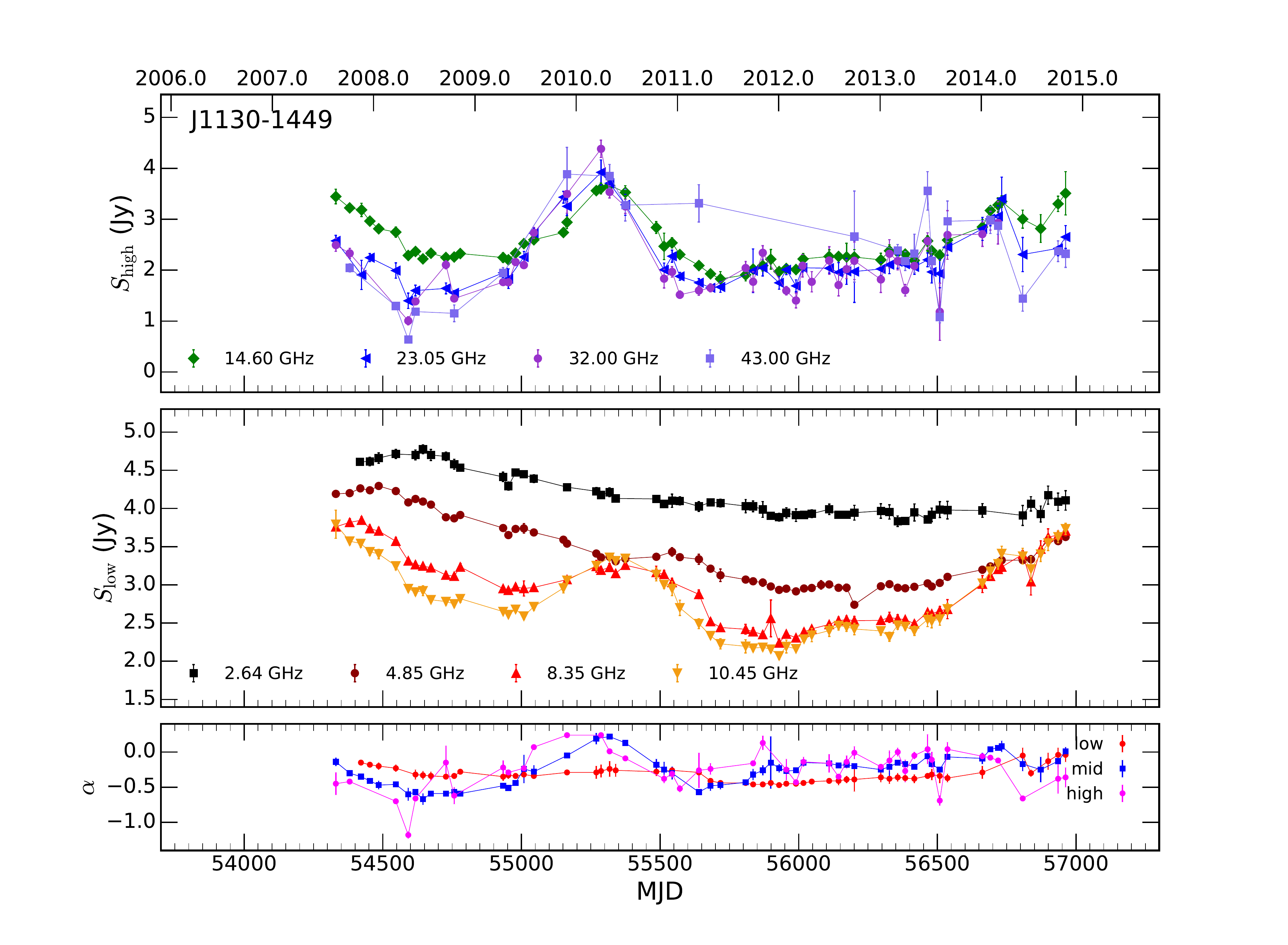}\\
\\[10pt]
\end{tabular}
\caption{Multi-frequency light curves for all the sources monitored by the \fg programme (``f'', ``s1'', ``s2'', ``old'') and the F-GAMMA-\textit{Planck} MoU. The lower panel in each frame shows the evolution of the low (2.64, 4.85 and 8.35~GHz) and mid-band (8.35, 10.45 and 14.6~GHz) and high-band (14.6, 23.05, 32, 43~GHz) spectral index. Only spectral index estimates from at least three frequencies are shown. Connecting lines have been used to guide the eye. }
\label{fig:sample_pg8}
\end{figure*}

\begin{figure*}[p]
\centering
\begin{tabular}{cc}
\includegraphics[trim=60pt 30pt 100pt 50pt  ,clip, width=0.49\textwidth,angle=0]{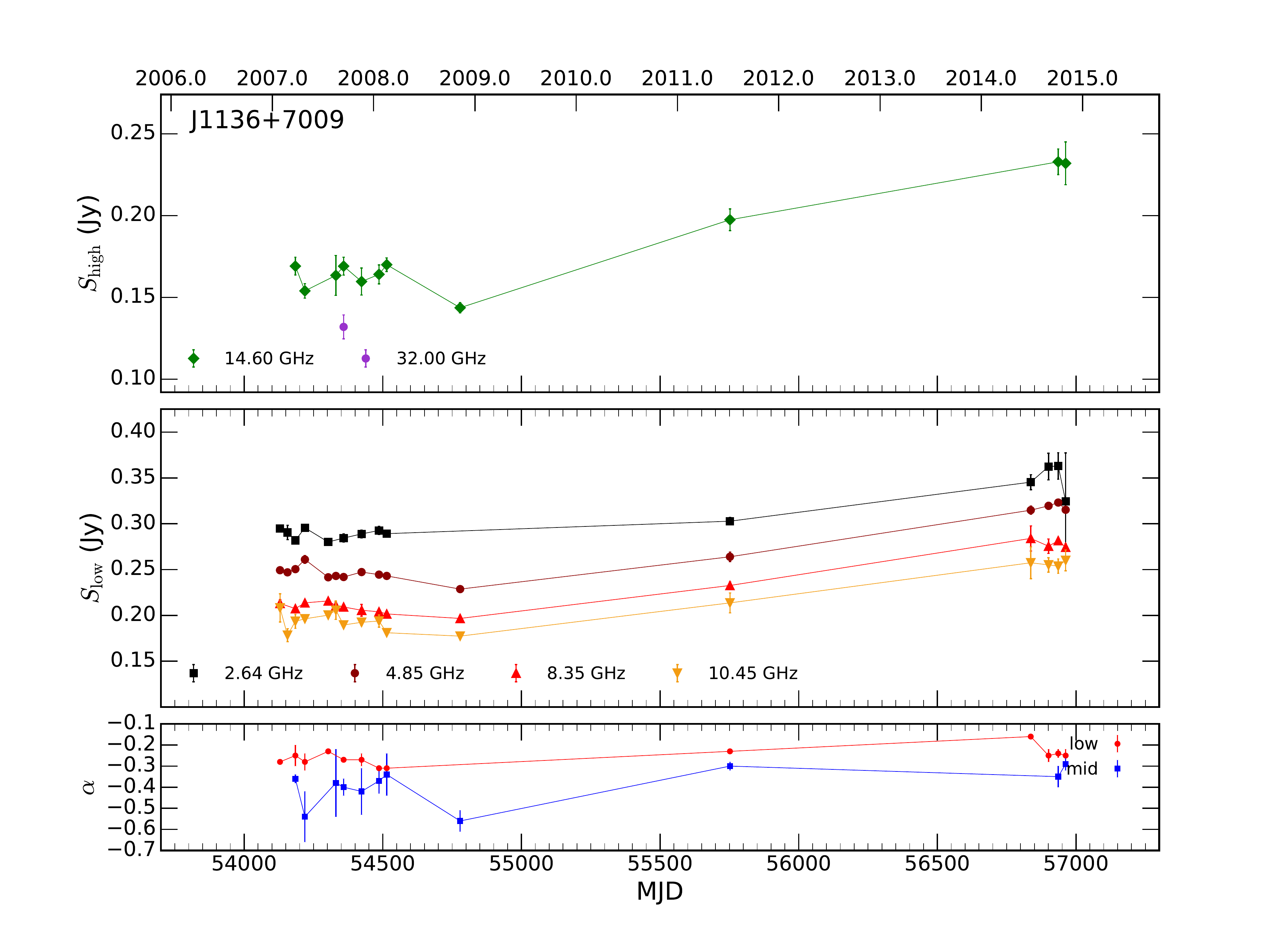}&
\includegraphics[trim=60pt 30pt 100pt 50pt  ,clip, width=0.49\textwidth,angle=0]{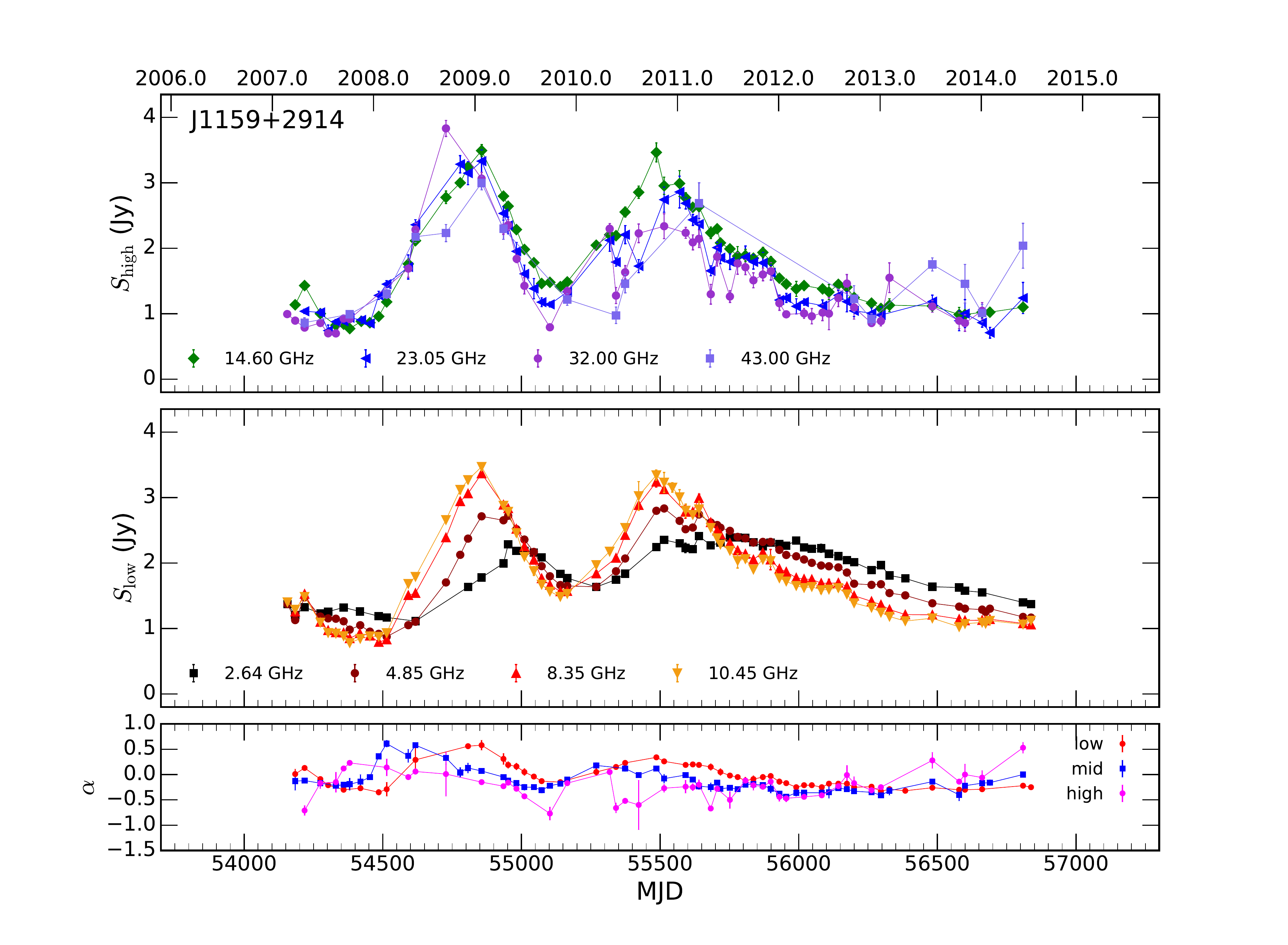}\\
\\[10pt]
\includegraphics[trim=60pt 30pt 100pt 50pt  ,clip, width=0.49\textwidth,angle=0]{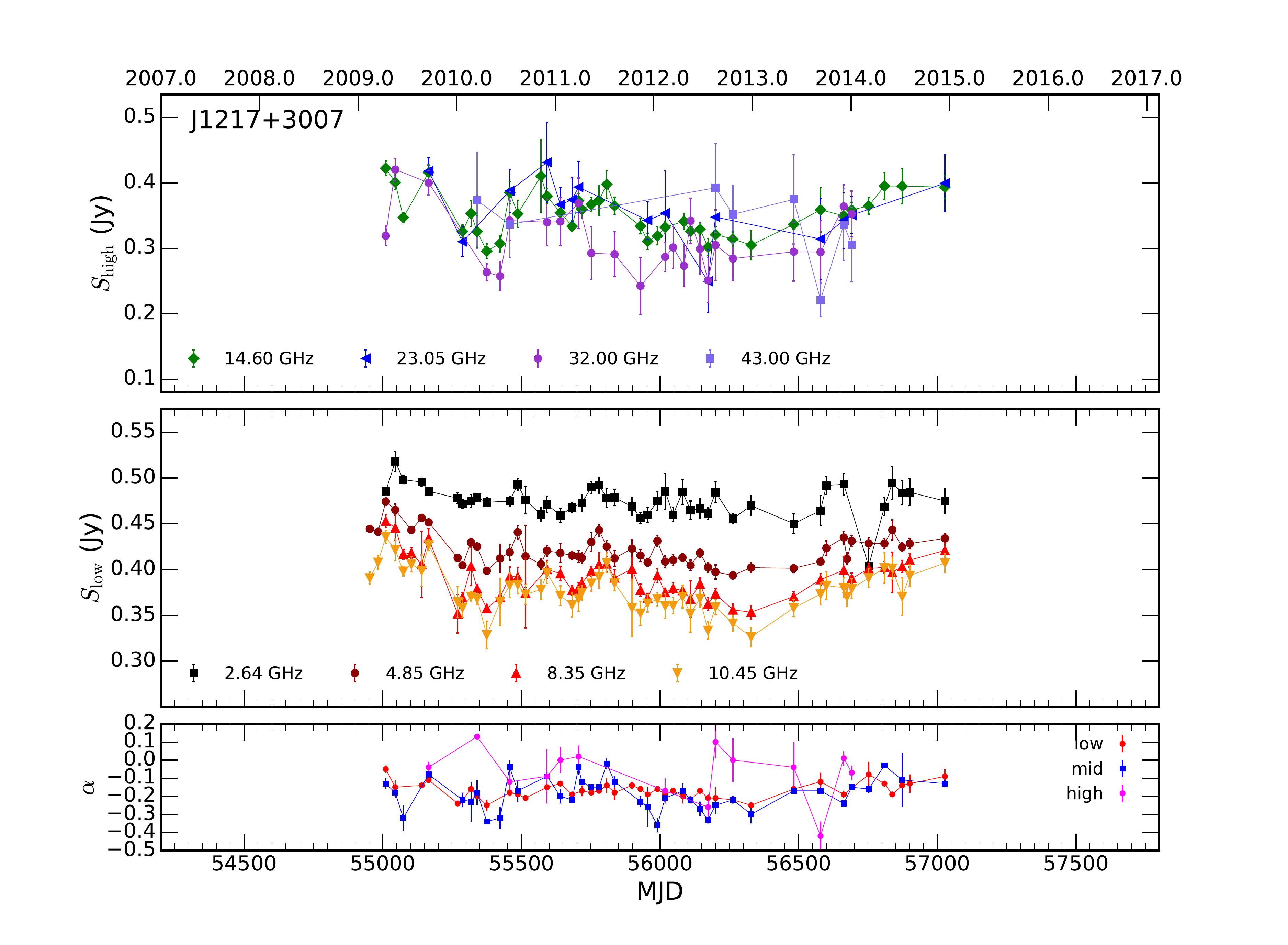}&
\includegraphics[trim=60pt 30pt 100pt 50pt  ,clip, width=0.49\textwidth,angle=0]{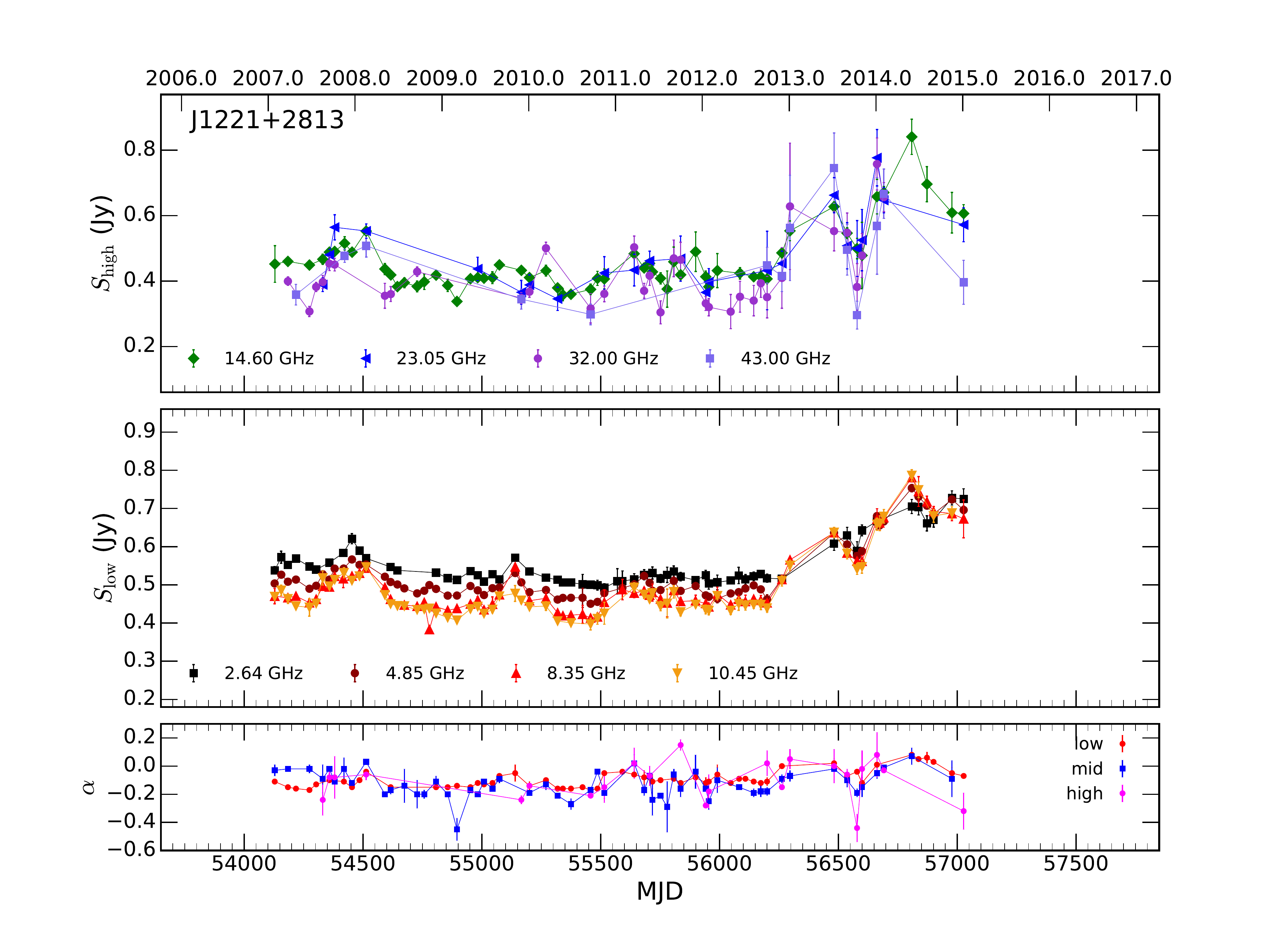}\\
\\[10pt]
\includegraphics[trim=60pt 30pt 100pt 50pt  ,clip, width=0.49\textwidth,angle=0]{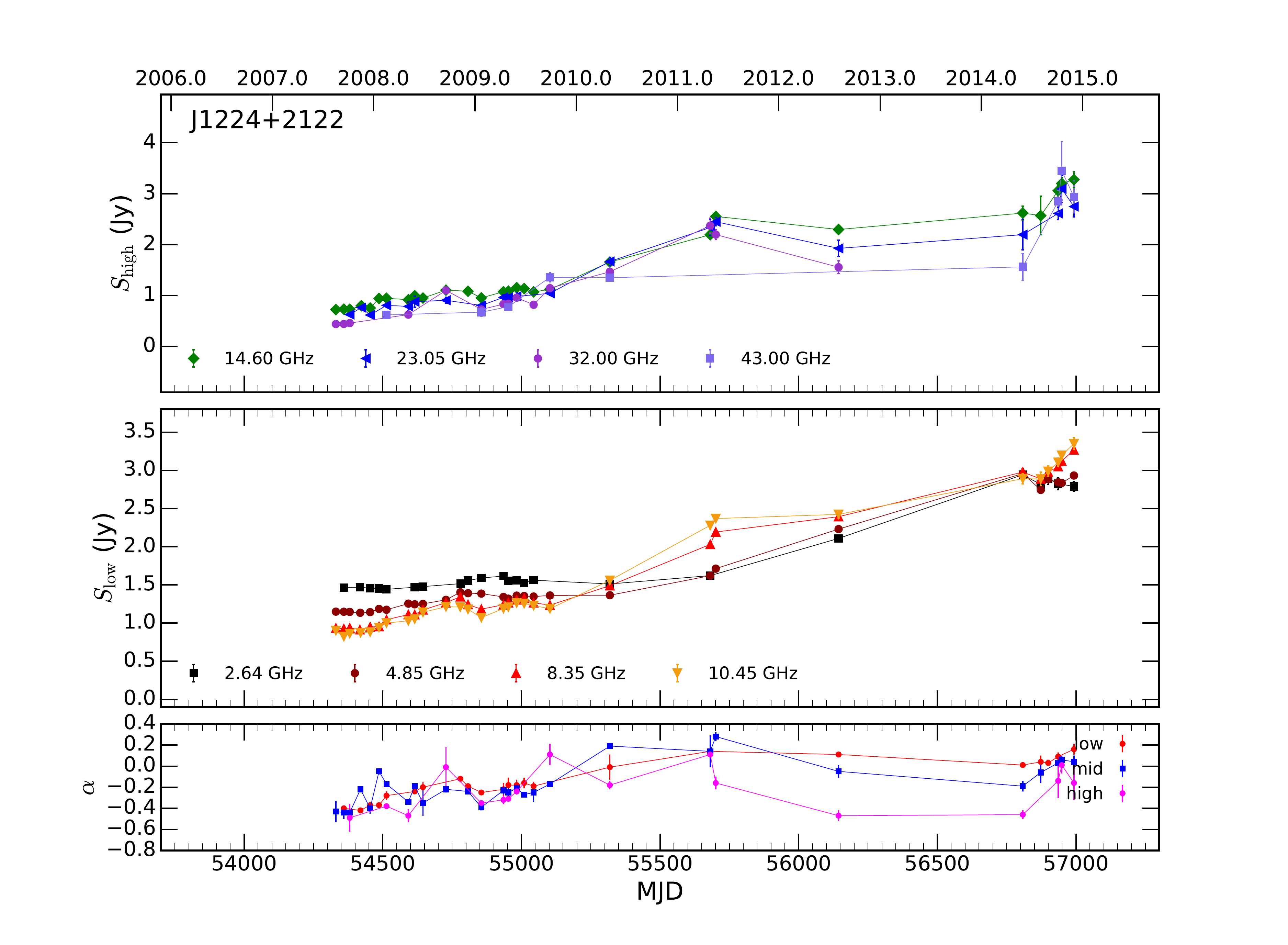}&
\includegraphics[trim=60pt 30pt 100pt 50pt  ,clip, width=0.49\textwidth,angle=0]{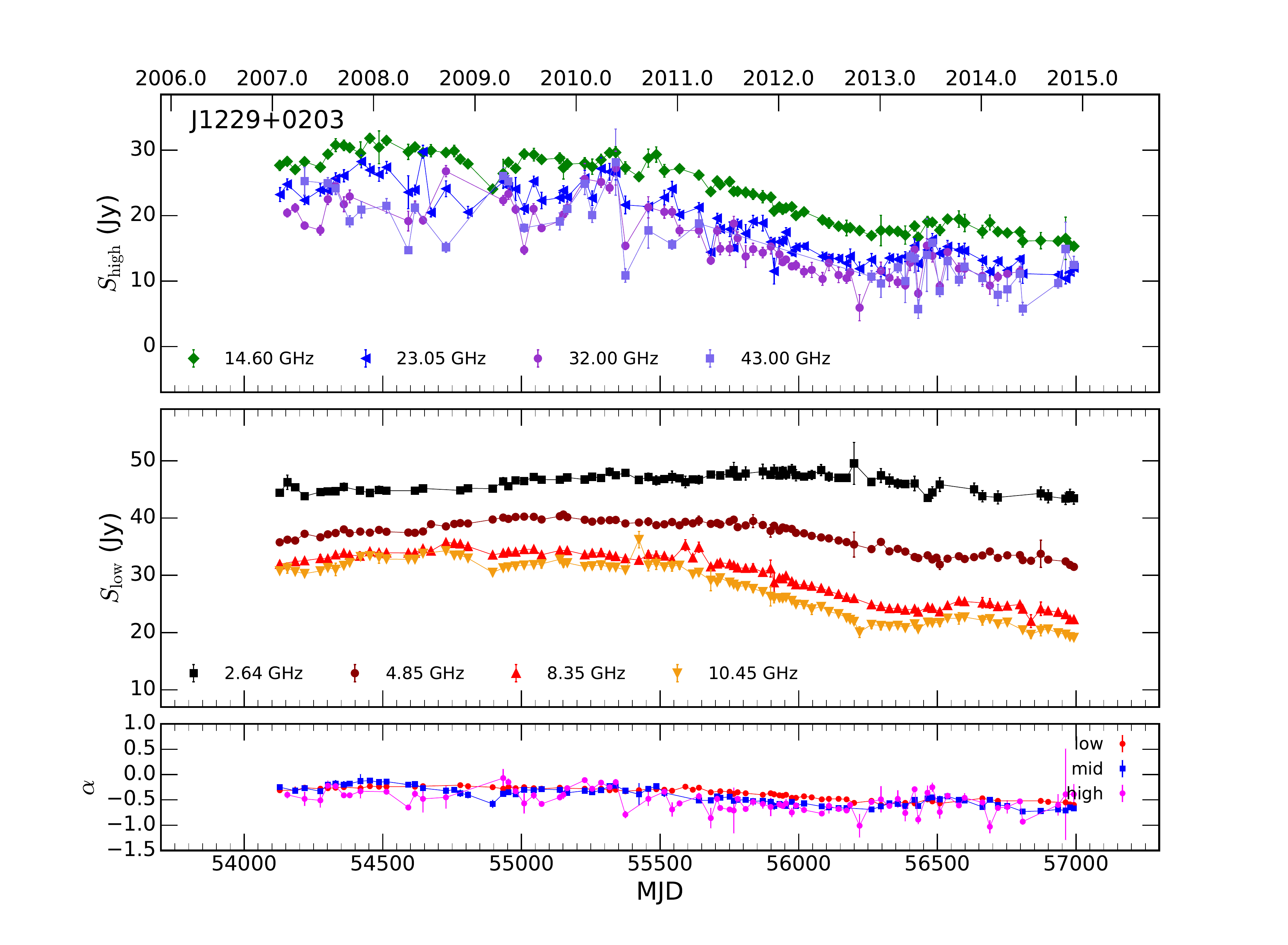}\\
\\[10pt]
\end{tabular}
\caption{Multi-frequency light curves for all the sources monitored by the \fg programme (``f'', ``s1'', ``s2'', ``old'') and the F-GAMMA-\textit{Planck} MoU. The lower panel in each frame shows the evolution of the low (2.64, 4.85 and 8.35~GHz) and mid-band (8.35, 10.45 and 14.6~GHz) and high-band (14.6, 23.05, 32, 43~GHz) spectral index. Only spectral index estimates from at least three frequencies are shown. Connecting lines have been used to guide the eye. }
\label{fig:sample_pg9}
\end{figure*}

\begin{figure*}[p]
\centering
\begin{tabular}{cc}
\includegraphics[trim=60pt 30pt 100pt 50pt  ,clip, width=0.49\textwidth,angle=0]{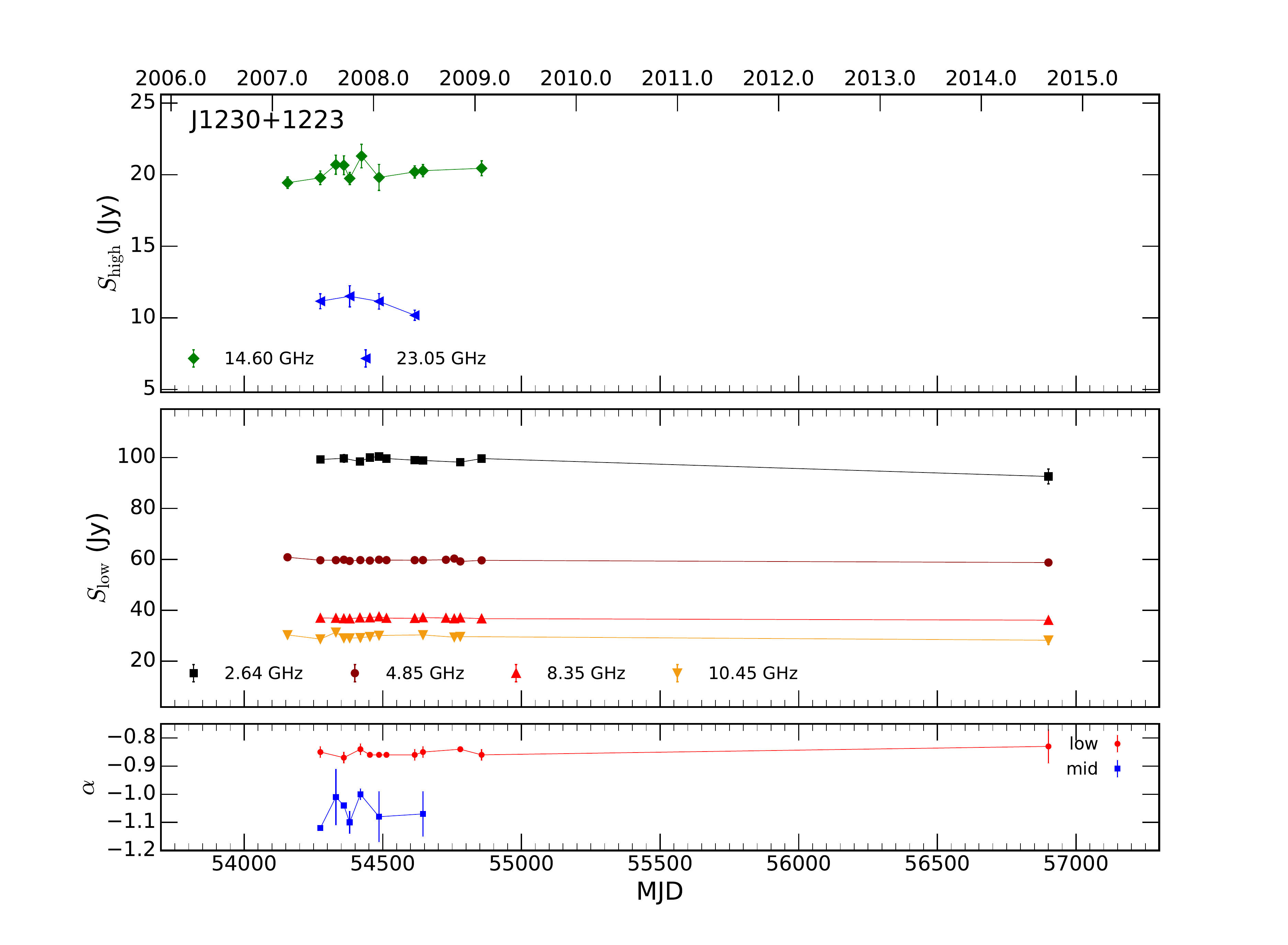}&
\includegraphics[trim=60pt 30pt 100pt 50pt  ,clip, width=0.49\textwidth,angle=0]{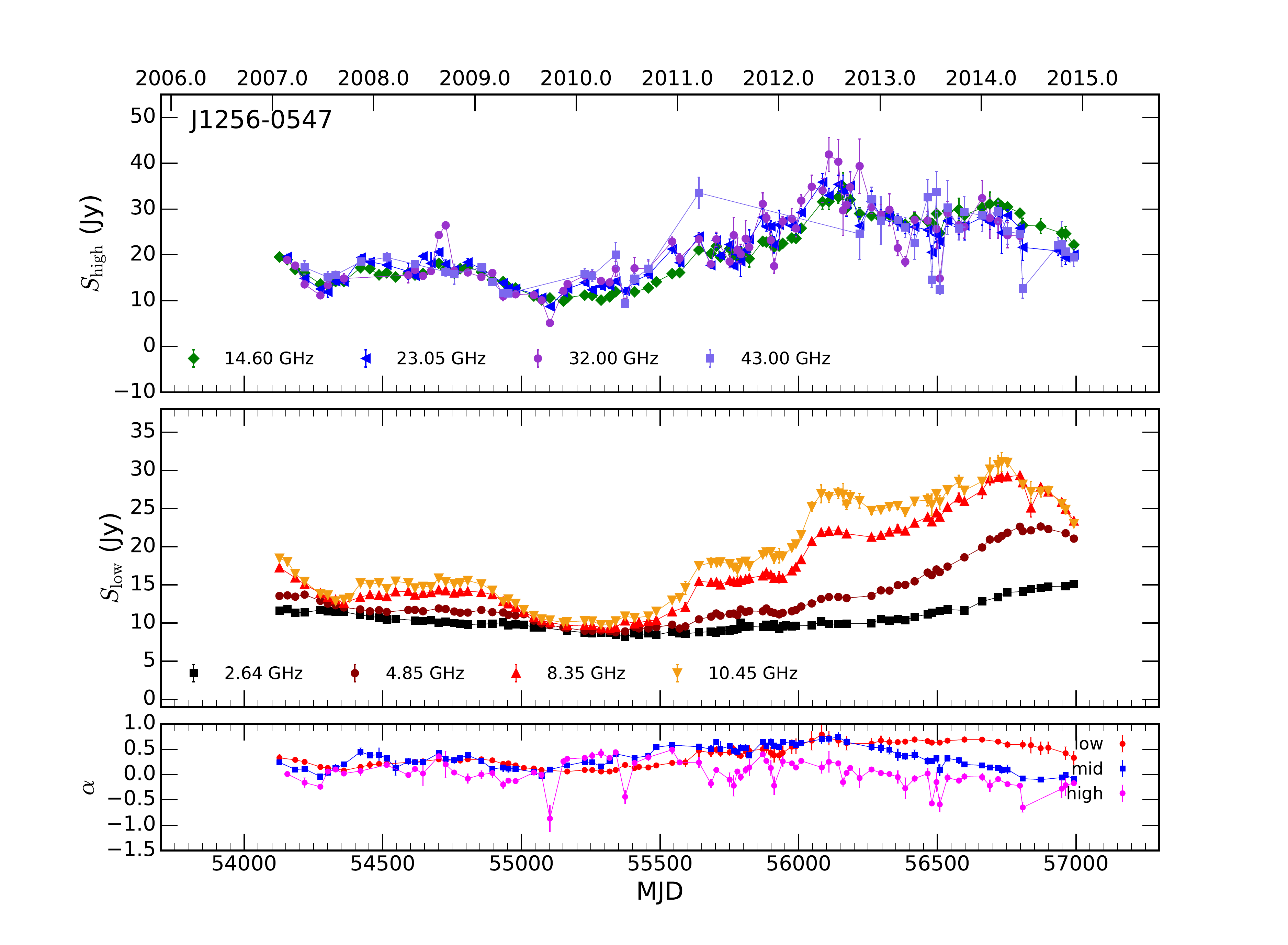}\\
\\[10pt]
\includegraphics[trim=60pt 30pt 100pt 50pt  ,clip, width=0.49\textwidth,angle=0]{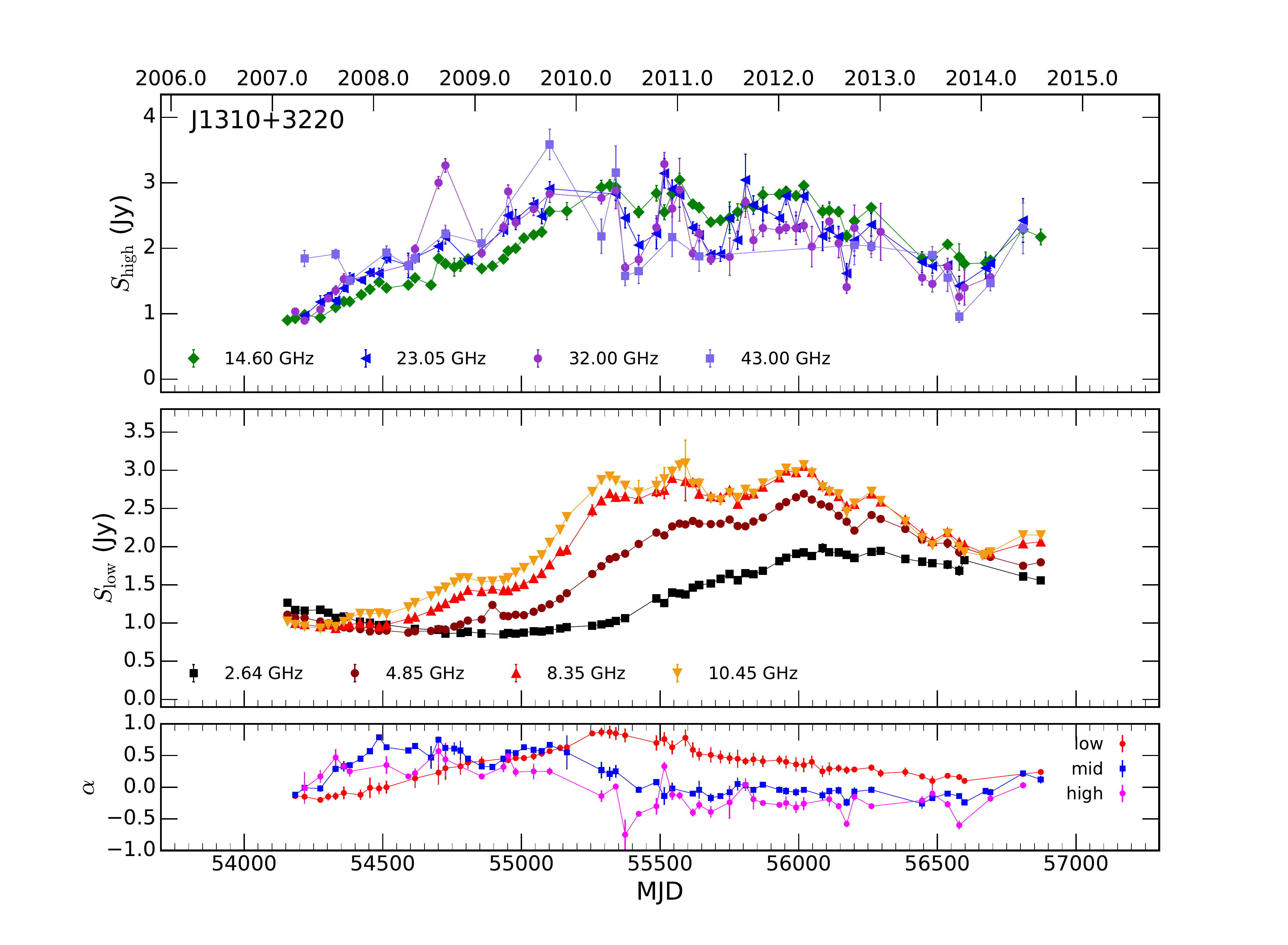}&
\includegraphics[trim=60pt 30pt 100pt 50pt  ,clip, width=0.49\textwidth,angle=0]{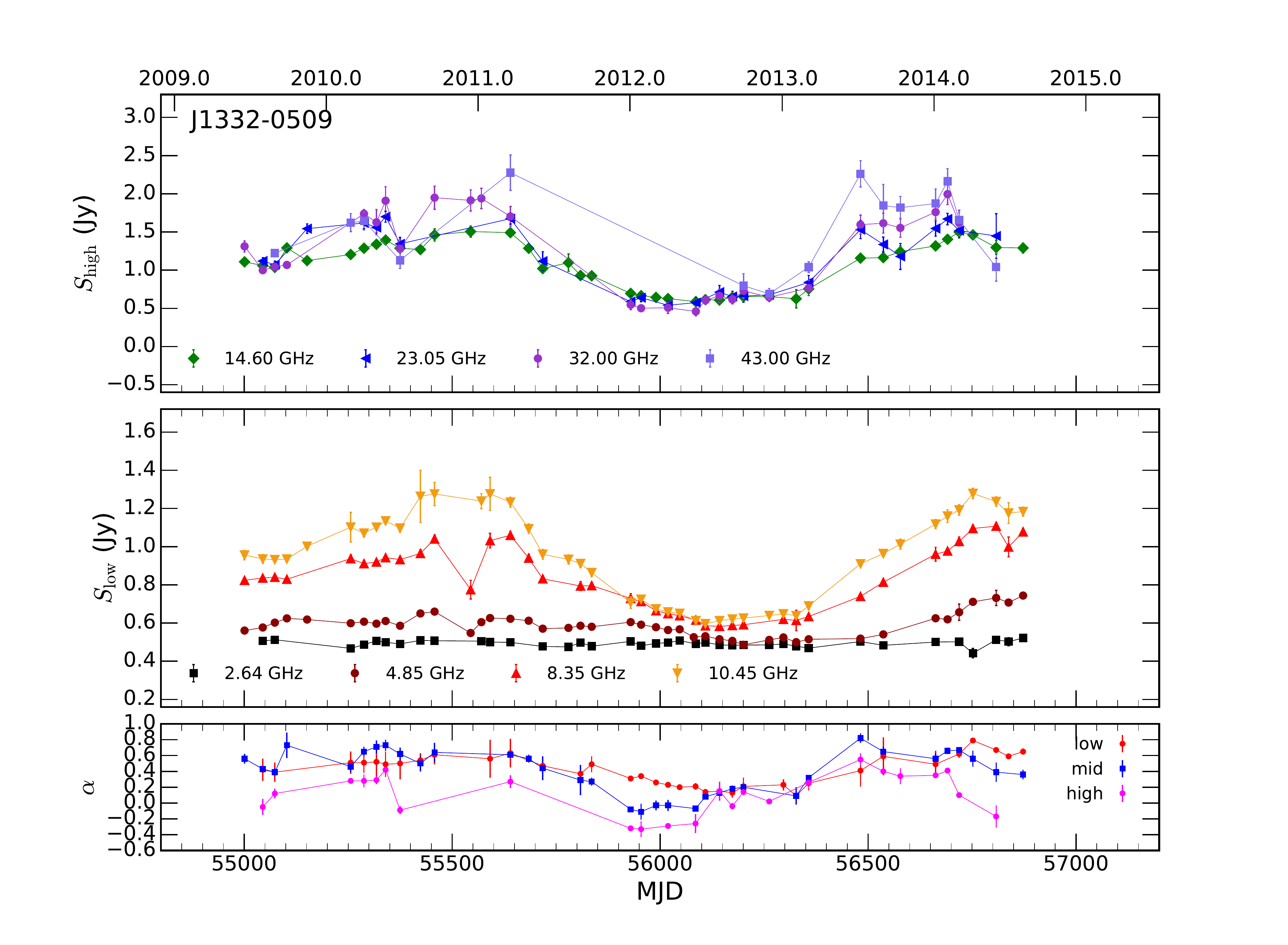}\\
\\[10pt]
\includegraphics[trim=60pt 30pt 100pt 50pt  ,clip, width=0.49\textwidth,angle=0]{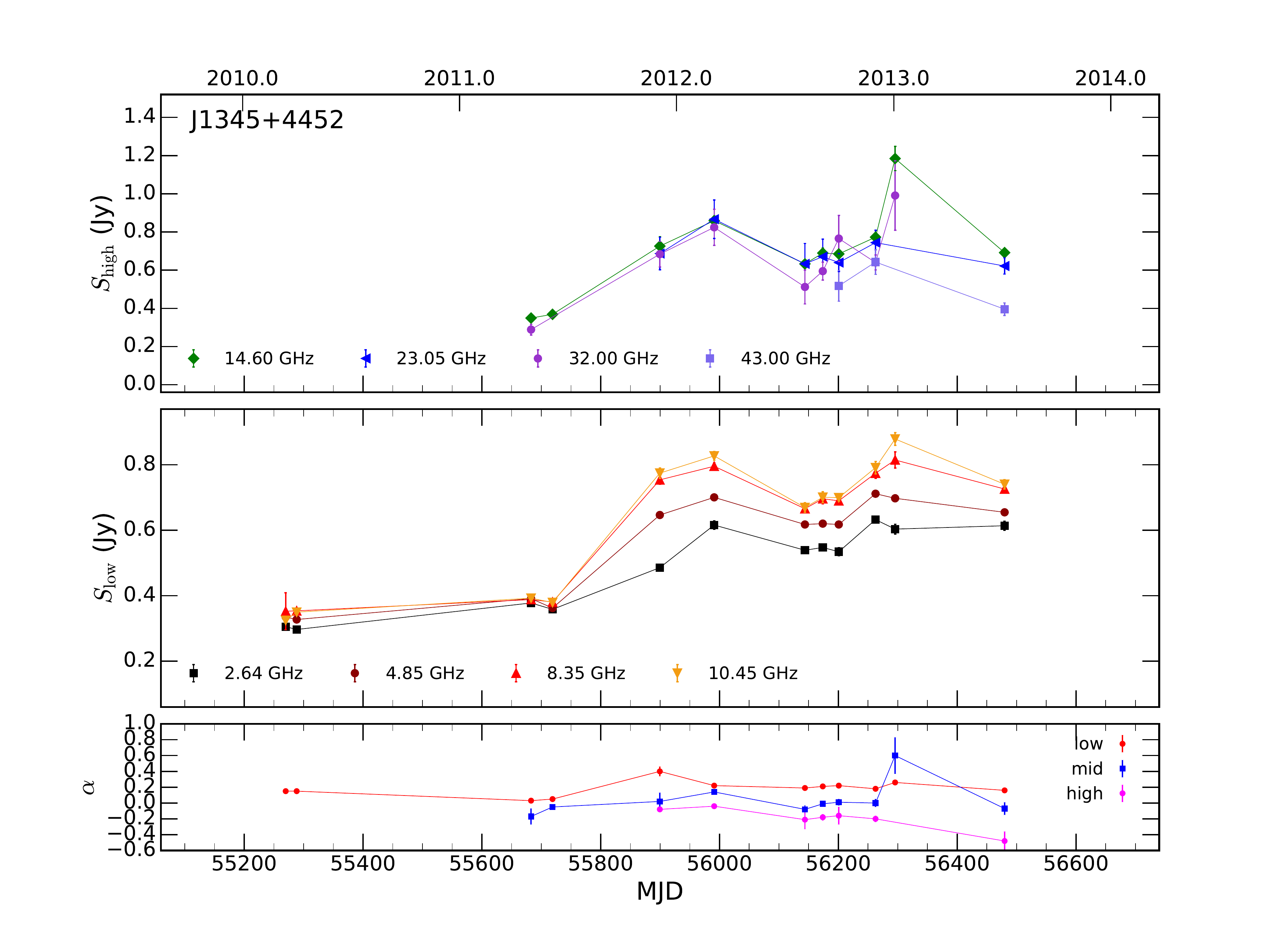}&
\includegraphics[trim=60pt 30pt 100pt 50pt  ,clip, width=0.49\textwidth,angle=0]{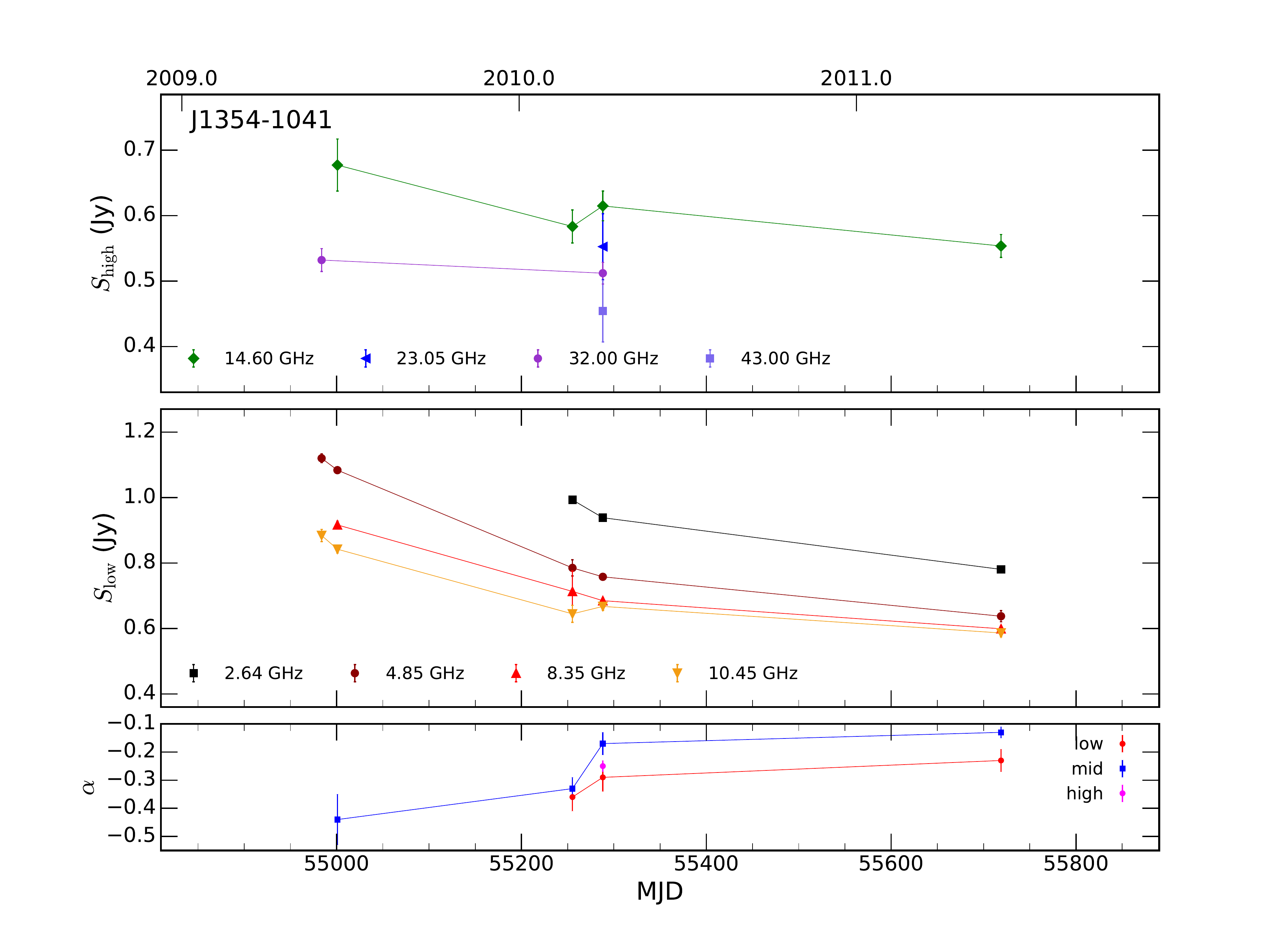}\\
\\[10pt]
\end{tabular}
\caption{Multi-frequency light curves for all the sources monitored by the \fg programme (``f'', ``s1'', ``s2'', ``old'') and the F-GAMMA-\textit{Planck} MoU. The lower panel in each frame shows the evolution of the low (2.64, 4.85 and 8.35~GHz) and mid-band (8.35, 10.45 and 14.6~GHz) and high-band (14.6, 23.05, 32, 43~GHz) spectral index. Only spectral index estimates from at least three frequencies are shown. Connecting lines have been used to guide the eye. }
\label{fig:sample_pg10}
\end{figure*}

\begin{figure*}[p]
\centering
\begin{tabular}{cc}
\includegraphics[trim=60pt 30pt 100pt 50pt  ,clip, width=0.49\textwidth,angle=0]{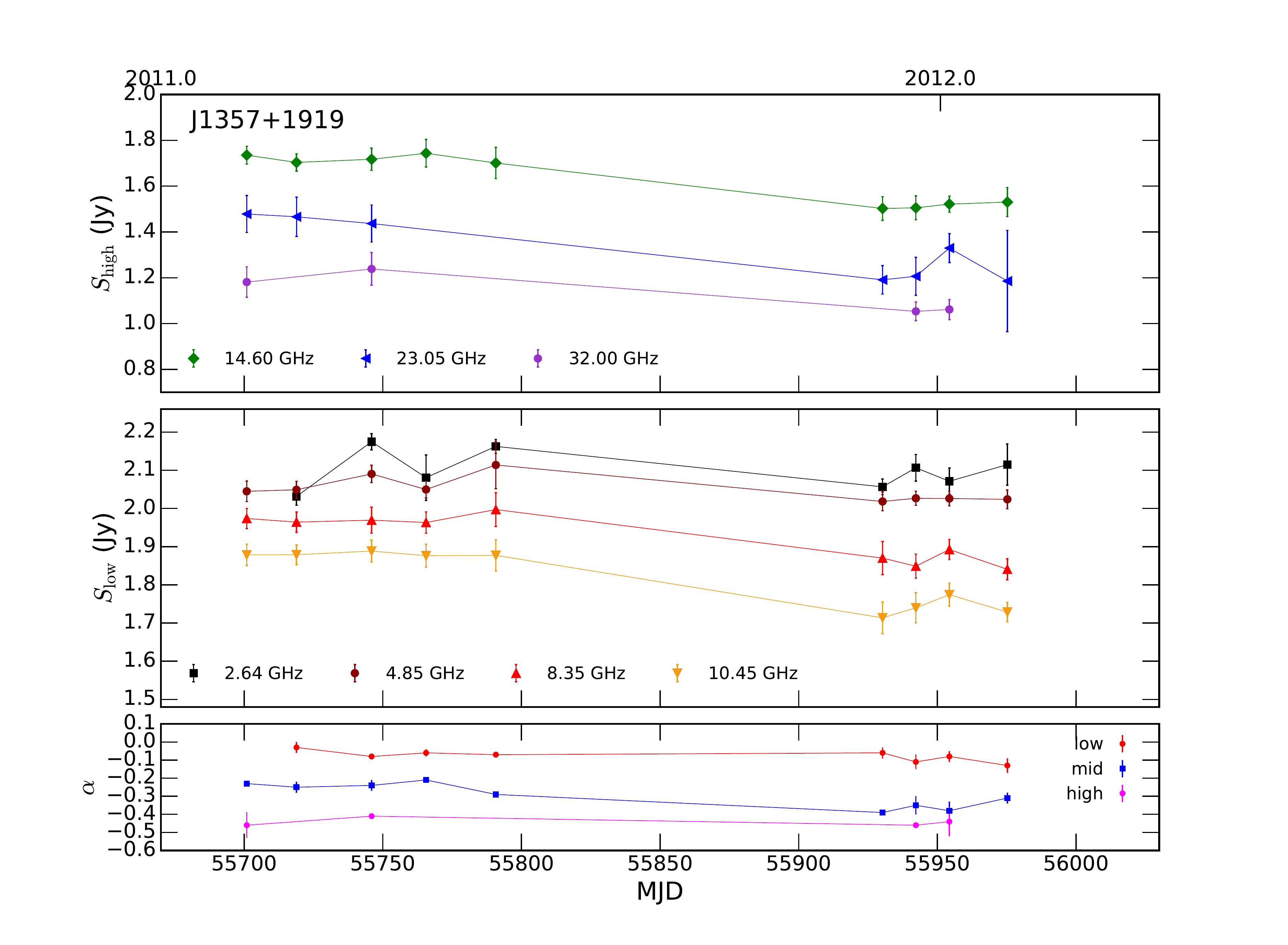}&
\includegraphics[trim=60pt 30pt 100pt 50pt  ,clip, width=0.49\textwidth,angle=0]{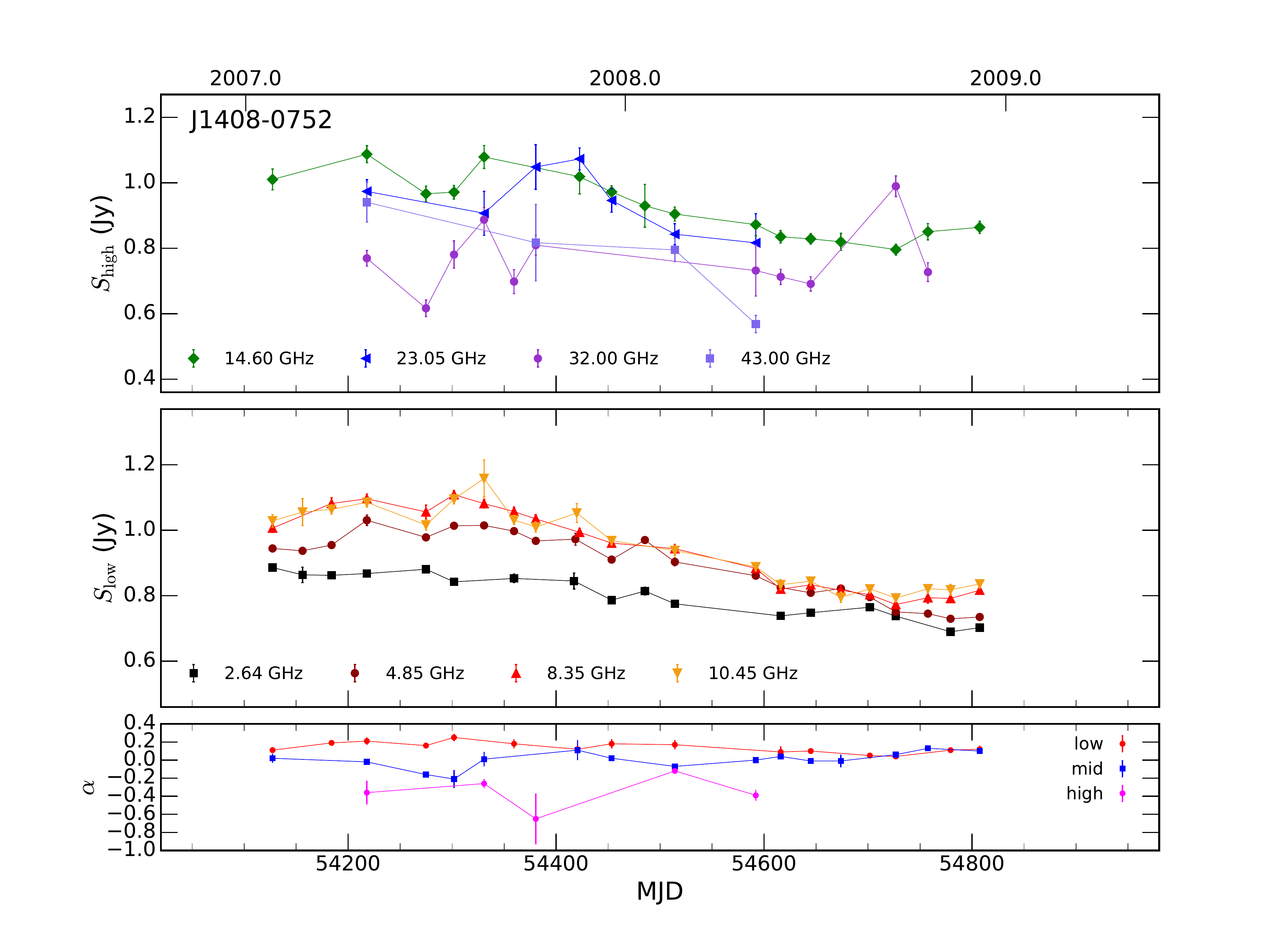}\\
\\[10pt]
\includegraphics[trim=60pt 30pt 100pt 50pt  ,clip, width=0.49\textwidth,angle=0]{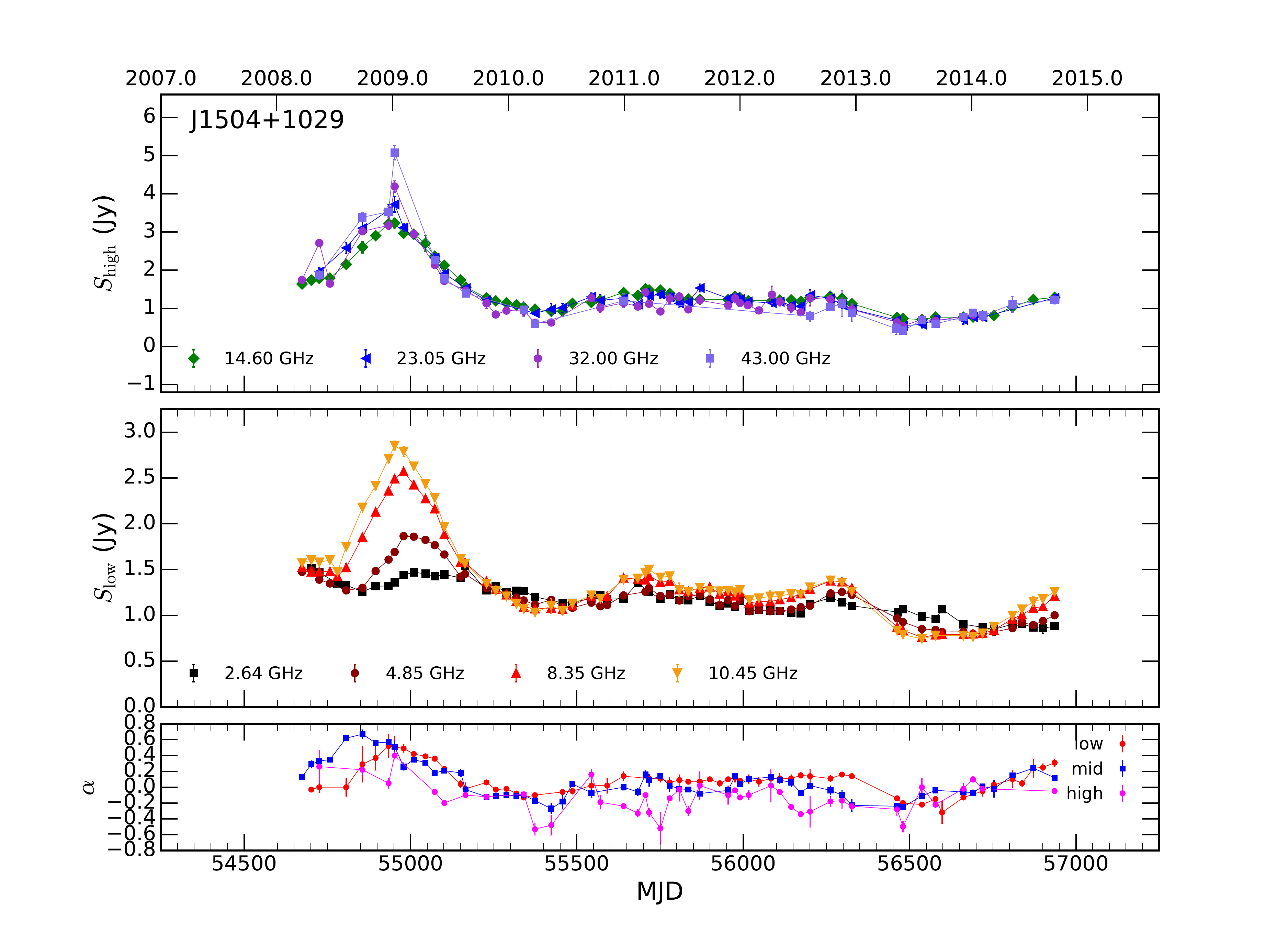}&
\includegraphics[trim=60pt 30pt 100pt 50pt  ,clip, width=0.49\textwidth,angle=0]{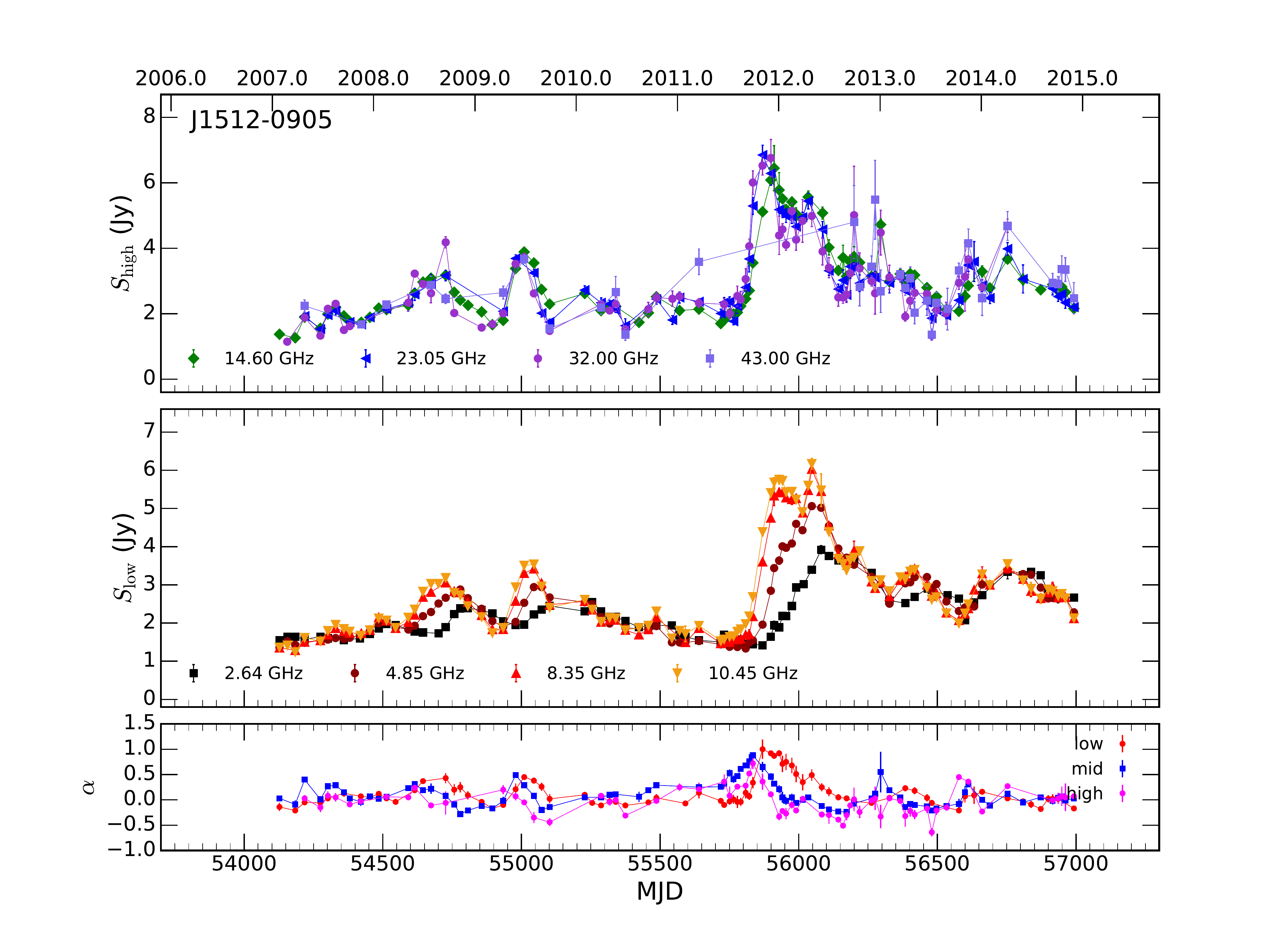}\\
\\[10pt]
\includegraphics[trim=60pt 30pt 100pt 50pt  ,clip, width=0.49\textwidth,angle=0]{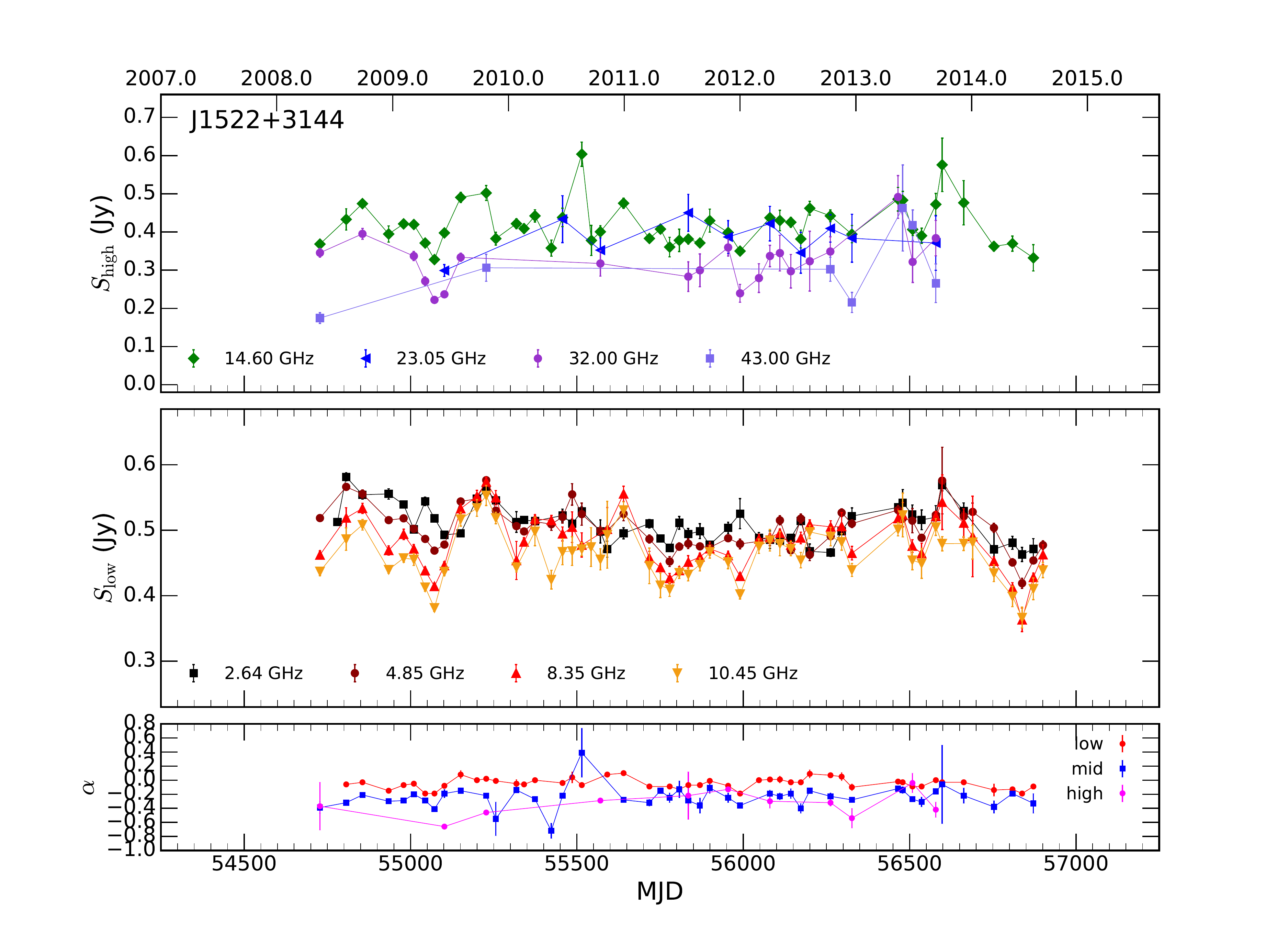}&
\includegraphics[trim=60pt 30pt 100pt 50pt  ,clip, width=0.49\textwidth,angle=0]{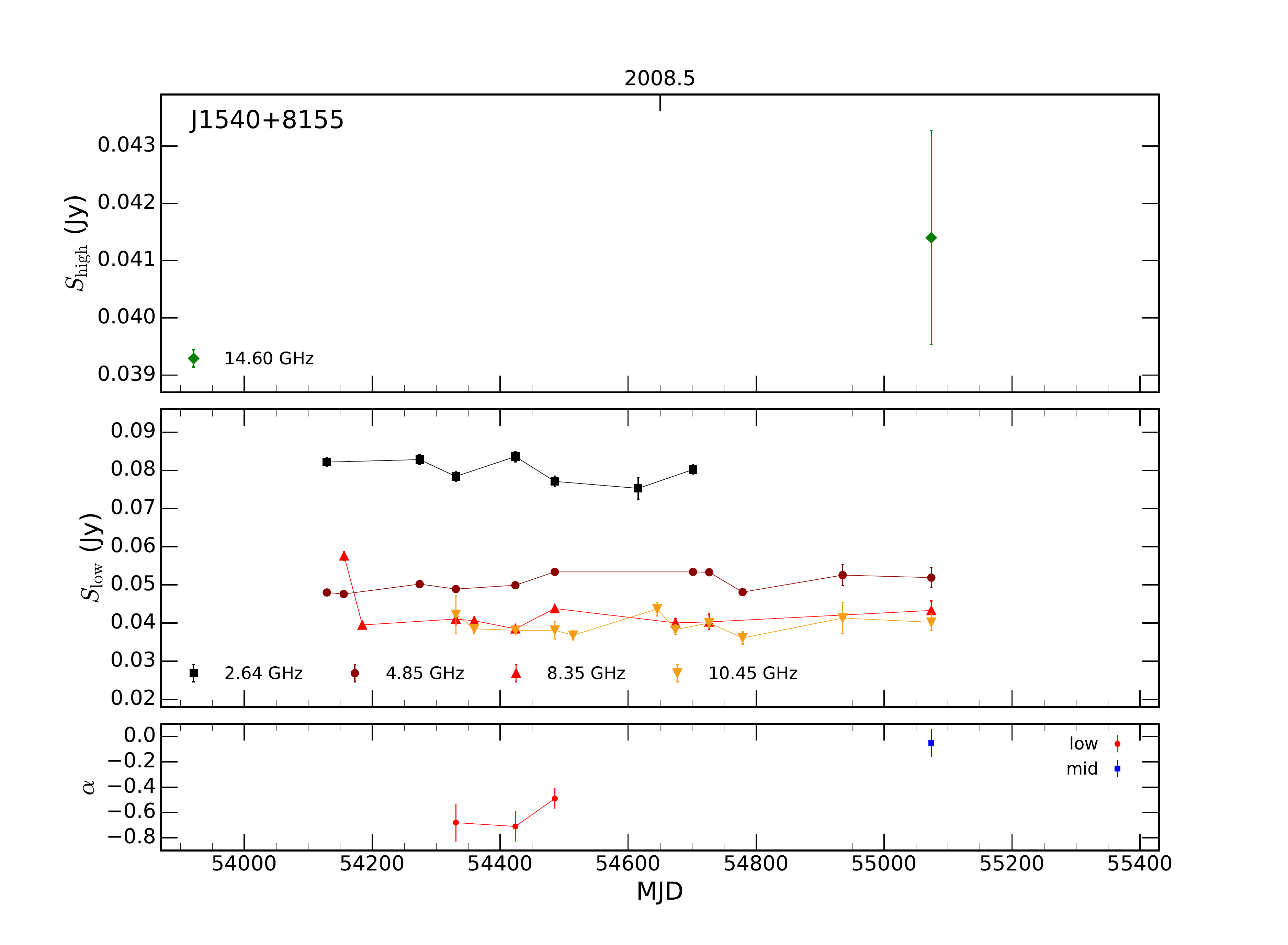}\\
\\[10pt]
\end{tabular}
\caption{Multi-frequency light curves for all the sources monitored by the \fg programme (``f'', ``s1'', ``s2'', ``old'') and the F-GAMMA-\textit{Planck} MoU. The lower panel in each frame shows the evolution of the low (2.64, 4.85 and 8.35~GHz) and mid-band (8.35, 10.45 and 14.6~GHz) and high-band (14.6, 23.05, 32, 43~GHz) spectral index. Only spectral index estimates from at least three frequencies are shown. Connecting lines have been used to guide the eye. }
\label{fig:sample_pg11}
\end{figure*}

\begin{figure*}[p]
\centering
\begin{tabular}{cc}
\includegraphics[trim=60pt 30pt 100pt 50pt  ,clip, width=0.49\textwidth,angle=0]{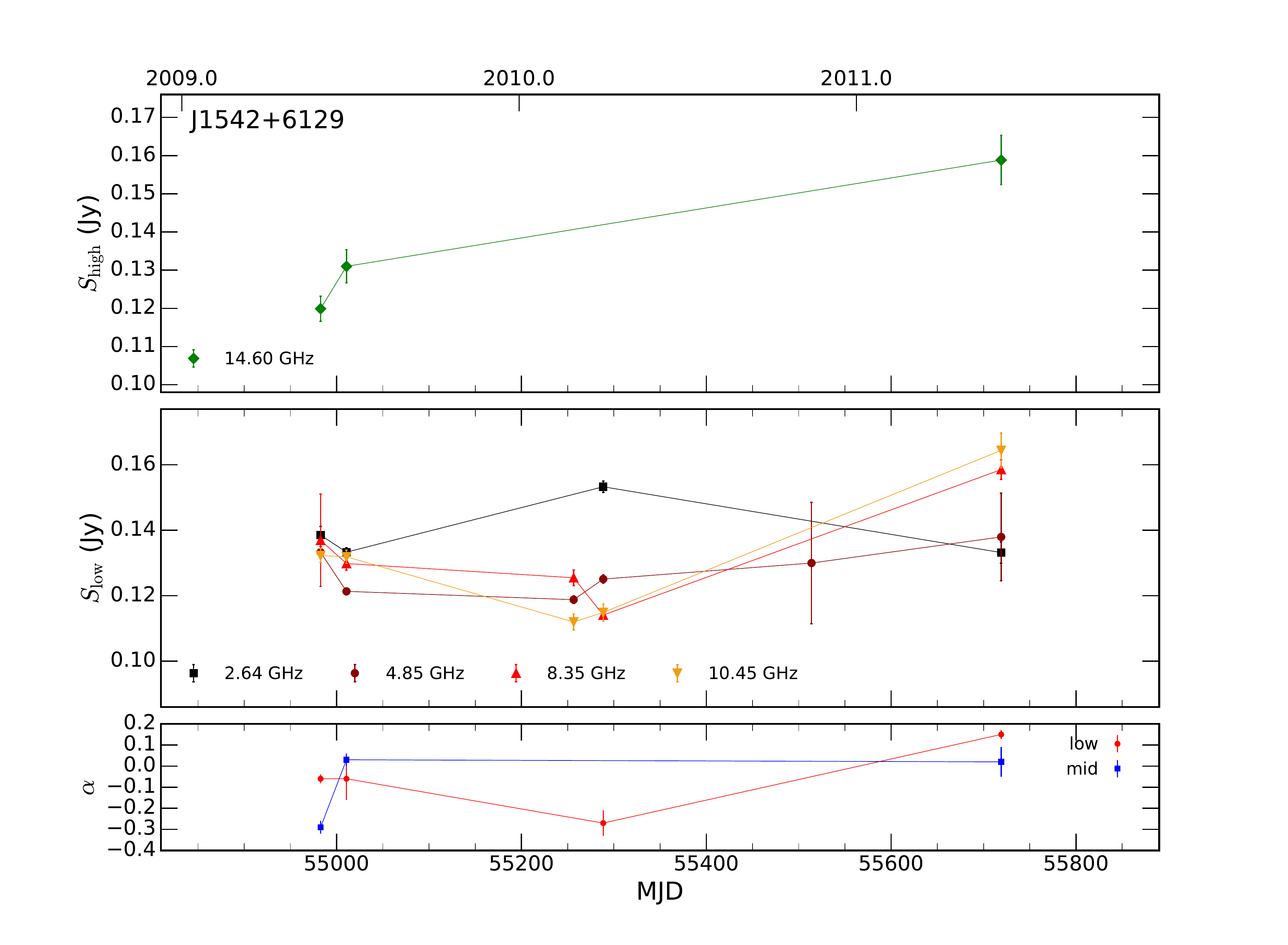}&
\includegraphics[trim=60pt 30pt 100pt 50pt  ,clip, width=0.49\textwidth,angle=0]{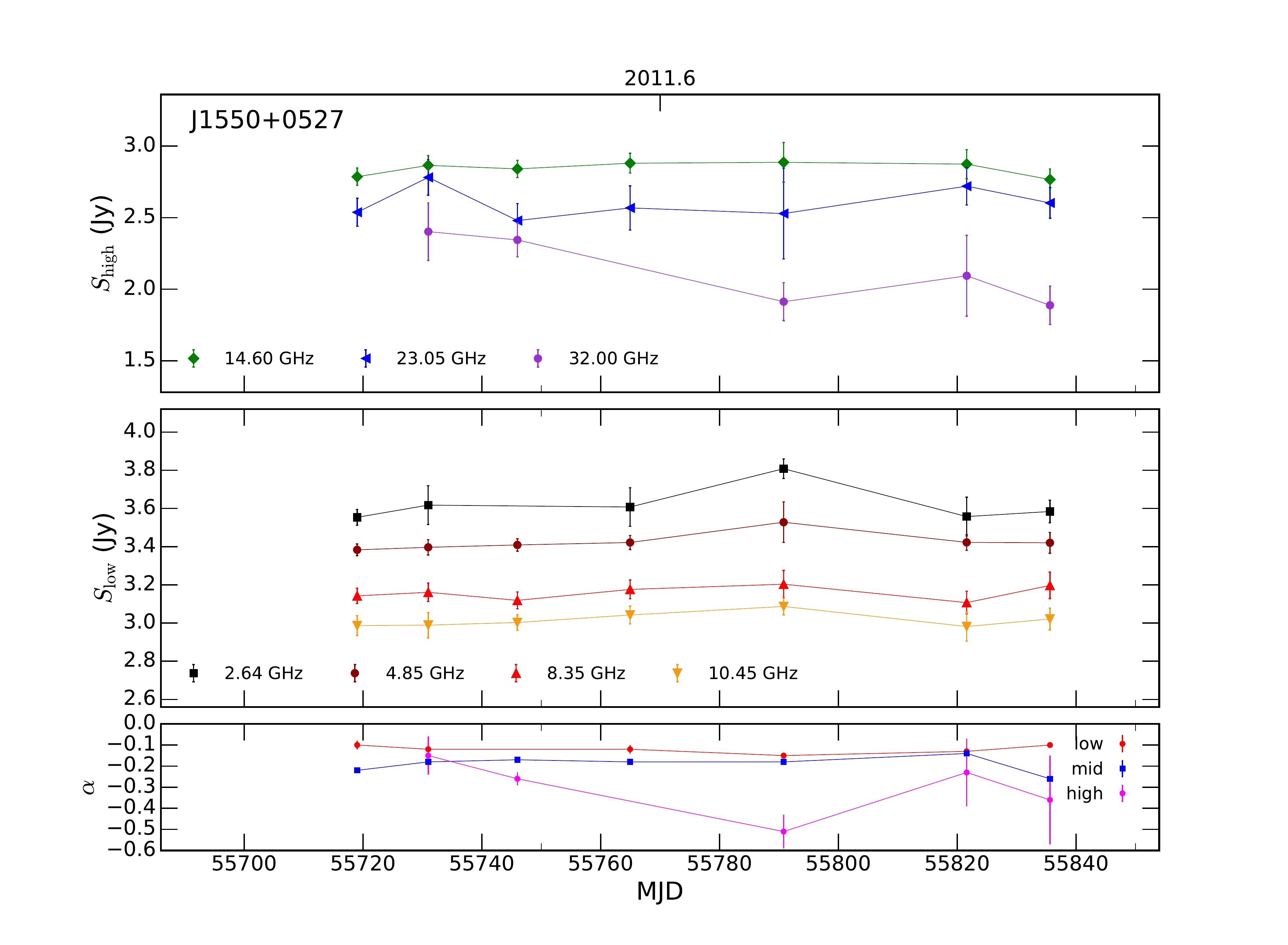}\\
\\[10pt]
\includegraphics[trim=60pt 30pt 100pt 50pt  ,clip, width=0.49\textwidth,angle=0]{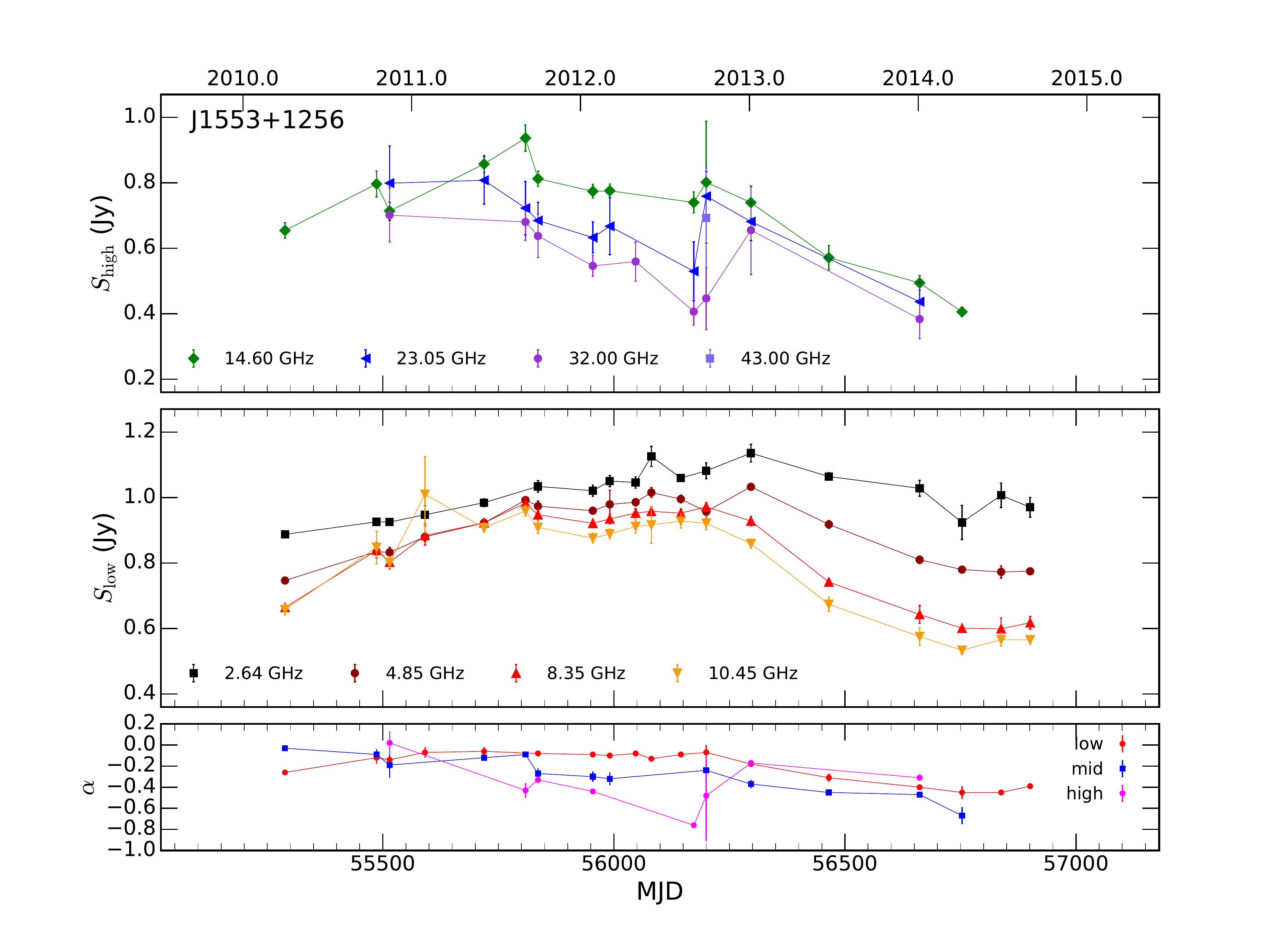}&
\includegraphics[trim=60pt 30pt 100pt 50pt  ,clip, width=0.49\textwidth,angle=0]{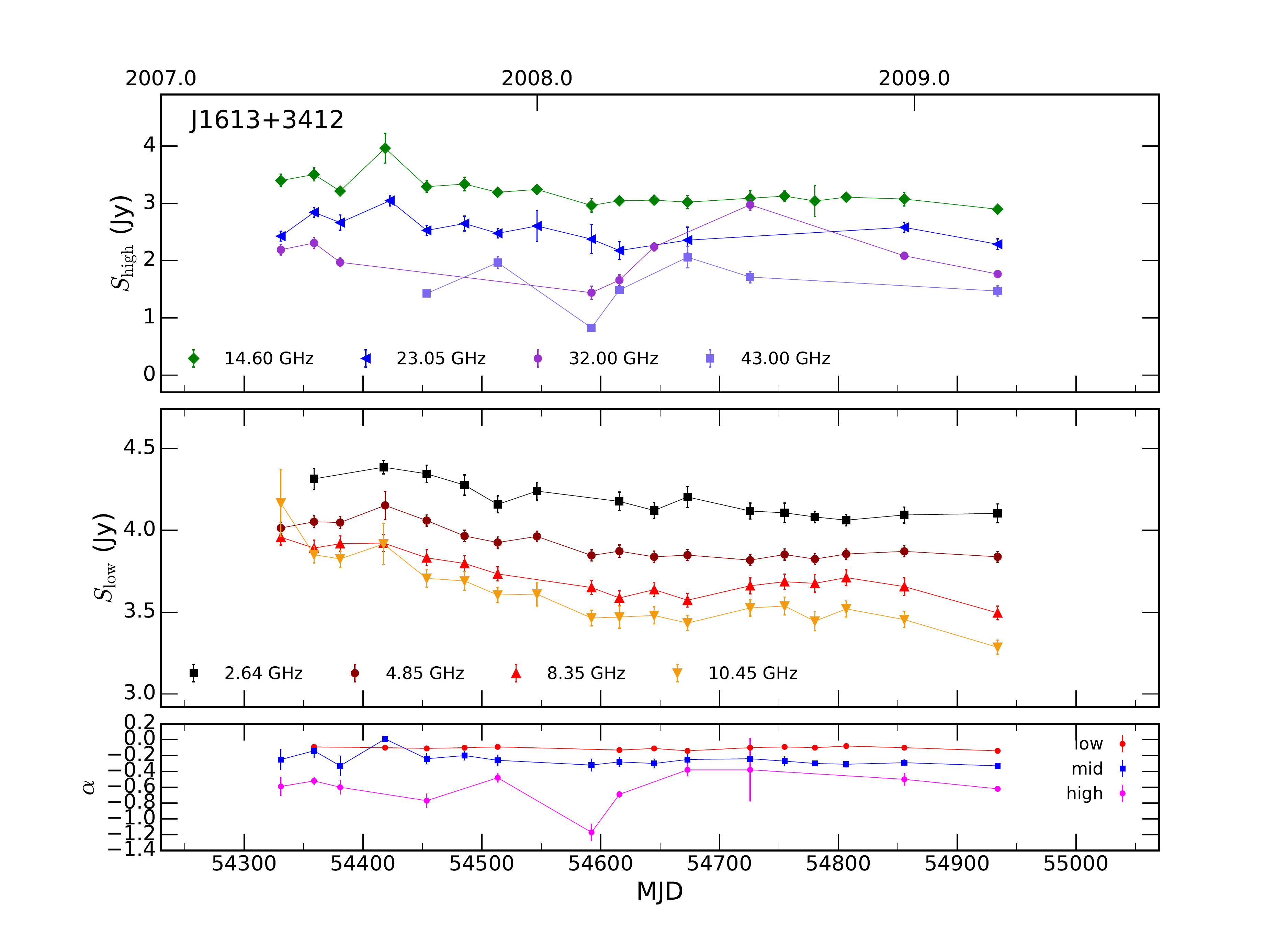}\\
\\[10pt]
\includegraphics[trim=60pt 30pt 100pt 50pt  ,clip, width=0.49\textwidth,angle=0]{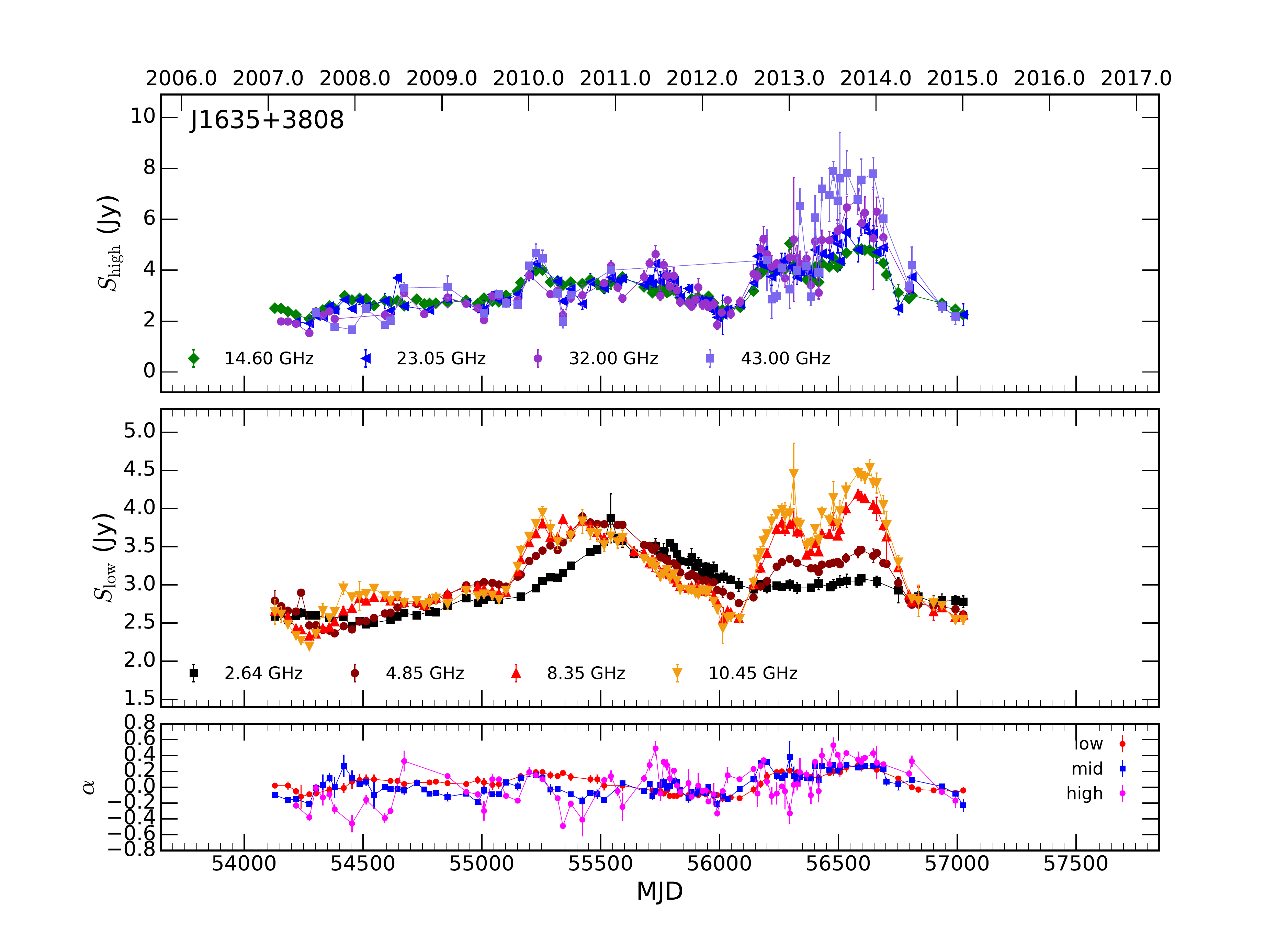}&
\includegraphics[trim=60pt 30pt 100pt 50pt  ,clip, width=0.49\textwidth,angle=0]{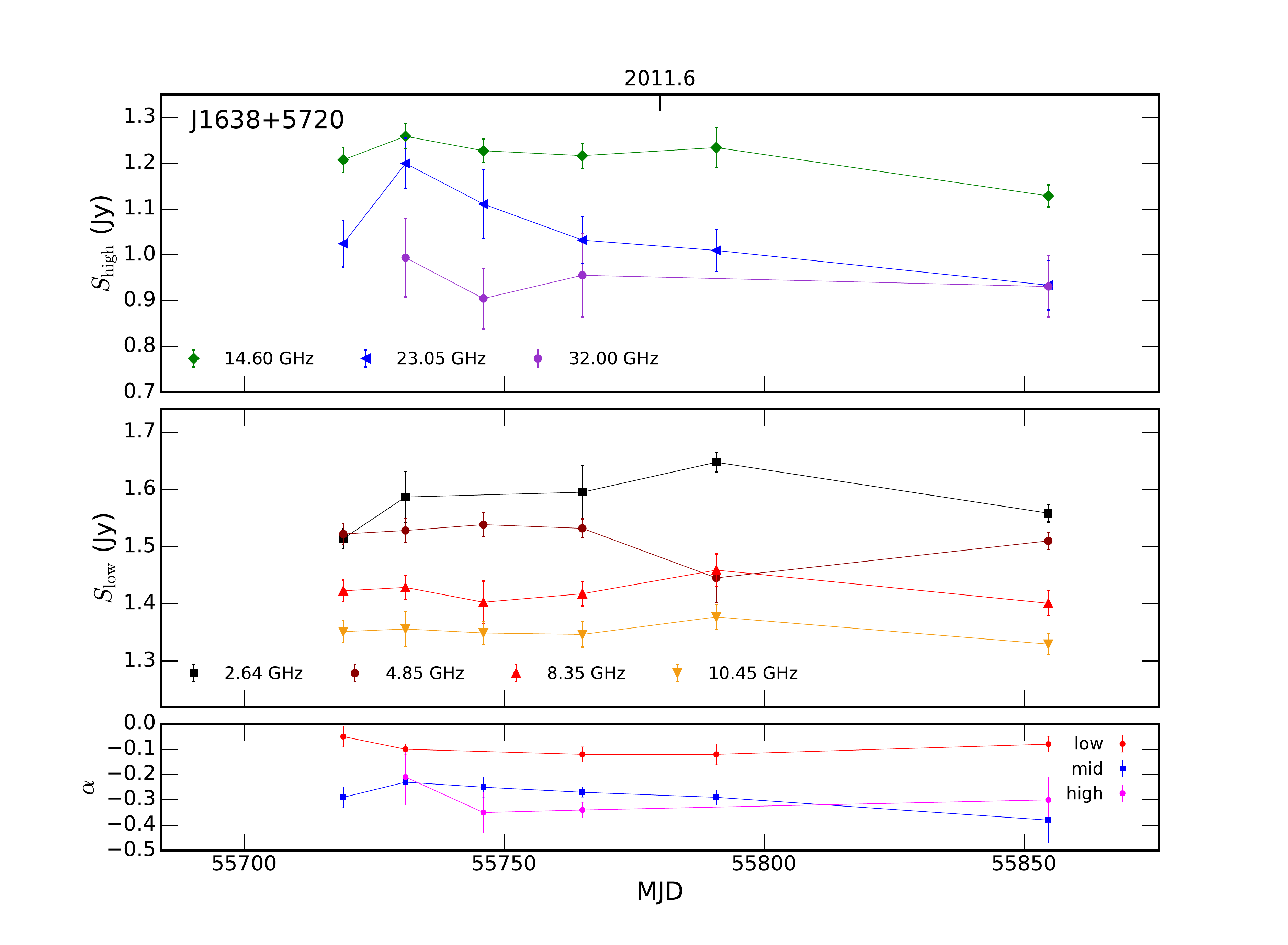}\\
\\[10pt]
\end{tabular}
\caption{Multi-frequency light curves for all the sources monitored by the \fg programme (``f'', ``s1'', ``s2'', ``old'') and the F-GAMMA-\textit{Planck} MoU. The lower panel in each frame shows the evolution of the low (2.64, 4.85 and 8.35~GHz) and mid-band (8.35, 10.45 and 14.6~GHz) and high-band (14.6, 23.05, 32, 43~GHz) spectral index. Only spectral index estimates from at least three frequencies are shown. Connecting lines have been used to guide the eye. }
\label{fig:sample_pg12}
\end{figure*}

\begin{figure*}[p]
\centering
\begin{tabular}{cc}
\includegraphics[trim=60pt 30pt 100pt 50pt  ,clip, width=0.49\textwidth,angle=0]{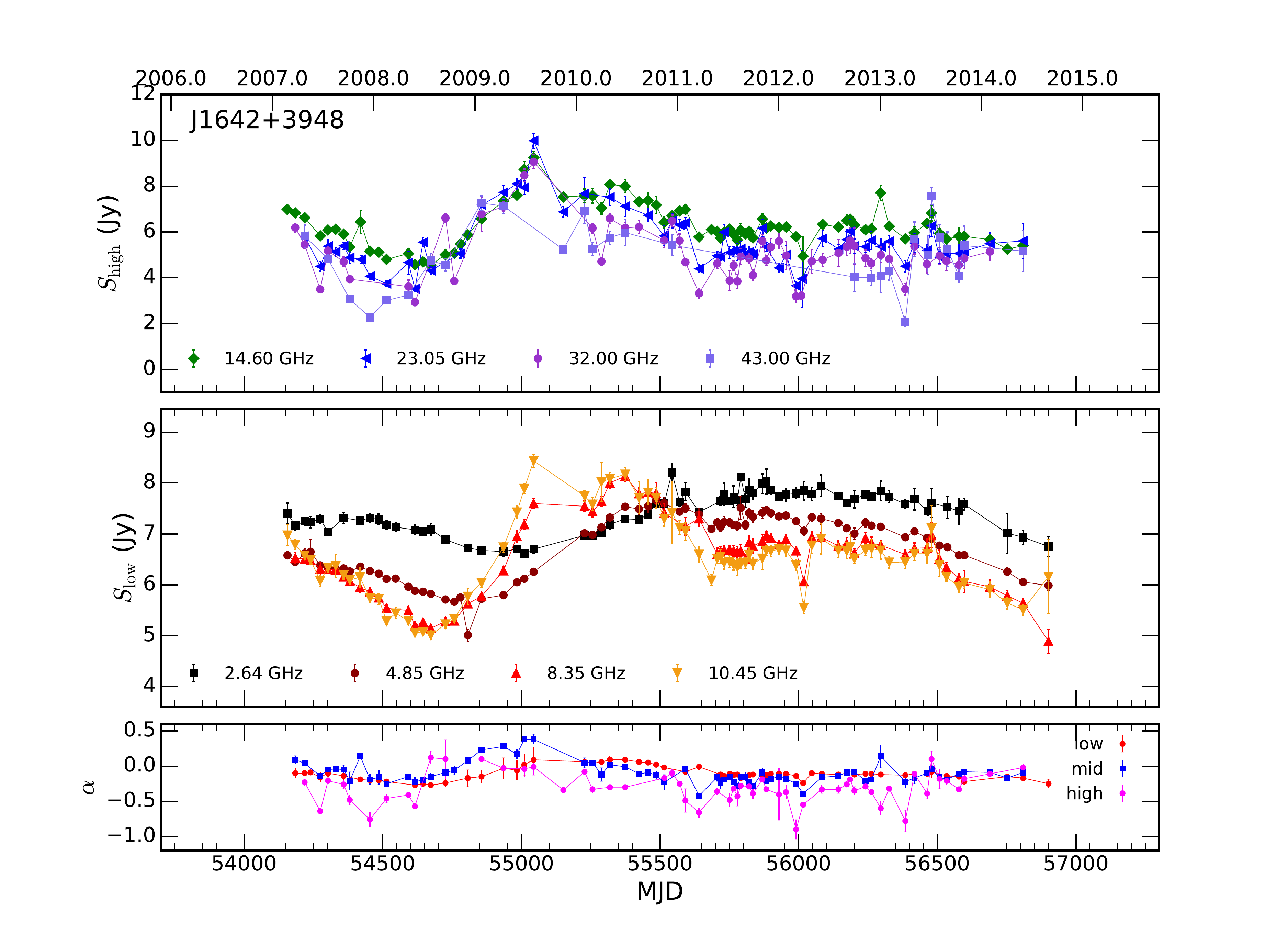}&
\includegraphics[trim=60pt 30pt 100pt 50pt  ,clip, width=0.49\textwidth,angle=0]{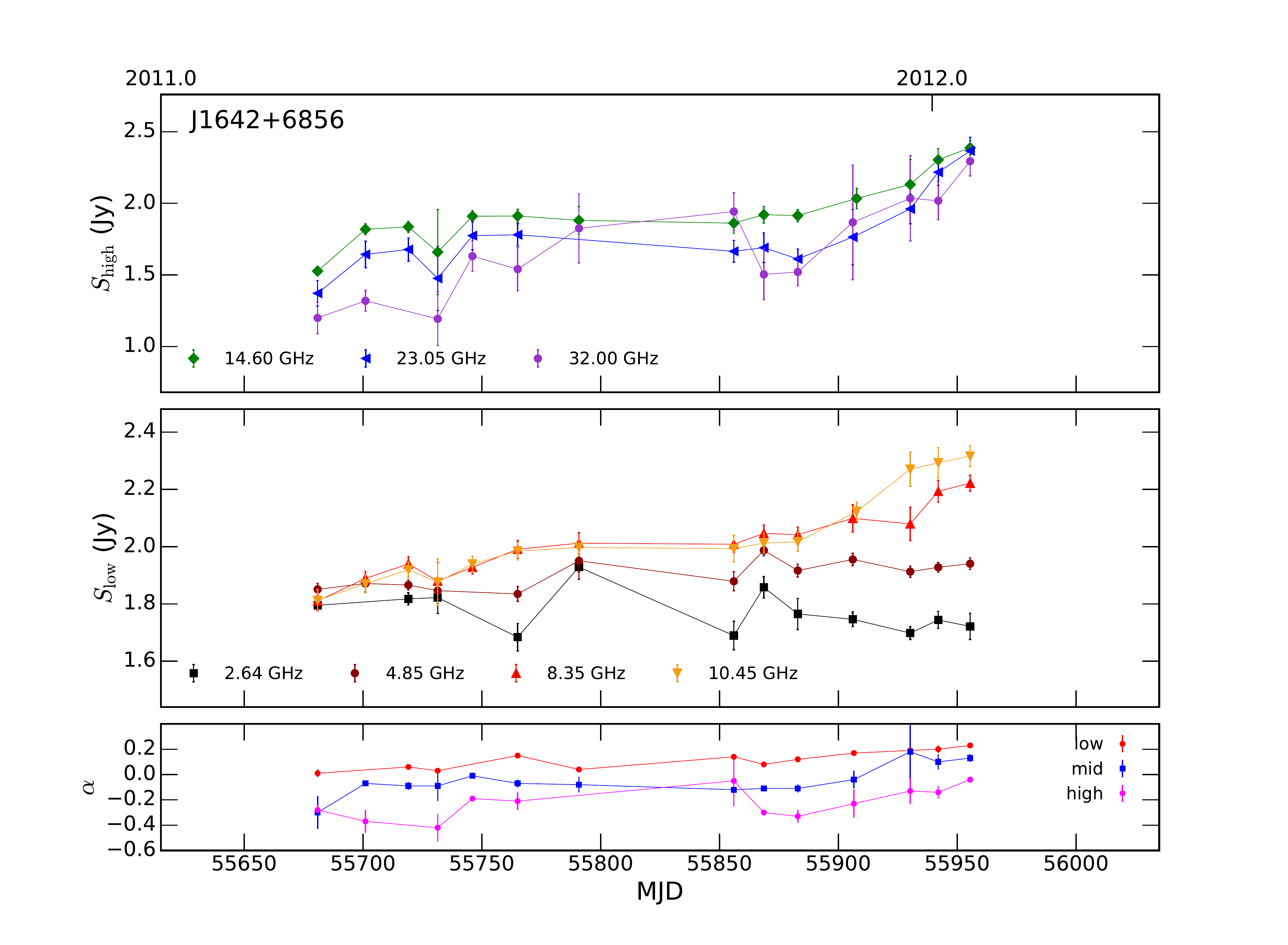}\\
\\[10pt]
\includegraphics[trim=60pt 30pt 100pt 50pt  ,clip, width=0.49\textwidth,angle=0]{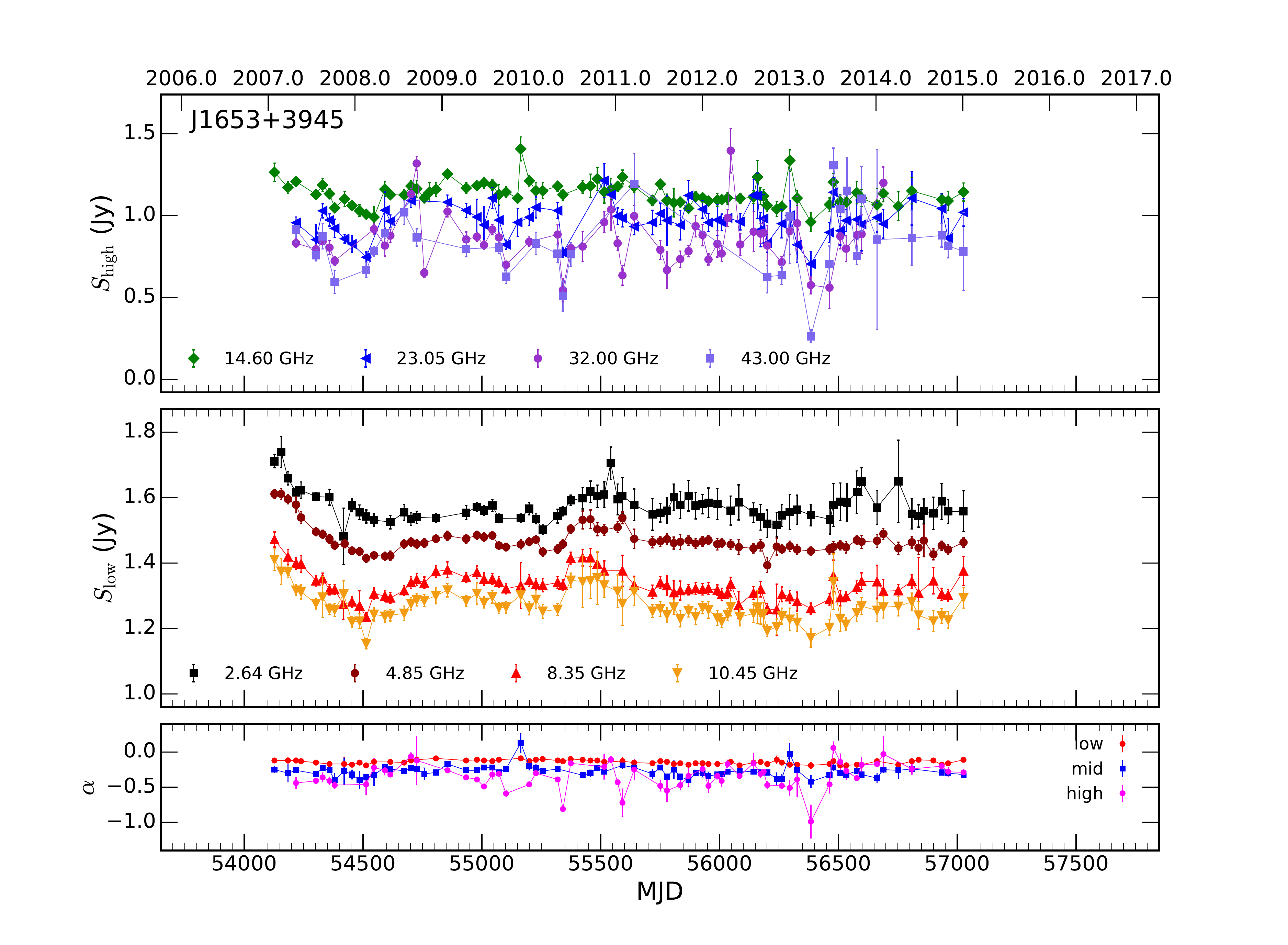}&
\includegraphics[trim=60pt 30pt 100pt 50pt  ,clip, width=0.49\textwidth,angle=0]{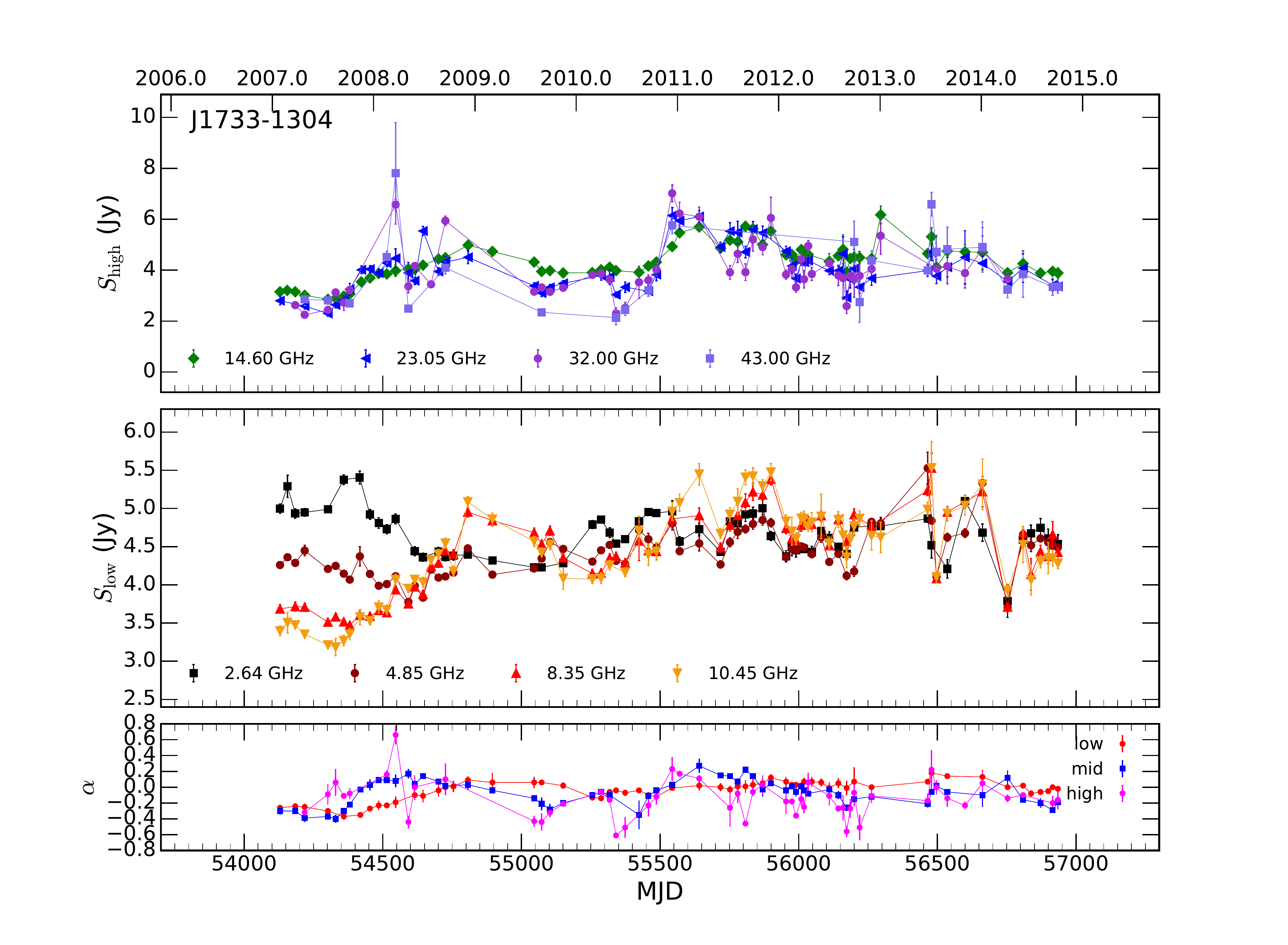}\\
\\[10pt]
\includegraphics[trim=60pt 30pt 100pt 50pt  ,clip, width=0.49\textwidth,angle=0]{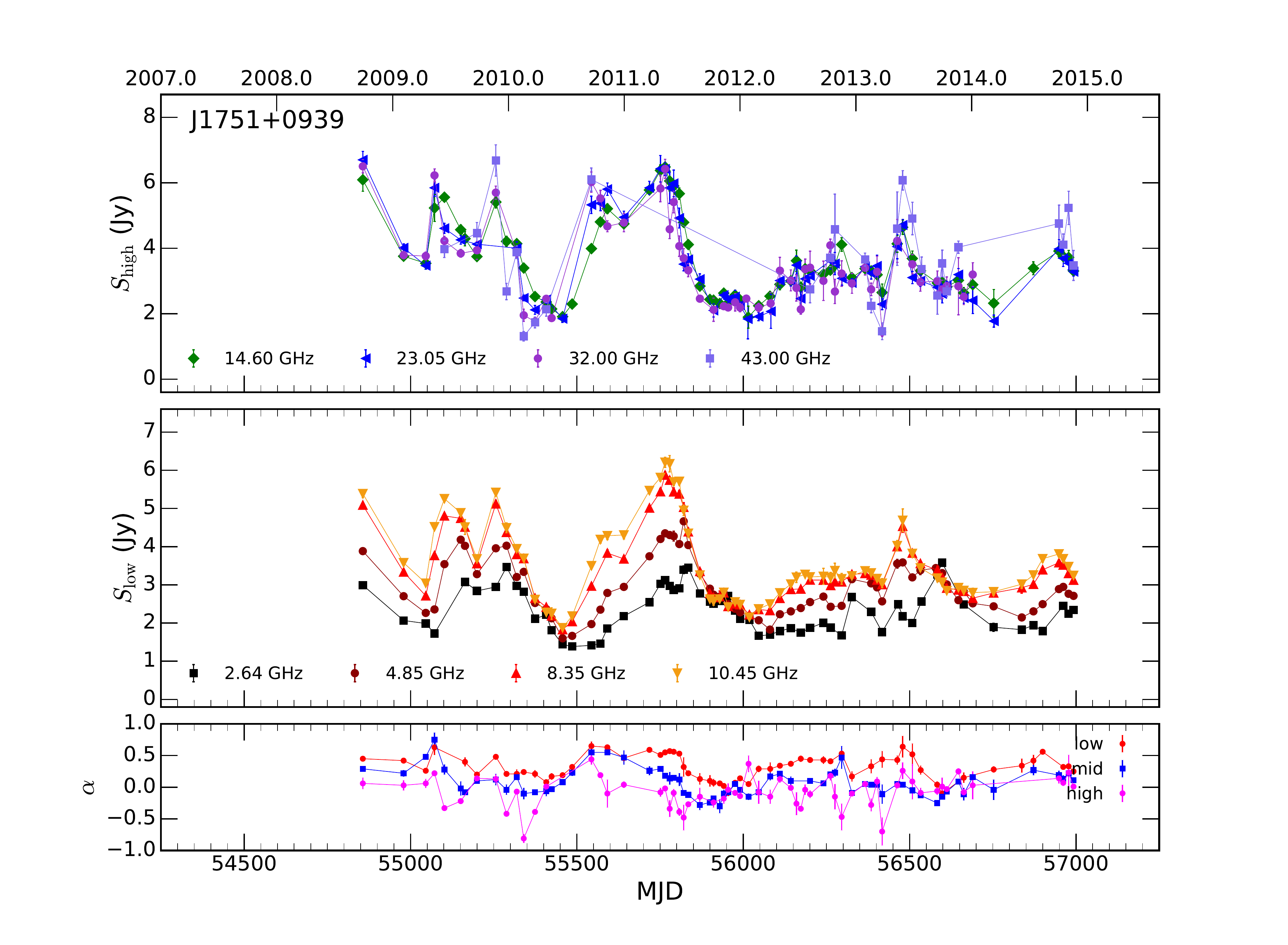}&
\includegraphics[trim=60pt 30pt 100pt 50pt  ,clip, width=0.49\textwidth,angle=0]{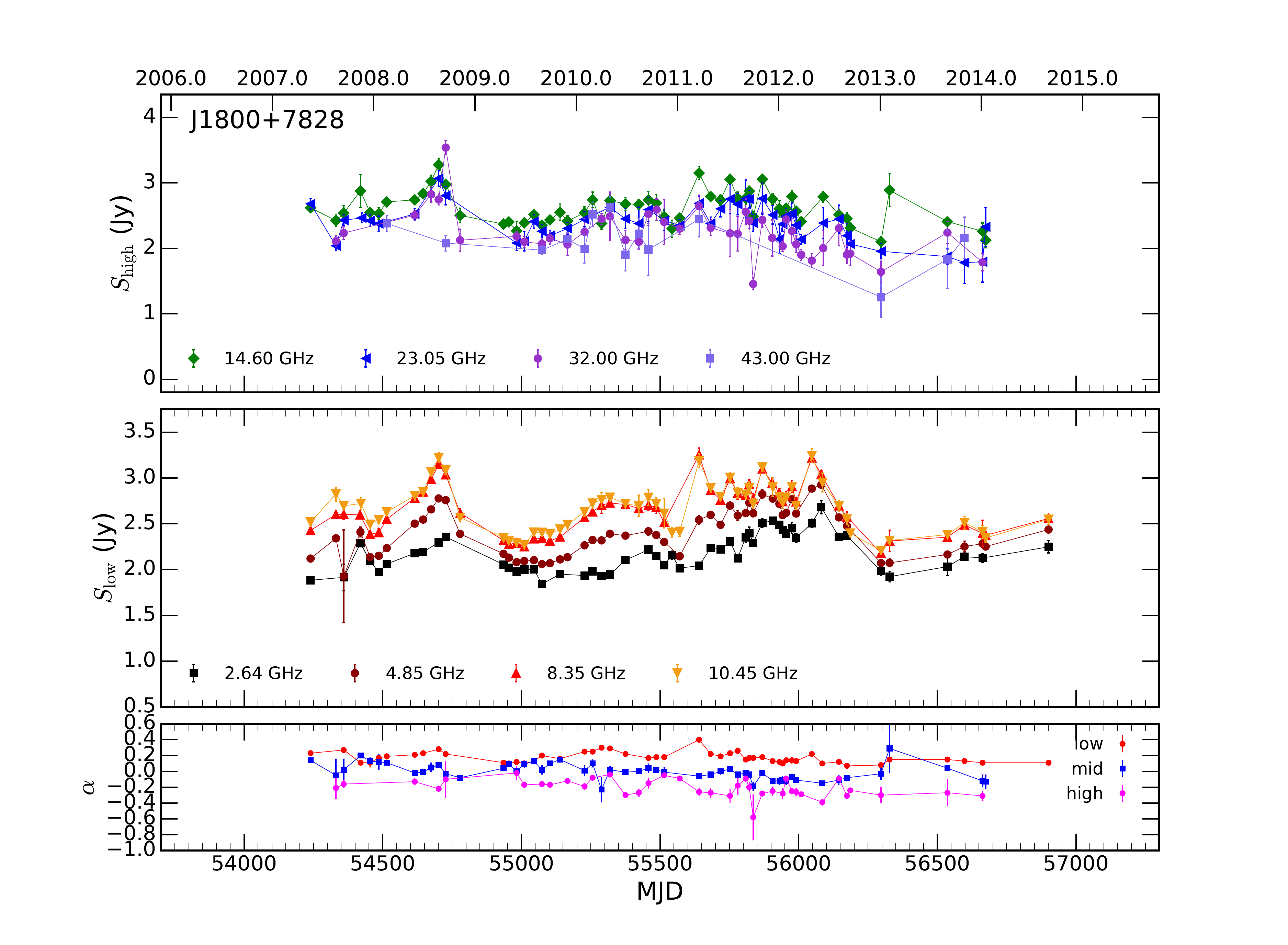}\\
\\[10pt]
\end{tabular}
\caption{Multi-frequency light curves for all the sources monitored by the \fg programme (``f'', ``s1'', ``s2'', ``old'') and the F-GAMMA-\textit{Planck} MoU. The lower panel in each frame shows the evolution of the low (2.64, 4.85 and 8.35~GHz) and mid-band (8.35, 10.45 and 14.6~GHz) and high-band (14.6, 23.05, 32, 43~GHz) spectral index. Only spectral index estimates from at least three frequencies are shown. Connecting lines have been used to guide the eye. }
\label{fig:sample_pg13}
\end{figure*}

\begin{figure*}[p]
\centering
\begin{tabular}{cc}
\includegraphics[trim=60pt 30pt 100pt 50pt  ,clip, width=0.49\textwidth,angle=0]{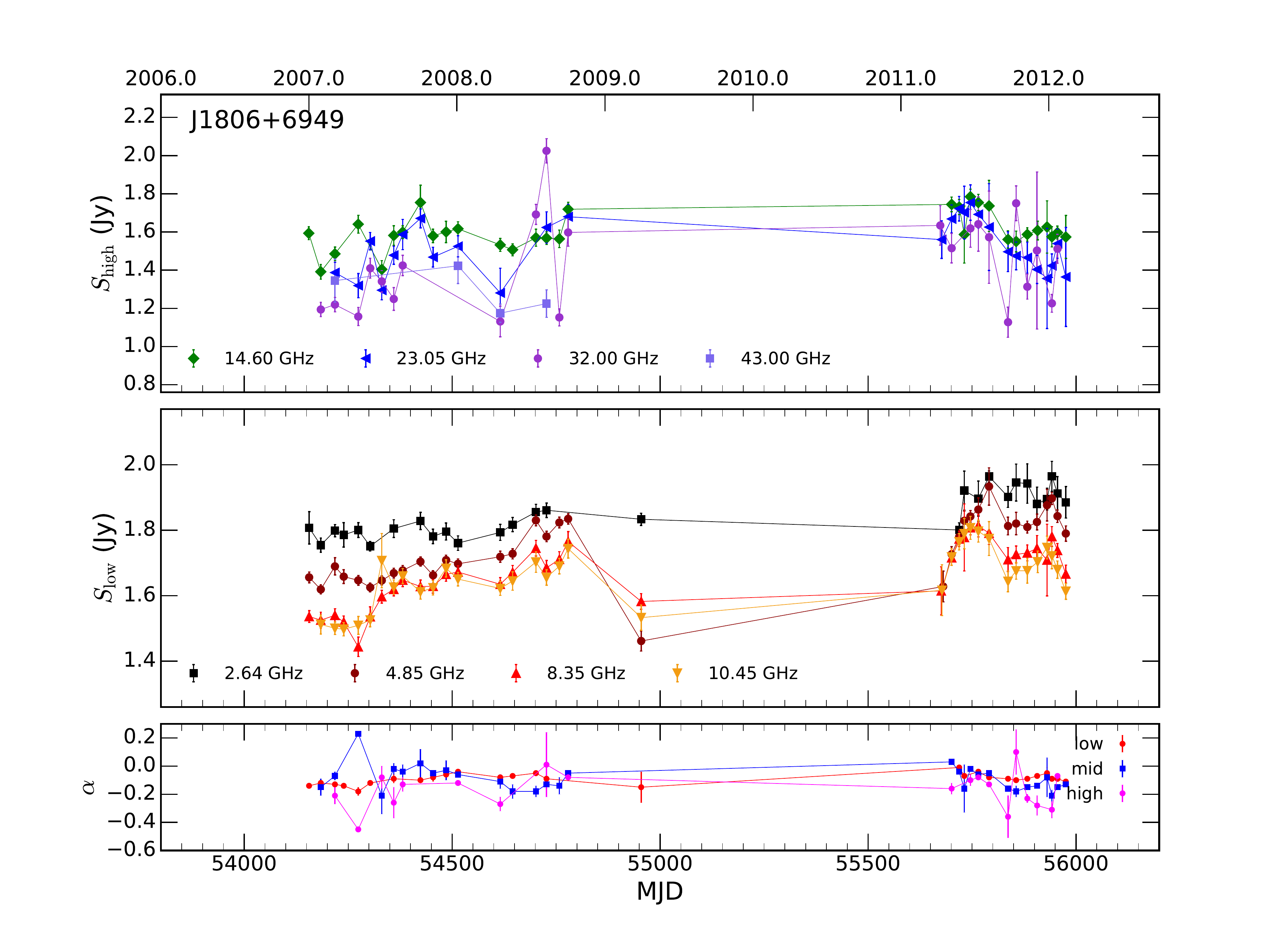}&
\includegraphics[trim=60pt 30pt 100pt 50pt  ,clip, width=0.49\textwidth,angle=0]{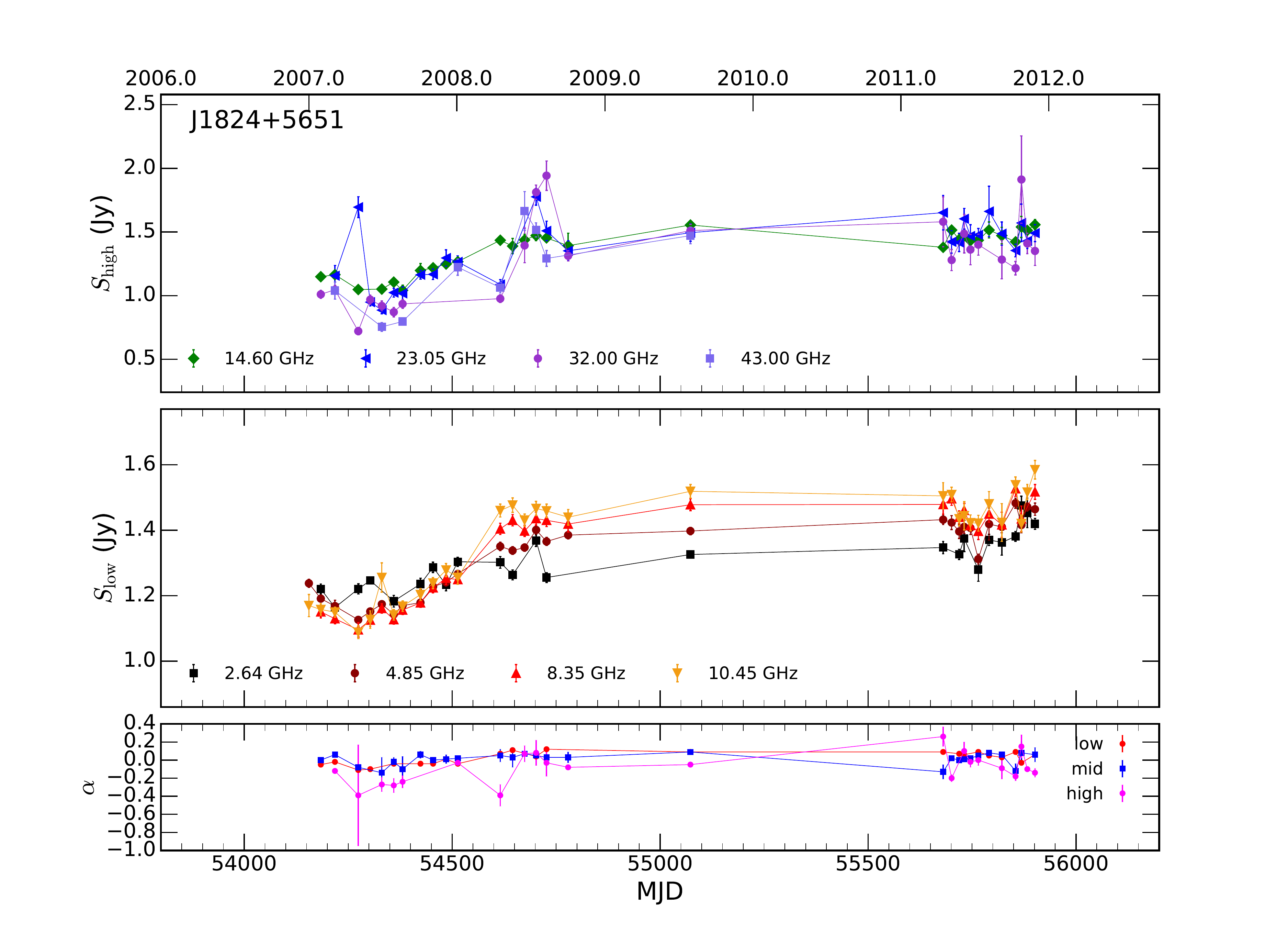}\\
\\[10pt]
\includegraphics[trim=60pt 30pt 100pt 50pt  ,clip, width=0.49\textwidth,angle=0]{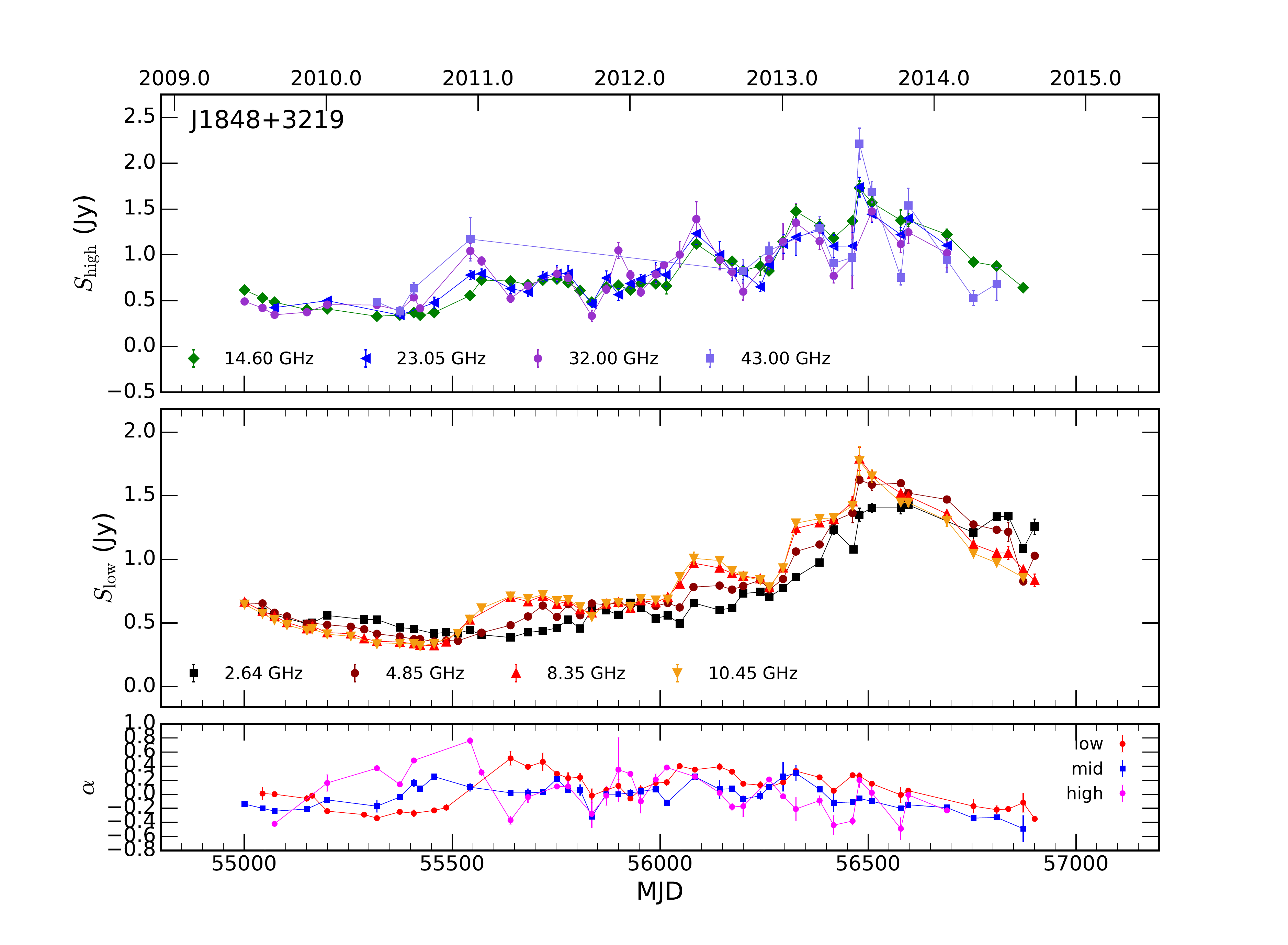}&
\includegraphics[trim=60pt 30pt 100pt 50pt  ,clip, width=0.49\textwidth,angle=0]{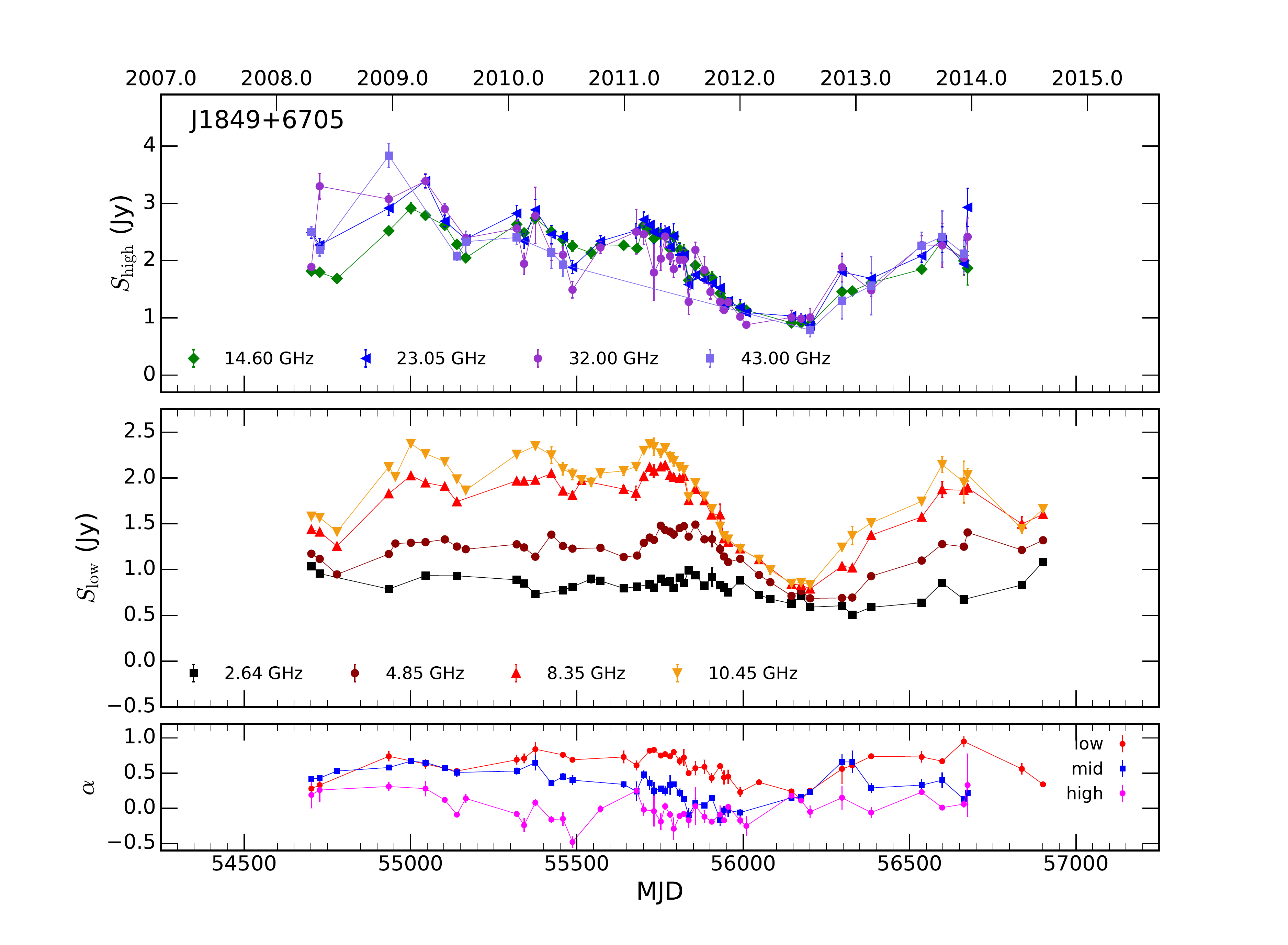}\\
\\[10pt]
\includegraphics[trim=60pt 30pt 100pt 50pt  ,clip, width=0.49\textwidth,angle=0]{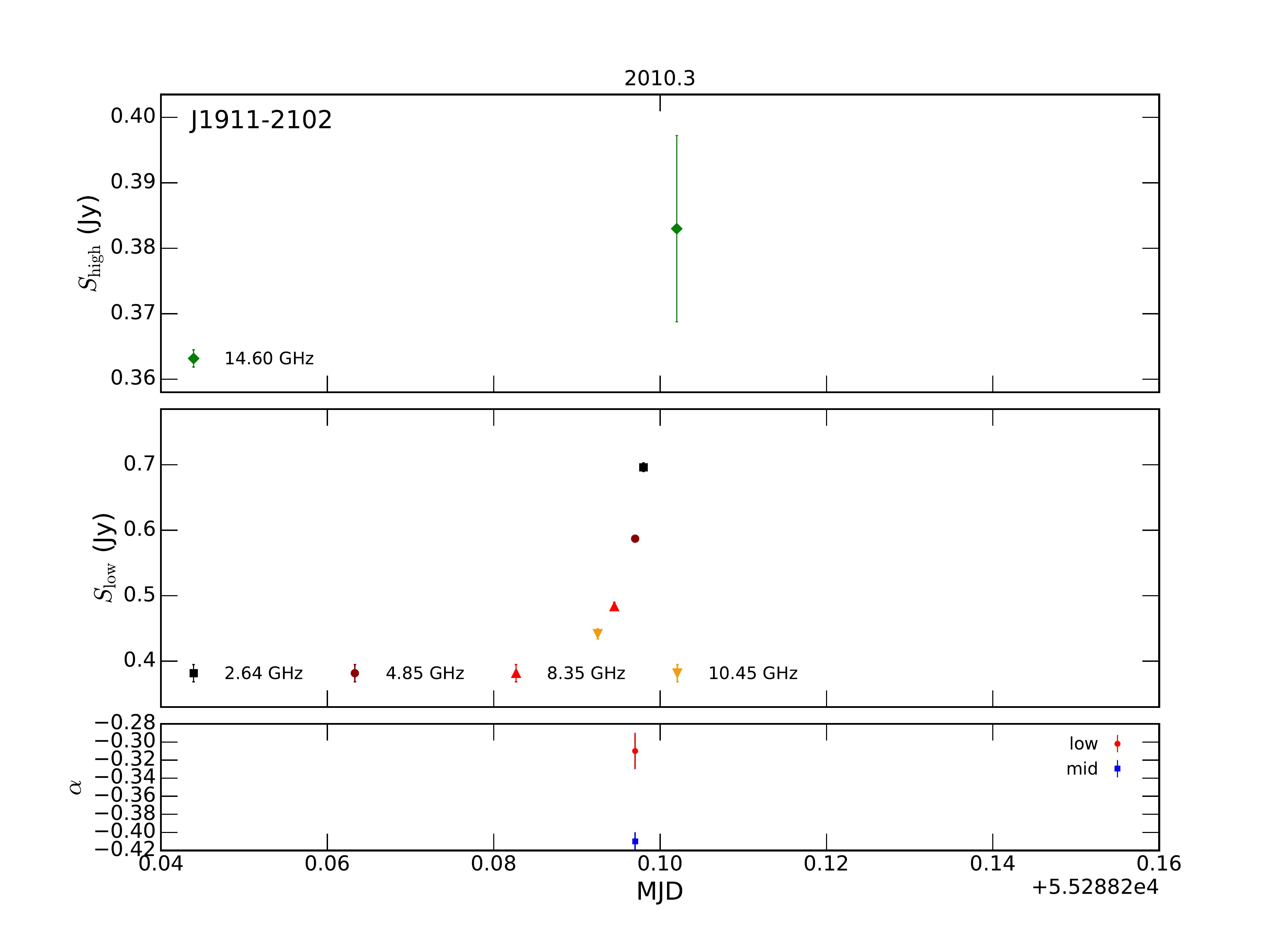}&
\includegraphics[trim=60pt 30pt 100pt 50pt  ,clip, width=0.49\textwidth,angle=0]{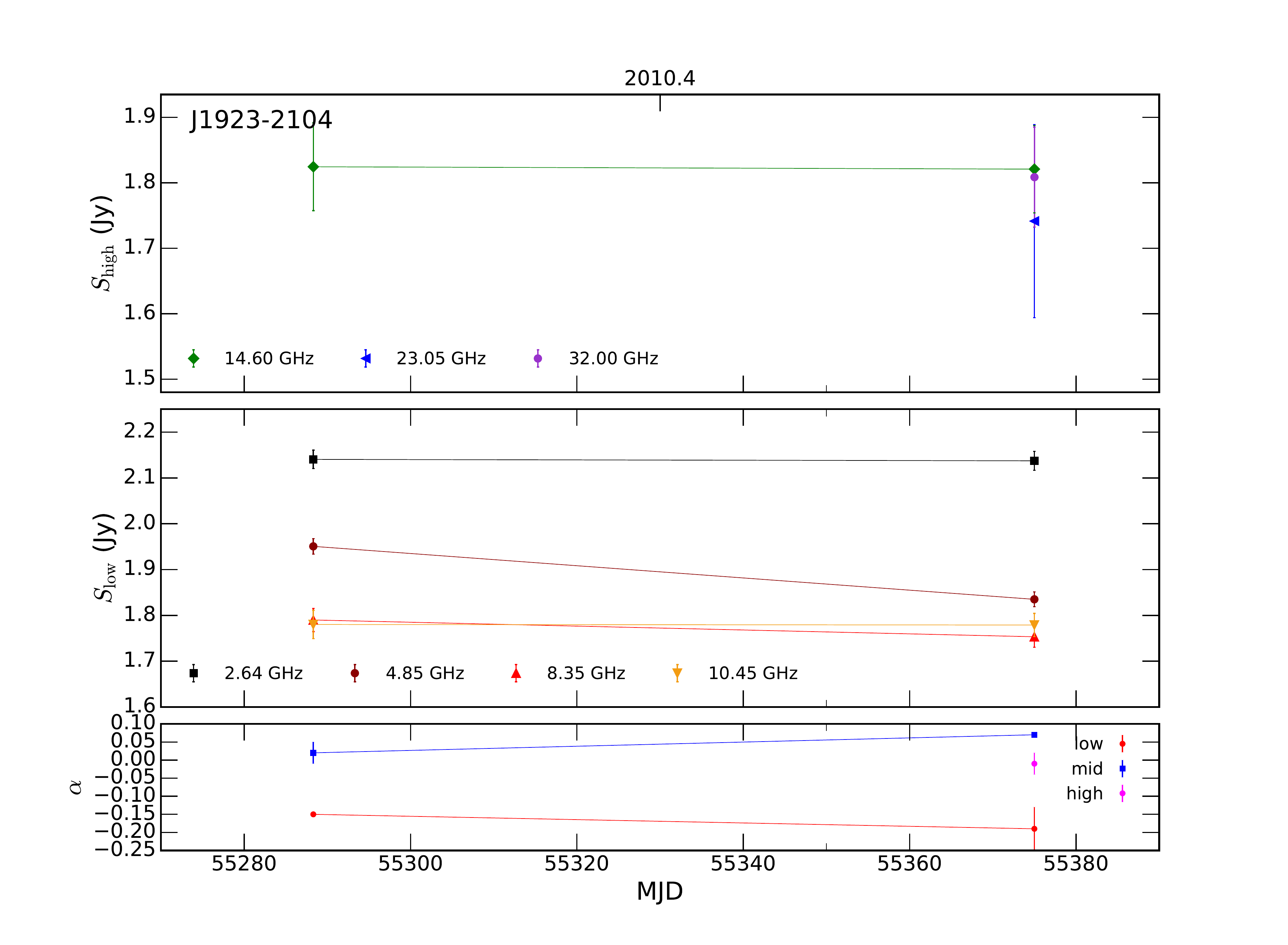}\\
\\[10pt]
\end{tabular}
\caption{Multi-frequency light curves for all the sources monitored by the \fg programme (``f'', ``s1'', ``s2'', ``old'') and the F-GAMMA-\textit{Planck} MoU. The lower panel in each frame shows the evolution of the low (2.64, 4.85 and 8.35~GHz) and mid-band (8.35, 10.45 and 14.6~GHz) and high-band (14.6, 23.05, 32, 43~GHz) spectral index. Only spectral index estimates from at least three frequencies are shown. Connecting lines have been used to guide the eye. }
\label{fig:sample_pg14}
\end{figure*}

\begin{figure*}[p]
\centering
\begin{tabular}{cc}
\includegraphics[trim=60pt 30pt 100pt 50pt  ,clip, width=0.49\textwidth,angle=0]{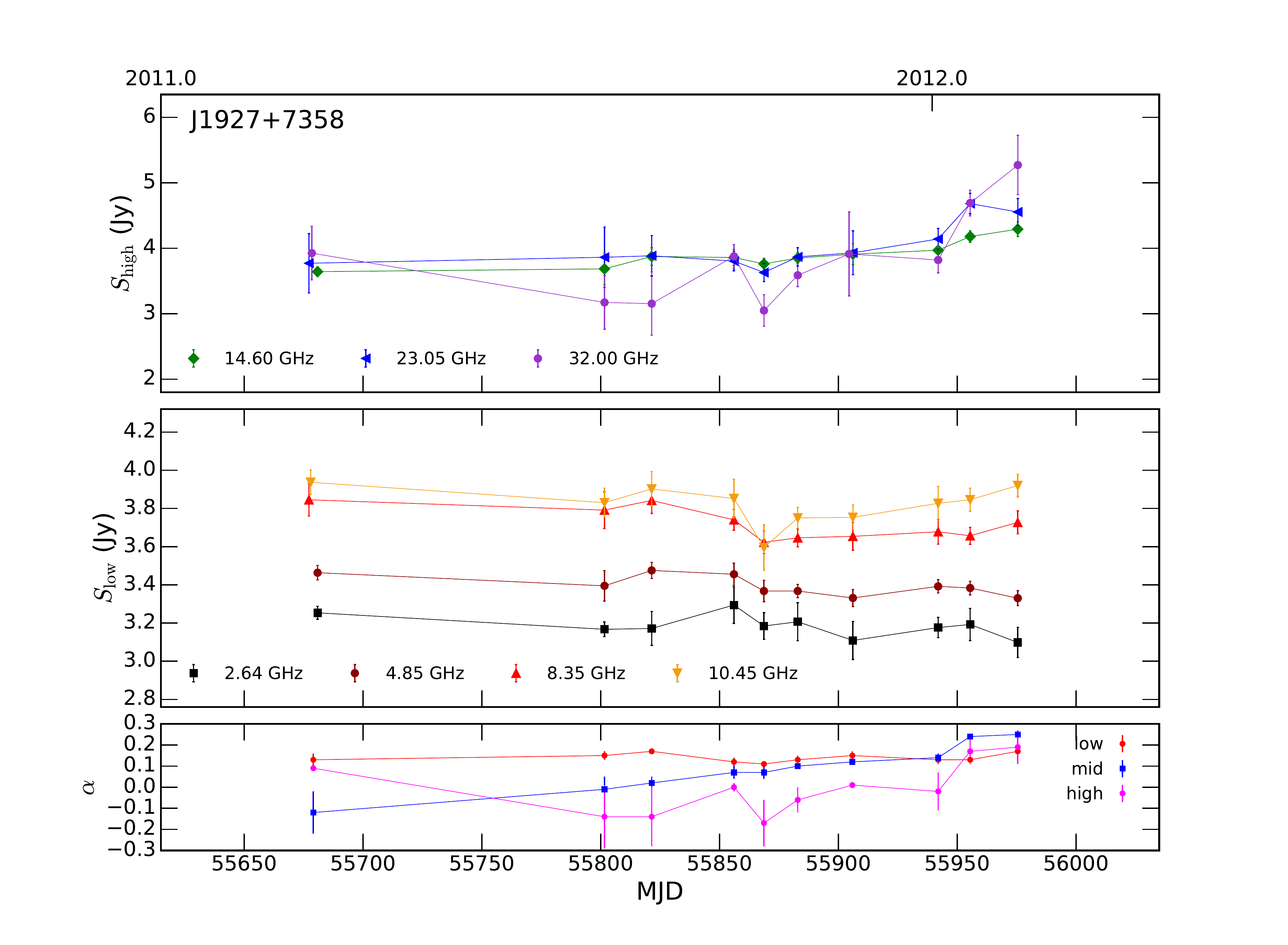}&
\includegraphics[trim=60pt 30pt 100pt 50pt  ,clip, width=0.49\textwidth,angle=0]{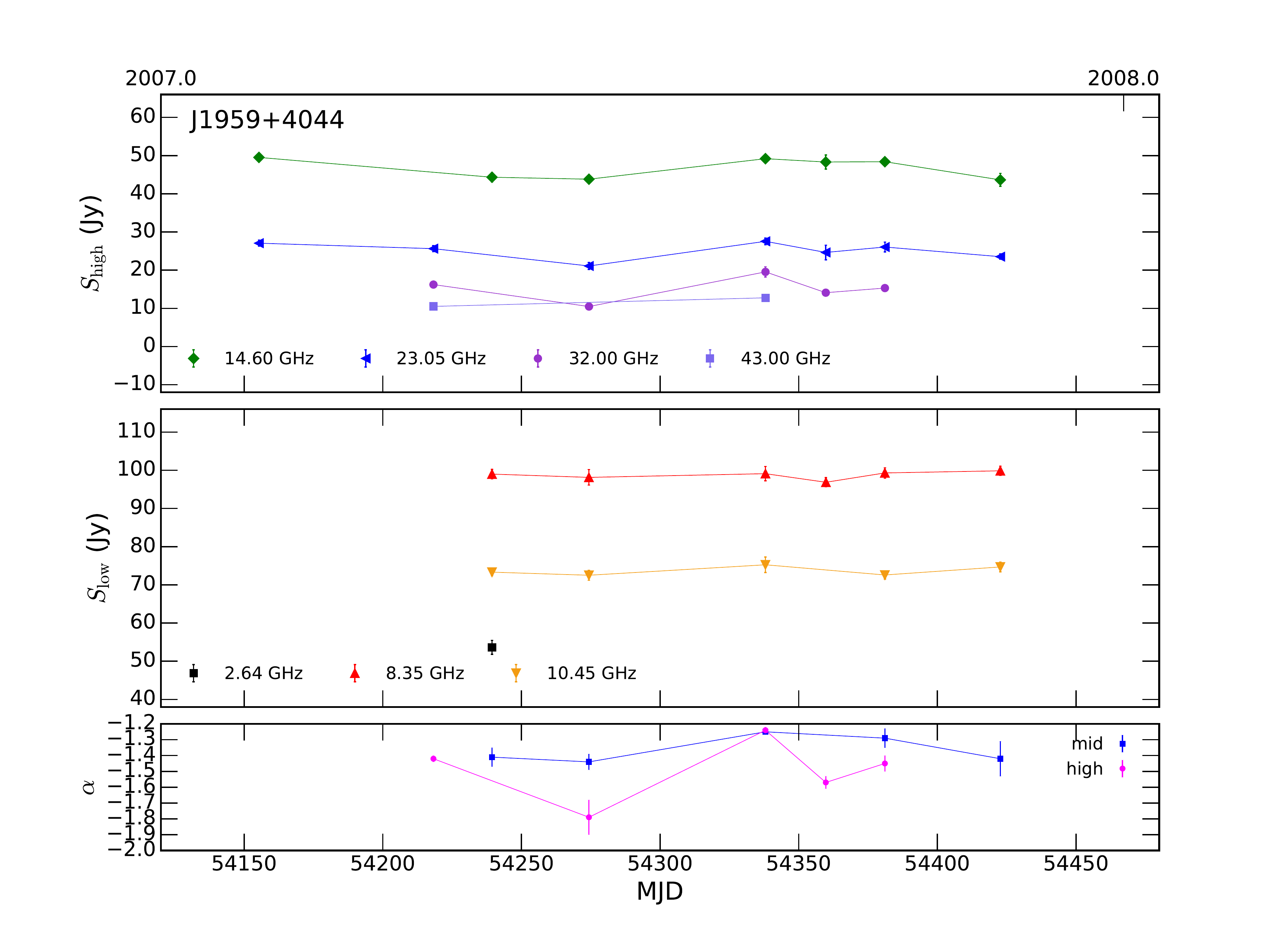}\\
\\[10pt]
\includegraphics[trim=60pt 30pt 100pt 50pt  ,clip, width=0.49\textwidth,angle=0]{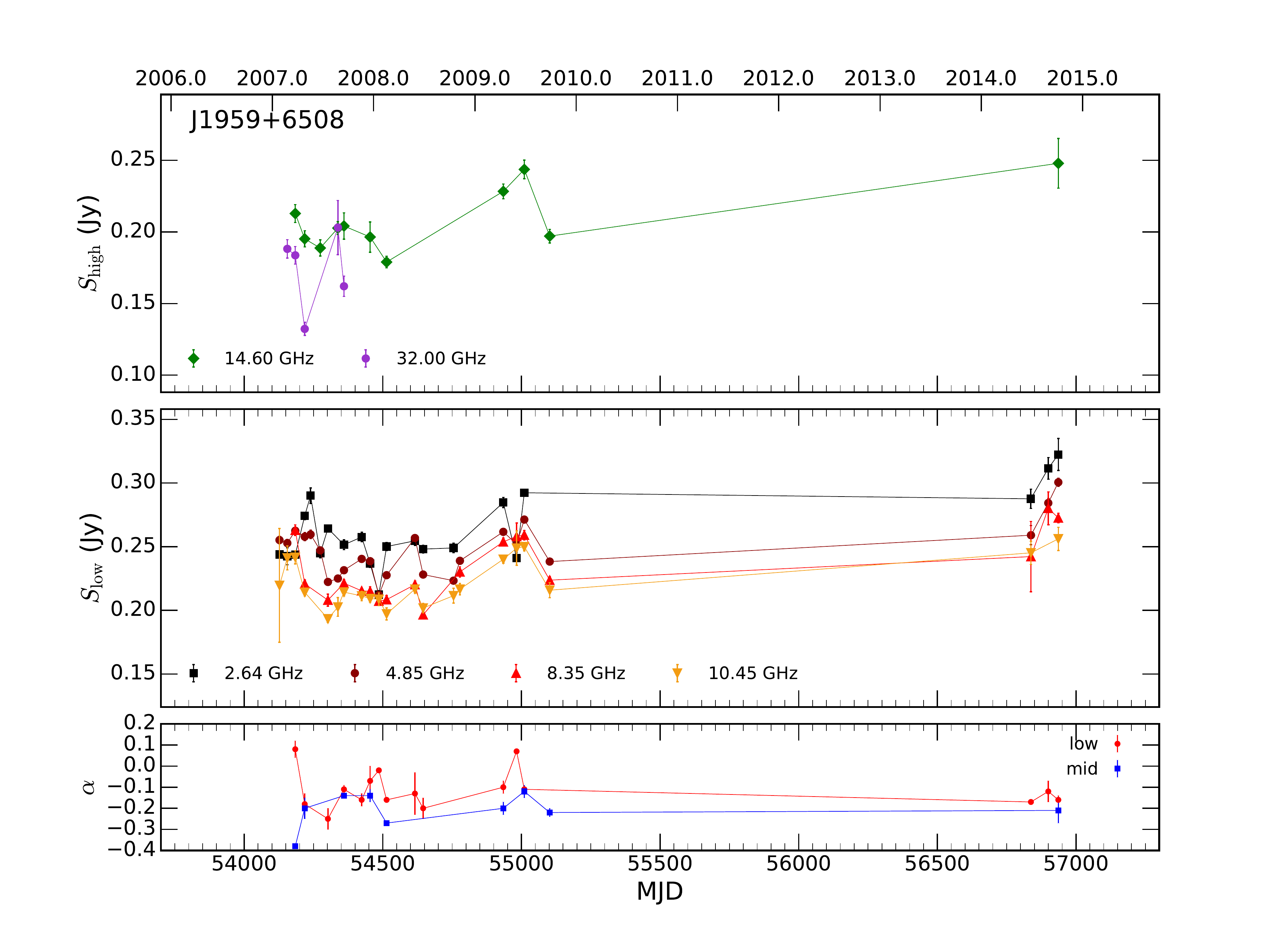}&
\includegraphics[trim=60pt 30pt 100pt 50pt  ,clip, width=0.49\textwidth,angle=0]{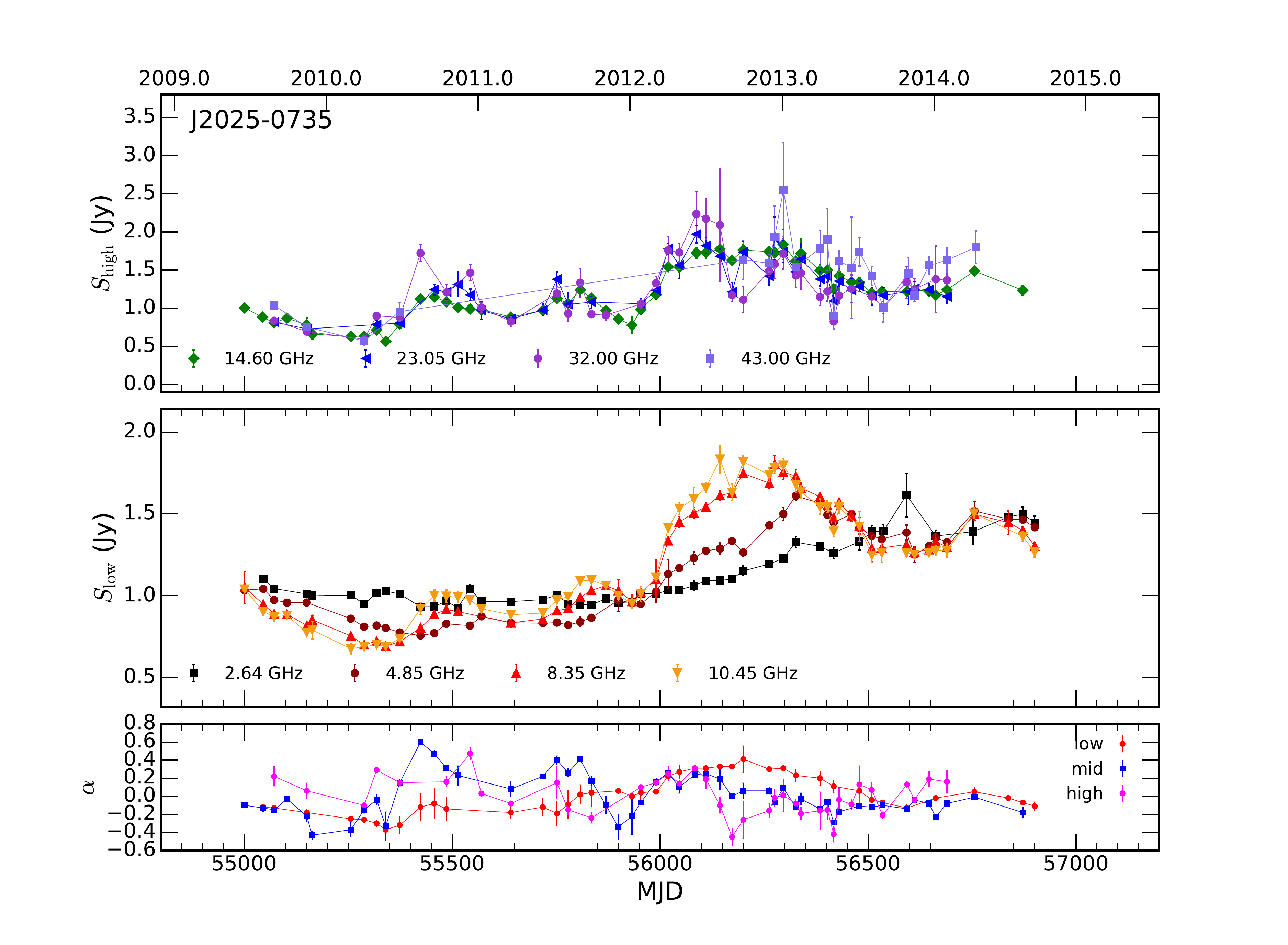}\\
\\[10pt]
\includegraphics[trim=60pt 30pt 100pt 50pt  ,clip, width=0.49\textwidth,angle=0]{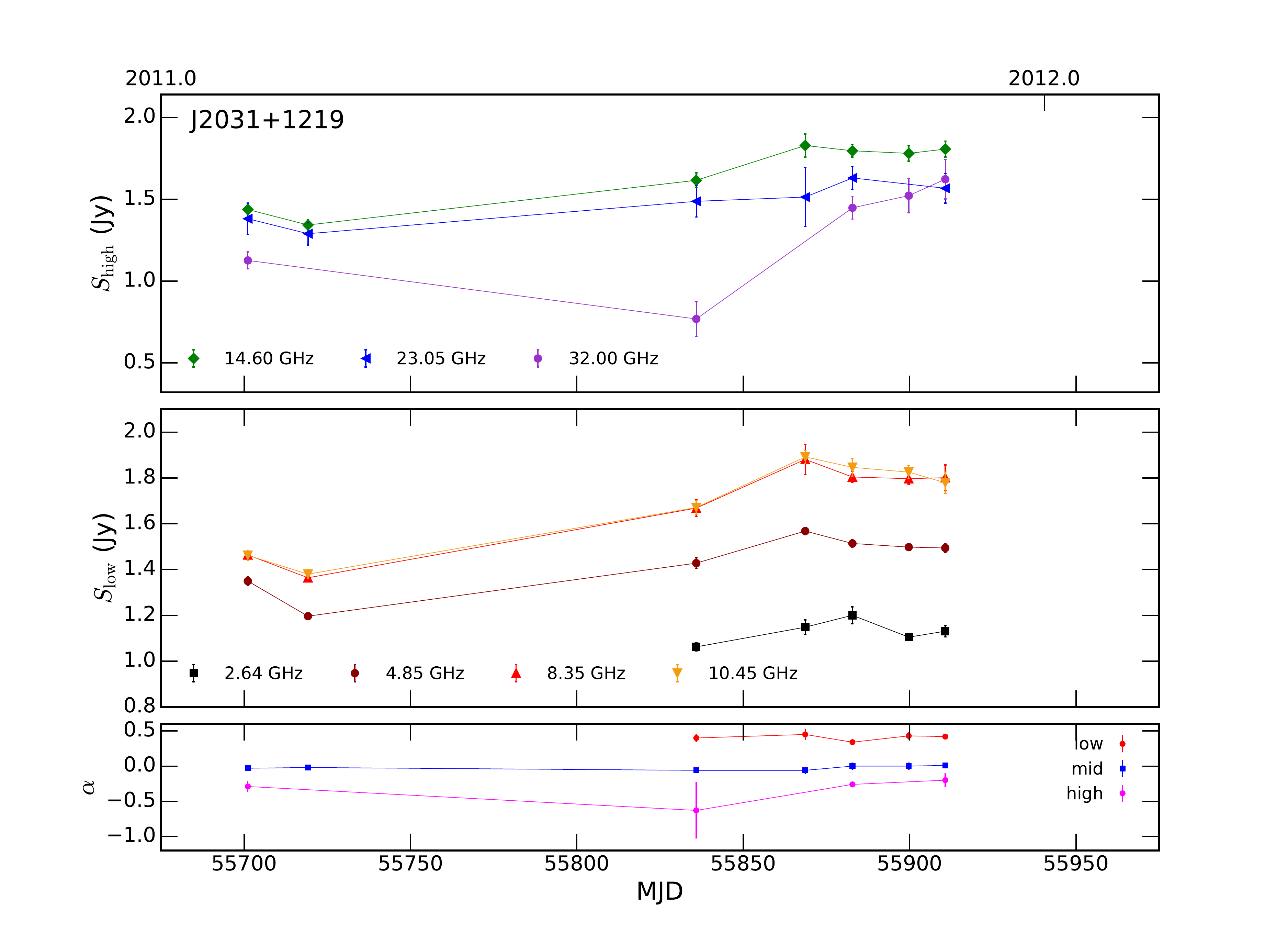}&
\includegraphics[trim=60pt 30pt 100pt 50pt  ,clip, width=0.49\textwidth,angle=0]{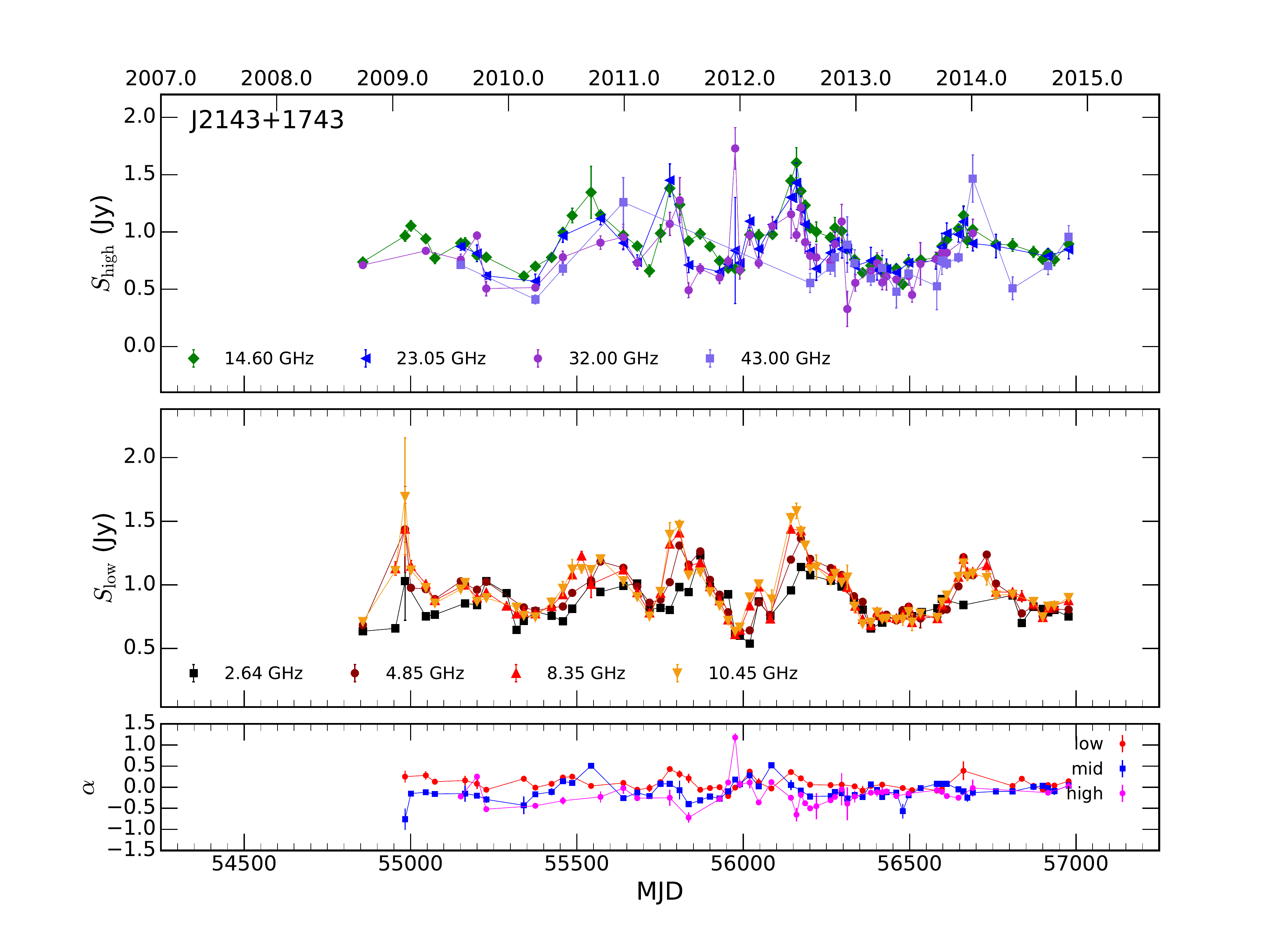}\\
\\[10pt]
\end{tabular}
\caption{Multi-frequency light curves for all the sources monitored by the \fg programme (``f'', ``s1'', ``s2'', ``old'') and the F-GAMMA-\textit{Planck} MoU. The lower panel in each frame shows the evolution of the low (2.64, 4.85 and 8.35~GHz) and mid-band (8.35, 10.45 and 14.6~GHz) and high-band (14.6, 23.05, 32, 43~GHz) spectral index. Only spectral index estimates from at least three frequencies are shown. Connecting lines have been used to guide the eye. }
\label{fig:sample_pg15}
\end{figure*}

\begin{figure*}[p]
\centering
\begin{tabular}{cc}
\includegraphics[trim=60pt 30pt 100pt 50pt  ,clip, width=0.49\textwidth,angle=0]{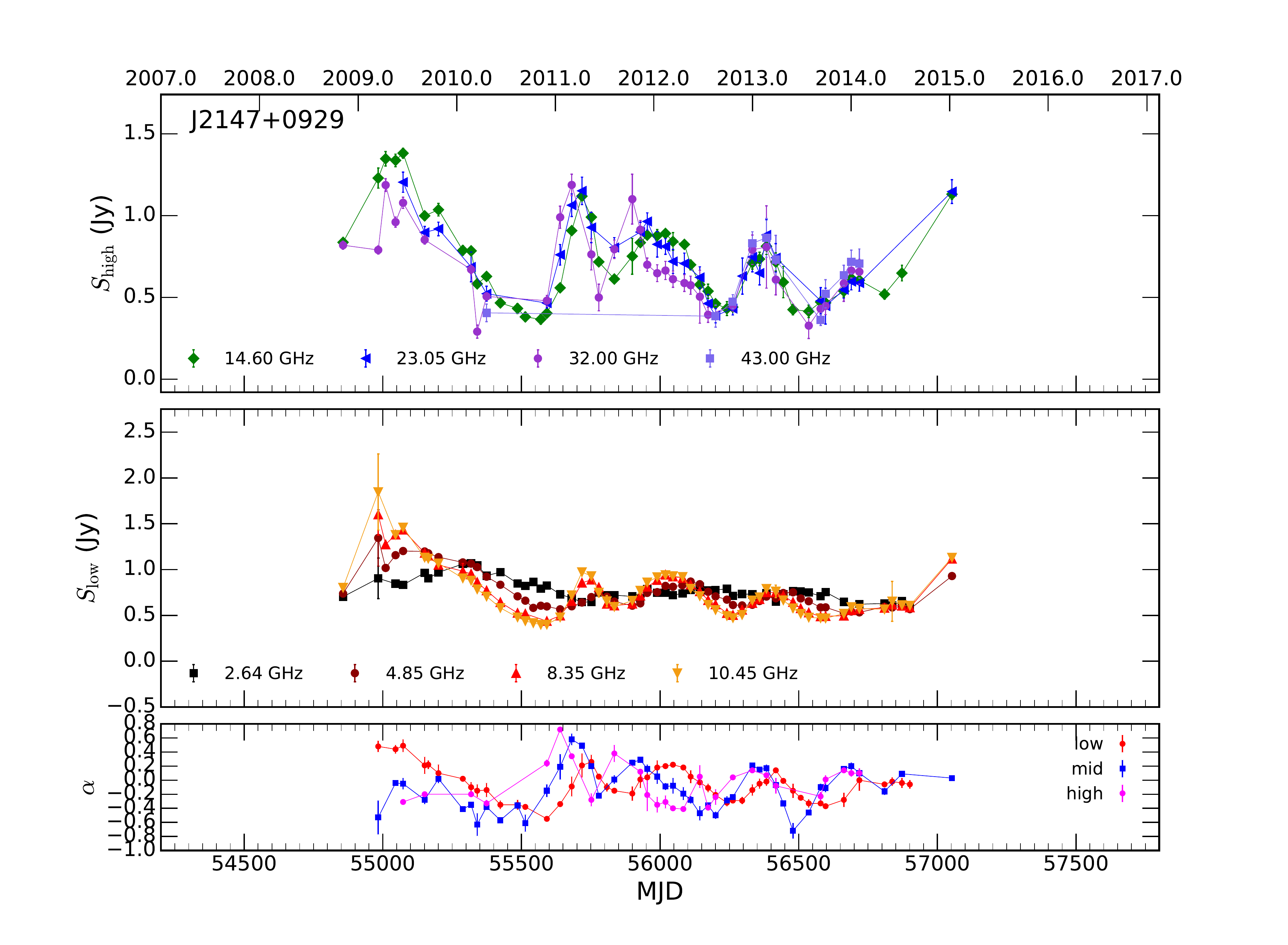}&
\includegraphics[trim=60pt 30pt 100pt 50pt  ,clip, width=0.49\textwidth,angle=0]{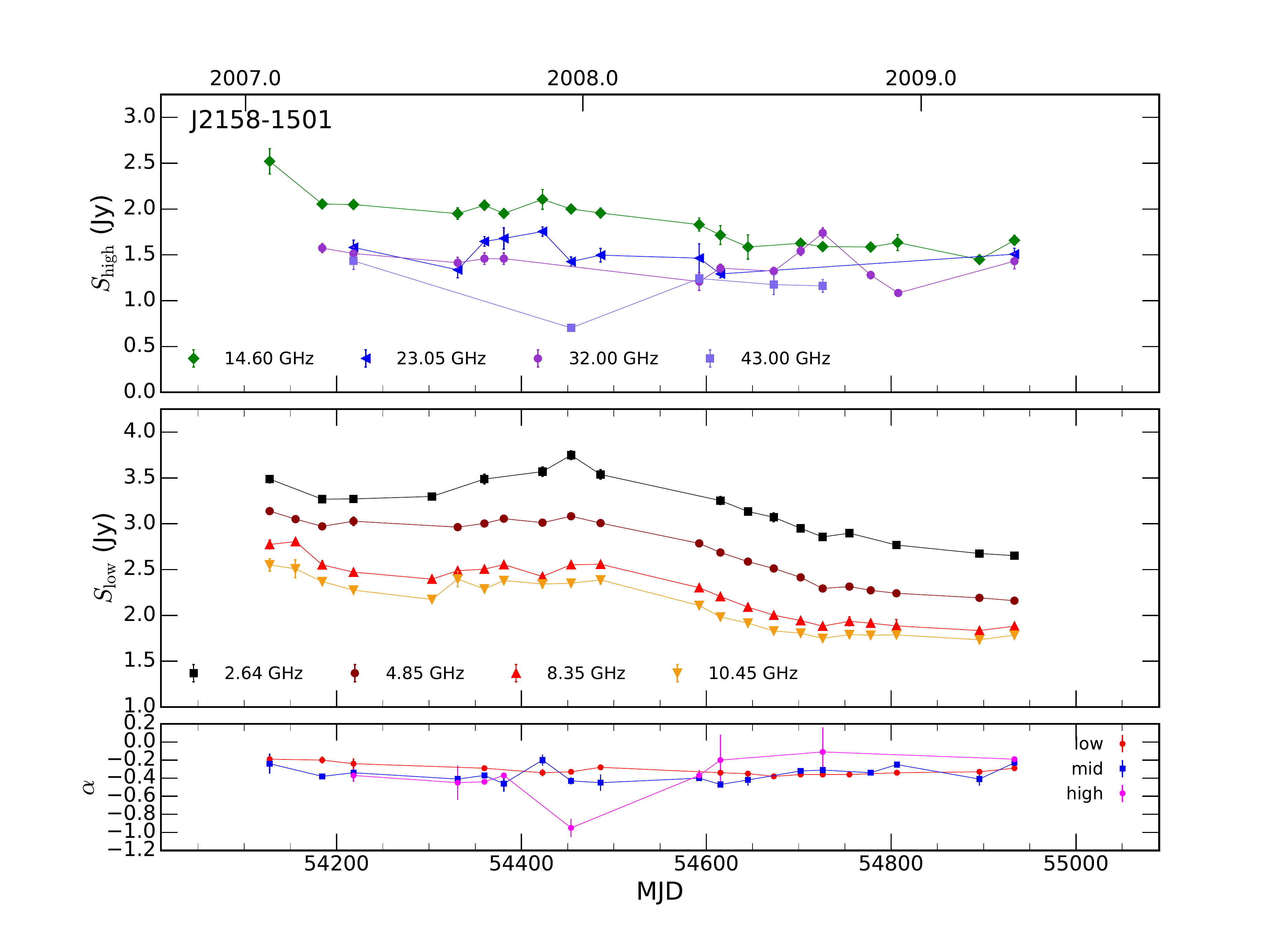}\\
\\[10pt]
\includegraphics[trim=60pt 30pt 100pt 50pt  ,clip, width=0.49\textwidth,angle=0]{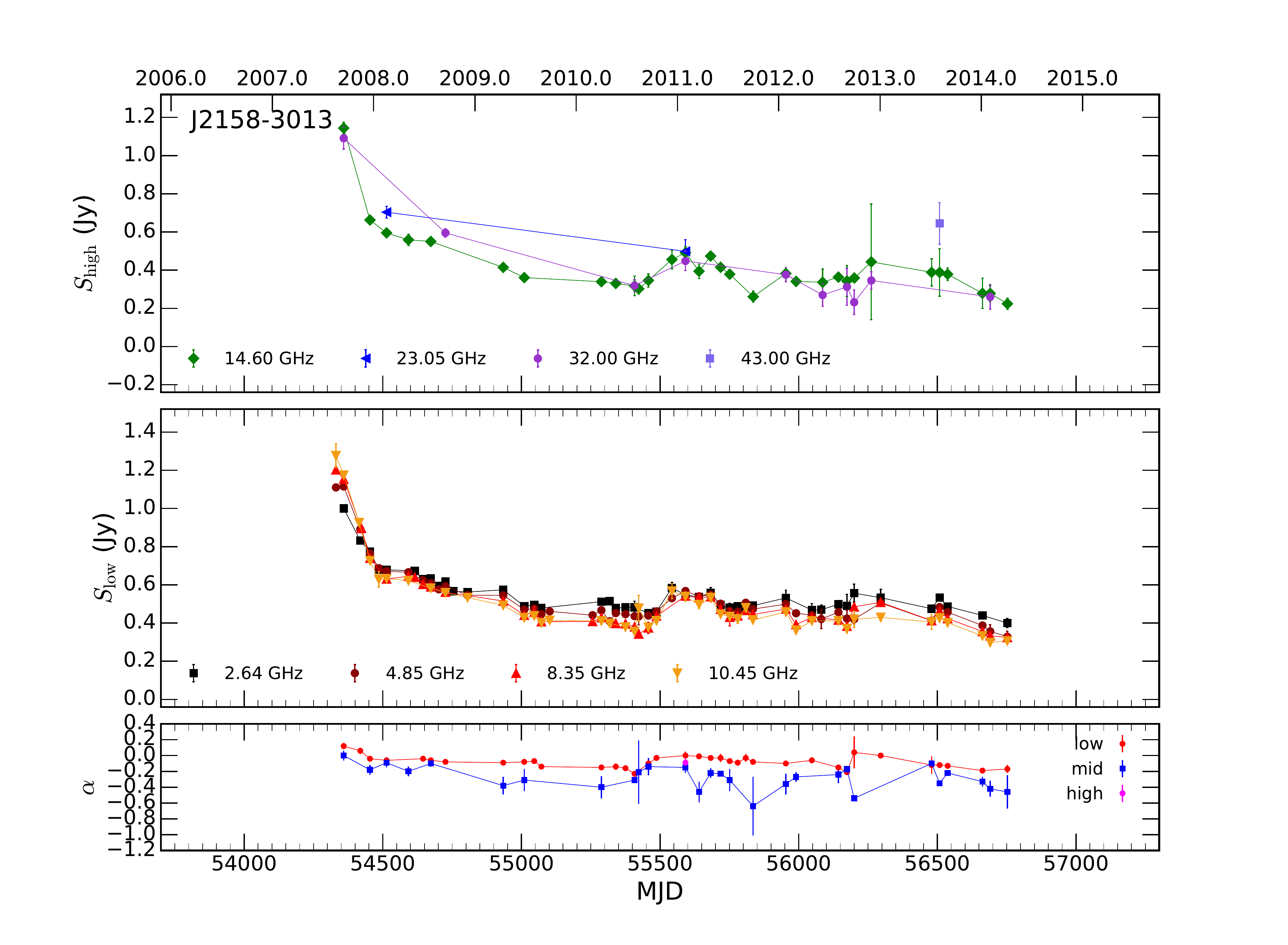}&
\includegraphics[trim=60pt 30pt 100pt 50pt  ,clip, width=0.49\textwidth,angle=0]{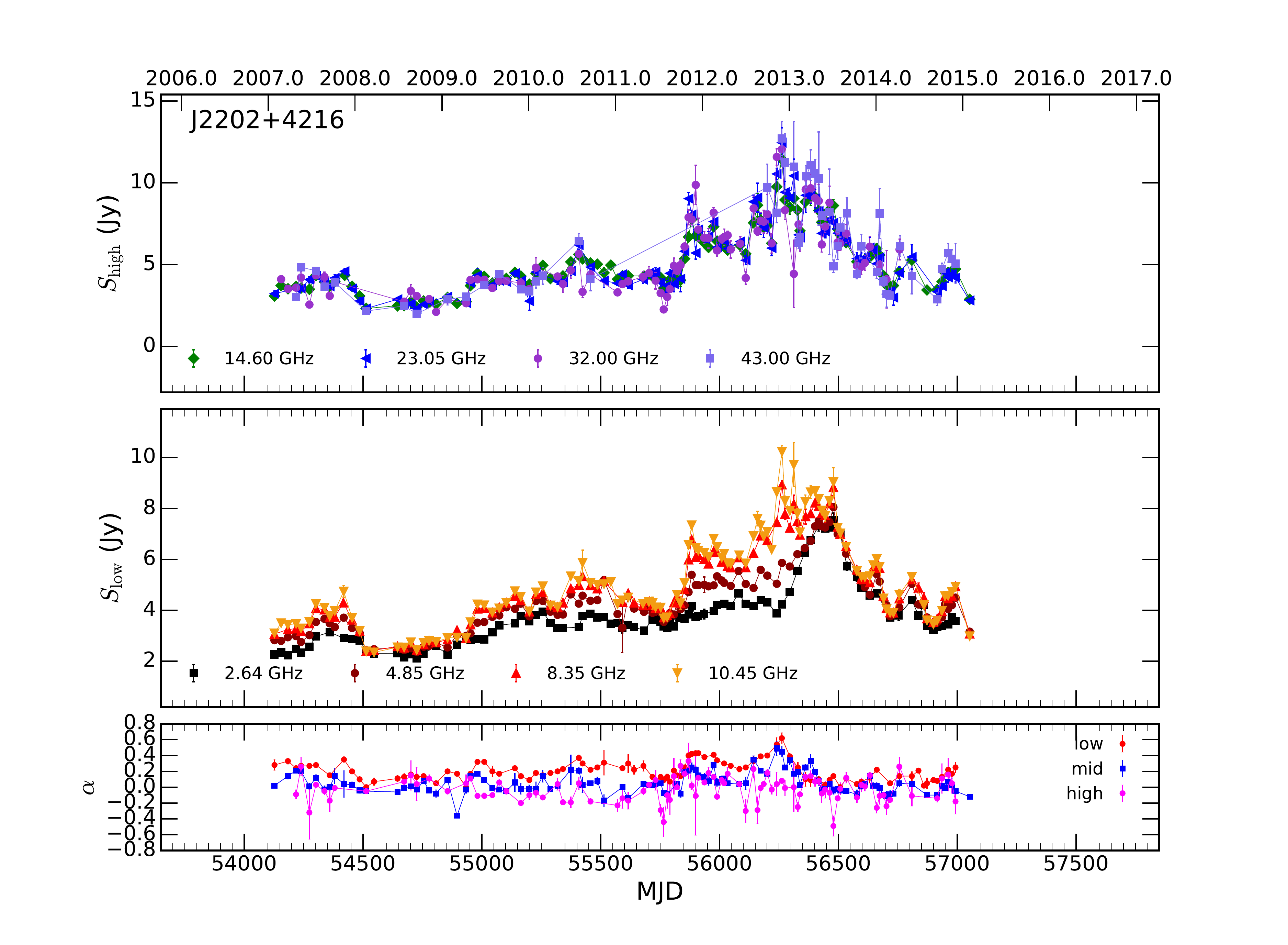}\\
\\[10pt]
\includegraphics[trim=60pt 30pt 100pt 50pt  ,clip, width=0.49\textwidth,angle=0]{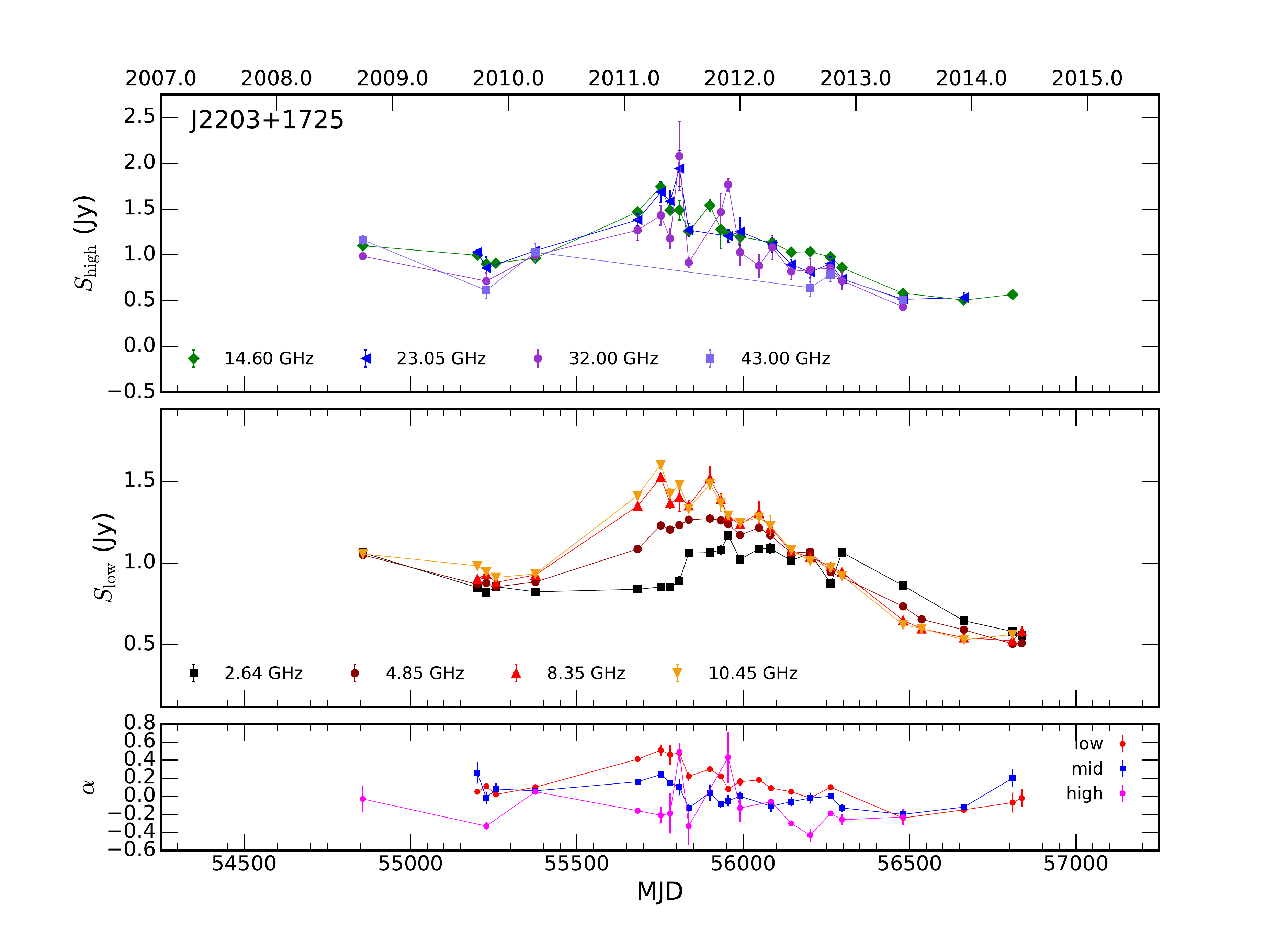}&
\includegraphics[trim=60pt 30pt 100pt 50pt  ,clip, width=0.49\textwidth,angle=0]{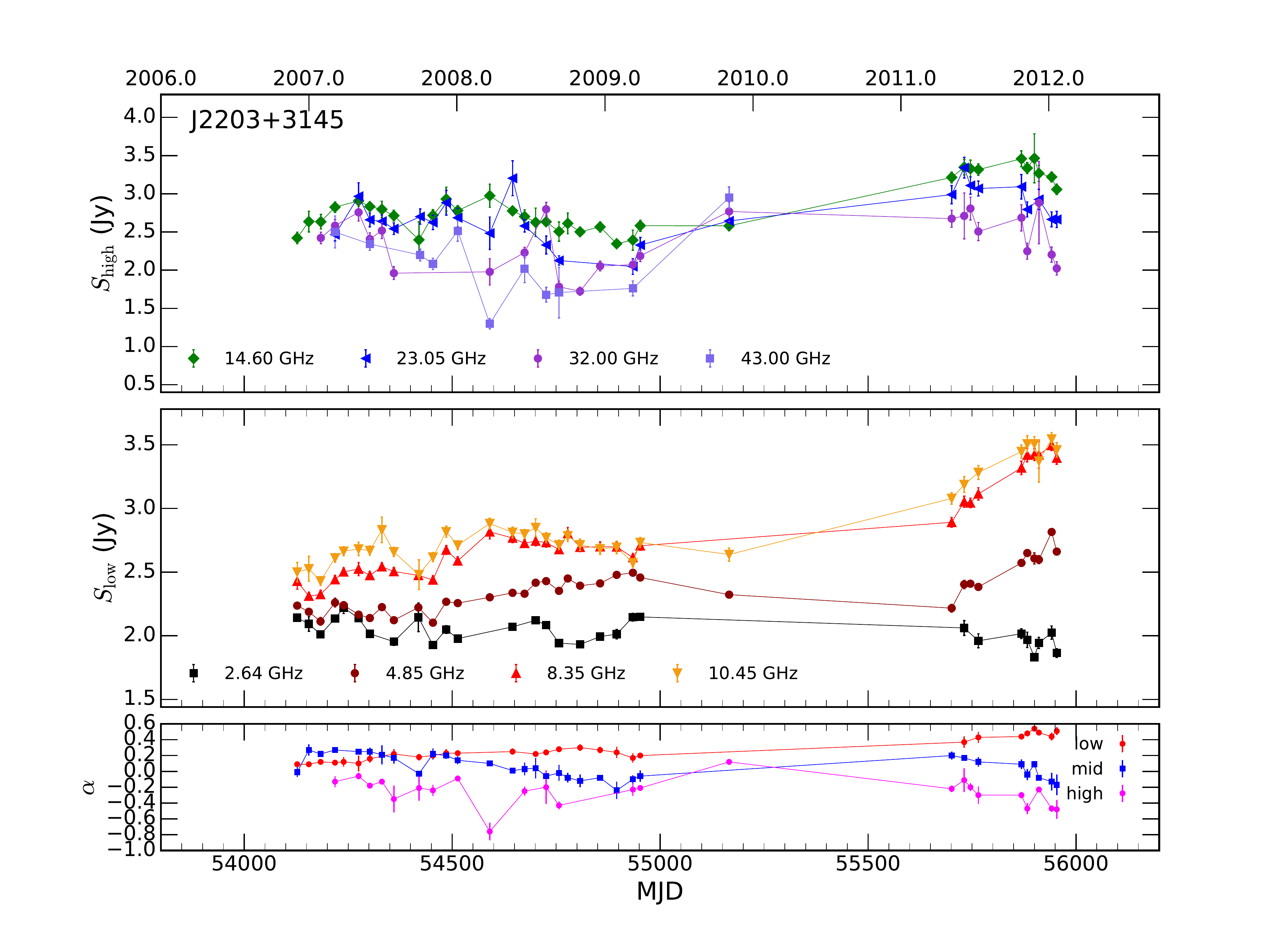}\\
\\[10pt]
\end{tabular}
\caption{Multi-frequency light curves for all the sources monitored by the \fg programme (``f'', ``s1'', ``s2'', ``old'') and the F-GAMMA-\textit{Planck} MoU. The lower panel in each frame shows the evolution of the low (2.64, 4.85 and 8.35~GHz) and mid-band (8.35, 10.45 and 14.6~GHz) and high-band (14.6, 23.05, 32, 43~GHz) spectral index. Only spectral index estimates from at least three frequencies are shown. Connecting lines have been used to guide the eye. }
\label{fig:sample_pg16}
\end{figure*}

\begin{figure*}[p]
\centering
\begin{tabular}{cc}
\includegraphics[trim=60pt 30pt 100pt 50pt  ,clip, width=0.49\textwidth,angle=0]{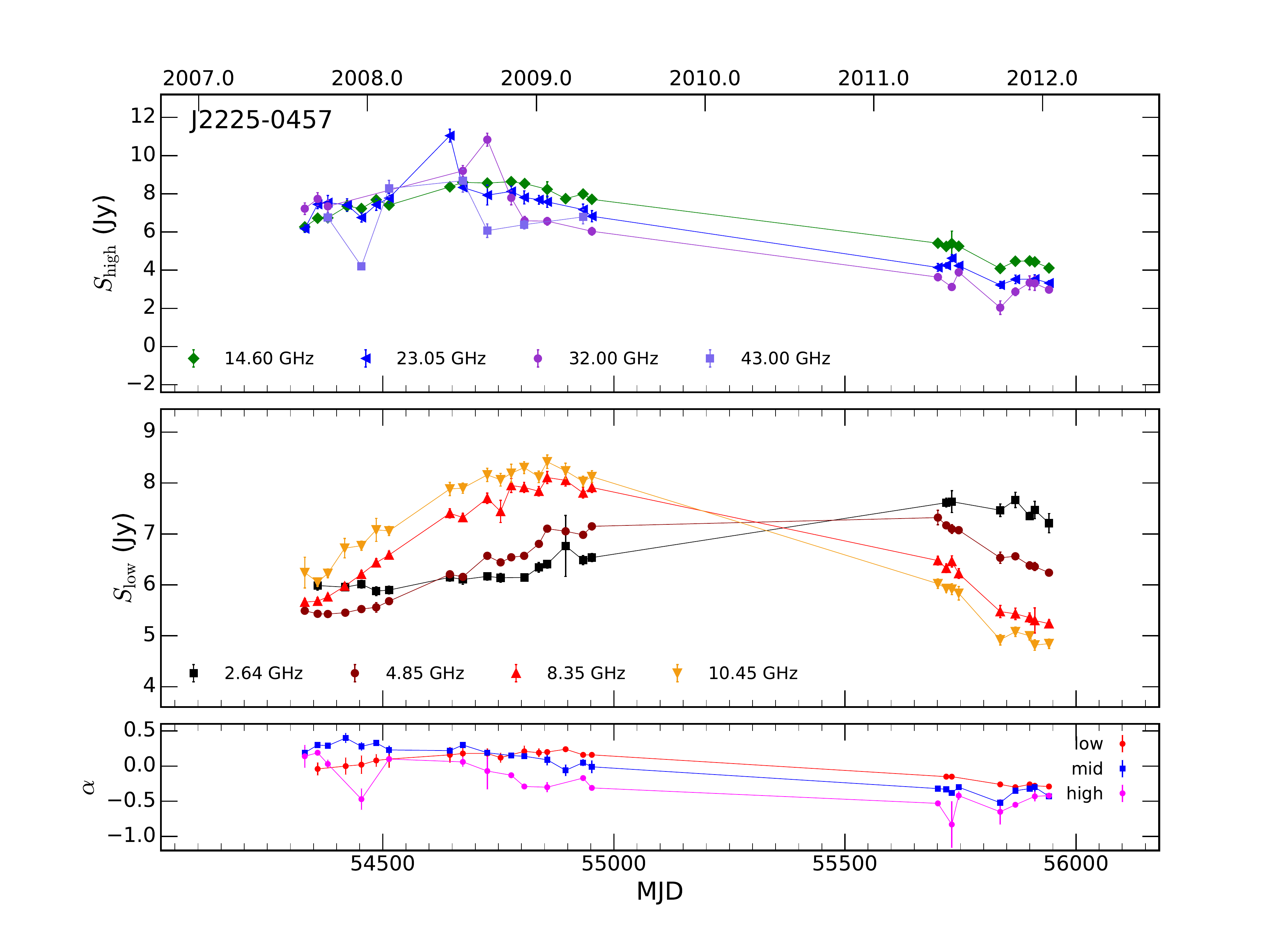}&
\includegraphics[trim=60pt 30pt 100pt 50pt  ,clip, width=0.49\textwidth,angle=0]{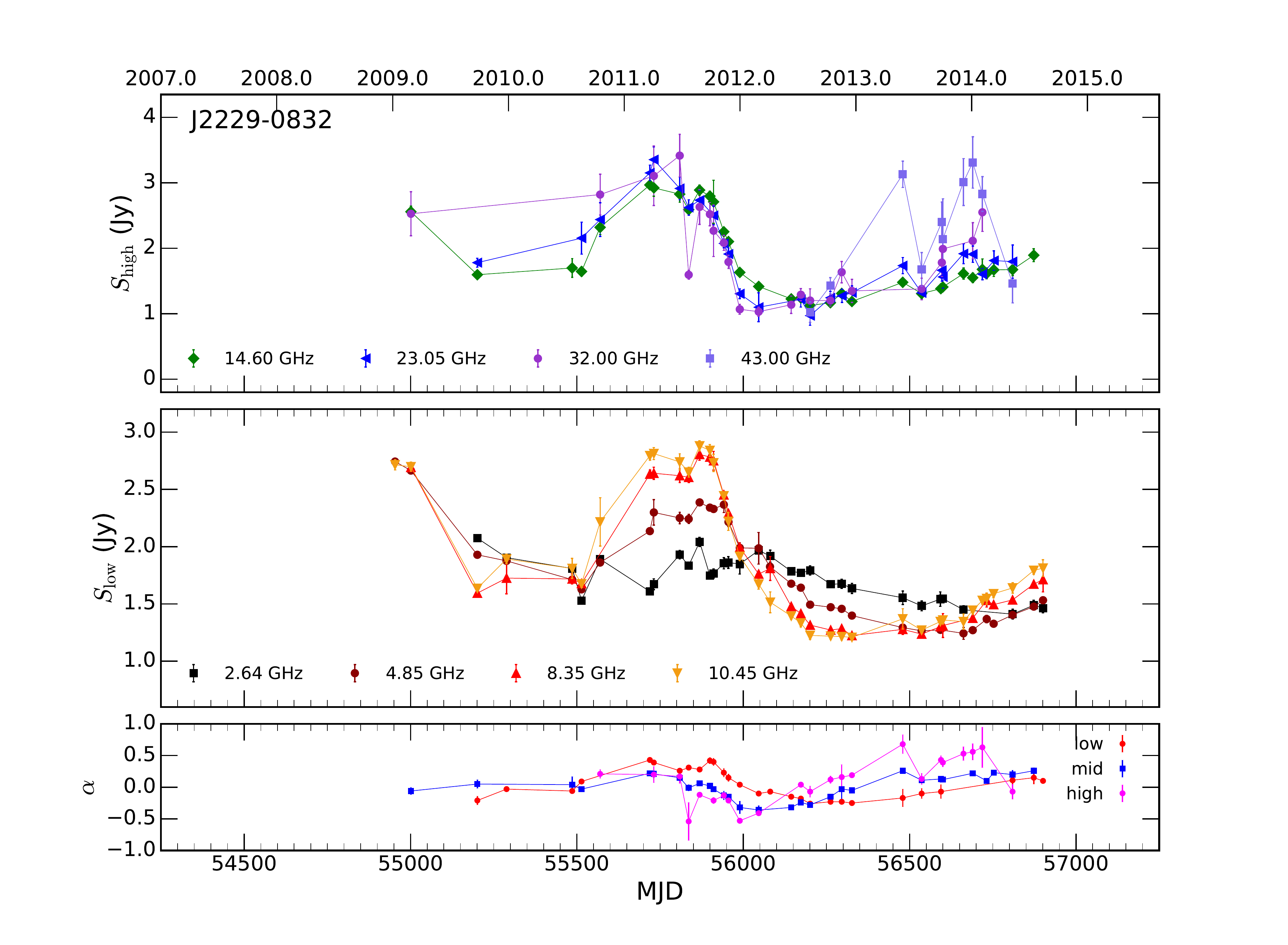}\\
\\[10pt]
\includegraphics[trim=60pt 30pt 100pt 50pt  ,clip, width=0.49\textwidth,angle=0]{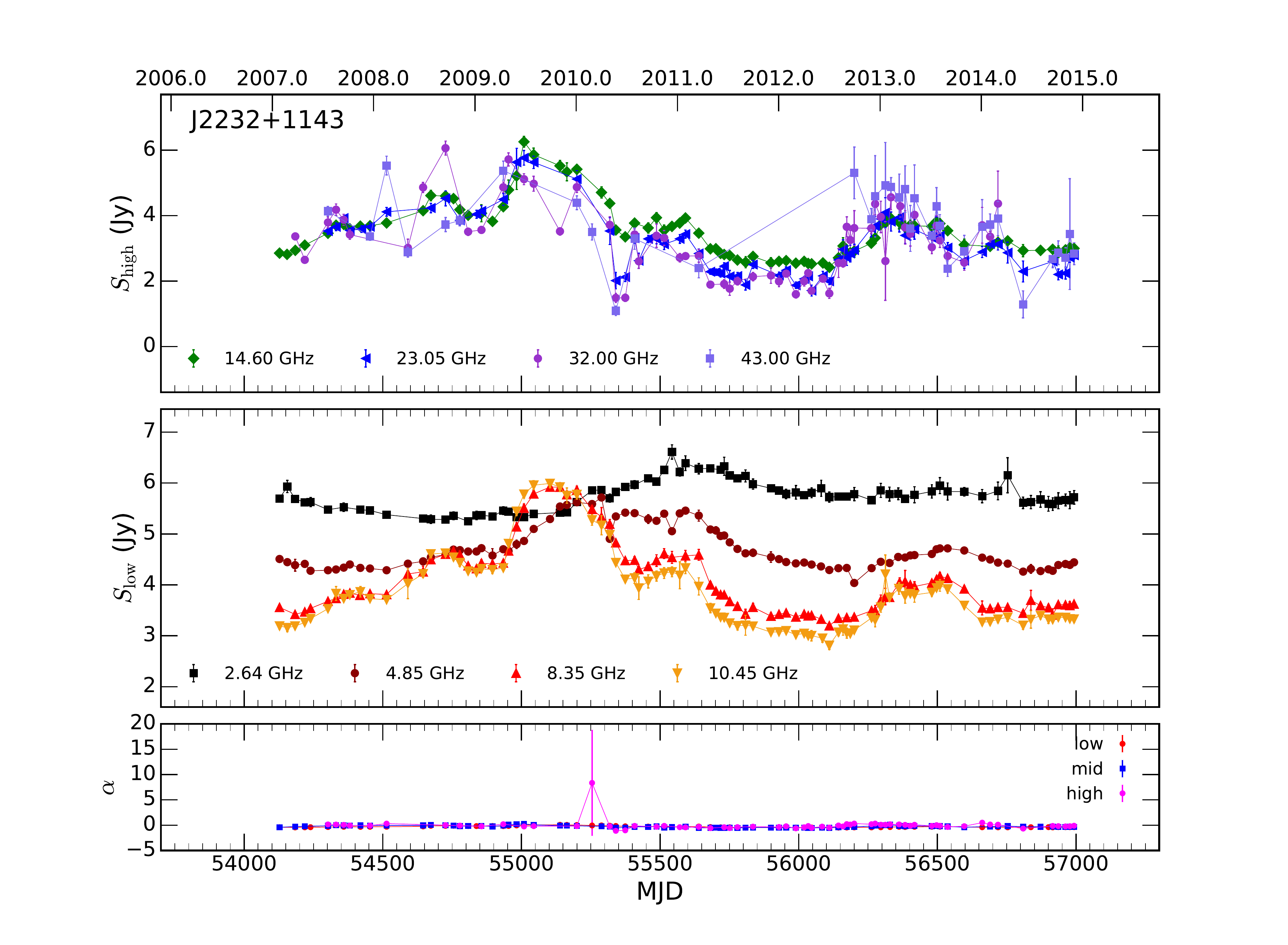}&
\includegraphics[trim=60pt 30pt 100pt 50pt  ,clip, width=0.49\textwidth,angle=0]{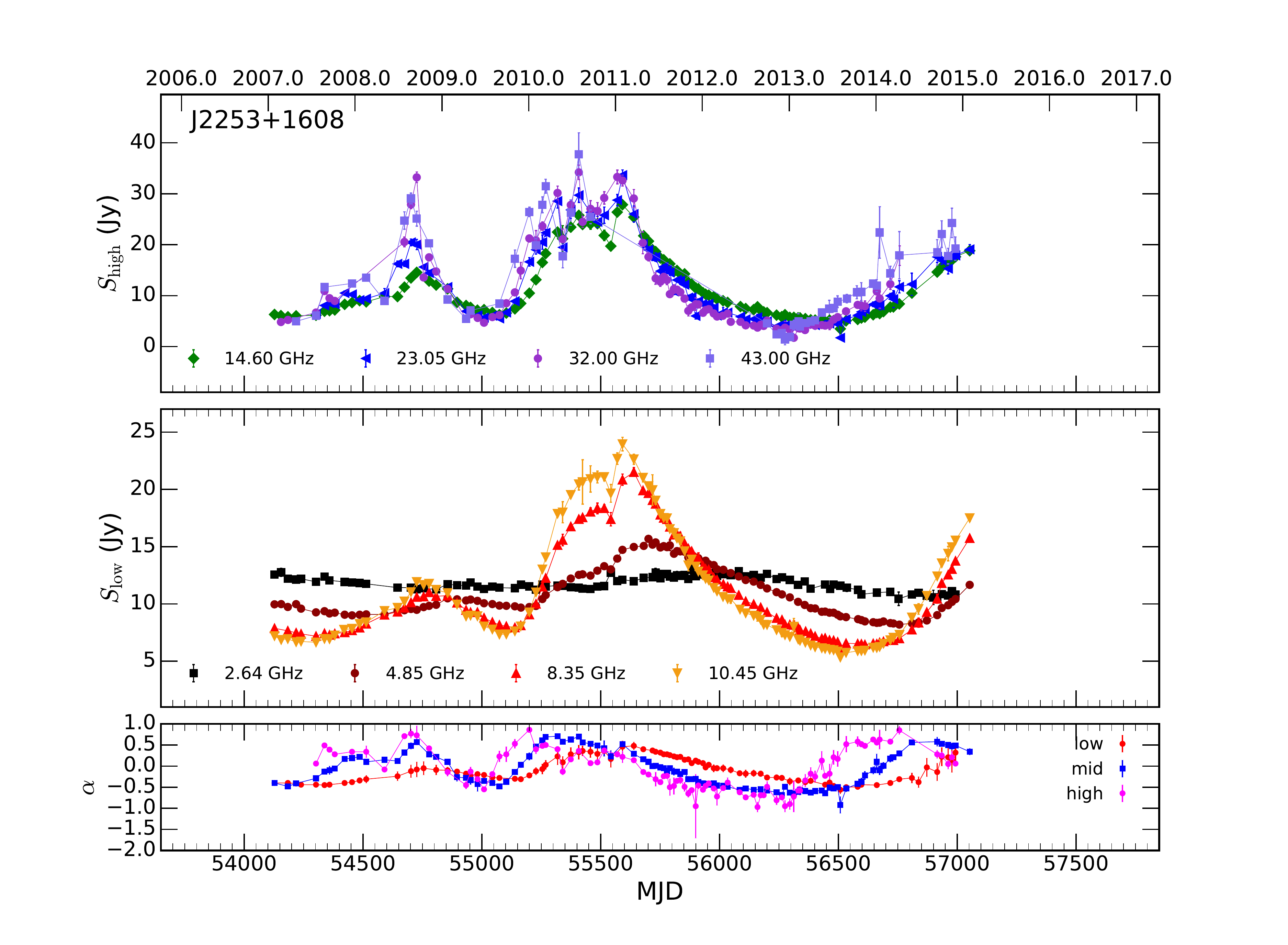}\\
\\[10pt]
\includegraphics[trim=60pt 30pt 100pt 50pt  ,clip, width=0.49\textwidth,angle=0]{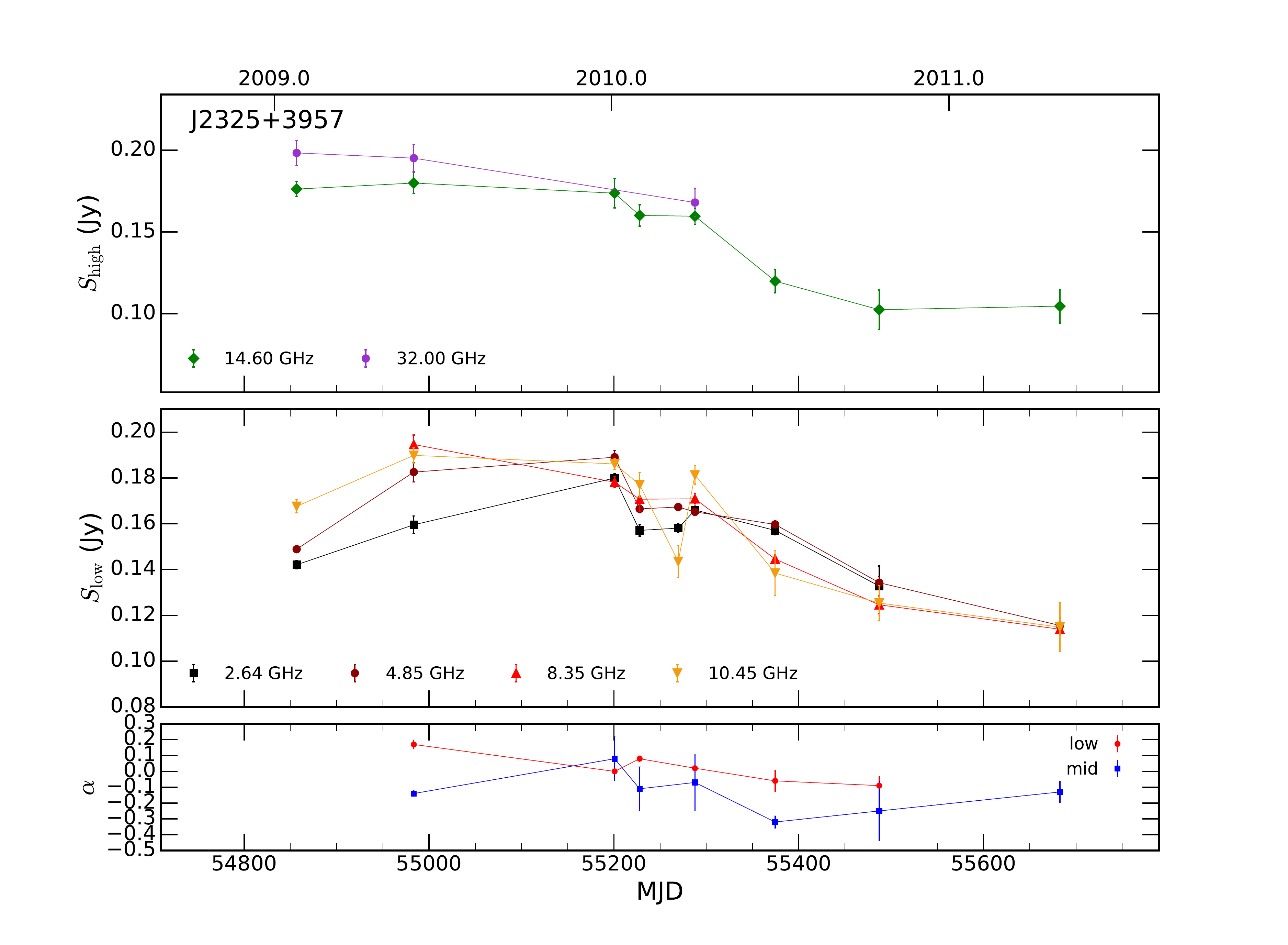}&
\includegraphics[trim=60pt 30pt 100pt 50pt  ,clip, width=0.49\textwidth,angle=0]{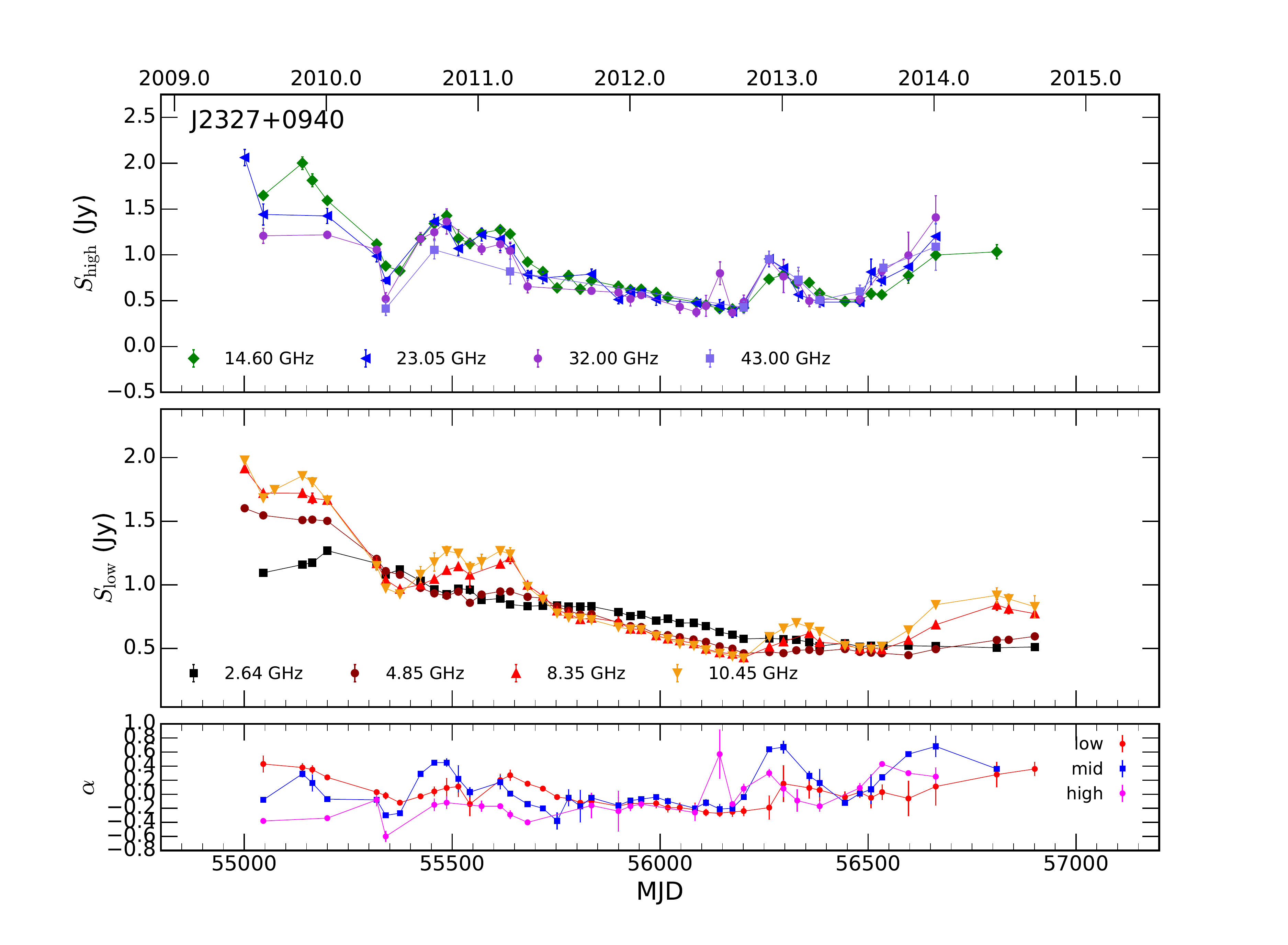}\\
\\[10pt]
\end{tabular}
\caption{Multi-frequency light curves for all the sources monitored by the \fg programme (``f'', ``s1'', ``s2'', ``old'') and the F-GAMMA-\textit{Planck} MoU. The lower panel in each frame shows the evolution of the low (2.64, 4.85 and 8.35~GHz) and mid-band (8.35, 10.45 and 14.6~GHz) and high-band (14.6, 23.05, 32, 43~GHz) spectral index. Only spectral index estimates from at least three frequencies are shown. Connecting lines have been used to guide the eye. }
\label{fig:sample_pg17}
\end{figure*}

\begin{figure*}[p]
\centering
\begin{tabular}{cc}
\includegraphics[trim=60pt 30pt 100pt 50pt  ,clip, width=0.49\textwidth,angle=0]{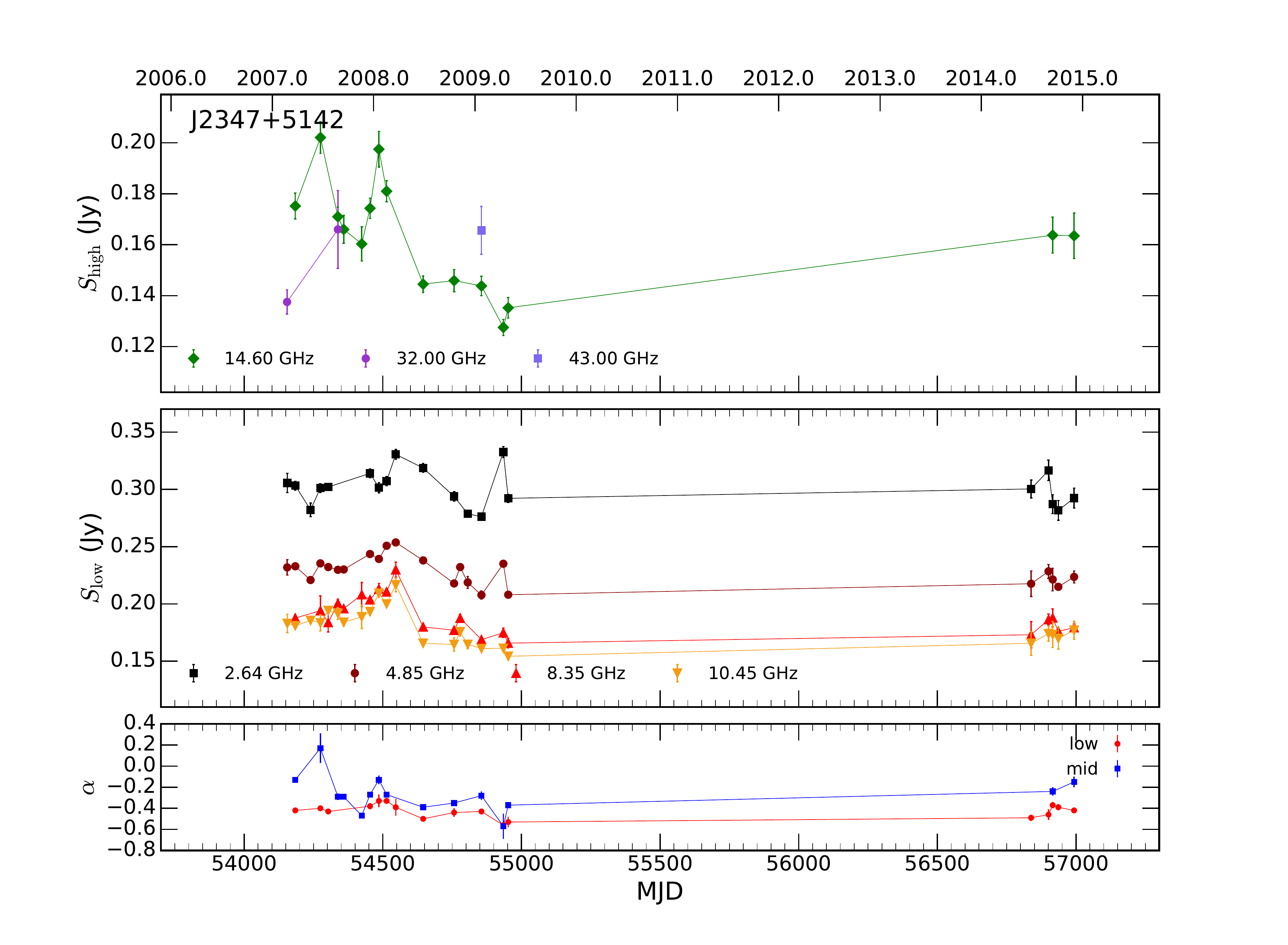}&
\includegraphics[trim=60pt 30pt 100pt 50pt  ,clip, width=0.49\textwidth,angle=0]{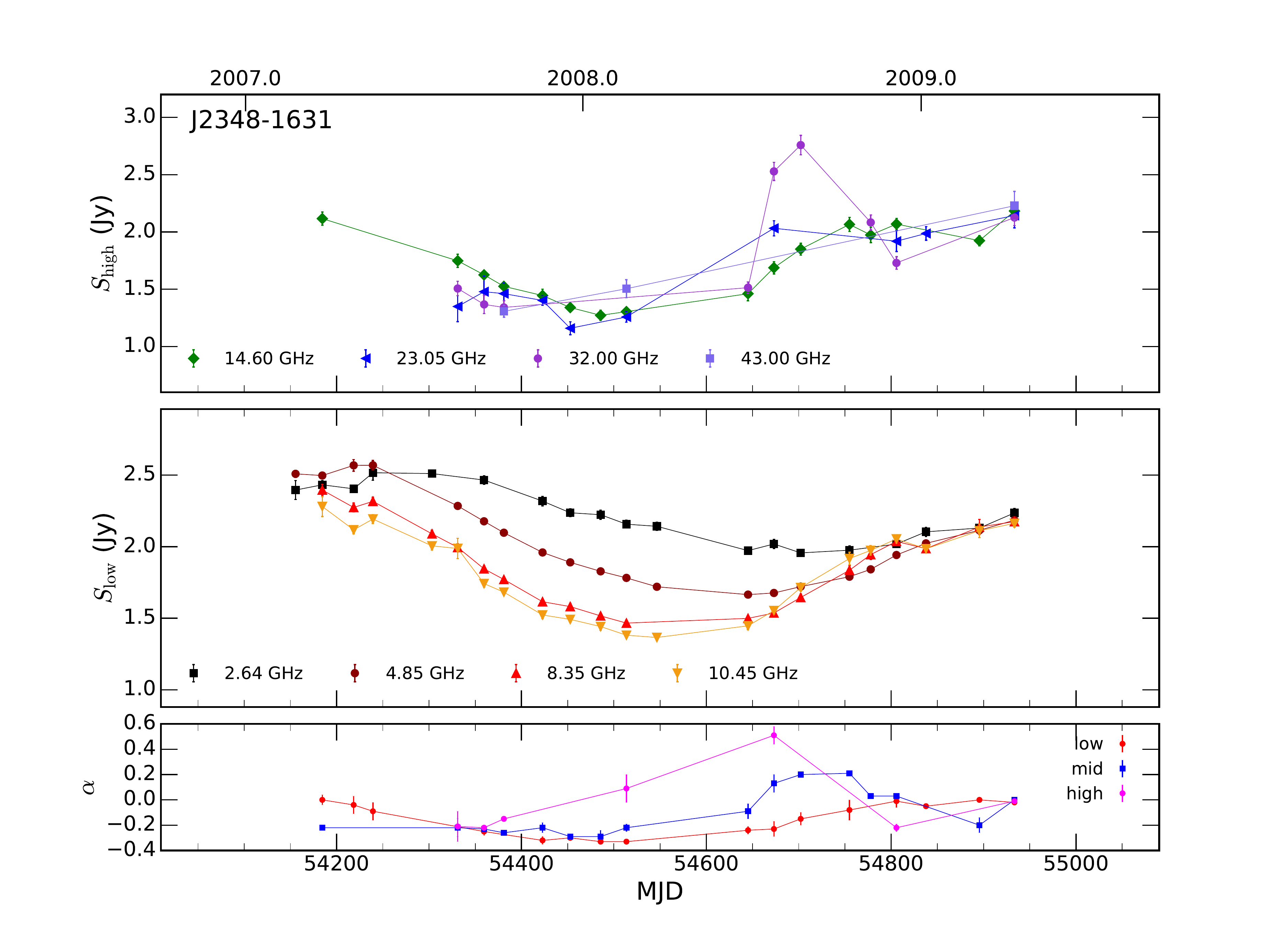}\\
\\[10pt]
\end{tabular}
\caption{Multi-frequency light curves for all the sources monitored by the \fg programme (``f'', ``s1'', ``s2'', ``old'') and the F-GAMMA-\textit{Planck} MoU. The lower panel in each frame shows the evolution of the low (2.64, 4.85 and 8.35~GHz) and mid-band (8.35, 10.45 and 14.6~GHz) and high-band (14.6, 23.05, 32, 43~GHz) spectral index. Only spectral index estimates from at least three frequencies are shown. Connecting lines have been used to guide the eye. }
\label{fig:sample_pg18}
\end{figure*}

Figures~\ref{fig:sample_pg1}-\ref{fig:sample_pg18} present the multi-frequency light curves for the faster variable sources in the \fg sample. {Some of their statistical moments are listed in table~\ref{tbl:data}. Finally, table~\ref{tbl:spinds} lists median and extreme values of the spectral index distributions of all the sources discussed in table~\ref{tab:sample} for which an adequate dataset was available (in each band, observations at least two frequencies are required). }

%%\red{ea-180527: NUMBER of sources to be processed:  65 I have here the "f" "s1" "s2" "old" as long as they have enough data. IF the "OLD " are  in the first paper consider exclduing them. IT MAY BE OBSCURING THE BEAUTY of F and S1 dunno. think and synchronise with catalogs}
%%%% -------------------------------------------------------------------------------------------------
% ea-180527: NUMBER of sources to be processed:  65 I have here the "f" "s1" "s2" "old" + "PLANCK" as long as they have enough data. IF the "OLD " are  in the first paper consider exclduing them
% NUMBER of sources to be processed:  90 + 14 PLANCK = 104
\clearpage
\clearpage
\clearpage
\clearpage
\clearpage
\clearpage
\clearpage
\clearpage
\clearpage
\clearpage
\clearpage
\clearpage
\clearpage
\clearpage
\clearpage
\clearpage
\clearpage
\clearpage
\end{appendix}
%==================================================================================================

\setcounter{table}{0}

 \clearpage
 \onecolumn
%\longtab{
\begin{longtable}{llrrllll}
\caption{\label{tab:sample} The list of sources that have been observed within the \fg programme }\\
\hline\hline
  \fg ID &Catalogue ID &\mc{1}{c}{RA (J2000)}         &\mc{1}{c}{DEC (J2000)}        &Class\tablefootmark{a} &\mc{1}{c}{$z$\tablefootmark{a}}  &Sample &Priority  \\
     &           &\mc{1}{c}{(hh:mm:ss)} &\mc{1}{c}{(dd:mm:ss)} &                       &                                 &       &          \\
\hline \\
\endfirsthead
\caption{continued.}\\
\hline\hline
  \fg ID &Catalogue ID &\mc{1}{c}{RA}         &\mc{1}{c}{DEC}        &Class\tablefootmark{a} &\mc{1}{c}{$z$\tablefootmark{a}}  &Sample &Priority \\
     &           &\mc{1}{c}{(hh:mm:ss)} &\mc{1}{c}{(dd:mm:ss)} &                       &                      &                      &     \\
\hline\\
\endhead\\
\hline
\endfoot
%
%\mc{8}{c}{\red{180523: this tables  has been checked with catalogs etc. In teh final database that IM did the QC mepoch\_new\_qc\_im-USEME theer are discrepancies. Must make sure that all the appropriate sources IN THAT folder are listed in the tables and then this table and that with spectral indices and fluxes are all synchronised.}}\\
\mc{8}{c}{
%\blue{180712: READY} 
Main monitored sample (source groups "f", "s1", "s2")}\\\\
J0050$-$0929  &PKS\,0048$-$097      &00:50:41.3  &$-$09:29:05.2  &BL Lac                         &0.635\tablefootmark{1}  &2       &f    \\ 
J0102$+$5824  &87GB\,0059$+$5808    &01:02:45.8  &$+$58:24:11.1  &Blazar Uncertain type          & 0.644                  &1,2     &s1   \\ 
J0136$+$4751  &BZQ\,J0136$+$4751    &01:36:58.6  &$+$47:51:29.1  &FSRQ                           &0.859                   &2       &s2   \\ 
J0217$+$0144  &PKS\,0215$+$015      &02:17:49.0  &$+$01:44:49.7  &FSRQ                           &1.715                   &1,2     &f    \\ 
J0221$+$3556  &B2\,0218$+$35        &02:21:05.5  &$+$35:56:13.9  &Blazar Uncertain type          &0.944                   &  2     &s2   \\ 
J0222$+$4302  &3C\,066A             &02:22:39.6  &$+$43:02:07.8  &BL Lac                         &0.444\tablefootmark{2}  &1,2     &f    \\ 
J0237$+$2848  &[HB89]\,0234$+$285   &02:37:52.4  &$+$28:48:09.0  &FSRQ                           &1.206                   &1,2     &s1   \\ 
J0238$+$1636  &[HB89]\,0235$+$164   &02:38:38.9  &$+$16:36:59.3  &BL Lac                         &0.940                   &1,2     &f    \\ 
J0241$-$0815  &NGC\,1052            &02:41:04.8  &$-$08:15:20.8  &Blazar Uncertain type          &0.005                   &1,2     &s1   \\ 
J0319$+$4130  &3C\,084              &03:19:48.2  &$+$41:30:42.1  &Blazar Uncertain type          &0.018                   &1,2     &f    \\ 
J0349$-$2102  &PKS\,0347$-$211      &03:49:57.9  &$-$21:02:47.7  &FSRQ                           &2.944                   &  2     &s2   \\ 
J0359$+$5057  &4C\,$+$50.11         &03:59:29.7  &$+$50:57:50.2  &FSRQ                           &1.512                   &1,2     &s1   \\ 
J0418$+$3801  &3C\,111              &04:18:21.3  &$+$38:01:35.8  &Sy 1\tablefootmark{3}          &0.049\tablefootmark{4} &1,2     &s1   \\ 
J0423$-$0120  &HB89]\,0420$-$014    &04:23:15.8  &$-$01:20:33.1  &FSRQ                           &0.916                   &1,2     &s1   \\ 
J0530$+$1331  &PKS\,0528$+$134      &05:30:56.4  &$+$13:31:55.1  &FSRQ                           &2.070                   &1,2     &f    \\ 
J0654$+$4514  &B3\,0650$+$453       &06:54:23.6  &$+$45:14:22.9  &FSRQ                           &0.928                   &  2     &f    \\ 
J0719$+$3307  &B2\,0716$+$33        &07:19:19.4  &$+$33:07:09.7  &FSRQ                           &0.779                   &  2     &s2   \\ 
J0721$+$7120  &S5\,0716$+$714       &07:21:53.4  &$+$71:20:36.4  &BL Lac                         &0.300\tablefootmark{5}  &1,2     &f    \\ 
J0730$-$1141  &PKS\,0727$-$11       &07:30:19.0  &$-$11:41:13.0  &FSRQ                           &1.589          &  2     &s1   \\ 
J0738$+$1742  &PKS\,0735$+$178      &07:38:07.4  &$+$17:42:19.0  &BL Lac                         &0.424          &1,2     &s1   \\ 
J0808$-$0751  &PKS\,0805$-$07       &08:08:15.5  &$-$07:51:09.9  &FSRQ                           &1.837          &  2     &s2   \\ 
J0818$+$4222  &B3\,0814$+$425       &08:18:16.0  &$+$42:22:45.4  &BL Lac                         &0.530          &1,2     &f    \\ 
J0824$+$5552  &BZQ\,J0824$+$5552    &08:24:47.2  &$+$55:52:42.7  &FSRQ                           &1.417          &  2     &s2   \\ 
J0841$+$7053  &S5\,0836$+$710       &08:41:24.4  &$+$70:53:42.2  &FSRQ                           &2.218          &1,2     &s1   \\ 
J0854$+$2006  &OJ\,$+$287           &08:54:48.9  &$+$20:06:30.6  &BL Lac                         &0.306          &1,2     &f    \\ 
J0920$+$4441  &S4\,0917$+$44        &09:20:58.3  &$+$44:41:53.9  &FSRQ                           &2.190          &  2     &f    \\ 
J0948$+$0022  &PMN\,J0948$+$0022    &09:48:57.3  &$+$00:22:25.6  &FSRQ                           &0.585          &  2     &f    \\ 
J0958$+$6533  &S4\,0954$+$658       &09:58:47.2  &$+$65:33:54.8  &BL Lac                         &0.367          &1,2     &s1   \\ 
J1104$+$3812  &MRK\,0421            &11:04:27.3  &$+$38:12:31.8  &BL Lac                         &0.030          &1,2     &f    \\ 
J1130$-$1449  &PKS\,1127$-$14       &11:30:07.1  &$-$14:49:27.4  &FSRQ                           &1.184          &1,2     &f    \\ 
J1159$+$2914  &4C\,$+$29.45         &11:59:31.8  &$+$29:14:43.8  &FSRQ                           &0.729          &1,2     &f    \\ 
J1217$+$3007  &BZB\,J1217$+$3007    &12:17:52.1  &$+$30:07:00.6  &BL Lac                         &0.130          &  2     &f    \\ 
J1221$+$2813  &W\,Com               &12:21:31.7  &$+$28:13:58.5  &BL Lac                         &0.102          &1,2     &f    \\ 
J1229$+$0203  &3C\,273              &12:29:06.7  &$+$02:03:08.6  &FSRQ                           &0.158          &1,2     &f    \\ 
J1256$-$0547  &3C\,279              &12:56:11.2  &$-$05:47:21.5  &FSRQ                           &0.536          &1,2     &f    \\ 
J1310$+$3220  &OP\,$+$313           &13:10:28.7  &$+$32:20:43.8  &Blazar Uncertain type          &0.997          &1,2     &f    \\ 
J1332$-$0509  &PKS\,1329$-$049      &13:32:04.3  &$-$05:09:42.9  &FSRQ                           &2.150          &  2     &f    \\ 
J1345$+$4452  &B3\,1343$+$451       &13:45:33.2  &$+$44:52:59.6  &FSRQ                           &2.534          &  2     &s2   \\ 
J1354$-$1041  &PKS\,1352$-$104      &13:54:46.5  &$-$10:41:02.7  &FSRQ                           &0.332          &  2     &s2   \\ 
J1504$+$1029  &PKS\,1502$+$106      &15:04:25.0  &$+$10:29:39.0  &FSRQ                           &1.839          &1,2     &f    \\ 
J1512$-$0905  &PKS\,1510$-$089      &15:12:50.5  &$-$09:05:59.8  &FSRQ                           &0.360          &1,2     &f    \\ 
J1522$+$3144  &B2\,1520$+$31        &15:22:10.0  &$+$31:44:14.4  &FSRQ                           &1.489          &1,2     &f    \\ 
J1542$+$6129  &BZB\,J1542$+$6129    &15:42:56.8  &$+$61:29:54.9  &BL Lac                         &0.117\tablefootmark{6}                &  2     &s2   \\ 
J1553$+$1256  &PKS\,1551$+$130      &15:53:32.7  &$+$12:56:51.7  &FSRQ                           &1.309          &  2     &s2   \\ 
J1635$+$3808  &4C\,$+$38.41         &16:35:15.5  &$+$38:08:04.5  &FSRQ                           &1.814          &1,2     &f    \\ 
J1642$+$3948  &3C\,345              &16:42:58.8  &$+$39:48:37.0  &FSRQ                           &0.593          &1,2     &s1   \\ 
J1653$+$3945  &MRK\,0501            &16:53:52.2  &$+$39:45:36.6  &BL Lac                         &0.033          &1,2     &f    \\ 
J1733$-$1304  &PKS\,1730$-$13       &17:33:02.7  &$-$13:04:49.5  &FSRQ                           &0.902          &1,2     &s1   \\ 
J1751$+$0939  &OT\,$+$081           &17:51:32.8  &$+$09:39:00.7  &BL Lac                         &0.322          &  2     &f    \\ 
J1800$+$7828  &S5\,1803$+$78        &18:00:45.7  &$+$78:28:04.0  &BL Lac                         &0.680          &1,2     &f    \\ 
J1848$+$3219  &B2\,1846$+$32A       &18:48:22.0  &$+$32:19:01.9  &FSRQ                           &0.798          &  2     &f    \\ 
J1849$+$6705  &S4\,1849$+$67        &18:49:16.1  &$+$67:05:41.7  &FSRQ                           &0.657          &  2     &f    \\ 
J1911$-$2102  &PMN\,J1911$-$2102    &19:11:53.9  &$-$21:02:43.8  &FSRQ                           &1.420          &  2     &s2   \\ 
J1923$-$2104  &PMN\,J1923$-$2104    &19:23:32.2  &$-$21:04:33.3  &FSRQ                           &0.874          &  2     &s2   \\ 
J2025$-$0735  &PKS\,2023$-$07       &20:25:40.6  &$-$07:35:52.0  &FSRQ                           &1.388          &  2     &f    \\ 
J2143$+$1743  &OX169                &21:43:35.5  &$+$17:43:48.0  &FSRQ                           &0.213          &  2     &s1   \\ 
J2147$+$0929  &PKS\,2144$+$092      &21:47:10.0  &$+$09:29:45.9  &FSRQ                           &1.113          &  2     &s1   \\ 
J2158$-$3013  &PKS\,2155$-$304      &21:58:52.0  &$-$30:13:32.0  &BL Lac                         &0.116          &1,2     &f    \\ 
J2202$+$4216  &BL~Lacertae          &22:02:43.3  &$+$42:16:40.0  &BL Lac                         &0.069          &1,2     &f    \\ 
J2203$+$1725  &PKS\,2201$+$171      &22:03:27.0  &$+$17:25:48.2  &FSRQ                           &1.076          &  2     &s2   \\ 
J2229$-$0832  &PKS\,2227$-$08       &22:29:40.1  &$-$08:32:54.4  &FSRQ                           &1.560          &  2     &s2   \\ 
J2232$+$1143  &CTA\,102             &22:32:36.4  &$+$11:43:50.9  &FSRQ                           &1.037          &1,2     &f    \\ 
J2253$+$1608  &3C\,454.3            &22:53:57.7  &$+$16:08:53.6  &FSRQ                           &0.859          &1,2     &f    \\ 
J2325$+$3957  &B3\,2322$+$396       &23:25:17.9  &$+$39:57:37.0  &BL Lac                         &\ldots         &2       &s2   \\ 
J2327$+$0940  &PKS\,325$+$093       &23:27:33.4  &$+$09:40:09.0  &FSRQ                           &1.843          &2       &s1   \\ 
\\\mc{8}{c}{
%\blue{180712: READY} 
Sources monitored until the sample revision (source group ``old'')} \\ \\
%
% ATTENTION the next are sources that were monitored and were interrupted after 2009.5. No dublicate with the  previous 65. Checked if all are in the log.xls Tjey are
%                                                                                                                                                                          
J0006$-$0623   &PKS\,0003$-$066     &00:06:13.9   &$-$06:23:35.3    &BL Lac                &0.347    &1   &\ldots \\     
J0303$+$4716   &4C\,$+$47.08        &03:03:35.2   &$+$47:16:16.3    &BL~Lac                &0.475\tablefootmark{7}    &1   &\ldots \\     
J0319$+$1845   &[HB89]\,0317$+$185  &03:19:51.8   &$+$18:45:34.2    &BL Lac galaxy dominated           &0.190    &1   &\ldots \\     
J0336$+$3218   &NRAO\,140           &03:36:30.1   &$+$32:18:29.3    &FSRQ                  &1.259    &1   &\ldots \\      
J0339$-$0146   &CTA\,26             &03:39:30.9   &$-$01:46:35.8    &FSRQ                  &0.850    &1   &\ldots \\                       
J0433$+$0521   &3C\,120             &04:33:11.1   &$+$05:21:15.6    &Blazar Uncertain type &0.033    &1   &\ldots \\      
J0750$+$1231   &PKS\,0748$+$126     &07:50:52.0   &$+$12:31:04.8    &FSRQ                  &0.889    &1   &\ldots \\                       
J0830$+$2410   &OJ\,248             &08:30:52.1   &$+$24:10:59.8    &FSRQ                  &0.939    &1   &\ldots \\     
J1041$+$0610   &PKS\,1038$+$064     &10:41:17.2   &$+$06:10:16.9    &FSRQ                  &1.264    &1   &\ldots \\     
J1128$+$5925   &TXS\,1125$+$596     &11:28:13.3   &$+$59:25:14.8    &FSRQ                  &1.795    &1   &\ldots \\     
J1136$+$7009   &MRK\,0180           &11:36:26.4   &$+$70:09:27.3    &BL~Lac                &0.045    &1   &\ldots \\     
J1224$+$2122   &PG\,1222$+$216      &12:24:54.5   &$+$21:22:46.4    &FSRQ                  &0.434    &1   &\ldots \\     
J1230$+$1223   &M\,087              &12:30:49.4   &$+$12:23:28.0    &LINER\tablefootmark{3}   &0.004\tablefootmark{8}    &1   &\ldots \\     
J1408$-$0752   &PKS\,B1406$-$076    &14:08:56.5   &$-$07:52:26.7    &FSRQ                  &1.494    &1   &\ldots \\     
J1540$+$8155   &1ES\,1544$+$820     &15:40:16.0   &$+$81:55:05.5    &BL~Lac                &0.000    &1   &\ldots \\     
J1613$+$3412   &1611$+$343          &16:13:41.1   &$+$34:12:47.9    &FSRQ                  &1.397    &1   &\ldots \\                      
J1806$+$6949   &3C\,371             &18:06:50.7   &$+$69:49:28.1    &BL~Lac                &0.046    &1   &\ldots \\      
J1824$+$5651   &4C\,$+$56.27        &18:24:07.1   &$+$56:51:01.5    &BL~Lac                &0.663    &1   &\ldots \\     
J1959$+$4044   &Cyg\,A              &19:59:28.4   &$+$40:44:02.1    &FRII\tablefootmark{9} &0.056\tablefootmark{10}    &1   &\ldots \\     
J1959$+$6508   &1ES\,1959$+$650     &19:59:59.9   &$+$65:08:54.7    &BL~Lac                &0.047    &1   &\ldots \\     
J2158$-$1501   &PKS\,2155$-$152     &21:58:06.3   &$-$15:01:09.3    &FSRQ                  &0.672    &1   &\ldots \\     
J2203$+$3145   &PKS\,2201$+$315     &22:03:15.0   &$+$31:45:38.3    &FSRQ                  &0.295    &1   &\ldots \\     
J2225$-$0457   &3C\,446             &22:25:47.3   &$-$04:57:01.4    &FSRQ                  &1.404    &1   &\ldots \\     
J2347$+$5142   &1ES\,2344$+$514     &23:47:04.8   &$+$51:42:17.9    &BL~Lac                &0.044    &1   &\ldots \\     
J2348$-$1631   &PKS\,2345$-$16      &23:48:02.6   &$-$16:31:12.0    &FSRQ                  &0.576    &1   &\ldots \\   
%                                      
% ea-180523: the follwoing sources are fond in the QC folder as old. I looked at all their LC and they indeed are old that is monitoried untoll 2009.5 or so        
%      
\\\mc{8}{c}{
%\blue{180715: READY} 
Sources observed as part of the F-GAMMA-\textit{Planck} MoU } \\ \\
J0108$+$0135   &PKS\,0106$+$01      &01:08:38.8   &$+$01:35:00.3    &FSRQ                    &2.099    &\ldots     &\ldots \\              
J0217$+$7349   &[HB89]\,0212$+$735  &02:17:30.8   &$+$73:49:32.5    &FSRQ                    &2.367    &\ldots     &\ldots \\       
J0321$+$1221   &PKS\,0321$+$1221    &03:21:53.1   &$+$12:21:14.0    &FSRQ\tablefootmark{11}  &2.662\tablefootmark{12}    &\ldots     &\ldots \\       
J0532$+$0732   &PMN\,J0532$+$0732   &05:32:39.0   &$+$07:32:43.3    &FSRQ                    &1.254    &\ldots     &\ldots \\       
J0607$-$0834   &[HB89]\,0605$-$085  &06:07:59.7   &$-$08:34:50.0    &FSRQ                    &0.870    &\ldots     &\ldots \\       
J0739$+$0137   &[HB89]\,0736$+$017  &07:39:18.0   &$+$01:37:05.0    &FSRQ                    &0.189    &\ldots     &\ldots \\       
J1058$+$0133   &PKS\,1055$+$01      &10:58:29.6   &$+$01:33:59.0    &Blazar Uncertain type   &0.890    &\ldots     &\ldots \\       
J1357$+$1919   &[HB89]\,1354$+$195  &13:57:04.4   &$+$19:19:07.4    &FSRQ                    &0.720    &\ldots     &\ldots \\       
J1550$+$0527   &PKS\,1548$+$056     &15:50:35.3   &$+$05:27:10.4    &FSRQ                    &1.422    &\ldots     &\ldots \\       
J1638$+$5720   &S4\,1637$+$57       &16:38:13.5   &$+$57:20:24.0    &FSRQ                    &0.751    &\ldots     &\ldots \\       
J1642$+$6856   &8C\,1642$+$690      &16:42:07.9   &$+$68:56:39.7    &FSRQ                    &0.751    &\ldots     &\ldots \\       
J1748$+$7005   &S4\,1749$+$70         &17:48:32.8   &$+$70:05:50.8    &BL~Lac                   &0.770                    &\ldots   &\ldots \\
J1927$+$7358   &8C\,1928$+$738      &19:27:48.5   &$+$73:58:01.6    &FSRQ                    &0.302    &\ldots     &\ldots \\       
J2031$+$1219   &PKS\,2029$+$121     &20:31:55.0   &$+$12:19:41.3    &Blazar Uncertain type   &1.215    &\ldots     &\ldots \\       
\\\mc{8}{c}{
%\blue{1807166:READY} 
Other sources observed within the \fg framework {as Targets of Opportunity}} \\\\ 
J0017$-$0512   &PMN\,J0017$-$0512     &00:17:35.8   &$-$05:12:41.6    &FSRQ                     &0.227                    &\ldots   &\ldots \\
J0033$-$1921   &1FGL\,J0033.5$-$1921  &00:33:34.4   &$-$19:21:32.9    &BL~Lac                   &0.610\tablefootmark{14}  &\ldots   &\ldots \\
J0051$-$0650   &PKS\,0048$-$071       &00:51:08.2   &$-$06:50:01.9    &FSRQ                     &1.975                    &\ldots   &\ldots \\
J0109$+$6134   &TXS\,0106$+$612       &01:09:46.3   &$+$61:33:30.5    &FSRQ\tablefootmark{15}   &0.783\tablefootmark{15}  &\ldots   &\ldots \\
J0112$+$2244   &TXS\,0109$+$224       &01:12:05.8   &$+$22:44:38.8    &BL~Lac                   &0.265                    &\ldots   &\ldots \\
J0136$+$3906   &B3\,0133$+$388        &01:36:32.5   &$+$39:06:00.0    &BL Lac\tablefootmark{17} &0.750\tablefootmark{14}  &\ldots   &\ldots \\
J0231$-$0110   &LQAC\,037$-$001\,022  &02:31:40.0   &$-$01:10:05.0    &BLAGN\tablefootmark{16}  &0.054\tablefootmark{16}  &\ldots   &\ldots \\
J0240$+$6113   &LSI$+$61\,303         &02:40:31.7   &$+$61:13:45.6    &High-mass X-ray binary\tablefootmark{17} &\ldots   &\ldots   &\ldots \\
J0255$+$0037   &PMN\,J0255$+$0037     &02:55:15.1   &$+$00:37:39.9    &Flat-Spec. Radio Source\tablefootmark{11} &1.015\tablefootmark{18} &\ldots  &\ldots  \\
J0257$+$0601   &3C\,75                &02:57:41.6   &$+$06:01:29.0    &FR\,I\tablefootmark{19}  &0.023\tablefootmark{20}                  &\ldots  &\ldots  \\
J0319$+$4134   &NGC\,1277             &03:19:51.5   &$+$41:34:25.0    &Group Member\tablefootmark{21}     &0.017\tablefootmark{20}        &\ldots  &\ldots  \\
J0442$-$0017   &NRAO\,190             &04:42:38.6   &$-$00:17:43.4    &FSRQ                     &0.845                    &\ldots   &\ldots \\
J0457$-$2324   &PKS\,0454$-$234       &04:57:03.2   &$-$23:24:52.0    &FSRQ                     &1.003                    &\ldots   &\ldots \\
J0507$+$6737   &1ES\,0502$+$675       &05:07:56.2   &$+$67:37:24.4    &BL~Lac                   &0.416                    &\ldots   &\ldots \\
J0510$+$1800   &PKS\,J0510$+$1800     &05:10:02.4   &$+$18:00:42.0    &FSRQ\tablefootmark{11}   &0.416\tablefootmark{13}  &\ldots   &\ldots \\       
J0521$+$2112   &RGB\,J0521$+$212      &05:21:46.0   &$+$21:12:51.5    &BL~Lac                   &\ldots                   &\ldots   &\ldots \\
J0632$+$0548   &HESS\,J0632+057       &06:32:59.2   &$+$05:48:00.8    &\ldots                   &\ldots                   &\ldots   &\ldots \\
J0648$+$1516   &GB6\,J0648$+$1516     &06:48:47.6   &$+$15:16:24.8    &BL~Lac                   &0.179                    &\ldots   &\ldots \\
J0710$+$5908   &TXS 0706+592          &07:10:30.1   &$+$59:08:20.2    &BL~Lac                   &0.125                    &\ldots   &\ldots \\
J0713$+$1935   &[WB92]\,0711$+$1940   &07:13:55.7   &$+$19:35:00.4    &FSRQ                     &0.540                    &\ldots   &\ldots \\
J0725$+$1425   &PKS\,0722$+$145       &07:25:16.8   &$+$14:25:13.7    &FSRQ                     &1.038                    &\ldots   &\ldots \\
J0906$+$0057   &\ldots                &09:06:24.0   &$+$00:57:58.0    &Sy 1\tablefootmark{3}    &0.070\tablefootmark{16}  &\ldots   &\ldots \\
J0909$+$2311   &RGB\,J0909$+$231      &09:09:00.6   &$+$23:11:12.9    &BL~Lac                   &0.223                    &\ldots   &\ldots \\
J0927$+$3902   &4C\,$+$39.25          &09:27:03.0   &$+$39:02:20.9    &FSRQ                     &0.695                    &\ldots   &\ldots \\
J0927$-$2034   &[HB89]\,0925$-$203    &09:27:51.8   &$-$20:34:51.2    &FSRQ                     &0.348                    &\ldots   &\ldots \\
J0959$+$0118   &PKS\,0956$+$015       &09:59:21.6   &$+$01:18:01.2    &\ldots                   &\ldots                   &\ldots   &\ldots \\
J1016$+$0513   &TXS\,1013$+$054       &10:16:03.1   &$+$05:13:02.3    &FSRQ                     &1.714                    &\ldots   &\ldots \\
J1033$+$6051   &S4\,1030$+$61         &10:33:51.4   &$+$60:51:07.3    &FSRQ                     &1.401                    &\ldots   &\ldots \\
J1044$+$0655   &PKS\,1042$+$071       &10:44:55.9   &$+$06:55:37.4    &QSO\tablefootmark{3}     &0.698\tablefootmark{22}  &\ldots   &\ldots \\
J1103$+$1158   &TXS\,1100$+$122       &11:03:03.5   &$+$11:58:16.6    &QSO\tablefootmark{3}     &0.912\tablefootmark{23}  &\ldots   &\ldots \\
J1120$+$0641   &ULAS\,J1120+0641      &11:20:01.5   &$+$06:41:24.3    &QSO\tablefootmark{24}    &7.085\tablefootmark{24}  &\ldots   &\ldots \\
J1211$+$3326   &\ldots                &12:11:32.8   &$+$33:26:25.0    &\ldots                   &\ldots                   &\ldots   &\ldots \\
J1220$+$3343   &3C\,270.1             &12:20:33.9   &$+$33:43:12.0    &FR II\tablefootmark{25}  &1.528\tablefootmark{26}  &\ldots   &\ldots \\
J1239$+$0443   &GB6\,J1239$+$0443     &12:39:32.8   &$+$04:43:05.2    &FSRQ                     &1.761                    &\ldots   &\ldots \\
J1312$+$4828   &GB6\,B1310$+$4844     &13:12:43.3   &$+$48:28:30.9    &Blazar Uncertain type    &0.501                    &\ldots   &\ldots \\
J1428$+$4240   &BZB\,J1428$+$4240     &14:28:32.6   &$+$42:40:20.6    &BL~Lac                   &0.129                    &\ldots   &\ldots \\
J1516$+$0015   &PKS\,1514$+$00        &15:16:40.2   &$+$00:15:01.9    &BL~Lac galaxy dominated  &0.052                    &\ldots   &\ldots \\
J1555$+$1111   &PG\,1553$+$113        &15:55:43.0   &$+$11:11:24.4    &BL~Lac                   &0.430\tablefootmark{28}  &\ldots   &\ldots \\
J1700$+$6830   &TXS\,1700$+$685       &17:00:09.3   &$+$68:30:07.0    &FSRQ                     &0.301                    &\ldots   &\ldots \\
J1719$+$1745   &OT\,129               &17:19:13.1   &$+$17:45:06.4    &BL~Lac                   &0.137\tablefootmark{29}  &\ldots   &\ldots \\
J1833$-$2104   &2FGL\,J1833.6-2104    &18:33:36.0   &$-$21:04:00.0    &\ldots                   &\ldots                   &\ldots   &\ldots \\
J1925$+$2106   &PKS\,B1923$+$210      &19:25:59.6   &$+$21:06:26.2    &BL~Lac                   &\ldots                   &\ldots   &\ldots \\
J2047$-$0246   &PMN\,J2047$-$0246     &20:47:45.7   &$-$02:46:05.0    &Flat-Spec. Radio Source\tablefootmark{11}  &\ldots &\ldots   &\ldots \\
J2102$+$4546   &V407\,Cyg             &21:02:09.9   &$+$45:46:33.6    &Symbiotic Star           &\ldots                   &\ldots   &\ldots \\
J2129$-$1538   &PKS\,2126$-$158       &21:29:12.2   &$-$15:38:41.0    &FSRQ                     &3.268                    &\ldots   &\ldots \\
J2136$+$0041   &PKS\,2134$+$004       &21:36:38.6   &$+$00:41:54.2    &FSRQ                     &1.941                    &\ldots   &\ldots \\
J2157$+$3127   &TXS\,2155$+$312       &21:57:28.8   &$+$31:27:01.4    &FSRQ                     &1.486                    &\ldots   &\ldots \\
J2212$+$2355   &PKS\,2209$+$236       &22:12:06.0   &$+$23:55:40.5    &Blazar Uncertain type    &1.125                    &\ldots   &\ldots \\
J2236$-$1433   &PKS\,2233$-$148       &22:36:34.1   &$-$14:33:22.2    &BL~Lac                   &0.325\tablefootmark{30}  &\ldots   &\ldots \\
J2314$+$2243   &RX\,J2314.9$+$2243        &23:14:55.7   &$+$22:43:25.0    &NLSy1\tablefootmark{27}  &0.169\tablefootmark{27}  &\ldots   &\ldots \\
J2358$+$1955   &PKS\,2356$+$196       &23:58:46.1   &$+$19:55:20.3    &FSRQ                     &1.066                    &\ldots   &\ldots \\%
\\\mc{8}{c}{
%\blue{180720-READY} 
The calibrators used for the F-GAMMA programme} \\ \\
3C\,48         &\ldots                &01:37:41.3   &$+$33:09:35.4    &FR I\tablefootmark{31}   &0.367\tablefootmark{32}  &1,2      &f      \\%
3C\,138        &\ldots                &05:21:09.9   &$+$16:38:22.0    &CSS\tablefootmark{33}    &0.759\tablefootmark{26}  &\ldots   &\ldots \\%
3C\,147        &\ldots                &05:42:36.1   &$+$49:51:07.2    &Sy 1.8\tablefootmark{3}  &0.545\tablefootmark{26}  &\ldots   &\ldots \\%
3C\,161        &\ldots                &06:27:10.0   &$-$05:53:07.0    &Quasar                   &\ldots                &1,2      &f      \\%
3C\,196        &\ldots                &08:13:36.0   &$+$48:13:02.2    &Sy 1.8\tablefootmark{3}  &0.871\tablefootmark{26}  &1,2      &f      \\%
3C\,286        &\ldots                &13:31:08.3   &$+$30:30:32.9    &Sy 1.5\tablefootmark{3}  &0.850\tablefootmark{26}  &1,2      &f      \\%
3C\,295        &\ldots                &14:11:20.7   &$+$52:12:09.0    &FR II\tablefootmark{25}  &0.461\tablefootmark{32}  &1,2      &f      \\%
NGC\,7027      &\ldots                &21:07:01.6   &$+$42:14:10.0    &Planetary Nebula         &\ldots                   &1,2      &f      \\%
DR\,21         &\ldots                &20:39:01.6,  &$+$42:19:38.0    &Star forming region      &\ldots                   &\ldots   &\ldots \\%
JUPITER        &\ldots                &\ldots       &\ldots           &Planet                   &\ldots                   &\ldots   &\ldots \\%
\end{longtable}
\tablefoot{Columns: (1) \fg source identifier; (2) survey
  name; (3), (4) targeted coordinates; (5) source class; (6) redshift; (7) \fg sample: ``1'' sources monitored until $\sim 2009.5$, ``2'' for sources monitored after 2009.5; (8) \fg priority group  \\
  \tablefoottext{a}{taken from \cite{2015ApSS.357...75M} unless explicitly said otherwise. The term ``FSRQ'' abbreviates their class ``QSO RLoud flat radio sp.'' and the term ``Blazar'' stands for ``Blazar Uncertain type''.}
}
\tablebib{(1) {\cite{2014MNRAS.439..690H}}; (2) {\cite{2011MNRAS.410.2556D}}; (3) {\cite{2006A&A...455..773V}};(4) {\cite{1991ApJS...75..297H}}; (5) {\cite{1996AJ....111.2187W}}; (6) {\cite{2011ApJ...740...98M}}; (7) \cite{1992ApJ...396..469H}; (8) \cite{2011MNRAS.413..813C};  (9) \cite{1984MNRAS.210..929L}; (10) \cite{1997ApJ...488L..15O}; (11) \cite{2007ApJS..171...61H}; (12) \cite{2013ApJ...767...14P}; (13) \cite{2014MNRAS.442.3329X}; (14) \cite{2015A&A...575A..21N}; (15) \cite{2010ApJ...718L.166V}; (16) \cite{2015MNRAS.449.3191Y}; (17) \cite{2009yCat....102025S}; (18) \cite{2010MNRAS.405.2302H}; (19) \cite{1999A&A...349...77C}; (20) \cite{2001ApJS..134..355M}; (21) \cite{1999AJ....118.2014W}; (22) \cite{2012A&A...544A..34P}; (23) \cite{2014ApJ...795...63F}; (24) \cite{2011Natur.474..616M}; (25) \cite{1983MNRAS.204..151L}; (26) \cite{2015ApJS..220....5M}; (27) \cite{2015A&A...578A..28B}; (28) \cite{2013A&A...554A..75S}; (29) \cite{2005ApJ...626...95S}; (30) \cite{1995AJ....110..880J}; (31) \cite{2000AJ....120.2269P}; (32) \cite{2011MNRAS.414..500H}; (33) \cite{1997ApJS..110..191d} .}
%}
 \clearpage
 \twocolumn

\setcounter{table}{9}

 \clearpage
 \onecolumn
%\longtab{
\begin{longtable}{lrrrrrrrrrr}
\caption{\label{tbl:data} Average flux densities and light curve parameters of all observed sources. Available online.}\\
\hline\hline
Source &\mc{1}{c}{$\nu$} &\mc{1}{c}{N} &\mc{1}{c}{$\langle S \rangle$} &\mc{1}{c}{$\hat{S}$} &\mc{1}{c}{SD}   &\mc{1}{c}{$S_\mathrm{min}$} &\mc{1}{c}{$S_\mathrm{max}$} &\mc{1}{c}{$\Delta t$} & \mc{1}{c}{Rate} & \mc{1}{c}{M} \\
       &\mc{1}{c}{(GHz)} &               &\mc{1}{c}{(Jy)}                &\mc{1}{c}{(Jy)}      &\mc{1}{c}{(Jy)} &\mc{1}{c}{(Jy)}             &\mc{1}{c}{(Jy)}             &\mc{1}{c}{(yr)}       &\mc{1}{c}{(d)}  &\mc{1}{c}{(yr$^{-1}$)}                    \\
%\cline{2-9} 
 %          &test  &test   &test &test   &test   & test  &test  &test     \\
\hline\\
\endfirsthead
\caption{continued.}\\
\hline\hline
Source &\mc{1}{c}{$\nu$} &\mc{1}{c}{N} &\mc{1}{c}{$\langle S \rangle$} &\mc{1}{c}{$\hat{S}$} &\mc{1}{c}{SD}   &\mc{1}{c}{$S_\mathrm{min}$} &\mc{1}{c}{$S_\mathrm{max}$} &\mc{1}{c}{$\Delta t$} & \mc{1}{c}{Rate} & \mc{1}{c}{Density} \\
       &\mc{1}{c}{(GHz)} &               &\mc{1}{c}{(Jy)}                &\mc{1}{c}{(Jy)}      &\mc{1}{c}{(Jy)} &\mc{1}{c}{(Jy)}             &\mc{1}{c}{(Jy)}             &\mc{1}{c}{(yr)}       &\mc{1}{c}{(d)}  &\mc{1}{c}{(yr$^{-1}$)}                \\
\hline\\
\endhead\\
\hline
\endfoot
%# Source          Freq..  N...  mean_S.  medianS  StD....  min_S..  max_S..  Dt.....  Rate.  density
%#                 (GHz)         (Jy)     (Jy)     (Jy)     (Jy)     (Jy)     (yr)    (d)    (1/yr)
% \blue{ea-180524:} 
 \mc{11}{c}{Main monitored sample (source groups ``f'', ``s1'', ``s2'') } \\ 
 \\
J0050$-$0929 &   2.64 &     48 &  0.623 &  0.634 &  0.150 &  0.282 &  0.896 &    5.1 &     39 &  9.5\\
      \ldots &   4.85 &     52 &  0.690 &  0.698 &  0.221 &  0.236 &  1.210 &    5.3 &     38 &    9.8     \\
      \ldots &   8.35 &     50 &  0.702 &  0.737 &  0.237 &  0.219 &  1.086 &    5.1 &     38 &    9.8     \\
      \ldots &  10.45 &     53 &  0.709 &  0.715 &  0.283 &  0.206 &  1.745 &    5.2 &     37 &   10.1     \\
      \ldots &  14.60 &     48 &  0.674 &  0.695 &  0.242 &  0.178 &  1.066 &    5.1 &     40 &    9.4     \\
      \ldots &  23.05 &     28 &  0.750 &  0.788 &  0.154 &  0.432 &  1.075 &    4.6 &     62 &    6.1     \\
      \ldots &  32.00 &     35 &  0.690 &  0.623 &  0.176 &  0.439 &  1.134 &    4.5 &     48 &    7.8     \\
      \ldots &  43.00 &     13 &  0.702 &  0.748 &  0.169 &  0.447 &  1.059 &    4.3 &    132 &    3.0     \\
	  \ldots & \ldots & \ldots & \ldots & \ldots & \ldots & \ldots & \ldots & \ldots & \ldots & \ldots \\
\\
  \mc{11}{c}{Sources monitored until the sample revision (source group ``old'')} \\ 
\\
J0006$-$0623 &   2.64 &     18 &  2.477 &  2.450 &  0.111 &  2.306 &  2.703 &    2.1 &     46 &  8.5\\
      \ldots &   4.85 &     21 &  2.337 &  2.305 &  0.131 &  2.200 &  2.709 &    2.1 &     39 &    9.9     \\
      \ldots &   8.35 &     21 &  2.205 &  2.146 &  0.117 &  2.061 &  2.452 &    2.1 &     37 &   10.2     \\
      \ldots &  10.45 &     21 &  2.145 &  2.110 &  0.101 &  2.033 &  2.422 &    2.1 &     37 &   10.2     \\
      \ldots &  14.60 &     19 &  2.060 &  2.030 &  0.106 &  1.926 &  2.346 &    2.1 &     43 &    8.9     \\
      \ldots &  23.05 &     14 &  1.855 &  1.898 &  0.214 &  1.421 &  2.305 &    2.0 &     55 &    7.2     \\
      \ldots &  32.00 &      9 &  1.958 &  1.792 &  0.274 &  1.655 &  2.440 &    2.1 &     94 &    4.4     \\
      \ldots &  43.00 &      7 &  1.894 &  1.956 &  0.333 &  1.489 &  2.441 &    1.5 &     92 &    4.6     \\
	  \ldots & \ldots & \ldots & \ldots & \ldots & \ldots & \ldots & \ldots & \ldots & \ldots & \ldots \\
\\
 \mc{11}{c}{Sources observed as part of the F-GAMMA-\textit{Planck} MoU} \\ 
\\
J0108$+$0135 &   2.64 &      8 &  3.859 &  3.879 &  0.176 &  3.598 &  4.062 &    0.7 &     37 & 11.4\\
      \ldots &   4.85 &      9 &  4.314 &  4.327 &  0.066 &  4.184 &  4.437 &    0.7 &     32 &   12.8     \\
      \ldots &   8.35 &      9 &  4.211 &  4.202 &  0.126 &  4.043 &  4.400 &    0.7 &     32 &   12.8     \\
      \ldots &  10.45 &      9 &  3.995 &  4.030 &  0.142 &  3.776 &  4.199 &    0.7 &     32 &   12.8     \\
      \ldots &  14.60 &      9 &  3.693 &  3.660 &  0.204 &  3.436 &  4.043 &    0.7 &     32 &   12.8     \\
      \ldots &  23.05 &      9 &  3.101 &  3.092 &  0.300 &  2.725 &  3.599 &    0.7 &     32 &   12.8     \\
      \ldots &  32.00 &      6 &  2.551 &  2.330 &  0.703 &  1.558 &  3.567 &    0.6 &     41 &   10.6     \\
	  \ldots & \ldots & \ldots & \ldots & \ldots & \ldots & \ldots & \ldots & \ldots & \ldots & \ldots \\
\\
 \mc{11}{c}{Other sources observed within the \fg framework {as Targets of Opportunity}}  \\ \\
% ea-1805390: empty files
% \mc{10}{l}{J1359$+$4011 NOT FOUND in the dictionary} \\
% ea-1805390: empty files
%\mc{11}{l}{NGC6572 NOT FOUND in the dictionary} \\
J0017$-$0512   &4.85    &1   &0.259  &0.259  &\ldots   &0.259   &0.259  &\ldots  &\ldots        &\ldots        \\
\ldots         &8.35    &1   &0.371  &0.371  &\ldots   &0.371   &0.371  &\ldots  &\ldots        &\ldots        \\
\ldots         &10.45   &1   &0.369  &0.369  &\ldots   &0.369   &0.369  &\ldots  &\ldots        &\ldots        \\
\ldots         &14.60   &1   &0.560  &0.560  &\ldots   &0.560   &0.560  &\ldots  &\ldots        &\ldots        \\
	  \ldots & \ldots & \ldots & \ldots & \ldots & \ldots & \ldots & \ldots & \ldots & \ldots & \ldots \\
\\
 \mc{11}{c}{The calibrators used for the F-GAMMA programme} \\ 
 \\
  3C48 &   2.64 &     99 &  9.595 &  9.575 &  0.083 &  9.417 &  9.869 &    7.9 &     30 & 12.5\\
      \ldots &   4.85 &    109 &  5.507 &  5.506 &  0.028 &  5.371 &  5.607 &    8.0 &     27 &   13.6     \\
      \ldots &   8.35 &    110 &  3.261 &  3.260 &  0.032 &  3.097 &  3.368 &    8.0 &     27 &   13.7     \\
      \ldots &  10.45 &    108 &  2.606 &  2.604 &  0.029 &  2.521 &  2.765 &    8.0 &     27 &   13.5     \\
      \ldots &  14.60 &    110 &  1.868 &  1.862 &  0.032 &  1.804 &  2.014 &    8.0 &     27 &   13.7     \\
      \ldots &  23.05 &     94 &  1.164 &  1.154 &  0.041 &  1.057 &  1.343 &    7.5 &     30 &   12.5     \\
      \ldots &  32.00 &     71 &  0.820 &  0.815 &  0.033 &  0.732 &  0.910 &    6.9 &     36 &   10.2     \\
      \ldots &  43.00 &     37 &  0.603 &  0.590 &  0.042 &  0.548 &  0.714 &    7.2 &     74 &    5.1     \\
	  \ldots & \ldots & \ldots & \ldots & \ldots & \ldots & \ldots & \ldots & \ldots & \ldots & \ldots \\
\end{longtable}
\tablefoot{Column description: 1: \fg source identifier; 2: observing frequency; 3: number of available measurements; 4: mean flux density; 5: median flux density; 6: flux density standard deviation; 7: minimum flux density; 8: maximum flux density; 9: light curve span; 10: mean span between consecutive measurements; 11: mean number of measurements in a  year.}
%}
 \clearpage
 \twocolumn

\clearpage
\onecolumn
%\longtab{
\begin{landscape}
\begin{longtable}{lrrrrrrcrrrrrrcrrrrrr}
\caption{\label{tbl:spinds} Spectral indices for all the sources observed in the \fg programme computed in three separate sub-bands. Available online. 
%\red{TO DO: check which sources stay if they are OK by looking at the frames. Also check if the spinds are correctly computed. \red{\bf ea$-$180524 ATTENTION: this table has dummy entries i.e that do not conatcin any dat in the "other sources" section. They will have AT the end be snchronised wit te other tables. This means that you keep it as is and then the Table 1 must have all the sources of this and the orher table. Will be the super table. Note that the inf values is because of 2 data points for fitiing a spec}}
}\\
\hline\hline
       &\mc{6}{c}{2.64, 4.85, 8.35~GHz } & &\mc{6}{c}{8.35, 10.4.5, 14.6~GHz} & &\mc{6}{c}{14.6, 23.0, 32, 43~GHz}\\
\cline{2-7}
\cline{9-13}
\cline{16-21}
Source &\mc{1}{c}{$\alpha_\mathrm{min}$} &\mc{1}{c}{$\alpha_\mathrm{max}$} &\mc{1}{c}{$\hat{\alpha}$} &\mc{1}{c}{$\langle \alpha\rangle $} &\mc{1}{c}{$\hat{\sigma}_\alpha$} &N & &\mc{1}{c}{$\alpha_\mathrm{min}$} &\mc{1}{c}{$\alpha_\mathrm{max}$} &\mc{1}{c}{$\hat{\alpha}$} &\mc{1}{c}{$\langle \alpha\rangle $} &\mc{1}{c}{$\hat{\sigma}_\alpha$} &N  & &\mc{1}{c}{$\alpha_\mathrm{min}$} &\mc{1}{c}{$\alpha_\mathrm{max}$} &\mc{1}{c}{$\hat{\alpha}$} &\mc{1}{c}{$\langle \alpha\rangle $} &\mc{1}{c}{$\hat{\sigma}_\alpha$} &N \\
%\cline{2-9} 
 %          &test  &test   &test &test   &test   & test  &test  &test     \\
\hline\\
\endfirsthead
\caption{continued.}\\
\hline\hline
       &\mc{6}{c}{2.64, 4.85, 8.35~GHz } & &\mc{6}{c}{8.35, 10.4.5, 14.6~GHz} & &\mc{6}{c}{14.6, 23.0, 32, 43~GHz}\\
\cline{2-7}
\cline{9-13}
\cline{16-21}
Source &\mc{1}{c}{$\alpha_\mathrm{min}$} &\mc{1}{c}{$\alpha_\mathrm{max}$} &\mc{1}{c}{$\hat{\alpha}$} &\mc{1}{c}{$\langle \alpha\rangle $} &\mc{1}{c}{$\hat{\sigma}_\alpha$} &N & &\mc{1}{c}{$\alpha_\mathrm{min}$} &\mc{1}{c}{$\alpha_\mathrm{max}$} &\mc{1}{c}{$\hat{\alpha}$} &\mc{1}{c}{$\langle \alpha\rangle $} &\mc{1}{c}{$\hat{\sigma}_\alpha$} &N  & &\mc{1}{c}{$\alpha_\mathrm{min}$} &\mc{1}{c}{$\alpha_\mathrm{max}$} &\mc{1}{c}{$\hat{\alpha}$} &\mc{1}{c}{$\langle \alpha\rangle $} &\mc{1}{c}{$\hat{\sigma}_\alpha$} &N \\
\hline\\
\endhead\\
\hline
\endfoot
% These are fg-f, S1 and S2 65 sources
 \mc{21}{c}{Main monitored sample (source groups ``f'', ``s1'', ``s2'')} \\ 
 \\
 J0050$-$0929  &$-$0.37  &$+$0.28  &$+$0.11  &$+$0.04  &0.02   & 52    &        &$-$0.55  &$+$0.43  &$-$0.08  &$-$0.09  &0.04   & 51      &        &$-$0.57  &$+$0.23  &$-$0.14  &$-$0.13  &0.08   & 36 \\
 \ldots        & \ldots  & \ldots  & \ldots  & \ldots  & \ldots&\ldots & \ldots & \ldots  & \ldots  & \ldots  & \ldots  & \ldots & \ldots & \ldots & \ldots  & \ldots  & \ldots  & \ldots  & \ldots & \ldots    \\
\\
% These are fg-o, 25 sources
 \mc{21}{c}{Sources monitored until the sample revision (source group ``old'')} \\ 
 \\
 J0006$-$0623  &$-$0.14  &$-$0.01  &$-$0.11  &$-$0.10  &0.01   & 22 &   &$-$0.47  &$-$0.03  &$-$0.11  &$-$0.13  &0.03   & 21 &   &$-$1.00  &$+$0.09  &$-$0.17  &$-$0.18  &0.08   & 14 \\
 \ldots        & \ldots  & \ldots  & \ldots  & \ldots  & \ldots&\ldots & \ldots & \ldots  & \ldots  & \ldots  & \ldots  & \ldots & \ldots & \ldots & \ldots  & \ldots  & \ldots  & \ldots  & \ldots & \ldots    \\
\\
% These are planck ONLY, 14 sources
 \mc{21}{c}{Sources observed as part of the F-GAMMA-\textit{Planck} MoU} \\ 
 \\
 J0108$+$0135  &$-$0.01  &$+$0.18  &$+$0.05  &$+$0.06  &0.08   &  9 &   &$-$0.37  &$-$0.13  &$-$0.25  &$-$0.24  &0.02   &  9 &   &$-$0.68  &$-$0.15  &$-$0.52  &$-$0.42  &0.15   &  9 \\
 \ldots        & \ldots  & \ldots  & \ldots  & \ldots  & \ldots&\ldots & \ldots & \ldots  & \ldots  & \ldots  & \ldots  & \ldots & \ldots & \ldots & \ldots  & \ldots  & \ldots  & \ldots  & \ldots & \ldots    \\
 \\
% These are other, 54 sources
 \mc{21}{c}{Other sources observed within the \fg framework {as Targets of Opportunity}} \\ 
 \\
J0017$-$0512   &$+$0.67  &$+$0.67  &$+$0.67   &$+$0.67  &\ldots &        1  &        &$+$0.71  &$+$0.71  &$+$0.71  &$+$0.71   &0.58    &        1         &        &\ldots   &\ldots   &\ldots   &\ldots   &\ldots  &\ldots        \\
 \ldots        & \ldots  & \ldots  & \ldots  & \ldots  & \ldots&\ldots & \ldots & \ldots  & \ldots  & \ldots  & \ldots  & \ldots & \ldots & \ldots & \ldots  & \ldots  & \ldots  & \ldots  & \ldots & \ldots    \\
\\
% These are cals, 10 sources
 \mc{21}{c}{The calibrators used for the F-GAMMA programme}  \\
\\
 3C48              &$-$0.98    &$-$0.90    &$-$0.93    &$-$0.94    &0.02     &   108   &    &$-$1.08    &$-$0.90    &$-$1.00    &$-$1.00    &0.01     &   111   &    &$-$1.24    &$-$0.77    &$-$1.05    &$-$1.04    &0.03     &    99   \\
 \ldots        & \ldots  & \ldots  & \ldots  & \ldots  & \ldots&\ldots & \ldots & \ldots  & \ldots  & \ldots  & \ldots  & \ldots & \ldots & \ldots & \ldots  & \ldots  & \ldots  & \ldots  & \ldots & \ldots    \\
\end{longtable}
\tablefoot{Column description: 1: \fg source identifier; 2, 8 and 14: minimum (softest) spectral index measured; 3, 9 and 15: maximum (hardest) spectral index measured; 4, 10 and 16: median spectra index; 5, 11 and 17: mean spectral index; 6,12 and 18: median spectral uncertainty; 7, 13 and 19: number of measured SEDs. In the cases of only two flux density measurements the spectral indices are still computed; however, the formal error propagation is not applicable and hence no median uncertainty is reported. }
\end{landscape}
%}
\clearpage
\twocolumn

\end{document}